\documentclass[titlepage, 11pt]{book}
\usepackage{amsmath,amssymb, diagrams}
\usepackage{epsfig}

\newcommand{\be}{\begin{equation}}
\newcommand{\ee}{\end{equation}}
\newcommand{\bea}{\begin{array}}
\newcommand{\ea}{\end{array}}
\newcommand{\beqa}{\begin{eqnarray}}
\newcommand{\eeqa}{\end{eqnarray}}

\newcommand{\bean}{\begin{eqnarray*}}
\newcommand{\eean}{\end{eqnarray*}}

\newcommand{\nn}{\nonumber}
\def\sqr#1#2{{\vcenter{\vbox{\hrule height.#2pt
        \hbox{\vrule width.#2pt height#1pt \kern#1pt
          \vrule width.#2pt}
        \hrule height.#2pt}}}}

\def\half{\frac{1}{2}}
\def\tr{\mathop{\rm Tr}\nolimits}
\def\BI{{\rm 1\!l}}
\def\CP{{\mathbb C}P}
\frontmatter

\title{Lectures on Fuzzy and Fuzzy SUSY Physics \thanks{SU-4252-819, 
DIAS-STP-05-12, IISc/CHEP/11/05}}
\author{A.P. Balachandran\thanks{e-mail:bal@phy.syr.edu}\\
{\it Department of Physics, Syracuse University, Syracuse NY,
13244-1130, USA}
\and
S. K\"{u}rk\c{c}\"{u}o\v{g}lu\thanks{e-mail:seckin@stp.dias.ie} \\
{\it Dublin Institute for Advanced Studies, 10 Burlington Road, Dublin 4,
Ireland}
\and
S. Vaidya\thanks{e-mail:vaidya@cts.iisc.ernet.in}\\
{\it Centre for High Energy Physics, Indian
Institute of Science, Bangalore, 560012, India.}
}

\date{November 2005}

\begin{document}

\maketitle

\thispagestyle{empty}

\vskip 8em

\begin{center}
{\Large
{\it Dedicated to Rafael Sorkin, \\ our
friend and teacher, \\
and a true and creative seeker of knowledge.}}
\end{center}

\newpage

\chapter{Preface}

One of us (Balachandran) gave a course of lectures on ``Fuzzy
Physics'' during spring, 2002 for students of Syracuse and Brown
Universities. The course which used video conferencing technology was
also put on the websites \cite{courseweb}. Subsequently A.P. Balachandran,
S. K\"{u}rk\c{c}\"{u}o\v{g}lu and S.Vaidya decided to edit the material and publish
them as lecture notes. The present book is the outcome of that effort.

The recent interest in fuzzy physics begins from the work of Madore
\cite{madore2, madore1} and others even though the basic mathematical
ideas are older and go back at least to Kostant and Kirillov
\cite{kostant} and Berezin \cite{berezin}. It is based on the fundamental observation
that coadjoint orbits of Lie groups are symplectic manifolds which can
therefore be quantized under favorable circumstances. When that can be
done, we get a quantum representation of the manifold. It is the fuzzy
manifold for the underlying ``classical manifold''. It is fuzzy
because no precise localization of points thereon is possible. The
fuzzy manifold approaches its classical version when the effective
Planck's constant of quantization goes to zero.

Our interest will be in compact simple and semi-simple Lie groups for
which coadjoint and adjoint orbits can be identified and are
compact as well. In such a case these fuzzy manifold is a finite-dimensional matrix
algebra on which the Lie group acts in simple ways.  Such fuzzy spaces
are therefore very simple and also retain the symmetries of their
classical spaces. These are some of the reasons for their attraction.

There are several reasons to study fuzzy manifolds. Our interest has
its roots in quantum field theory (qft). Qft's require regularization
and the conventional nonperturbative regularization is lattice
regularization. It has been extensively studied for over thirty years. 
It fails to preserve space-time symmetries of quantum fields. It also has
problems in dealing with topological subtleties like instantons, and
can deal with index theory and axial anomaly only
approximately. Instead fuzzy physics does not have these problems. So
it merits investigation as an alternative tool to investigate qft's.

A related positive feature of fuzzy physics, 
is its ability to deal with supersymmetry(SUSY) in a
precise manner \cite{GKP1, fuzzyS, seckin1, seckinthesis}. 
(See however,\cite{caterall}). Fuzzy SUSY models are also 
finite-dimensional matrix models amenable to numerical work, 
so this is another reason for our attraction to this field.

Interest in fuzzy physics need not just be utilitarian. Physicists
have long speculated that space-time in the small has a discrete
structure. Fuzzy space-time gives a very concrete and interesting
method to model this speculation and test its consequences. There are
many generic consequences of discrete space-time, like CPT and
causality violations, and distortions of the Planck spectrum.  Among
these must be characteristic signals for fuzzy physics, but they
remain to be identified.

These lecture notes are not exhaustive, and reflect the research
interests of the authors. It is our hope that the interested reader
will be able to learn about the topics we have not covered with the
help of our citations. 

{\bf Acknowledgements} The work of A.P.B. was supported by DOE under grant number
DE-FG02-85ER40231. S.K. acknowledges financial support from Irish 
Research Council Science Engineering and Technology(IRCSET) under
the postdoctoral fellowship program.

\tableofcontents

\mainmatter

%%%%%%%%%%%%%%%%%%%%%%%%%%%%%%%%%%%%%%%%%%%%%%%%%%%%%%%%%%%%%%%%%%%%%%%%%%%%%%%%%%
\chapter{Introduction}

We can find few fundamental physical models amenable to exact
treatment. Approximation methods like perturbation theory are
necessary and are part of our physics culture.

Among the important approximation methods for quantum field theories
(qft's) are strong coupling methods based on lattice discretization of
underlying space-time or perhaps its time-slice. They are among the
rare effective approaches for the study of confinement in QCD and for
non-perturbative regularization of qft's. They enjoyed much popularity
in their early days and have retained their good reputation for
addressing certain fundamental problems.

One feature of naive lattice discretizations however can be
criticized. They do not retain the symmetries of the exact theory
except in some rough sense. A related feature is that topology and
differential geometry of the underlying manifolds are treated only
indirectly, by limiting the couplings to ``nearest neighbors''. Thus
lattice points are generally manipulated like a trivial topological
set, with a point being both open and closed. The upshot is that these
models have no rigorous representation of topological defects and
lumps like vortices, solitons and monopoles. The complexities in the
ingenious solutions for the discrete QCD $\theta$-term \cite{ref1}
illustrate such limitations. There do exist radical attempts to
overcome these limitations using partially ordered sets \cite{ref2},
but their potentials are yet to be adequately studied.

As mentioned in the preface, a new approach to discretization, under the
name of ``fuzzy physics'' inspired by non-commutative geometry (NCG),
is being developed since a few years. The key remark here is that when
the underlying space-time or spatial cut can be treated as a phase
space and quantized, with a parameter ${\hat \hbar}$ assuming the role of
$\hbar$, the emergent quantum space is fuzzy, and the number of
independent states per (``classical'') unit volume becomes finite. We
have known this result after Planck and Bose introduced such an
ultraviolet cut-off and quantum physics later justified
it. A ``fuzzified'' manifold is expected to be ultraviolet finite, and
if the parent manifold is compact too, supports only finitely many
independent states. The continuum limit is the semi-classical $\hat
h\rightarrow 0$ limit. This unconventional discretization of classical
topology is not at all equivalent to the naive one, and we shall see
that it does significantly overcome the previous criticisms.

There are other reasons also to pay attention to fuzzy spaces, be they
space-times or spatial cuts. There is much interest among string
theorists in matrix models and in describing D-branes using
matrices. Fuzzy spaces lead to matrix models too and their ability to
reflect topology better than elsewhere should therefore evoke our
curiosity. They let us devise new sorts of discrete models and are
interesting from that perspective. In addition,as mentioned in the
preface, it has now been discovered that when open strings end on
D-branes which are symplectic manifolds, then the branes can become
fuzzy. In this way one comes across fuzzy tori, ${\mathbb C}P^N$ and
many such spaces in string physics.

The central idea behind fuzzy spaces is discretization by
quantization. It does not always work. An obvious limitation is that
the parent manifold has to be even dimensional. If it is not, it has
no chance of being a phase space. But that is not all. Successful use
of fuzzy spaces for qft's requires good fuzzy versions of the
Laplacian, Dirac equation, chirality operator and so forth, and their
incorporation can make the entire enterprise complicated. The torus
$T^2$ is compact, admits a symplectic structure and on quantization
becomes a fuzzy, or a non-commutative torus. It supports a finite number
of states if the symplectic form satisfies the Dirac quantization
condition. But it is impossible to introduce suitable derivations
without escalating the formalism to infinite dimensions.

But we do find a family of classical manifolds elegantly escaping
these limitations. They are the co-adjoint orbits of Lie groups.  For
semi-simple Lie groups, they are the same as adjoint orbits. It is a
theorem that these orbits are symplectic. They can often be quantized
when the symplectic forms satisfy the Dirac quantization
condition. The resultant fuzzy spaces are described by linear
operators on irreducible representations (IRR's) of the group. For
compact orbits, the latter are finite-dimensional. In addition, the
elements of the Lie algebra define natural derivations, and that helps
to find Laplacian and the Dirac operator. We can even define chirality
with no fermion doubling and represent monopoles and instantons. (See
chapters 5, 6 and 8). These orbits therefore are altogether well-adapted for
QFT's.

Let us give examples of these orbits:
\begin{itemize}
\item{$S^2 \simeq {\mathbb C}P^1 $}: This is the orbit of $SU(2)$ through the Pauli matrix
$\sigma_3$ or any of its multiples $\lambda\,\sigma_3$ ($\lambda\neq
0$). It is the set $\{\lambda\,g\,\sigma_3\,g^{-1}\,:\, g\in
SU(2)\}$. The symplectic form is $j\,d\,{\rm cos}\,\theta\wedge d\phi$
with $\theta,\phi$ being the usual $S^2$ coordinates. Quantization
gives the spin $j$ $SU(2)$ representations.
\item{${\mathbb C}P^2$}: ${\mathbb C}P^2$ is of particular interest
being of dimension 4. It is the orbit of $SU(3)$ through the
hypercharge $Y=1/3\,\,{\rm diag} (1,1,-2)$ (or its multiples):
\begin{equation}
{\mathbb C}P^2:\{g\,Y\,g^{-1}\,:\, g\in SU(3)\}.
\end{equation}
The associated representations are symmetric products of $3$'s or
${\bar 3}$'s.

In a similar way ${\mathbb C}P^N$ are adjoint orbits of $SU(N+1)$ for
any $N \leq 3$. They too can be quantized and give rise to fuzzy spaces.

\item{$SU(3)/[U(1)\times U(1)]$}:
This 6-dimensional manifold is the orbit of $SU(3)$ through
$\lambda_3={\rm diag}(1,-1,0)$ and its multiples. These orbits give
all the IRR's containing a zero hypercharge state.
\end{itemize}

In this book, we focus on the fuzzy spaces emerging from quantizing
$S^2$. They are called the fuzzy spheres $S_F^2$ and depend on the
integer or half integer $j$ labelling the irreducible representations
of $SU(2)$. Physics on  $S_F^2$ is treated in detail. Scalar and gauge
fields, the Dirac operator, instantons, index theory, and the so-called
UV-IR mixing \cite{Raamsdonk, ydri1, vaidya1, chu, dolan, sachin1, sachin2} are all covered. 
Supersymmetry can be elgantly discretized in the approach of fuzzy physics by replacing the Lie
algebra $su(2)$ of $SU(2)$ by the superalgebras $osp(2,1)$ and
$osp(2,2)$. Fuzzy supersymmetry is also discussed here including its
instanton and index theories. We also briefly discuss the fuzzy spaces
associated with  ${\mathbb C}P^N \, (N \leq 2)$. These spaces,
especially  ${\mathbb C}P^2$, are of physical interest. We refer to
the literature \cite{GrosseCP2, fuzzycp2, BDBJ, Nair, Julieta} for their more exhaustive
treatment.   

Fuzzy physics draws from many techniques and notions developed in the
context of noncommutative geometry. There are excellent books and
reviews on this vast subject some of which we include in the
bibliography \cite{madore1, connes, Landi, Varilly, Szabo:2001kg, Douglas:2001ba}.

%\end{document}
%\begin{document}

\chapter{Fuzzy Spaces}

In the present chapter, we approach the problem of quantization 
of classical manifolds like $S^2$ and ${\mathbb C}P^N$ using harmonic
oscillators. The method is simple and transparent, and enjoys
generality too. The point of departure in this approach is the
quantization of complex planes. We focus on quantizing ${\mathbb C}^2$ 
and its associated $S^2$ first. We will consider other
manifolds later in the chapter.

\section{Fuzzy ${\mathbb C}^2$}    

The two-dimensional complex plane ${\mathbb C}^2$ has coordinates
$z=(z_1, z_2)$ where $z_i \in {\mathbb C}$. We want to quantize
${\mathbb C}^2$ turning it into fuzzy ${\mathbb C}^2 \equiv {\mathbb
C}^2_F$.

This is easily accomplished. After quantization, $z_i$ become harmonic
oscillator annihilation operators $a_i$ and $z_i^*$ become their
adjoint. Their commutation relations are
\begin{equation}
\lbrack a_i \, a_j \rbrack = \lbrack a_i^\dagger  \, a_j^\dagger
\rbrack = 0 \,, \quad \quad \lbrack a_i \, a_j^\dagger   \rbrack =
\tilde{\hbar} \delta_{ij} \,, 
\end{equation}
where the $\tilde{\hbar}$ need not be the ``Planck's constant$/ 2 \pi$''. The
classical manifold emerges as $\tilde{\hbar} \rightarrow 0$. We set
the usual Planck's constant $\hbar$ to $1$ hereafter unless otherwise stated.

In the same way, we can quantize ${\mathbb C}^{N+1}$ for any $N$ using
an appropriate number of oscillators and that gives us fuzzy ${\mathbb
C}P^N$ as we shall later see.

\section{Fuzzy $S^3$ and Fuzzy $S^2$} 

There is a well-known descent chain from ${\mathbb C}^2$ to the
$3$-sphere $S^3$ and thence to $S^2$. Our tactics to obtain fuzzy $S^2
\equiv S^2_F$ is to quantize this chain, obtaining along the way fuzzy
$S^3 \equiv S_F^3$.

Let us recall this chain of manifolds. Consider ${\mathbb C}^2$ with
the origin removed, ${\mathbb C}^2 \setminus \{0\}$. As $z \neq 0$,
$\frac{z}{|z|}$ with $|z| = \big( \sum |z_i|^2 \big)^{\frac{1}{2}}$
makes sense here. Since $\big |\frac{z}{|z|} \big |$ is normalized to $1$,
$\frac{z}{|z|} = 1$, it gives the $3$-sphere $S^3$. Thus we have the
fibration
\begin{equation}
{\mathbb R} \rightarrow {\mathbb C}^2 \setminus \{0\} \rightarrow S^3
= \left \langle \xi =\frac{z}{|z|} \right \rangle \,, \quad z \rightarrow \frac{z}{|z|} \,.     
\end{equation}

Now $S^3$ is a $U(1)$-bundle (``Hopf fibration'') \cite{topology1} over $S^2$. If $\xi
\in S^3$, then $\vec{x}(\xi) = \xi^\dagger \vec{\tau} \xi$ (where $\tau_i
\,, i =1,2,3$ are the Pauli matrices) is invariant under the $U(1)$ action
$\xi \rightarrow \xi e^{i \theta}$ and is a real normalized
three-vector:
\begin{equation}
\vec{x}(\xi)^* =\vec{x}(\xi) \,, \quad \quad \vec{x}(\xi) \cdot
\vec{x}(\xi) =1 \,. 
\end{equation}
So $\vec{x}(\xi) \in S^2$ and we have the Hopf fibration
\begin{equation}
U(1) \rightarrow S^3 \rightarrow S^2 \,, \quad \quad \xi \rightarrow
\vec{x}(\xi) \,. 
\end{equation}
Note that $ \vec{x}(\xi) = \frac{1}{|z|} z^* \vec{\tau} z \frac{1}{|z|}$. 

The fuzzy $S^3$ is obtained by replacing $\frac{z_i}{|z|}$ by $a_i
\frac{1}{\sqrt{\hat{N}}}$ where $\hat{N}= a_j^\dagger a_j$ is the
number operator:
\begin{equation}
\frac{z_i}{|z|} \rightarrow a_i \frac{1}{\sqrt{\widehat{N}}} \,, \quad
\frac{z_i^*}{|z|} \rightarrow \frac{1}{\sqrt{\widehat{N}}} a_i^\dagger \,,
\quad \widehat{N} = a^\dagger_j a_j \,, \widehat{N} \neq 0 \,.  
\label{eq:secop1}
\end{equation}
The quantum condition $\hat{N} \neq 0$ means that the vacuum is
omitted from the Hilbert space, so that it is the orthogonal
complement of the vacuum in Fock space. This omission is like the
deletion of $0$ from ${\mathbb C}^2$.

There is a problem with this omission as $a_i
\frac{1}{\sqrt{\widehat{N}}}$ and its polynomials will create it from any
$\widehat{N} =n$ state. For this reason, and because $a_i
\frac{1}{\sqrt{\widehat{N}}}$ and its adjoint need the
infinite-dimensional Fock space to act on and do not give
finite-dimensional models for $S_F^3$, we will not dwell on this
space.

\section{The Fuzzy Sphere $S_F^2$}

The problems of $S_F^3$ melt away for $S_F^2$. Quantization of $S^2$
gives $S_F^2$ with $x_i(\xi)$ becoming the operator $x_i$:
\begin{equation}
x_i(\xi) \rightarrow x_i = \frac{1}{\sqrt{\hat{N}}}
a^\dagger \vec{\tau} a \frac{1}{\sqrt{\hat{N}}}  
=\frac{1}{\hat{N}}  a^\dagger \vec{\tau} a \,, \quad \quad \widehat{N} \neq 0 \,.
\label{eq:qz1}
\end{equation}
Since 
\begin{equation} 
\lbrack x_i \,, \hat{N} \rbrack = 0 \,,
\label{eq:com1}
\end{equation}  
we can restrict $x_i$ to the subspace ${\cal H}_n$ of the Fock space
where ${\hat N}=n \, (\neq 0)$. This space is $(n+1)$-dimensional and
is spanned by the orthogonal vectors
\begin{equation}
\frac{(a_1^\dagger)^{n_1}}{\sqrt{n_1!}}
\frac{(a_2^\dagger)^{n_2}}{\sqrt{n_2!}} |0 \rangle \equiv |n_1 \, n_2 \rangle
\,, \quad  n_1+n_2=n \,.
\label{eq:state1}
\end{equation}
$x_i$ act irreducibly on this space and generate the full matrix
algebra $Mat(n+1)$.

The $SU(2)$ angular momentum operators $L_i$ are given by the
Schwinger construction:
\begin{equation}
L_i = a^\dagger \frac{\tau_i}{2} a \,, \quad \lbrack L_i \,, L_j
\rbrack = i \epsilon_{ijk} L_k \,. 
\label{eq:sch1}
\end{equation}
$a_i^\dagger$ transform as spin $\frac{1}{2}$ spinors and (\ref{eq:state1})
spans the $n$-fold symmetric product of these spinors. It has angular
momentum $\frac{n}{2}$:
\begin{equation}
L_iL_i \big|_{{\cal H}_n} = \frac{n}{2} \big (\frac{n}{2}+1 \big)
{\bf 1} \big|_{{\cal H}_n} \,.
\end{equation}
Since 
\begin{equation}
x_i  \big|_{{\cal H}_n} = \frac{2}{n} L_i  \big|_{{\cal H}_n} \,,
\end{equation}
we find
\begin{eqnarray}
\lbrack x_i \,, x_j \rbrack  \big|_{{\cal H}_n} &=& \frac{2}{n} i
\epsilon_{ijk} x_k  \big|_{{\cal H}_n} \,, \nonumber \\ 
\big (\sum x_i^2 \big)  \big|_{{\cal H}_n} &=& \big ( 1+ \frac{2}{n}
\big) {\bf 1} \big|_{{\cal H}_n} \,. 
\label{eq:fs1}
\end{eqnarray}
$S_F^2$ has radius $\big ( 1+ \frac{1}{n} \big)^{\frac{1}{2}}$ which
becomes $1$ as $n \rightarrow \infty$.

We generally write the equations in (\ref{eq:fs1}) as $\lbrack x_i \,,
x_j \rbrack = \frac{2}{n} i \epsilon_{ijk} x_k \,, \big (\sum
x_i^2 \big) = \big (1+ \frac{2}{n} \big)$, omitting the indication of
${\cal H}_n$. $S_F^2$ should have an additional label $n$, but that too is usually
omitted. The $x_i$'s are seen to commute in the naive continuum limit
$n \rightarrow \infty$ giving back the commutative algebra of
functions on $S^2$.

The fuzzy sphere $S_F^2$ is a ``quantum'' object. It has wave
functions which are generated by $x_i$ restricted to ${\cal H}_n$. Its
Hilbert space is $Mat(n+1)$ with the scalar product
\begin{equation}
(m_1, m_2) = \frac{1}{n+1} Tr m_1^\dagger m_2 \,, \quad \quad m_i \in
  Mat(n+1) \,. 
\label{eq:scalarp1}
\end{equation}
We denote $Mat(n+1)$ with this scalar product also as $Mat(n+1)$.

\section{Observables of $S_F^2$}

The observables of $S_F^2$ are associated with linear operators on $Mat(n+1)$. We can
associate two linear operators $\alpha^L$ and $\alpha^R$ to each
$\alpha \in Mat(n+1)$. They have left- and right-actions on
$Mat(n+1)$;
\begin{equation}
\alpha^L m = \alpha m \,, \quad \quad \alpha^R m = m \alpha \,, \quad
\quad  \forall \quad m \in Mat(n+1) \, 
\end{equation}
and fulfill
\begin{equation}
(\alpha \beta)^L = \alpha^L \beta^L \,, \quad \quad (\alpha \beta)^R =
  \beta^R \alpha^R \,. 
\end{equation}
Such left- and right- operators commute:
\begin{equation} 
\lbrack\alpha^L \,, \beta^R \rbrack = 0 \,, \quad \quad \forall \quad
\alpha \,, \beta \in  Mat(n+1) \,. 
\end{equation} 
We denote the two commuting matrix algebras of left- and right- operators 
by $ Mat_{L,R}(n+1)$.
%the algebra ${\cal A}$ of
%operators of $S_F^2$ from which observables are constructed is their
%tensor product over ${\mathbb C}$: \be {\cal A} = Mat_L (n+1)
%\otimes_{{\mathbb C}} Mat_R(n+1) \,.  \ee 
$Mat(n+1)$ is generated by $a_i^\dagger a_j$ with the understanding that their
domain is ${\cal H}_n$. Accordingly, $Mat_{L,R}(n+1)$ are generated by $(a_i^\dagger a_j)^{L,R}$.

We can also define operators $a_i^{L,R}$, $(a_j^\dagger)^{L,R}$:
\begin{eqnarray}  
\bea {cc}
a_i^L m = a_i m \,, \quad \quad & a_i^R m = m a_i \\
a_j^{\dagger L} m = a_j^\dagger  m \,, \quad \quad & a_j^{\dagger R} m
= m a_j^\dagger \,.    
\ea
\end{eqnarray} 
They are operators changing $n$:
\begin{eqnarray} 
a_i^{L,R} &:& \quad \quad {\cal H}_n \rightarrow  {\cal
  H}_{n-1} \nonumber \\ 
a_j^{\dagger L, R} &:& \quad \quad  {\cal H}_n
  \rightarrow  {\cal H}_{n+1} \, .
\label{eq:shift1}
\end{eqnarray} 
Such operators are important for discussions of bundles.(See chapter 5)

With the help of these operators, we can write 
\begin{equation}  
(a_i^\dagger a_j)^L = a_i^{\dagger L} a_j^L \,, \quad \quad
  (a_i^\dagger a_j)^R = a_j^R a_i^{\dagger R} \,. 
\end{equation} 

Of particular interest are the three angular momentum operators
\begin{equation} 
L_i^L \,, L_i^R \,, \quad \quad {\cal L}_i = L_i^L - L_i^R \,.
\label{eq:am1}
\end{equation} 
Of these, ${\cal L}_i$ annihilates ${\bf 1}$ as does the continuum
orbital angular momentum. It is the fuzzy sphere angular momentum
approaching the orbital angular momentum of $S^2$ as $n \rightarrow
\infty$: 
\begin{equation}  
{\cal L}_i \rightarrow -i \big(\vec{x}(\xi) \wedge \vec{\nabla} \big)
\equiv -i \epsilon_{ijk} x(\xi)_j \frac{\partial}{\partial x(\xi)_k} \quad
\mbox{as} \quad n \rightarrow \infty.
\label{eq:difop1}
\end{equation} 

\section{Diagonalizing  ${\cal L}_i$}

We have $\sum(L_i^L)^2 = \sum (L_i^R)^2 = \frac{n}{2}
\big(\frac{n}{2}+1 \big)$ so that orbital angular momentum is the sum
of two angular momenta with values $\frac{n}{2}$. Hence the spectrum
of ${\cal L}^2$ is 
\be
\langle \ell (\ell +1) : \ell \in \{0,1,2,..., n \} \rangle.
\ee
A function $f$ in $C^\infty(S^2)$ has the expansion
\begin{equation} 
f = \sum a_{\ell m} Y_{\ell m} \,
\end{equation}   
in terms of the spherical harmonics. The spectrum of orbital angular
momentum is thus $\langle \ell (\ell +1) : \ell \in \{0,1,2,\ldots, n,,\ldots \} \rangle$.

The spectrum of ${\cal L}^2$ is thus precisely that of the continuum
orbital angular momentum cut off at $n$. There is no distortion of
eigenvalues upto $n$.

The eigenstates $T^{\ell}_m \,, m \in \{-\ell, -\ell+1,...,\ell \}$ of
${\cal L}^2$ are known as polarization operators \cite{moskalev}. They are eigenstates
of ${\cal L}_3$ and also orthonormal: 
\begin{eqnarray} 
{\cal L}^2 T^{\ell}_m &=& \ell(\ell+1) T^{\ell}_m \,, \nonumber \\ 
{\cal L}_3 T^{\ell}_m &=& m T^{\ell}_m \,, \nonumber \\ 
\big( T^{\ell^\prime}_{m^\prime} \,, T^{\ell}_m \big) &=& \delta_{\ell
  \ell^\prime} \delta_{m^\prime m} \,. 
\end{eqnarray} 

\section{Scalar Fields on $S_F^2$}

We will be brief here as they are treated in detail in chapter 4. A
complex scalar field $\Phi$ on $S^2$ is a power series in the
coordinate functions $m_i:= x_i$, 
\begin{equation}  
\Phi = \sum a_{i_1 \ldots i_n} m_{i_1}
\cdots m_{i_n} \,.
\label{eq:sf1}
\end{equation} 
(Note again that $\vec{m} \cdot \vec{m} = 1$)
The Laplacian on $S^2$ is $\Delta : = - ( -i \vec{x} \wedge
\vec{\nabla} )^2$ and a simple Euclidean action is 
\begin{equation}  
{\cal S} = - \int \frac{d \Omega}{4 \pi} \Phi^* \Delta \Phi \, \quad
\quad d \Omega = d \cos(\theta) d \psi \,.
\label{eq:ac1}
\end{equation} 
We can simplify (\ref{eq:ac1}) by the expansion
\begin{equation}  
\Phi(\vec{x}) = \sum \Phi_{\ell m} Y_{\ell m}(\vec{m}) \,.
\end{equation} 
Then since $\Delta Y_{\ell m}(\vec{m}) = - \ell (\ell+1) Y_{\ell
m}(\vec{m})$, and $\int \frac{d \Omega}{4 \pi} Y_{\ell^\prime
m^\prime}(\vec{m})^* Y_{\ell m}(\vec{m}) = \delta_{\ell \ell^\prime}
\delta_{m m^\prime}$,
\begin{equation}  
{\cal S} = \sum \ell (\ell+1) \Phi_{\ell m}^* \Phi_{\ell m} \,.
\label{eq:acf1}
\end{equation} 

From (\ref{eq:sf1}), we infer that the fuzzy scalar field $\psi$ is a
power series in the matrices $x_i$ and is hence itself a matrix.  The
Euclidean action replacing (\ref{eq:ac1}) is
\begin{equation} 
S = ({\cal L}_i \psi \,, {\cal L}_i \psi) = - (\psi \,, \Delta \psi) \,.
\end{equation} 

On expanding $\psi$ according to 
\begin{equation} 
\psi = \sum_{\ell \leq n+1} \psi_{\ell m} T^{\ell}_m \,,
\end{equation}   
this reduces to 
\begin{equation}  
S = \sum_{\ell \leq n}  \ell (\ell+1)|\psi_{\ell m}|^2 \,.
\end{equation}  

\section{Holstein-Primakoff Construction}

There is an interesting construction of $L_i$ for fixed $n$ using
just one oscillator due to Holstein and Primakoff. We outline this
construction here \cite{Holstein}

In brief, since ${\hat N}$ commutes with $L_i$, we can eliminate $a_2$
from $L_i$ and restrict $L_i$ to ${\cal H}_n$ without spoiling their
commutation relations. The result is the Holstein-Primakoff
construction.

We now give the details. (\ref{eq:secop1}) gives the following polar
decomposition of $a_2$:
\begin{equation} 
a_2 = U \sqrt{N-a_1^\dagger a_1} \,, \quad \quad U^\dagger U = U
U^\dagger = {\bf 1} \,,
\end{equation} 
where we choose the positive square root:
\begin{equation}  
\sqrt{N-a_1^\dagger a_1} \geq 0 \,. 
\end{equation} 
We can understand $U$ better by examining the action of $a_2$ on the
orthonormal states (\ref{eq:state1}) spanning ${\cal H}_n$. We find
\begin{eqnarray} 
a_2 | n_1 \,, n_2 \rangle &=& \sqrt{n_2} | n_1 \,, n_2 - 1 \rangle \nonumber \\
&=&  U \sqrt{N-a_1^\dagger a_1} | n_1 \,, n_2 \rangle \nonumber \\
&=& \sqrt{n_2} U | n_1 \,, n_2 \rangle 
\end{eqnarray} 
or
\begin{equation}  
U | n_1 \,, n_2 \rangle = | n_1 \,, n_2 - 1 \rangle 
\end{equation} 
Thus if
\begin{equation} 
A^\dagger = a_1^\dagger U \,, \quad \quad A = U^\dagger a_1 \,,
\label{eq:polard1}
\end{equation} 
then
\begin{eqnarray}  
A^\dagger  | n_1 \,, n_2 \rangle &=&  \sqrt{n_1 + 1} | n_1 + 1 \,, n_2
- 1 \rangle \,,\nonumber \\  
A  | n_1 \,, n_2 \rangle &=&  \sqrt{n_1} | n_1 - 1 \,, n_2 - 1 \rangle \,,
\end{eqnarray} 
and
\begin{gather} 
\lbrack A \,, A^\dagger \rbrack = {\bf 1} \,, \quad \lbrack
A^\dagger \,, A^\dagger \rbrack = \lbrack A \,, A \rbrack = 0 \,, \\
\lbrack A \,, N \rbrack = \lbrack A^\dagger \,, N \rbrack = 0 \,.
\label{eq:osal1}
\end{gather}    
$a_2$ vanishes on $|n \,, 0 \rangle$ and $U$ and $A^\dagger$ are
undefined on that vector. That is $A$ and $A^\dagger$ can not be
defined on ${\cal H}_n$. In any case the oscillator algebra of
(\ref{eq:osal1}) has no finite-dimensional representation. But this
is not the case for $L_i$. We have
\begin{eqnarray} 
L_+ &=& L_1 + i L_2 = a_1^\dagger a_2 = A^\dagger \sqrt{N - A^\dagger
  A} \nonumber \\ 
L_- &=& L_1 - i L_2 = a_2^\dagger a_1 = \sqrt{N - A^\dagger A} \, A
\nonumber \\ 
L_3 &=& a_1^\dagger a_1 - a_2^\dagger a_2 = A^\dagger A - N 
\label{eq:HPC}
\end{eqnarray} 
On ${\cal H}_n$ (\ref{eq:HPC}) gives the Holstein-Primakoff
realization of the $SU(2)$ Lie algebra for angular momentum 
$\frac{n}{2}$.

\section{${\mathbb C}P^N$ and Fuzzy ${\mathbb C}P^N$}

$S^2$ is ${\mathbb C}P^1$ as a complex manifold. The additional structure
for ${\mathbb C}P^1$ as compared to $S^2$ is only the complex
structure. So we can without great harm denote $S^2$ and $S_F^2$ also
as ${\mathbb C}P^1$ and ${\mathbb C}P^1_F$. In chapter 3 we will
in fact consider the complex structure and its quantization.

Generalizations of ${\mathbb C}P^1$ and ${\mathbb C}P^1_F$ are
${\mathbb C}P^N$ and ${\mathbb C}P^N_F$. They are
associated with the groups $SU(N+1)$.

Classically ${\mathbb C}P^N$ is the complex projective space of
complex dimension $N$. It can described as follows.  Consider
the $(2 N +1)$-dimensional sphere
\begin{equation} 
S^{2 N +1} = \Big \langle \xi = \big ( \xi_1 \,, \xi_2 \cdots ,
\xi_{N +1} \big) : \, \xi_i \in {\mathbb C} \,, |\xi|^2 : = \sum
|\xi_i|^2 =1 \Big \rangle \,.
\end{equation} 
It admits the $U(1)$ action
\begin{equation}  
\xi \, \rightarrow \, e^{i \theta} \xi \,.
\label{eq:u1ac}
\end{equation} 
${\mathbb C}P^N$ is the quotient of $S^{2 N +1}$ by this
action giving rise to the fibration
\begin{equation} 
U(1) \, \rightarrow \, S^{2 N +1} \, \rightarrow \, {\mathbb C}P^N \,.
\end{equation} 
If $\lambda_i$ are the Gell-Mann matrices of $SU(N + 1)$, a
point of ${\mathbb C}P^N$ is
\begin{equation} 
\vec{X}(\xi) = \xi^\dagger \vec{\lambda} \xi \,, \quad \xi \in  S^{2
  N +1} \,. 
\end{equation} 
For $N=1$, these become the previously constructed structures.

There is another description of $ S^{2 N +1}$ and ${\mathbb
C}P^N$. $SU(N + 1)$ acts transitively on $S^{2 N
+1}$ and the stability group at $(1 \,, \vec{0})$ is
\begin{equation} 
SU(N) = \left \langle u \in SU(N +1) : \, u = 
\left (
\begin{array}{cc}
1 & 0 \\
0 & {\hat u} \\
\end{array}
\right ) 
\right \rangle \,.
\end{equation} 
Hence 
\begin{equation} 
S^{2 N +1} = SU(N+1) \big /  SU(N) \,.
\end{equation} 

Consider the equivalence class
\begin{equation} 
\langle (1 \,, \vec{0} ) \rangle = \big \langle (e^{i \theta} \,,
\vec{0}) \big |e^{i \theta} \in U(1) \big \rangle
\label{eq:eqc1}
\end{equation} 
of all elements connected to $(1 \,, \vec{0})$ by the $U(1)$ action
(\ref{eq:u1ac}). Its orbit under $SU(N+1)$ is ${\mathbb
C}P^N$. The stability group of (\ref{eq:eqc1}) is
\begin{equation} 
S \big \lbrack  U(1) \times U(N) \big \rbrack = \left \lbrack
\upsilon \in SU(N+1) : \upsilon = 
\left (
\begin{array}{cc}
e^{i \theta} & 0 \\
0 & {\hat \upsilon} \\
\end{array}
\right )
\right \rbrack \,.
\end{equation} 
Thus
\begin{equation} 
{\mathbb C}P^N = SU(N+1) \big / S \big \lbrack  U(1)
\times U(N) \big \rbrack \,. 
\end{equation} 
$S \big \lbrack U(1) \times U(N) \big \rbrack$ is commonly
denoted as $U(N)$. The two groups are isomorphic.

To obtain ${\mathbb C}P^N_F$, we think of $S^{2 N +1}$
as a submanifold of ${\mathbb C}^{N +1} \setminus \{0 \}$:
\begin{equation} 
S^{2 N +1} = \Big \langle \xi =\frac{z}{|z|} \,, z = (z_1 \,,
 z_2 \,, \cdots, z_{N+1}) \in {\mathbb C}^{N +1} \setminus \{0
 \} \Big \rangle \,.
\end{equation} 

Just as before, we can quantize ${\mathbb C}^{N +1}$ by
replacing $z_i$ by annihilation operators $a_i$ and $z_i^*$ by
$a_i^\dagger$:
\begin{equation} 
\lbrack a_i \,, a_j \rbrack = \lbrack a_i^\dagger \,, a_j^\dagger
\rbrack =0 \,, \quad \lbrack a_i \,, a_j^\dagger \rbrack = \delta_{ij}
\,.
\end{equation} 
With
\begin{equation} 
{\widehat N} = a^\dagger_i a_i 
\end{equation} 
as the number operator, the quantized $\xi$ is given by the correspondence
\begin{equation} 
\xi_i = \frac{z_i}{|z|} \longrightarrow a_i \frac{1}{\sqrt{{\widehat N}}}
\,, \quad N \neq 0 \,.
\end{equation} 
Then as in (\ref{eq:qz1}), we get the ${\mathbb C}P^N_F$ coordinates
\begin{equation} 
X_i(z) \longrightarrow X_i = \frac{1}{{\widehat N}} a^\dagger \lambda_i a
\,, \quad N \neq 0 \,.
\end{equation} 

The rest of the discussion follows that of ${\mathbb C}P^1$ with
$SU(N+1)$ replacing $SU(2)$. Because of (the analogue of)
(\ref{eq:com1}), $X_i$ can be restricted to ${\cal H}_n$, the subspace
of the Fock space with ${\widehat N} = n$. It is spanned by the orthonormal vectors
\begin{equation} 
\prod_{i=1}^{N+1} \frac{(a_i^\dagger)^{n_i}}{\sqrt{n_i!}} |0
\rangle := | n_1 \, n_2 \,, \cdots \,, N+1 \rangle
\label{eq:states2}
\,, \quad \quad \sum n_i = n \,,
\end{equation} 
and is of dimension
\begin{equation} 
M = _{N+1+n} C_n = \frac{(N+n)!}{n! N!} \,.
\label{eq:dim1}
\end{equation} 

The $SU(N+1)$ angular momentum operators are given by a
generalized Schwinger construction :
\begin{equation} 
L_i = a^\dagger \frac{\lambda_i}{2} a \,, \quad \quad \lbrack L_i \,,
L_j \rbrack = i f_{ijk} L_k \,. 
\end{equation} 
$a_i^\dagger$ transform by the unitary irreducible representation (UIR)
$(N+1)$ of $SU(N+1)$ and (\ref{eq:states2}) span the
space of $n$ fold symmetric product of ther UIR's $(N+1)$ of  $SU(N+1)$. It carries a
UIR of dimension (\ref{eq:dim1}) and the quadratic Casimir operator
\begin{equation} 
\sum L_i^2 = \frac{N}{2} \left ( \frac{n^2}{N+1} +n \right) {\bf 1}
\label{eq:casimir1}
\end{equation} 
Its remaining Casimir operators are fixed by (\ref{eq:casimir1}). As before 
\begin{eqnarray} 
X_i  \big|_{{\cal H}_n} &=& \frac{2}{n} L_i \big|_{{\cal H}_n} \,, \nonumber \\
\lbrack X_i \,, X_j \rbrack  \big|_{{\cal H}_n} &=& \frac{2}{n} i f_{ijk} X_k  \big|_{{\cal H}_n}  \,, \nonumber \\  
\big (\sum X_i^2 \big ) \big|_{{\cal H}_n} &=& \left (\frac{2 N}{N+1}
+ \frac{2 N}{n} \right) {\bf 1} \big|_{{\cal H}_n}
\end{eqnarray} 
The ``size'' of ${\mathbb C}P^N_F$ is measured by the ``radius'' 
$\sqrt{ \left ( \frac{2 N}{N+1} + \frac{2 N}{n} \right)}$.
In the $N \rightarrow \infty$ limit, the $X_i$'s also commute and
generate $C^\infty({\mathbb C}P^N)$.

The wave function of ${\mathbb C}P^N_F$ are polynomials in
$X_i$, that is they are elements of $Mat(M)$, with a scalar product like
(\ref{eq:scalarp1}). As before, for each $\alpha \in Mat(M)$, we have
two observables $\alpha_{L, R}$ and they constitute the matrix algebras
$M_{L,R}(M)$.

The discussions leading up to (\ref{eq:shift1}) and (\ref{eq:am1}) can
be adapted also to ${\mathbb C}P^N_F$. As for
(\ref{eq:difop1}), it generalizes to
\begin{equation} 
{\cal L}_i \longrightarrow -i f_{ijk} X(\xi)_j \frac{\partial}{\partial
X(\xi)_k} \,.
\end{equation} 
Diagonalization of ${\cal L}_i$ involves the reduction of the product
of the UIR's of $SU(N+1)$ given by $L_i^L$ and its
complex conjugate given by $L_i^R$ to their irreducible
components. The corresponding polarization operators can also in
principle be constructed.

The scalar field action (\ref{eq:acf1}) generalizes easily to
${\mathbb C}P^N_F$.

\section{The ${\mathbb C}P^N$ Holstein-Primakoff Construction}    

The generalization of this construction to ${\mathbb C}P^N$ and
$SU(N+1)$ is due to Sen \cite{Sen}.

Consider for specificity $N =2$ and $SU(3)$ first. $SU(3)$ has
$3$ oscillators $a_1 \,, a_2 \,, a_3$. There are also the $SU(2)$
algebras with generators
\begin{equation} 
\sum_{i=j=1}^2 a_i^\dagger \Big (\frac{\vec{\sigma}}{2} \Big)_{ij} a_j
\,, \quad \quad \sum_{i,j =2}^3 a_i^\dagger \Big
(\frac{\vec{\sigma}}{2} \Big)_{ij} a_j \,,
\end{equation} 
acting on the indices $1,2$ and $2,3$ respectively, of $a$'s and
$a^\dagger$'s. Taking their commutators, we can generate the full
$SU(3)$ Lie algebra.

We will eliminate $a_2 \,, a_2^\dagger$ from both these sets using the
previous Holstein-Primakoff construction. In that way, we will obtain
the $SU(3)$ Holstein-Primakoff construction.

As previously we write the polar decompositions
\begin{equation} 
a_2 = U_2 \sqrt{N_2} \,, \quad a_2^\dagger = \sqrt{N_2} U_2^\dagger
\,, \quad N_2 = a_2^\dagger a_2 \,, \quad U_2^\dagger U_2 = {\bf 1}
\,.
\end{equation} 
The oscillators act on the Fock space $ \oplus_N {\cal H}_N$ spanned
by (\ref{eq:states2}) for $N=2$. The actions of $U_2$ and
\begin{equation} 
A_{12}^\dagger = a_1^\dagger U_2 \,, \quad A_{12} = U_2^\dagger a_1 \,,
\end{equation} 
follow (\ref{eq:polard1}). They do not affect $n_3$. Using
(\ref{eq:HPC}), we can write the $SU(2)$ generators acting on $(12)$
indices as
\begin{eqnarray} 
I_+ &=& a_1^\dagger a_2 = A_{12}^\dagger \sqrt{N_2} \,, \nonumber \\
I_- &=& a_2^\dagger a_1 = \sqrt{N_2} A_{12} \,, \nonumber \\
I_3 &=& \frac{1}{2} \big( a_1^\dagger a_1 - a_2^\dagger a_2 \big) =
\frac{1}{2} \big ( A_{12}^\dagger A_{12} - N_2 \big) \,.  
\label{eq:HPC2}
\end{eqnarray} 
We follow the $I \,, U \,, V$ spin notation of $SU(3)$ in particle
physics \cite{Lipkinbook}. They are connected by Weyl reflections.  

In a similar manner, the $SU(3)$ generators acting on $23$ indices are
constructed from
\begin{equation} 
A_{32}^\dagger = a_3^\dagger U_2 \,, A_{32} = U_2^\dagger a_3 \,,
\end{equation} 
and read
\begin{eqnarray} 
U_+ &=& a_3^\dagger a_2 = A_{32}^\dagger \sqrt{N_2} \nonumber \\
U_- &=& a_2^\dagger a_3 = \sqrt{N_2} A_{32} \nonumber \\
U_3 &=& \frac{1}{2} \big( a_2^\dagger a_2 - a_3^\dagger a_3 \big) =
\frac{1}{2} \big ( N_2 - A_{32}^\dagger A_{32} \big) \,.  
\label{eq:HPC3}
\end{eqnarray} 

In a UIR of $SU(3)$, the total number operator $N = N_1 + N_2 + N_3$
is fixed. Acting on on ${\cal H}_n$, it becomes $n$. Keeping this in
mind, we now substitute
\begin{equation} 
N_2 = N -N_1 -N_3 = N - A_{12}^\dagger A_{12} -  A_{32}^\dagger A_{32} 
\end{equation} 
in (\ref{eq:HPC2}) and (\ref{eq:HPC3}) to eliminate the second
oscillator. That gives
\begin{eqnarray} 
\bea{ccc}
I_+ = A_{12}^\dagger \sqrt{N-N_1-N_3} \,, \quad & I_- =
\sqrt{N-N_1-N_3} A_{12} \,, \quad & I_3 = N_1+ \frac{N_3}{2}-
\frac{N}{2} \\ 
U_+ = A_{32}^\dagger \sqrt{N-N_1-N_3} \,, \quad & U_- =
\sqrt{N-N_1-N_2} A_{32} \,, \quad & U_3 = N_3+ \frac{N_1}{2}-
\frac{N}{2} 
\ea 
\end{eqnarray} 
These operators and their commutators generate the full $SU(3)$ Lie
algebra when restricted to ${\cal H}_n$. That is the $SU(3)$
Holstein-Primakoff construction.

If the restriction to ${\cal H}_n$ is not made, $N$ is a new operator
and we get instead the $U(3)$ Lie algebra with $N$ generating its
central $U(1)$.

The Holstein-Primakoff construction for ${\mathbb C}P^N$ is
much the same. One introduces $N+1$ oscillators $a_i \,,
a_i^\dagger (i \in \lbrack 1\,, \cdots N \rbrack)$ with which
$SU(N+1)$ Lie algebra can be realized using the Schwinger
construction. The $SU(N+1)$ UIR's we get therefrom are
symmetric products of the fundamental representation $(N+1)$. The
number operator $N= a^\dagger \cdot a$ has a fixed value in one such
UIR. Next $a_2 \,,a_2^\dagger$ are eliminated from $SU(N+1)$
generators in favor of $N$ and the remaining operators to obtain the
generalized Holstein-Primakoff construction.

$SU(N+1)$ is of rank $N$, and we can realize its Lie
algebra with $N$ oscillators. There is a similar result in
quantum field theory where with the help of the vertex operator
construction, a (simply laced) rank $N$, Lie algebra can be
realized with $N$ scalar fields on $S^1 \times {\mathbb R}$
valued on $S^1$ \cite{Goddard}. This resemblance perhaps is not
an accident.

%\end{document}

%\begin{document}

\chapter{Star Products}

\section{Introduction}

The algebra of smooth functions on a manifold ${\cal M}$ under 
point-wise multiplication is commutative. 
In deformation quantization \cite{defqu}, this point-wise product 
is deformed to a non-commutative (but still associative) product 
called the $*$-product. It has a central role in many discussions 
of non-commutative geometry. It has been fruitfully used in quantum 
optics for a long time.

The existence of such deformations was understood many years ago by 
Weyl, Wigner, Groenewold and Moyal \cite{weyl, Groenewold:kp, Moyal:sk}. They noted that if there is a 
linear injection (one-to-one map) $\psi$ of an algebra ${\cal A}$ into 
smooth functions ${\mathbb C}^\infty({\cal M})$ on a manifold 
${\cal M}$, then the product in ${\cal A}$ can be transported to the 
image $\psi({\cal A})$ of ${\cal A}$ in ${\mathbb C}^
\infty({\cal M})$ using the map. That is then a $*$-product.

Let us explain this construction with greater completeness and 
generality \cite{BDBJ}. For concreteness we can consider
${\cal A}$ to be an algebra of bounded operators on a Hilbert 
space closed under the hermitian conjugation of $*$. It is then an
example of a $*$-algebra.

More generally, ${\cal A}$ can be a generic ``$*$-algebra', 
that is an algebra closed under an anti-linear involution:
\be
a \,, b \in {\cal A} \,, \, \lambda \in {\mathbb C} \, 
\Rightarrow a^* \,, b^* \in {\cal A} \,, (a b)^* = b^* a^* \,,
(\lambda a )^* = \lambda^* a^* \,.
\ee

A two-sided ideal ${\cal A}_0$ of ${\cal A}$ is a subalgebra 
of ${\cal A}$ with the property
\be
a_0 \in {\cal A}_0 \Rightarrow \alpha a_0 \, \, \mbox{and} \, 
\, a_0 \alpha \in {\cal A}_0 \,, \, \forall \alpha \in {\cal A} \,.
\ee
That is ${\cal A}{\cal A}_0\,, {\cal A}_0{\cal A} \subseteq {\cal A}_0$. 
A two-sided $*$-ideal ${\cal A}_0$ by definition is itself closed under
$*$ as well.

An element of the quotient ${\cal A}/{\cal A}_0$ is the equivalence class
\be
\lbrace \alpha + {\cal A}_0 \subset {\cal A} \rbrace =  \big \lbrace
\lbrack \alpha + a_0 \rbrack \big| a_0 \in {\cal A}_0 \big \rbrace \,.
\ee
If ${\cal A}_0$ is a two-sided ideal, ${\cal A}/ {\cal A}_0$ is itself 
an algebra with the sum and the product
\beqa
(\alpha + {\cal A}_0) + (\beta + {\cal A}_0) &=& 
\alpha + \beta + {\cal A}_0 \,, \nonumber \\
(\alpha + {\cal A}_0) (\beta + {\cal A}_0) &=& \alpha \beta + {\cal A}_0
\eeqa
If ${\cal A}_0$ is a two-sided $*$-ideal, then  ${\cal A}/ {\cal A}_0$ 
is a $*$-algebra with the $*$-operation
\begin{equation} 
(\alpha + {\cal A}_0)^*= \alpha^* + {\cal A}_0 \,.
\end{equation} 

We note that the product and $*$ are independent of the choice 
of the representatives $\alpha \,, \beta$ from the equivalence 
classes $\alpha + {\cal A}_0$ and $\beta + {\cal A}_0$ because 
${\cal A}_0$ is a two-sided ideal. So they make sense for 
${\cal A}/{\cal A}_0$.

Let $C^\infty({\cal M})$ denote the complex-valued smooth functions on
a manifold ${\cal M}$. Complex conjugation $-$(bar) is defined on
these functions. It sends a function $f$ to its complex conjugate
${\bar f}$.

We consider the linear maps
\begin{gather}  
\psi: {\cal A} \, \longrightarrow C^\infty({\cal M}) \, \\
\psi \left (\sum \lambda_i a_i \right ) = \sum \lambda_i \psi(a_i) 
\,, \quad a_i \in {\cal A} \,, \quad \lambda_i \in {\mathbb C} \,. 
\end{gather} 

The kernel of such a map is the set of all $\alpha \in {\cal A}$ 
for which $\psi({\alpha})$ is the zero
function $0$ (Its value is zero at all points of ${\cal M}$):
\begin{equation}  
Ker \, \psi = \langle \alpha_0 \in {\cal A} \big | \psi(\alpha_0) =0 \rangle \,.
\label{eq:kernel1}
\end{equation}  
$\psi$ descends to a linear map, called $\Psi$, from ${\cal A}/ Ker \,
\psi = \{\alpha + Ker \, \psi : \alpha \in {\cal A} \}$ to $C^\infty({\cal
M})$:
\begin{equation} 
\Psi (\alpha + Ker \, \psi) = \psi(\alpha) \,
\end{equation}  
$\psi(\alpha)$ does not depend on the choice of the representative
$\alpha$ from $\alpha + Ker \, \psi$ because of (\ref{eq:kernel1}).
Clearly $\Psi$ is an injective map from ${\cal A}/Ker \, \psi$ to
$C^\infty({\cal M})$.

If $Ker \, \psi$ is also a two sided ideal, $\Psi$ is a linear map
from the algebra ${\cal A}/ Ker \, \psi$ to $C^\infty({\cal M})$.
Using this fact, we define a product, also denoted by $*$, on
$\Psi({\cal A}/ Ker \, \psi)= \psi({\cal A}) \subseteq C^\infty({\cal
M})$ :
\begin{equation} 
\Psi( \alpha + Ker \, \psi) * \Psi( \beta + Ker \, \psi) = \Psi
\big((\alpha + Ker \, \psi) \,(\beta + Ker \, \psi)\big) \,.
\end{equation} 
or 
\begin{equation} 
\psi( \alpha ) * \psi( \beta ) = \psi(\alpha \beta) \,.
\end{equation}         
With this product, $\psi({\cal A})$ is an algebra $(\psi({\cal A})\,,
*)$ isomorphic to ${\cal A}/ Ker \,\psi$. (The notation means that
$\psi({\cal A})$ is considered with product $*$ and not say point-wise
product).

We assume that ${\cal A}/ Ker \, \psi$ is a $*$-algebra and that 
$\Psi$ preserves the stars on ${\cal A}/ Ker \, \psi$ and 
$ C^\infty({\cal M})$, the $*$ on the latter being complex conjugation
denoted by bar:
\begin{eqnarray} 
\Psi \big ( (\alpha + Ker \, \psi)^* \big ) &=& 
\overline{\Psi(\alpha + Ker \, \psi)} \,, \nonumber \\    
\psi(\alpha^*) &=& \overline{\psi(\alpha)} \,.
\end{eqnarray} 
Such $\psi$ and $\Psi$ are said to be $*$-morphisms from ${\cal A}$
and ${\cal A}/Ker \, \psi$ to $(\psi({\cal A})\,, *)$. The two
algebras ${\cal A}/ Ker \, \psi$ and $(\psi({\cal A})\,, *)$ are
$*$-isomorphic.

\vskip 1em

{\it Remark:} Star ($*$) occurs with two meanings.
\begin{enumerate}
\item It refers to involution on algebras in the phrase $*$-morphism.
\item It refers to the new product on functions in  $(\psi({\cal A})\,, *)$.
\end{enumerate}
These confusing notations, designed to keep the reader alert, are
standard in the literature.

The above is the general framework. In applications, we encounter more 
than one linear bijection (one-to-one, onto map) from an a
algebra ${\cal A}$ to $ C^\infty({\cal M})$ and that produces 
different-looking $*$'s on $C^\infty({\cal M})$ and algebras
$(C^\infty({\cal M}) \,, *)$, $(C^\infty({\cal M}) \,, *^\prime)$
etc. As they are $*$-isomorphic to ${\cal A}$, they are mutually
$*$-isomorphic as well. A simple example we encounter below is 
$C^\infty({\mathbb C})$ with Moyal- and coherent-state-induced
$*$-products. These algebras are $*$-isomorphic.

\section{Properties of Coherent States} 
 
It is useful to have the Campbell-Baker-Hausdorff (CBH) formula
written down. It reads
\begin{equation}  
e^{\hat A} e^{\hat B} = e^{{\hat A} +{\hat B}} e^{\frac{1}{2} \lbrack
{\hat A} \,, {\hat B} \rbrack}
\end{equation} 
for two operators ${\hat A}\,, {\hat B}$ if
\begin{equation}  
\lbrack {\hat A} \,, \lbrack {\hat A} \,, {\hat B} \rbrack \rbrack =
\lbrack {\hat B} \,, \lbrack {\hat A} \,, {\hat B} \rbrack \rbrack = 0
\,.
\end{equation} 

For one oscillator with annihilation-creation operators
$a$,$a^\dagger$, the coherent state
\begin{equation} 
|z \rangle = e^{z a^\dagger - {\bar z} a} |0 \rangle =
 e^{-\frac{1}{2}|z|^2} e^{z a^\dagger} |0 \rangle \,, \quad z \in
 {\mathbb C}
\end{equation} 
has the properties
\begin{equation} 
a |z \rangle = z |z \rangle \,; \quad \quad \langle z^\prime | z
\rangle = e^{\frac{1}{2} |z - z^\prime|^2} \,.
\label{eq:csp1}
\end{equation} 

The coherent states are overcomplete, with the resolution of identity
\begin{equation} 
{\bf 1} = \int \frac{d^2 z}{\pi} |z \rangle \langle z|  \,, \quad d^2 z
= dx_1 dx_2 \,,\quad \mbox{where} \quad \, z=\frac{x_1 + i x_2}{\sqrt2} \,.
\label{eq:overcomp}
\end{equation} 
The factor $\frac{1}{\pi}$ is easily checked: $Tr \, {\bf 1} |0
\rangle \langle 0| = 1$ while $\int d^2z | \langle 0 | z \rangle|^2$
is $\pi$ in view of (\ref{eq:csp1}).

A central property of coherent states is the following: an operator
${\hat A}$ is determined just by its diagonal matrix elements
\begin{equation} 
A ( z \,, {\bar z}) = \langle z| {\hat A} | z \rangle \,,
\end{equation} 
that is by its ``symbol'' $A$, a function on ${\mathbb C}$ with values 
$A ( z \,, {\bar z}) = \langle z| {\hat A} | z \rangle$
\footnote{The $\bar{z}$ argument in $A(z \,, {\bar z})$ 
is redundant. It is there to emphasize that $A$ is 
not necessarily a holomorphic function of the complex variable $z$.}.
An easy proof uses analyticity \cite{perelomov}. ${\hat A}$ is certainly determined by 
the collection of all its matrix elements 
$\langle {\bar \eta} | {\hat A} | \xi \rangle$ or equally by
\begin{equation}      
e^{\frac{1}{2}(|\eta|^2+|\xi|^2)} \langle {\bar \eta}| {\hat A}| \xi
\rangle = \langle 0 | e^{\eta a} {\hat A} e^{\xi a^\dagger} | 0
\rangle \,.
\end{equation} 
The right hand side (at least for appropriate ${\cal A}$) is seen to
be a holomorphic function of $\eta$ and $\xi$, or equally well of
\begin{equation} 
u = \frac{\eta +\xi}{2} \,, \quad v = \frac{\eta - \xi}{2i} \,.
\end{equation} 
Holomorphic functions are globally determined by their values for real
arguments. Hence the function ${\tilde A}$ defined by
\begin{equation} 
{\tilde A}(u,v) = \langle 0| e^{\eta a^\dagger} {\hat A} e^{\xi
  a^\dagger} | 0 \rangle 
\end{equation} 
is globally determined by its values for $u,v$ real or ${\eta} = {\bar
\xi}$. Thus $\langle \xi | {\hat A} | \xi \rangle$ determines ${\hat
A}$ as claimed.

There are also explicit formulas for ${\hat A}$ in terms of $\langle
\xi | {\bar A} | \xi \rangle$\cite{GarnikAlSasha}.

\section{The Coherent State or Voros $*$-product on the Moyal Plane}

As indicated above, we can map an operator $\hat A$ to a function $A$
using coherent states as follows:
\begin{equation} 
{\hat A} \longrightarrow A \,, \quad A(z \,, {\bar z}) = \langle z | {\hat
  A} |z \rangle. 
\end{equation} 
This map is linear and also bijective by the previous remarks and
induces a product $*_C$ on functions ($C$ indicating ``coherent
state''). With this product, we get an algebra $(C^\infty({\mathbb C})
\,, *_C)$ of functions. Since the map ${\hat A} \rightarrow A$ has the
property ${\hat A}^* \rightarrow A^* \equiv {\bar A}$, this map is a $*$-morphism from
operators to $(C^\infty({\mathbb C}) \,, *_C)$.

Let us get familiar with this new function algebra. 

The image of $a$ is the function $\alpha$ where $\alpha(z\,,{\bar z})
=z$. The image of $a^n$ has the value $z^n$ at $(z \,, {\bar z})$, so by definition,
\begin{equation} 
\alpha *_C \alpha \ldots *_C \alpha (z \,, {\bar z}) = z^n \,.
\end{equation} 

The image of $a^* \equiv a^\dagger$ is ${\bar \alpha}$ where ${\bar
\alpha}(z, {\bar z}) = {\bar z}$ and that of $(a^*)^n$ is ${\bar
\alpha} *_C {\bar \alpha} \cdots *_C {\bar \alpha}$ where
\begin{equation} 
{\bar \alpha} *_C {\bar \alpha} \cdots *_C {\bar \alpha}(z\,, {\bar
  z}) = {\bar z}^n \,. 
\end{equation} 

Since $\langle z | a^* a | z \rangle = {\bar z} z$ and $\langle z | a
a^* | z \rangle = {\bar z} z + 1$, we get
\begin{equation} 
{\bar \alpha} *_C \alpha = {\bar \alpha} \alpha \,, \quad \quad
\alpha *_C {\bar \alpha} = \alpha {\bar \alpha} + {\bf 1} \,, 
\end{equation}     
where $ {\bar \alpha} \alpha =  \alpha {\bar \alpha}$ is the pointwise
product of $\alpha$ and ${\bar \alpha}$, and ${\bf 1}$ is the constant
function with value $1$ for all $z$. 

For general operators ${\hat f}$, the construction proceeds as
follows. Consider 
\begin{equation} 
: e^{\xi a^\dagger - {\bar \xi} a}:
\end{equation} 
where the normal ordering symbol $: \cdots :$ means as usual that
$a^\dagger$'s are to be put to the left of $a$'s. Thus  
\begin{eqnarray} 
: a a^\dagger a^\dagger a : &=& a^\dagger a^\dagger a a \,, \nonumber \\
: e^{\xi a^\dagger - {\bar \xi} a}: &=& e^{\xi a^\dagger} e^{-{\bar
      \xi} a} \,. 
\end{eqnarray} 
Hence
\begin{equation} 
\langle z | :e^{\xi a^\dagger - {\bar \xi} a}: |z \rangle = e^{\xi
  {\bar z} - {\bar \xi} z} \,. 
\end{equation}
 
Writing ${\hat f}$ as a Fourier transform,
\begin{equation} 
{\hat f} = \int \frac{d^2 \xi}{\pi} : e^{\xi a^\dagger - {\bar \xi}
  a}: {\tilde f}(\xi \,, {\bar \xi}) \,, \quad \quad {\tilde f} 
(\xi \,, {\bar \xi}) \in {\mathbb C} \,,
\end{equation} 
its symbol is seen to be
\begin{equation} 
f = \int \frac{d^2 \xi}{\pi} e^{\xi {\bar z} - {\bar \xi} z} {\tilde
  f}(\xi \,, {\bar \xi}) \,. 
\end{equation}     
This map is invertible since $f$ determines ${\tilde f}$. 

Consider also the second operator
\begin{equation} 
{\hat g} = \int \frac{d^2 \eta}{\pi} : e^{\eta a^\dagger - {\bar \eta}
  a}: {\tilde g}(\eta \,, \bar {\eta}) \,, 
\end{equation} 
and its symbol
\begin{equation} 
g = \int \frac{d^2 \eta}{\pi} e^{\eta {\bar z} - {\bar \eta} z}
{\tilde g}(\eta \,, \bar {\eta}) \,. 
\end{equation} 
The task is to find the symbol $f *_C g$ of ${\hat f}{\hat g}$.

Let us first find 
\begin{equation} 
e^{\xi {\bar z} - {\bar \xi} z} *_C  e^{\eta {\bar z} - {\bar \eta} z} \,.
\end{equation} 
We have
\begin{equation} 
:e^{\xi a^\dagger - {\bar \xi} a}: \,  : e^{\eta a^\dagger - {\bar
      \eta} a}: = : e^{\xi a^\dagger - {\bar \xi} a} \,  
e^{\eta a^\dagger - {\bar \eta} a}: e^{-{\bar \xi}{\eta}}
\end{equation} 
and hence
\begin{eqnarray} 
e^{\xi {\bar z} - {\bar \xi} z} *_C e^{\eta {\bar z} - {\bar \eta} z}
&=& e^{-{\bar \xi} \eta} e^{\xi {\bar z} - {\bar \xi} z} \,  
e^{\eta {\bar z} - {\bar \eta} z} \nonumber \\
&=& e^{\xi {\bar z} - {\bar \xi} z} e^{{\overleftarrow \partial}_z \,
  {\overrightarrow \partial}_{\bar z}} 
e^{\eta {\bar z} - {\bar \eta} z} \,.
\label{eq:exp1}
\end{eqnarray} 
The bidifferential operators $\big ({\overleftarrow \partial}_z \,
{\overrightarrow \partial}_{\bar z} \big )^k \,, (k= 1,2,...)$ have
the definition
\begin{equation}  
\alpha \big ({\overleftarrow \partial}_z \, {\overrightarrow
  \partial}_{\bar z} \big )^k \beta \, (z \,, {\bar z}) =  
\frac{\partial^k \alpha (z \,, {\bar z})}{\partial z^k}
\frac{\partial^k \beta (z \,, {\bar z})}{\partial {\bar z}^k} \,.
\end{equation} 
The exponential in (\ref{eq:exp1}) involving them can be defined
using the power series.

$f *_C g$ follows from (\ref{eq:exp1}):
\begin{equation} 
f *_C g \,(z \,, {\bar z}) = \big ( f  e^{{\overleftarrow \partial}_z
  \, {\overrightarrow \partial}_{\bar z}} g \big ) (z \,, {\bar z})  \,.    
\label{eq:csstar1}
\end{equation} 
(\ref{eq:csstar1}) is the coherent state $*$-product \cite{Voros}

We can explicitly introduce a deformation parameter $\theta > 0 $ in
the discussion by changing (\ref{eq:csstar1}) to
\begin{equation} 
f *_C g \, (z \,, {\bar z}) = \big ( f  e^{ \theta \, {\overleftarrow
    \partial}_z \, {\overrightarrow \partial}_{\bar z}} g \big )  
(z \,, {\bar z}) \,.   
\label{eq:csstar2}
\end{equation}  
After rescaling $z^\prime = \frac{z}{\sqrt{\theta}}$,
(\ref{eq:csstar2}) gives (\ref{eq:csstar1}). As $z^\prime$ and ${\bar
z}^\prime$ after quantization become $a \,, a^\dagger$, $z$ and ${\bar
z}$ become the scaled oscillators $a_\theta \,, a_\theta^\dagger$:
\begin{equation} 
\lbrack a_\theta \,, a_\theta \rbrack = \lbrack a_\theta^\dagger  \,,
a_\theta^\dagger \rbrack = 0 \,, \quad  \lbrack a_\theta \,,
a_\theta^\dagger \rbrack = \theta \,. 
\label{eq:tetacom}
\end{equation} 
(\ref{eq:tetacom}) is associated with the Moyal plane with Cartesian
coordiante functions $x_1 \,, x_2$. 
If $a_\theta = \frac{x_1 + i x_2}{\sqrt2} \,, 
a_\theta^\dagger = \frac{x_1 - i x_2}{\sqrt2}$,
\begin{equation}  
\lbrack x_i \,, x_j \rbrack = i \theta \varepsilon_{ij} \,, \quad
\varepsilon_{ij} = - \varepsilon_{ji} \,, \quad \varepsilon_{12} = 1 \,.
\label{eq:deform1}
\end{equation} 

The Moyal plane is the plane ${\mathbb R}^2$, but with its function
algebra deformed in accordance with (\ref{eq:deform1}). The deformed
algebra has the product (\ref{eq:csstar2}) or equivalently the Moyal
product derived below.

\section{The Moyal Product on the Moyal Plane}

We get this by changing the map ${\hat f} \rightarrow f$ from
operators to functions. For a given function $f$, the operator ${\hat
f}$ is thus different for the coherent state and Moyal $*$'s. The
$*$-product on two functions is accordingly also different.

\subsection{The Weyl Map and the Weyl Symbol}

The Weyl map of the operator
\begin{equation}  
{\hat f} = \int \frac{d^2 \xi}{\pi} {\tilde f}(\xi \,, {\bar \xi})
e^{\xi a^\dagger - {\bar \xi} a} \,, 
\label{eq:Weyl1}
\end{equation} 
to the function $f$ is defined by 
\begin{equation} 
f(z\,,{\bar z}) = \int \frac{d^2 \xi}{\pi} {\tilde f}(\xi \,, 
{\bar \xi}) e^{\xi {\bar z} - {\bar \xi} z} \,. 
\label{eq:Weyl2}
\end{equation} 
(\ref{eq:Weyl2}) makes sense since ${\tilde f}$ is fully determined by
${\hat f}$ as follows: 
\begin{equation} 
\langle z| {\hat f} | z \rangle = \int \frac{d^2 \xi}{\pi} {\tilde
  f}(\xi \,, \bar {\xi}) e^{-\frac{1}{2} \xi {\bar \xi} } 
e^{\xi {\bar z} - {\bar \xi} z} \,.       
\end{equation} 
${\tilde f}$ can be calculated from here by Fourier transformation.

The map is invertible since ${\tilde f}$ follows from $f$ by Fourier
transform of (\ref{eq:Weyl2}) and ${\tilde f}$ fixes ${\hat f}$ by
(\ref{eq:Weyl1}). $f$ is called the {\it Weyl symbol} of ${\hat f}$.

As the Weyl map is bijective, we can find a new $*$ product, call it
$*_W$, between functions by setting $f*_W g = \quad \mbox{Weyl Symbol of} \quad
{\hat f}{\hat g}$.

For
\begin{equation} 
{\hat f} =  e^{\xi a^\dagger - {\bar \xi} a} \,, \quad {\hat g} =
e^{\eta a^\dagger - {\bar \eta} a} \,, 
\end{equation} 
to find $f*_W g$, we first rewrite ${\hat f}{\hat g}$ according to
\begin{equation} 
{\hat f}{\hat g} = e^{\frac{1}{2}(\xi {\bar \eta} - {\bar \xi} \eta)}  e^{(\xi +
  \eta) a^\dagger - ({\bar \xi} +{\bar \eta}) a} \,.  
\end{equation} 
Hence
\begin{eqnarray} 
f*_W g \,(z\,, {\bar z}) &=& e^{\xi {\bar z}-{\bar \xi} z}  
e^{\frac{1}{2}(\xi {\bar \eta} - {\bar \xi} \eta)} 
e^{\eta {\bar z}-{\bar \eta} z}
\nonumber \\     
&=& f e^{\frac{1}{2} \big ( {\overleftarrow \partial}_z \,
{\overrightarrow \partial}_{\bar z} - {\overleftarrow \partial}_{\bar
  z} \, {\overrightarrow \partial}_z
\big )} g \, (z \,,{\bar z}) \,. 
\label{eq:Weyl3}
\end{eqnarray} 
Multiplying by ${\tilde f}(\xi \,, {\bar \xi})$, ${\tilde g}(\eta \,,
{\bar \eta})$ and integrating, we get (\ref{eq:Weyl3}) for arbitrary
functions:
\begin{equation} 
f*_W g \, (z\,, {\bar z}) = \Big ( f e^{\frac{1}{2} \big (
  {\overleftarrow \partial}_z \, {\overrightarrow \partial}_{\bar z} - 
{\overleftarrow \partial}_{\bar z} \, {\overrightarrow \partial}_z
\big )} g \Big ) (z \,,{\bar z}) \,. 
\end{equation} 
Note that 
\begin{equation} 
{\overleftarrow \partial}_z \, {\overrightarrow \partial}_{\bar z}
-{\overleftarrow \partial}_{\bar z} \, {\overrightarrow \partial}_z 
= i ( {\overleftarrow \partial}_1 \, {\overrightarrow \partial}_2
-{\overleftarrow \partial}_2 \, {\overrightarrow \partial}_1 ) 
= i \varepsilon_{ij}  {\overleftarrow \partial}_i \, {\overrightarrow
  \partial}_j  \,. 
\end{equation} 

Introducing also $\theta$, we can write the $*_W$-product as
\begin{equation} 
f *_W g = f e^{i \frac{\theta}{2} \varepsilon_{ij}  {\overleftarrow \partial}_i
  \, {\overrightarrow \partial}_j} g \,. 
\label{eq:Weyl4}
\end{equation} 
By (\ref{eq:deform1}), $\theta \varepsilon_{ij} = \omega_{ij}$ fixes
the Poisson brackets, or the Poisson structure on the Moyal
plane.(\ref{eq:Weyl4}) 
is customarily written as
\begin{equation} 
f *_W g = f  e^{\frac{i}{2} \omega_{ij}  {\overleftarrow \partial}_i \,
  {\overrightarrow \partial}_j} g \, .
\end{equation} 
using the Poisson structure. (But we have not cared to position
the indices so as to indicate their tensor nature and to write  
$\omega^{ij}$.)

\section{Properties of $*$-Products}

A $*$-product without a subscript indicates that it can be either a $*_C$ or a $*_W$.

\subsection{Cyclic Invariance}

The trace of operators has the fundamental property
\begin{equation} 
Tr {\hat A} {\hat B} = Tr {\hat B} {\hat A}
\end{equation} 
which leads to the general cyclic identities
\begin{equation} 
Tr \, {\hat A}_1 \ldots {\hat A}_n = Tr \, {\hat A}_n {\hat A}_1
\ldots {\hat A}_{n-1} \,. 
\label{eq:ctr1}
\end{equation} 
We now show that
\begin{equation} 
Tr \, {\hat A} {\hat B} = \int \frac{d^2 z}{\pi} \,  A * B \, (z \,,
{\bar z}) \,, \quad \quad * = *_C \quad \mbox{or} \quad *_W \,. 
\label{eq:ctr2}
\end{equation}   
(The functions on R.H.S. are different for $*_C$ and $*_W$ if ${\hat
  A} \,, {\hat B}$ are fixed). From this follows the analogue 
of (\ref{eq:ctr1}):
\begin{equation} 
\int \frac{d^2 z}{\pi} \, \big ( A_1 * A_2 * \cdots * A_n) \, (z \,, {\bar
  z} \big ) = \int \frac{d^2 z}{\pi} \big ( A_n * A_1 * \cdots * 
A_{n-1}) \, (z \,, {\bar z} \big ) \,. 
\label{eq:ctr3}
\end{equation} 

For $*_C$, (\ref{eq:ctr2}) follows from (\ref{eq:overcomp}).

The coherent state image of $ e^{\xi a^\dagger - {\bar \xi} a} $ is
the function with value
\begin{equation} 
e^{\xi {\bar z} - {\bar \xi} z} e^{-\frac{1}{2}{\bar \xi}{\xi}}
\label{eq:csf1}
\end{equation} 
at $z$, with a similar correspondence if $\xi \rightarrow \eta$. So
\begin{equation} 
Tr  \, e^{\xi a^\dagger - {\bar \xi} a} \, e^{\eta a^\dagger - {\bar
    \eta} a} = \int {\frac{d^2 z}{\pi}} \, \Big ( 
e^{\xi {\bar z} - {\bar \xi} z} e^{-\frac{1}{2}{\bar \xi}{\xi}} \Big)
    \Big ( e^{\eta {\bar z} - {\bar \eta} z}  
e^{-\frac{1}{2}{\bar \eta}{\eta}} \Big ) e^{-{\bar \xi}{\eta}}
\end{equation} 
The integral produces the $\delta$-function
\begin{equation} 
\prod_i 2 \delta (\xi_i + \eta_i) \,, \quad \quad  \xi_i = \frac{\xi_1 +
  \xi_2}{\sqrt{2}} \,, \quad \eta_i = \frac{\eta_1 + \eta_2} {\sqrt{2}} \,.
\end{equation} 
We can hence substitute $ e^{- \big ( \frac{1}{2}{\bar \xi}{\xi} +
  \frac{1}{2}{\bar \eta}{\eta} + {\bar \xi}{\eta} \big)}$ by 
$e^{\frac{1}{2} (\xi {\bar \eta} - {\bar \xi} \eta)}$ and get
(\ref{eq:ctr2}) for Weyl $*$ for these exponentials and so for general 
functions by using  (\ref{eq:Weyl1}).

\subsection{A Special Identity for the Weyl Star}

The above calculation also gives, the identity 
\begin{equation}  
\int \frac{d^2 z}{\pi} A *_W B \, (z \,, {\bar z}) = \int \frac{d^2
  z}{\pi} A (z \,, {\bar z}) \, B \, (z \,, {\bar z}) \,.  
\end{equation}  
That is because 
\begin{equation} 
\prod_i \delta(\xi_i + \eta_i) \, e^{\frac{1}{2} (\xi {\bar \eta} -
{\bar \xi} \eta)} = \prod_i \, \delta(\xi_i + \eta_i) \,.  
\end{equation}  
In (\ref{eq:ctr3}), $A$ and $B$ in turn can be Weyl $*$-products of
other functions. Thus in integrals of Weyl $*$-products of functions,
one $*_W$ can be replaced by the pointwise (commutative) product:
\begin{eqnarray} 
&&\int \frac{d^2 z}{\pi} \big ( A_1 *_W A_2 *_W \cdots A_K \big ) *_W
  ( B_1 *_W B_2 *_W \cdots B_L \big ) \, (z \,, {\bar z})  
\nonumber \\
&& \quad \quad \quad \quad = \int \frac{d^2 z}{\pi} \big ( A_1 *_W A_2
*_W \cdots A_K \big ) \,( B_1 *_W B_2 *_W  \cdots B_L \big ) \, (z \,,
{\bar z}) \,.  
\end{eqnarray} 
This identity is frequently useful.

\subsection{Equivalence of $*_C$ and $*_W$}

For the operator
\begin{equation} 
{\hat A} = e^{\xi a^\dagger -{\bar \xi} a} \,,
\end{equation} 
the coherent state function $A_C$ has the value (\ref{eq:csf1}) at
$z$, and the Weyl symbol $A_W$ has the value
\begin{equation}  
A_W(z \,, {\bar z}) = e^{\xi {\bar z} - {\bar \xi} z} \,.
\end{equation} 

As both $\big ( C^\infty({\mathbb R}^2) \,, *_C \big )$ and $\big (
C^\infty({\mathbb R}^2) \,, *_W \big )$ are isomorphic to the
operator algebra, they too are isomorphic. The isomorphism is
established by the maps 
\be
A_C \longleftrightarrow A_W
\ee
and their extension via Fourier transform to all operators and 
functions ${\hat A} \,, A_{C \,, W}$.

Clearly
\begin{gather} 
A_W = e^{-\frac{1}{2} \partial_z \partial_{\bar z}} A_C \,, \quad 
A_C = e^{\frac{1}{2} \partial_z \partial_{\bar z}} A_W  \,, \nonumber \\
A_C *_C B_C \longleftrightarrow A_W *_W B_W \,.  
\end{gather}   

The mutual isomorphism of these three algebras is a $*$-isomorphism since 
$({\hat A} {\hat B})^\dagger \longrightarrow {\bar B}_{C \,, W} *_{C \,, W} {\bar A}_{C \,, W}$. 

\subsection{Integration and Tracial States}

This is a good point to introduce the ideas of a state and a tracial
state on a $*$-algebra ${\cal A}$ with unity ${\bf 1}$.

A state $\omega$ is a linear map from ${\cal A}$ to ${\mathbb C}$,
$\omega (a) \in {\mathbb C}$ for all $a \in {\cal A}$ with the
following properties:
\begin{eqnarray} 
\omega(a^*) &=& \overline{\omega(a)} \,, \nonumber \\  
\omega (a^*a) & \geq & 0 \,, \nonumber \\
\omega({\bf 1}) & = & 1 \,.
\label{eq:ts1}
\end{eqnarray} 

If ${\cal A}$ consists of operators on a Hilbert space and $\rho$ is a
density matrix, it defines a state $\omega_\rho$ via 
\begin{equation}  
\omega_\rho (a) =  Tr (\rho a) \,.
\label{eq:ts2}
\end{equation} 

If $\rho = e^{- \beta H}/ Tr (e^{-\beta H})$ for a Hamiltonian
$H$, it gives a Gibbs state via (\ref{eq:ts2}).

Thus the concept of a state on an algebra ${\cal A}$ generalizes the
notion of a density matrix. There is a remarkable construction, the
Gel'fand- Naimark-Segal (GNS) construction which shows how to
associate any state with a rank-$1$ density matrix \cite{Haag}. 
%We consider this construction in chapter {\bf [....].}.

A state is {\it tracial} if it has cyclic invariance \cite{Haag}:
\begin{equation} 
\omega (ab) = \omega (ba) \,.
\label{eq:tracial1}
\end{equation} 
The Gibbs state is not tracial, but fulfills an identity generalizing
(\ref{eq:tracial1}). It is a Kubo-Martin-Schwinger (KMS) state \cite{Haag}. 

A positive map $\omega^\prime$ is in general an unnormalized state: It must
fulfill all the conditions that a state fulfills, but is not obliged
to fulfill the condition $\omega^\prime({\bf 1}) = 1$.

Let us define a positive map $\omega^\prime$ on $(C^\infty(\mathbb R^2) \,, *)$ ($*
= *_C \, \mbox{or} \, *_W$) using integration:
\begin{equation} 
\omega^\prime(A) = \int \frac{d^2 z}{\pi} \, {\hat A}
(z \,, {\bar z}) \,.
\end{equation}
It is easy to verfy that $\omega^\prime$ fulfills the properties of a positive map.
 
A {\it tracial} positive map $\omega^\prime$ also has the cyclic
invariance (\ref{eq:tracial1}).

The cyclic invariance (\ref{eq:tracial1}) of $\omega^\prime (A * B)$ means that
it is a tracial positive map. 

\subsection{The $\theta$-Expansion}

On introducing $\theta$, we have (\ref{eq:csstar2}) and 
\begin{equation} 
f *_W g (z \,, {\bar z}) =  f e^{\frac{\theta}{2} \big (
  {\overleftarrow \partial}_z \, {\overrightarrow \partial}_{\bar z} - 
{\overleftarrow \partial}_{\bar z} \, {\overrightarrow \partial}_z
  \big )} g \, (z \,,{\bar z}) \,. 
\end{equation}   
The series expansion in $\theta$ is thus 
\begin{equation} 
f *_C g \, (z \,, {\bar z}) = f g \, (z \,, {\bar z}) + \theta \,
\frac{\partial f}{\partial z} (z \,, {\bar z}) \frac{\partial g} 
{\partial {\bar z}} (z \,, {\bar z}) + {\cal O} (\theta^2) \,,
\end{equation} 
\begin{equation} 
f *_W g \, (z \,, {\bar z}) =  f g (z \,, {\bar z}) + \frac{\theta}{2}
\Big ( \frac{\partial f}{\partial z} \frac{\partial g} 
{\partial {\bar z}} - \frac{\partial f}{\partial {\bar z}}
\frac{\partial g} {\partial z} \Big ) \, (z \,, {\bar z}) +  {\cal O} (\theta^2) \,.
\end{equation} 

Introducing the notation 
\begin{equation} 
\lbrack f \,, g \rbrack_* = f * g - g * f \,, \quad *=*C \quad
\mbox{or} \quad *_W \,, 
\label{eq:ps1}
\end{equation} 
We see that
\begin{eqnarray} 
\lbrack f \,, g \rbrack_{*_C} &=& \theta \Big ( \frac{\partial
  f}{\partial z} \frac{\partial g} {\partial {\bar z}} -  
\frac{\partial f}{\partial {\bar z}} \frac{\partial g} {\partial z}
\Big ) (z \,, {\bar z}) +  {\cal O} (\theta^2) \,, \nonumber \\ 
\lbrack f \,, g \rbrack_{*_W} &=& \theta \Big ( \frac{\partial
  f}{\partial z} \frac{\partial g} {\partial {\bar z}} -  
\frac{\partial f}{\partial {\bar z}} \frac{\partial g} {\partial z}
\Big ) (z \,, {\bar z}) +  {\cal O} (\theta^2) \,,       
\end{eqnarray} 
We thus see that
\begin{equation} 
\lbrack f \,, g \rbrack_* = i \theta \{f \,, g \}_{P.B.} + {\cal O} (\theta^2) \,,
\label{eq:ps2}
\end{equation} 
where $\{f \,, g \}$ is the Poisson Bracket of $f$ and $g$ and the
${\cal O}(\theta^2)$ term depends on $*_{C \,, W}$. Thus the $*$-product is
an associative product which to leading order in the deformation parameter
(``Planck's'' constant) $\theta$ is compatible with the rules of
quantization of Dirac. We can say that with the $*$-product, we have
deformation quantization of the classical commutative algebra of
functions.

But it should be emphasized that even to leading order in $\theta$, $f
*_C g$ and $f *_W g$ do not agree. Still the algebras $\big (
C^\infty({\mathbb R}^2 \,, *_C) \big )$ and $\big ( C^\infty({\mathbb
R}^2 \,, *_W) \big )$ are $*$-isomorphic.

Suppose we are given a Poisson structure on a manifold $M$ with
Poisson bracket $\{. \,, .\}$. Then Kontsevich (\cite{kontsevich}) has given the
$*$-product $f * g$ as a formal power series in $\theta$ such that
(\ref{eq:ps2}) holds.

\section{The $*$-Product for the Fuzzy Sphere}

Star products for K\"{a}hler manifolds have been known for a long time. The
approach we take here was initiated by Pre\v{s}najder, it produces
particularly compact expressions.

Let $P_n$ be the orthogonal projection operator to the subspace with
$N=n$. The fuzzy sphere algebra is then the algebra with elements $P_n
\gamma (a_i^\dagger a_j) P_n$ where $\gamma$ is any polynomial in
$(a_i^\dagger a_j)$. As any such polynomial commutes with $N$, if
$\gamma$ and $\delta$ are two of these polynomials,
\be
P_n \gamma(a_i^\dagger a_j) P_n P_n \delta (a_i^\dagger a_j) P_n = 
P_n \gamma (a_i^\dagger a_j) \delta (a_i^\dagger a_j) P_n
\ee

This algebra, more precisely, is the orthogonal direct sum $Mat(n+1)
\oplus 0$ where $Mat(n+1)$ acts on the ${\widehat N}=n$ subspace and is the fuzzy
sphere. But the extra $0$ here is entirely harmless.

\subsection{The Coherent State $*$-Product $*_C$}

There are now two oscillators $a_1 \,, a_2$, so the coherent states
are labeled by two complex variables, being
\begin{equation} 
|Z_1 \,, Z_2 \rangle = e^{Z a^\dagger - {\bar Z} a} | 0 \rangle \,,
\quad  \quad Z = (Z_1 \,, Z_2) \,. 
\end{equation}     
We use capital $Z$'s for unnormalized $Z$'s and $z$'s for normalized
ones: $z = \frac{Z}{|Z|} \,, |Z|^2 = \sum |Z_i|^2$. 

The normalized coherent states $|z \rangle_n$ for $S_F^2$, as one can
guess, are obtained by projection from $| Z \rangle$, 
\begin{equation} 
|z \rangle_n = \frac{P_n | Z \rangle }{| \langle P_n | Z \rangle |}
 = \frac{\big ( \sum_i z_i a^\dagger_i \big )^n }{\sqrt{n !}} | 0 \rangle \,.
\end{equation} 
where we have used 
\begin{equation} 
P_n | Z \rangle =  \frac{( Z_i a^\dagger_i)^n}{n!} | 0 \rangle \,.
\end{equation} 
They are called Perelomov coherent states \cite{perelomov}

For an operator $P_n {\hat A} P_n$, the coherent state symbol has the
value 
\be
\langle Z | P_n {\hat A} P_n | Z \rangle = e^{-|z|^2} \frac{| z |^{2n}}{n!}
\langle z | {\hat A} | z \rangle_n
\ee
at $Z$. By a previous result, the
diagonal coherent state expectation values $\langle z | P_n {\hat A}
P_n | z \rangle_n$ determines $P_n {\hat A} P_n$ uniquely and there is
a $*$-product for $S_F^2$. We call it a $*_C$-product in analogy to
the notation used before.

We can find it explicitly as follows \cite{chiral, BDBJ, seckin1}. For
$n=1$ (spin $\frac{n}{2} = \frac{1}{2})$, a basis for $2 \times 2$
matrices is
\begin{equation} 
\big \lbrace \sigma_A : \sigma_0 = {\bf 1} \,, \sigma_i \quad (i =
1,2,3) \quad = \quad \mbox{Pauli Matrices} \,, \quad Tr \sigma_A  
\sigma_B = 2 \delta_{AB} \big \rbrace \,. 
\end{equation} 

Let
\be
| i \rangle = a_i^\dagger | 0 \rangle \,, \quad i = 1, 2
\ee
be an orthonormal vector for $n=1$. A general operator is  
\begin{equation} 
{\hat F} = f_A {\hat \sigma}_A \,, \quad {\hat \sigma}_A = a^\dagger
  \sigma_A a \big |_{n=1} \,, \quad f_A \in {\mathbb C} \,.
\end{equation} 
and $ {\hat \sigma}_A| i \rangle = | j \rangle (\sigma_A)_{ji}$.
In above by $a^\dagger \sigma_A a \big |_{n=1}$, we mean the restriction of 
$a^\dagger \sigma_A a$ to the subspace with $n=1$.  

Call the coherent state symbol of ${\hat \sigma}_A$ for $n=1$ as $\chi_A$:
\begin{equation} 
\chi_A (z) = \langle z| {\hat \sigma}_A |z \rangle \,, \quad \quad \chi_0 (z) =
1 \,, \quad \quad \chi_i = {\bar z} \sigma_i  z \,, \quad i =1,2,3 \,.
\end{equation} 
%The $\chi_i$ here are the same as those in section 2.4.

The $*$-product for $n=1$ now follows: 
\begin{equation} 
\chi_A *_C \chi_B (z) = \langle z| {\hat \sigma}_A {\hat \sigma}_B | z \rangle_1 \,. 
\end{equation} 
Write
\begin{equation} 
\sigma_A \sigma_B = \delta_{AB} + E_{ABi} \sigma_i 
\label{eq:chi}
\end{equation} 
to get
\begin{eqnarray} 
\chi_A *_C \chi_B (z) &=& \delta_{AB} +  E_{ABi} \chi_i (z) \nonumber \\
&:=& \chi_A (z) \chi_B(z) + {\cal K}_{AB} (z) \,.
\label{eq:chi1}
\end{eqnarray}
 
Let us use the notation
\begin{equation}  
n_i = \chi_i (z) \,, \quad \quad n_0 =1 \,.
\end{equation} 
$\vec{n}$ is the coordinate on $S^2$: $\vec{n} \cdot \vec{n} =1$. Then
(\ref{eq:chi1}) is
\begin{equation} 
n_A *_C n_B (z) = n_A n_B + K_{AB}(n) \,, \quad \quad {\cal K}_{AB}(z)
:= K_{AB}(n) \,. 
\end{equation}
 
This $K_{AB}$ has a particular significance for complex
analysis. Since $\chi_0 (z) = 1$, $\chi_0(z) * \chi_A = \chi_0 \chi_A$
by (\ref{eq:chi1}) and
\begin{equation} 
K_{0A} = 0 \,.
\end{equation} 
The components $K_{ij}(n)$ of $K$ can be calculated from
(\ref{eq:chi}), (\ref{eq:chi1}). Let $\theta(\alpha)$ be the spin $1$
angular momentum matrices:
\begin{equation} 
\theta(\alpha)_{ij} = -i \varepsilon_{\alpha i j} \,.
\end{equation} 
Then 
\begin{eqnarray} 
K_{ij}(\vec{n}) &=& \frac{\{ \vec{\theta} \cdot \vec{n} \, (
  \vec{\theta} \cdot \vec{n} - 1 ) \}_{ij}}{2} \nonumber \\  
\vec{\theta} \cdot \vec{n} &:=& \theta(\alpha) n_\alpha \,.
\end{eqnarray} 
The eigenvalues of $\vec{\theta} \cdot \vec{n}$ are $\pm 1 \,, 0$ and
$K_{ij}(\vec{n})$ is the projection operator to the eigenspace $\vec{\theta} \cdot \vec{n} = -1$,
\begin{equation} 
K(\vec{n})^2 = K(\vec{n}) \,.
\end{equation} 
It is related to the complex structure of $S^2$ in the projective module
picture treated in chapter 5.

The vector space for angular momentum $\frac{n}{2}$ is the $n$-fold
symmetric tensor product of the spin-$\frac{1}{2}$ vector spaces. The
general linear operator on this space can be written as
\begin{equation} 
{\widehat F} = f_{A_1 A_2 \cdots A_n} {\hat \sigma}_{A_1} \otimes {\hat \sigma}_{A_2}
\otimes \cdots {\hat \sigma}_{A_n}  
\end{equation} 
where $f$ is totally symmetric in its indices. Its symbol is thus
\begin{equation} 
F(\vec{n}) = f_{A_1 A_2 \cdots A_n} n_{A_1} \otimes n_{A_2} \otimes
\cdots n_{A_n} \,, \quad \quad n_0 : = 1 \,. 
\end{equation} 
The symbol of another operator
\begin{equation} 
{\widehat G} = g_{B_1 B_2 \cdots B_n} {\hat\sigma}_{B_1} \otimes {\hat
  \sigma}_{B_2} \otimes \cdots {\hat \sigma}_{B_n} \,,  
\end{equation} 
where $g$ is symmetric in its indices, is 
\begin{equation} 
G(\vec{n}) = g_{B_1 B_2 \cdots B_n} n_{B_1} \otimes n_{B_2} \otimes
\cdots n_{B_n} \,. 
\end{equation} 
Since
\begin{equation}  
{\widehat F}{\widehat G} =  f_{A_1 A_2 \cdots A_n} \,  g_{B_1 B_2 \cdots B_n}
\, \sigma_{A_1} \, \sigma_{B_1} \otimes \sigma_{A_2} \,   
\sigma_{B_2} \otimes \cdots \otimes  \sigma_{A_n} \, \sigma_{B_n} \,,
\end{equation} 
we have that 
\begin{equation} 
F * G (\vec{n}) =  f_{A_1 A_2 \cdots A_n} \,  g_{B_1 B_2 \cdots B_n}
\, \prod_i \big ( n_{A_i} \, n_{B_i} + K_{A_i B_i} \big)  
\end{equation} 
or
\begin{multline} 
F * G (\vec{n}) = F G (\vec{n}) + \sum_{m=1}^n \frac{n!}{m! (n-m)!}
f_{A_1 A_2 \cdots A_m A_{m+1} \cdots A_n}  n_{A_{m+1}} \,  
n_{A_{m+2}} \cdots n_{A_n} \, \\
\times K_{A_1 B_1} (\vec{n}) \,  K_{A_2 B_2}
(\vec{n}) \, \cdots  K_{A_m B_m} (\vec{n})         
g_{B_1 B_2 \cdots B_m B_{m+1} \cdots B_n} n_{B_{m+1}} \, n_{B_{m+2}} \cdots n_{B_n} \,. 
\end{multline} 
Now as $f$ and $g$ are symmetric in indices, there is the expression
\begin{equation} 
\partial_{A_1} \partial_{A_2} \cdots \partial_{A_m} F (\vec{n}) =
\frac{n!}{(n-m)!} f_{A_1 A_2 \cdots A_m A_{m+1} \cdots A_n}  
n_{A_{m+1}} \, n_{A_{m+2}} \cdots n_{A_n}
\label{eq:diff1}
\end{equation} 
for $F$ and a similar expression for $G$. Hence 
\begin{multline} 
F *_C G (\vec{n}) = \sum_{m=0}^n  \frac{(n-m)!}{m! n!} \big (
\partial_{A_1} \partial_{A_2} \cdots \partial_{A_m} F \big ) \,  
(\vec{n}) \\
\times K_{A_1 B_1} (\vec{n}) \,  K_{A_2 B_2} (\vec{n})
\, \cdots  K_{A_m B_m} (\vec{n}) \, \big 
( \partial_{B_1} \partial_{B_2} \cdots \partial_{B_m} G \big ) \, (\vec{n}) \,.
\label{eq:spheres1}
\end{multline} 
which is the final answer. Here the $m=0$ terms is to be understood as
$FG(\vec{n})$, the pointwise product of $F$ and $G$ evaluated at
$\vec{n}$. This formula was first given in \cite{chiral}. It was derived by
a similar method.

Differentiating on $n_A$ ignoring the constraint $\vec{n}
\cdot \vec{n} = 1$ is justified in the final answer
(\ref{eq:spheres1}) (although not in (\ref{eq:diff1}), since
$K_{AB}(\vec{n}) \partial_A (\vec{n} \cdot \vec{n}) = K_{AB}(\vec{n})
\partial_B (\vec{n} \cdot \vec{n}) = 0$.  (\ref{eq:diff1}) being only
an intermediate step on the way to (\ref{eq:spheres1}), this sloppiness
is immaterial.

For large $n$, (\ref{eq:spheres1}) is an expansion in powers of
$\frac{1}{n}$, the leading term giving the commutative product. Thus
the algebra $S_F^2$ is in some sense a deformation of the commutative
algebra of functions $C^\infty(S^2)$. But as the maximum angular
momentum in $F$ and $G$ is $n$, we get only the spherical harmonics
$Y_{\ell m} \,, \ell \in \{0,1, \cdots \,n \}$ in their expansion.
For this reason, $F$ and $G$ span a finite-dimensional subspace of
$C^\infty(S^2)$ and $S_F^2$ is not properly a deformation of the
commutative algebra $C^\infty(S^2)$.

\subsection{The Weyl $*$-Product $*_W$}

The Weyl $*$-products are characterized by the special identity
described before. For this reason they are very convenient for use in
loop expansions in quantum field theory (see chapter 4).

A simple way to find $*_M$ is to find it via its connection to
$*_C$. For this purpose let us consider   
\begin{equation} 
Tr ({\hat T}^\ell_m)^\dagger {\hat T}_{m^\prime}^{\ell^\prime} = 
\frac{n+1}{4 \pi} \int d \Omega \big \lbrack T_n(\ell)^{\frac{1}{2}} 
\overline{Y}_{\ell m} \big \rbrack *_C \big \lbrack 
T_n(\ell^\prime)^{\frac{1}{2}} Y_{\ell^\prime m^\prime} \big  
\rbrack (\vec{x}) \,,
\label{eq:weylfs}
\end{equation} 
where 
\begin{equation} 
\langle z ,n | T^\ell_m | z , n \rangle = 
T_n(\ell)^{\frac{1}{2}} Y_{\ell m} ({\hat n}) \,. 
\label{eq:tlm}
\end{equation} 
The factor $T_n(\ell)^{\frac{1}{2}}$ is independent of $m$ by 
rotational invariance. It is real as shown by complex conjugating
(\ref{eq:tlm}) and using
\begin{equation} 
(T^\ell_m)^\dagger = (-1)^m T_{-m}^\ell \,, \quad {\bar Y}_{\ell
    m}({\hat n}) 
= (-1)^m Y_{\ell \,, - m} (\vec{n}) \,.
\end{equation} 
It can be chosen to be positive as well. We shall evaluate it later.

The normalization of $T_m^\ell$ and $Y_{\ell m}$ are
\begin{equation} 
Tr (T_m^\ell)^\dagger T_{m^\prime}^{\ell^\prime} = \int d \Omega 
\overline{Y}_{\ell m}(\vec{x}) Y_{\ell^\prime m^\prime}
(\vec{x}) = \delta_{\ell \ell^\prime} \delta_{m m^\prime} \,.
\end{equation} 
Hence using (\ref{eq:weylfs})  
\begin{equation} 
\delta_{\ell \ell^\prime} \delta_{m m^\prime} = \int d \Omega 
\overline{Y}_{\ell m}(\vec{x}) Y_{\ell^\prime m^\prime}
(\vec{x}) = \frac{n+1}{4 \pi} \int d \Omega \big
(T_n(\ell)^{\frac{1}{2}} 
\overline{Y}_{\ell m} \big) (\vec{x}) *_C 
\big(T_n(\ell^\prime)^{\frac{1}{2}} 
Y_{\ell^\prime m^\prime} \big ) (\vec{x}) \,. 
\label{eq:ws1}
\end{equation} 
Equation (\ref{eq:ws1}) suggests that the fuzzy sphere algebra 
$(S_F^2 \,, *_M)$ with the Weyl-Moyal product $*_M$ is obtained from 
the fuzzy sphere algebra $(S_F^2 \,, *_C)$ with the coherent state 
$*_C$ product from the map
\begin{gather} 
\chi :  (S_F^2 \,, *_C) \longrightarrow  (S_F^2 \,, *_W) \nonumber \\ 
\chi \left (\sqrt{\frac{n+1}{4 \pi}} T_n(\ell)^{\frac{1}{2}} Y_{\ell m} \right) = Y_{\ell m} 
\label{eq:cwmap1}
\end{gather} 
The induced $*$, call it for a moment as $*^\prime$, on the image of $\chi$ is 
\begin{equation} 
Y_{\ell m} *^\prime Y_{\ell^\prime m^\prime} = \chi \left (
\sqrt{\frac{n+1}{4 \pi}} T_n(\ell)^{\frac{1}{2}} Y_{\ell m} *_C
\sqrt{\frac{n+1}{4 \pi}} T_n(\ell^\prime)^{\frac{1}{2}} Y_{\ell^\prime m^\prime} \right )\,.
\label{eq:ws2}
\end{equation} 
For the evaluation of (\ref{eq:ws2}), $Y_{\ell m} *_C Y_{\ell^\prime
  m^\prime}$ 
has to be written as a series in $Y_{\ell^{\prime 
\prime} m^{\prime \prime}}$ and $\chi$ applied to it term-by-term. 
We will not need its full details here.

Now replace $Y_{\ell m}$ by $\overline{Y}_{\ell m}$ and integrate. 
As $\chi$ commutes with rotations, only the angular momentum $0$
component of $\sqrt{\frac{n+1}{4 \pi}} T_n(\ell)^{\frac{1}{2}} 
\overline{Y}_{\ell m} *_C \sqrt{\frac{n+1}{4 \pi}} T_n(\ell^\prime)
^{\frac{1}{2}} Y_{\ell^\prime m^\prime}$ contributes to the integral. 
This component is $\delta_{\ell \ell^\prime} \delta_{m m^\prime} 
\overline{Y}_{00} *_C Y_{00} = \delta_{\ell \ell^\prime} \delta_{m m^\prime} 
\frac{1}{4 \pi}$. Using (\ref{eq:cwmap1}), for $\ell =0$ and the value 
$T_n(0)^{\frac{1}{2}} = \sqrt{\frac{4 \pi}{n+1}}$ to be 
derived below, we get
\be 
\int d \Omega \overline{Y}_{\ell m} *^\prime Y_{\ell^\prime m^\prime} 
= \delta_{\ell \ell^\prime} \delta_{m m^\prime}
=\int d \Omega \overline{Y}_{\ell m} Y_{\ell^\prime m^\prime} \,.
\ee 
Hence $*^\prime$ enjoys the special identity characterizing the 
Weyl-Moyal product for the basis of functions in our algebra 
and hence for all functions. $*^\prime$ is the Weyl-Moyal product $*_M$.

$T_n$ is a function ${\cal T}_n$ of $\ell ( \ell +1)$. The latter is 
the eigenvalue of ${\cal L}^2$, the square of angular momentum.
The map $\chi$ can hence be defined directly on all functions $\alpha$ by
\begin{equation} 
\chi (\alpha) = \sqrt{\frac{n+1}{4 \pi}} {\cal T}_n({\cal L}^2)
^{\frac{1}{2}} \alpha 
\end{equation} 
where R.H.S. can be calculated for example by expanding 
$\alpha$ in spherical harmonics.

The evaluation of $T_n^{\frac{1}{2}}(\ell)$ can be done as follows. 
It is enough to compare the two sides of (\ref{eq:tlm})
for $m = \ell$. For $m = \ell$,
\begin{equation} 
Y_{\ell \ell}(\vec{x}) = \frac{\sqrt{(2 \ell +1)!}}{\ell !} 
\, {\bar z}_2^\ell z_1^\ell 
\label{eq:yll}
\end{equation} 
%{\bf we need to check this formula}

The operator $T_\ell^\ell$ being the highest weight state commutes 
with $L_+ = a_2^\dagger a_1$ while $\lbrack L_3 \,, T_\ell^\ell
\rbrack = \ell \, T_\ell^\ell$. Hence in terms of $a_i$ and $a_j^\dagger$,
\begin{equation} 
T_\ell^\ell = N_\ell a_2^{\dagger \ell} a_1^\ell 
\end{equation} 
where the constant $N_\ell$ is to be fixed by the condition
\begin{equation} 
Tr (T_\ell^\ell)^\dagger T_\ell^\ell =1 \,.
\end{equation} 
Evaluating L.H.S. in the basis $\frac{(a_1^\dagger)^{n_1} 
(a_2^\dagger)^{n_2}}{\sqrt{ n_1! n_2 !}} | 0 \rangle \,, n_1 +n_2 =
n+1$, we get after a choice of sign,
\begin{equation} 
N_\ell = \sqrt{\frac{4 \pi}{n+1}} \frac{(n-\ell)! (n+1)! \sqrt{(2 \ell
    +1)!}}{n! \ell! (n + \ell +1)!}  
\end{equation} 
and 
\begin{equation} 
T_\ell^\ell = \sqrt{\frac{4 \pi}{n+1}} \frac{(n-\ell)! (n+1)! \sqrt{(2 \ell
    +1)!}}{n! \ell! (n + \ell +1)!} a_2^{\dagger \ell} a_1^\ell \,. 
\label{eq:tll}
\end{equation} 
Inserting (\ref{eq:tll}) in (\ref{eq:tlm}) and using 
(\ref{eq:yll}), we get, after a short calculation, 
\begin{equation} 
T_n(\ell)^{\frac{1}{2}} = \sqrt{\frac{4 \pi}{n+1}} 
\frac{n! {(n+1)}!}{(n - \ell)! (n + \ell + 1)!} 
\end{equation} 
which gives $T_n(0)^{\frac{1}{2}} = \sqrt{\frac{4 \pi}{n+1}}$ 
as claimed earlier.

%\end{document}

\chapter{Scalar Fields on the Fuzzy Sphere}

The free Euclidean action for the fuzzy sphere for a scalar field is
\begin{equation} 
S_0 = \frac{1}{n+1}\text{Tr} \left[-\frac{1}{2}[L_i, \hat{\phi}]
  [L_i,\hat{\phi}] + \frac{\mu^2}{2} \hat{\phi}^2 \right]
\label{freeS}
\end{equation} 
where we will now hat all operators or $(n+1) \times (n+1)$
matrices. 

As we saw in chapter 2, the scalar field can be
expanded in terms of the polarization tensors $\hat{T}^{\ell}_m$:
\begin{equation} 
\hat{\phi} = \sum_{\ell,m} \phi_{\ell m} \hat{T}^{\ell}_m
\end{equation} 
where $\phi_{\ell m}$ are complex numbers. For concreteness, we will
restrict our attention to hermitian scalar fields
$\hat{\phi}^\dagger=\hat{\phi}$. Since
$(\hat{T}^{\ell}_m)^\dagger=(-1)^m \hat{T}^\ell_m$, this implies that
$\bar{\phi}_{\ell, m} = (-1)^m \phi_{\ell, -m}$.

In terms of $\phi_{\ell m}$'s, the action (\ref{freeS}) is
\begin{equation} 
S_0 = \sum_{\ell,m}^{n+1} \frac{|\phi_{\ell m}|^2}{2}(\ell(\ell+1) +
  \mu^2) = \sum_{\ell=0}^{n+1} \frac{\phi_{\ell, 0}^2}{2}(\ell(\ell+1)
  + \mu^2) + 2\sum_{\ell=0}^{n+1} \sum_{m=1}^\ell \frac{|\phi_{\ell
  m}|^2}{2}(\ell(\ell+1) + \mu^2)   
\end{equation} 

The generating function for correlators in this model is 
\begin{equation} 
Z_0(\hat{J})={\cal N}_0 \int D \hat{\phi} e^{-S_0 +\frac{1}{n+1}{\rm Tr}
  \hat{J} \hat{\phi}}
\end{equation} 
where $\hat{J}$, the ``external current'' is an $(n+1) \times (n+1)$
hermitian matrix. Also
\begin{equation} 
{\cal N}_0 = \left[ \int D \hat{\phi} e^{-S_0} \right]^{-1}
\end{equation} 
is the usual normalization chosen so that 
\begin{equation} 
Z_0(0) =1
\end{equation} 
while 
\begin{equation} 
D\hat{\phi} = \prod_{\ell \leq n/2} \frac{d \phi_{\ell 0}}{\sqrt{2
    \pi}} \prod_{m \geq 1} \frac{d\bar{\phi}_{\ell m} d\phi_{\ell
    m}}{2 \pi i}\,.
\end{equation} 

Let us write 
\begin{equation} 
\hat{J} = \sum_{\ell, m} J_{\ell m} \hat{T}^{\ell}_{m} \,.
\end{equation} 
Then 
\begin{equation} 
 {\rm Tr} \hat{J}\hat{\phi} = \sum_{\ell,m} \bar{J}_{\ell m} \phi_{\ell
  m} = \sum_{\ell=0}^{n+1}J_{\ell 0} \phi_{\ell 0} +\sum_{\ell}
  \sum_{m\geq 1}^\ell (\bar{J}_{\ell m} \phi_{\ell m} + J_{\ell
  m}\bar{\phi}_{\ell m})
\label{sourceterm}
\end{equation} 
and
\begin{eqnarray} 
\lefteqn{Z_0(\hat{J}) = {\cal N}_0 \int d\hat{\phi} \exp \left[
  \sum_{\ell}\left( \frac{-\phi_{\ell, 0}^2}{2}(\ell(\ell+1) + \mu^2)
  + J_{\ell 0} \phi_{\ell 0}\right) + \right. } \nonumber \\
&& \left.\sum_{\ell=0}^{n+1} \sum_{m=1}^\ell -|\phi_{\ell
  m}|^2 (\ell(\ell+1) + \mu^2) + \bar{J}_{\ell m} \phi_{\ell m} + J_{\ell
  m}\bar{\phi}_{\ell m} \right]
\label{part1}
\end{eqnarray}  
It is a product of Gaussians. Substituting
\begin{equation} 
\phi_{\ell m} = \chi_{\ell m} + \frac{J_{\ell m}}{\ell(\ell+1)+\mu^2}
\end{equation} 
and fixing ${\cal N}_0$ by the condition $Z(0)=1$, we get
\begin{equation}
Z_0(\hat{J}) = \prod_{\ell m} \exp \left[\frac{\bar{J}_{\ell m} J_{\ell
    m}}{2[\ell(\ell+1)+\mu^2]}\right] = \exp \left[ {\rm Tr}
    \frac{1}{2}\hat{J}^\dagger \frac{1}{(-\Delta + \mu^2)} \hat{J} \right]
\label{part2}
\end{equation} 

Using (\ref{part1}) and (\ref{part2}) we can compute all correlators
(Schwinger functions) of $\phi$'s. For example,
\begin{equation} 
\langle {\bar \phi}_{\ell' m'}  \phi_{\ell m} \rangle :={\cal N}_0 \int D
\hat{\phi} {\bar \phi}_{\ell' m'} \phi_{\ell m} e^{-S} = \left .
\frac{\partial^2 Z_0(\hat{J})}{\partial
  J_{\ell' m'} \partial {\bar J}_{\ell m} } \right|_{J=0} = \frac{\delta_{\ell'
    \ell}\delta_{m'm}}{\ell(\ell+1)\mu^2} 
\label{propagator}
\end{equation} 
All the correlators of $\hat{\phi}$ follow from
(\ref{propagator}). For instance
\begin{equation} 
\langle \hat{\phi}^2 \rangle = \sum_{\ell,m,\ell',m'}
\hat{T}^{\ell' \dagger }_{m'} \hat{T}^{\ell}_m \langle {\bar \phi}_{\ell' m'}
\phi_{\ell m} \rangle = \sum_{\ell, m} \frac{\hat{T}^{\ell}_m
  \hat{T}^{\dagger \ell}_m}{\ell (\ell+1) + \mu^2}
\end{equation} 

From this follow the correlators under the coherent state or Weyl
maps. The latter (or working with matrices) is more convenient for
current purposes. We have not given $*_W$ explicitly earlier for
$S^2_F$. But we will give the needed details here.

The image $\phi_W$ under the Weyl map of $\hat{\phi}$ has been defined
earlier using the coherent state symbol $\phi_c$ of
$\hat{\phi}$, $\phi_c(z)$ being $\langle z |\hat{\phi}|z
\rangle$. Since $\hat{T}^{\ell}_m$ becomes $Y^{\ell}_m$ under the Weyl
map, we get, using $\bar{Y}^{\ell}_m = (-1)^m Y^{\ell}_m$, and
dropping the subscript $W$,
\begin{equation} 
\langle \phi(\vec{x}) \phi(\vec{x'}) \rangle \equiv G_n (\vec{x}, \vec{x'}) =
  \sum_{\ell=0}^n \sum_{m=-\ell}^\ell \frac{Y^{\ell}_m(\vec{x})
  \bar{Y}^{\ell}_m(\vec{x'})}{\ell(\ell+1)+\mu^2} = \sum_{\ell=0}^n
  \sum_{m=-\ell}^\ell (-1)^m \frac{Y^{\ell}_{m}(\vec{x})
  Y^{\ell}_{-m} (\vec{x'})}{\ell(\ell+1) + \mu^2} \,. 
\end{equation} 
So as 
\begin{equation} 
(-1)^m = (-1)^{-m},
\end{equation} 
\begin{equation} 
G_n (\vec{x}, \vec{x'}) = G_n(\vec{x'}, \vec{x}) \,.
\end{equation} 
The symmetry of $G_n$ is important for calculations.

\section{Loop Expansion}

There is a standard method to develop the loop expansion in the
presence of interactions. Suppose the partition function is
\begin{eqnarray} 
Z(\hat{J}) &=& {\cal N} \int D \hat{\phi} e^{-S + \frac{1}{n+1}{\rm Tr}
  \hat{J} \hat{\phi}},\\
S &=& S_0 + \frac{1}{n+1}\frac{\lambda}{4!}  {\rm Tr} \hat{\phi}^4 :=
  S_0 + S_I, \quad \lambda >0, \label{fullS} \\ 
{\cal N} &=& \left[ \int D \hat{\phi} e^{-S}\right] \Rightarrow Z(0) =1.
\end{eqnarray} 

Let
\be
V (\ell_1 m_1 ; \ell_2 m_2; \ell_3 m_3 ;\ell_4 m_4) =  {\rm Tr} \left( \hat{T}^{\ell_1}_{m_1}
\hat{T}^{\ell_2}_{m_2} \hat{T}^{\ell_3}_{m_3} \hat{T}^{\ell_4}_{m_4} 
\right)  \,.
\ee
We can further abbreviate L.H.S. as follows:
\be
V (\ell_1 m_1 ; \ell_2 m_2; \ell_3 m_3 ;\ell_4 m_4) := V(1234) \,.
\ee
Now since 
\begin{eqnarray} 
S_I &=& \frac{1}{n+1}\frac{\lambda}{4!} {\rm Tr} \left( \hat{T}^{\ell_1}_{m_1}
\hat{T}^{\ell_2}_{m_2} \hat{T}^{\ell_3}_{m_3} \hat{T}^{\ell_4}_{m_4}
\right) \phi_{\ell_1 m_1} \phi_{\ell_2 m_2} \phi_{\ell_3 m_3}
\phi_{\ell_4 m_4} , \\
&\equiv& \frac{\lambda}{4!} V(l_1,m_1;l_2,m_2;l_3,m_3;l_4,m_4;j)
\phi_{\ell_1 m_1} \phi_{\ell_2 m_2}\phi_{\ell_3 m_3}\phi_{\ell_4 m_4}, \\ 
&\equiv& \frac{\lambda}{4!} V(1234) \phi_{\ell_1 m_1} \phi_{\ell_2
  m_2}\phi_{\ell_3 m_3}\phi_{\ell_4 m_4}   
\label{Sint2} 
\end{eqnarray} 
we can write, using (\ref{sourceterm}),
\begin{gather} 
Z(\hat{J}) = {\cal N} \exp \left[-\frac{\lambda}{4!}V(1234)
      \frac{\partial}{\partial \bar{J}_{\ell_1 m_1}} 
      \frac{\partial}{\partial \bar{J}_{\ell_2 m_2}}
      \frac{\partial}{\partial \bar{J}_{\ell_3 m_3}}
      \frac{\partial}{\partial \bar{J}_{\ell_4 m_4}} \right]\int D
      \hat{\phi} e^{-S_0 + \frac{1}{n+1}{\rm Tr}\hat{J}\hat{\phi}}
      \nonumber \\   
= \frac{{\cal N}}{{\cal N}_0} \exp \left[-\frac{\lambda}{4!} V(1234)
      \frac{\partial}{\partial \bar{J}_{\ell_1 m_1}}
      \frac{\partial}{\partial \bar{J}_{\ell_2 m_2}}
      \frac{\partial}{\partial \bar{J}_{\ell_3 m_3}}  
      \frac{\partial}{\partial \bar{J}_{\ell_4 m_4}} \right] \nonumber
      \\
      \exp \left[\frac{1}{2}\sum_{\ell,m}\bar{J}_{\ell
      m}\frac{1}{-\Delta_\ell + \mu^2} J_{\ell m}\right],\nonumber \\ 
(-\Delta_\ell + \mu^2)^{-1} = \frac{1}{\ell(\ell+1)+\mu^2}
\label{fullZ}
\end{gather} 

Even before proceeding to calculate the one-loop two-point function,
one can see that the interaction $V(1234)$ in (\ref{Sint2}) has
invariance only under cyclic permutation of its factors $\ell_i, m_i$
and is not
invariant under transpositions of adjacent factors. This means that we
have to take care to distingiush between ``planar'' and ``non-planar''
graphs while doing perturbation theory as we shall see later below.

The function $V(1234)$ may be conveniently written as
\begin{eqnarray}         
\lefteqn{V(1234) = (n+1)\prod_{i=1}^{4}(2\ell_i
+1)^{1/2} \times} \nonumber \\
&& \sum_{l,m}^{l=n} \left\{ \begin{array}{ccc}
                                \ell_1 & \ell_2 & l \\
                                \frac{n}{2}   & \frac{n}{2}   & \frac{n}{2}
                          \end{array} \right\} \left\{ \begin{array}{ccc}
                                   \ell_3 & \ell_4 & l \\
                             \frac{n}{2}   & \frac{n}{2}   & \frac{n}{2}
                                                \end{array} \right\}
(-1)^m C_{m_1 m_2 m}^{\ell_1\;\; \ell_2\;\; \ell} C_{m_3 m_4
  -m}^{\ell_3\;\; \ell_4\;\; \ell}. 
\label{quarticint}
\end{eqnarray}
The $C_{m_1 m_2 m}^{l_1\;\; l_2\;\; l}$ are the Clebsch-Gordan (C-G)
coefficients and the objects with 6 entries within brace brackets are
the $6j$ symbols. Although less obvious, from the R.H.S of
(\ref{quarticint}), it too still has cyclic
symmetry, as can be verified using properties of $6j$ symbols and C-G
coefficients.

The loop expansion of $Z(J)$ is its power series expansion in
$\lambda$. By differentiating it with respect to the currents followed
by setting them zero, we can generate the loop expansion of
correlators. The $K$-loop term is the $\lambda^K$-th term, the zero
loop being referred to as the tree term. We can write
\begin{equation} 
Z(J) = \sum_0^\infty \lambda^K z_K(J)
\end{equation} 
where $\lambda^K z_K(J)$ is the $K$-loop term. 

The factor ${\cal N}/{\cal N}_0$ contributes multiplicative vacuum
fluctuation diagrams to the correlation functions. It is a common
factor to {\it all} correlators, and is a phase in Minkowski (real
time) regime. 

\section{The One-Loop Two-Point Function}

Of particular interest is the one-loop two-point function where one
can see a ``non-planar'' graph unique to noncommutative theories.

Expanding its numerator and denominator to $O(\lambda)$, we get for
$Z({\hat J})$, 
\begin{equation} 
Z({\hat J}) \approx \frac{\left(1-\frac{\lambda}{4!}V(1234)
      \frac{\partial}{\partial \bar{J}_{\ell_1 m_1}} 
      \frac{\partial}{\partial \bar{J}_{\ell_2 m_2}}
      \frac{\partial}{\partial \bar{J}_{\ell_3 m_3}}
      \frac{\partial}{\partial \bar{J}_{\ell_4 m_4}}\right)\exp
      \left[\frac{1}{2}\sum_{\ell,m}\bar{J}_{\ell 
      m}\frac{1}{-\Delta_\ell + \mu^2} J_{\ell m}\right]}{1-\frac{\lambda}{4!}
V(1234) \langle \phi_{\ell_1 m_1} \phi_{\ell_2 m_2} \phi_{\ell_3
  m_3}\phi_{\ell_4 m_4} \rangle} \,.
\label{ZJ}
\end{equation} 
Here, the argument $i \in (1,2,3,4)$ in $V(1234)$ is to be interpreted
as $\ell_i m_i$ and $\ell_i, m_i$ are to be summed over.
Also the denominator comes from expanding ${\cal N}$ as
power series in $\lambda$:
\begin{equation} 
{\cal N} ={\cal N}(\lambda):=\sum_{K=0}^\infty \lambda^K {\cal N}_K \,.
\end{equation} 
This contributes disconnected diagrams, two of which are planar and
one is non-planar. The disconnected diagrams are precisely cancelled
by other terms of (\ref{fullZ}) as we shall see.

The $O(\lambda)$ term of (\ref{fullZ}) or (\ref{ZJ}) is $\lambda z_1 (J)$ where 
\begin{equation} 
z_1(J) = \left[ \frac{{\cal N}_1}{{\cal N}_0} - \frac{1}{4!}V(1234)
      \frac{\partial}{\partial \bar{J}_{\ell_1 m_1}}
      \frac{\partial}{\partial \bar{J}_{\ell_2 m_2}}
      \frac{\partial}{\partial \bar{J}_{\ell_3 m_3}}
      \frac{\partial}{\partial \bar{J}_{\ell_4 m_4}} \right]
      \exp\left(\frac{1}{2}\sum_{\ell,m}\bar{J}_{\ell m}
      \frac{1}{-\Delta_\ell + \mu^2}J_{\ell m} \right),
\label{np1}
\end{equation} 
The two-point function follows by differentiation as in
(\ref{propagator}). 

Expanding the exact two-point function $\langle \phi_{\ell m}
\bar{\phi}_{\ell' m'} \rangle$ in powers of $\lambda$, 
\begin{equation} 
\langle \phi_{\ell m} \bar{\phi}_{\ell' m'} \rangle = \langle \phi_{\ell m}
\bar{\phi}_{\ell' m'} \rangle_0 + \lambda \langle \phi_{\ell m}
\bar{\phi}_{\ell' m'} \rangle_1 + \ldots
\end{equation} 
we get 
\begin{eqnarray}  
\langle \phi_{\ell m}\bar{\phi}_{\ell' m'} \rangle_1 &=&
\left. \frac{\partial}{\partial\bar{J}_{\ell m} \partial J_{\ell' m'}}
z_1(J) \right|_{J=0} \nonumber \\
&=& \frac{{\cal N}_1}{{\cal N}_0} \langle \phi_{\ell m} \bar{\phi}_{\ell' m'}
\rangle_0 - \frac{\partial}{\partial \bar{J}_{\ell m}} 
\frac{\partial}{\partial J_{\ell' m'}} \frac{\partial}{\partial
  J_{\ell_1 m_1}} \frac{\partial}{\partial J_{\ell_2 m_2}}
\frac{\partial}{\partial \bar{J}_{\ell_3 m_3}}
\frac{\partial}{\partial \bar{J}_{\ell_4 m_4}} \left[\frac{\lambda}{4!}V(1234)
  \right. \nonumber \\ 
&& \left.\exp\left(\frac{1}{2}\sum_{\ell,m}\bar{J}_{\ell m}
      \frac{1}{-\Delta_\ell + \mu^2}J_{\ell m} \right)\right]_{J=0}
\label{oneloop}
\end{eqnarray} 
(\ref{oneloop}) has both disconnected and connected diagrams. We
briefly examine them.

\vskip 1em

\noindent {\it i. Disconnected Diagrams:} 
\vskip 1em
They come when the differentiations $\frac{\partial}{\partial J_{\ell'
    m'}}, \frac{\partial}{\partial \bar{J}_{\ell m}}$ both hit the
same factor in the product of (\ref{oneloop}) to produce the free
propagator. There are three such terms, two of which are planar
diagrams and one non-planar diagram. These add up to $-{\cal
  N}/{\cal N}_0 [ -\Delta_{\ell}+\mu^2]^{-1} \delta_{\ell \ell'}
\delta_{m m'}$:
\begin{equation} 
\langle \phi_{\ell m}\bar{\phi}_{\ell' m'} \rangle^D_1 = -
\frac{{\cal N}_1}{{\cal N}_0} [\Delta_\ell + \mu^2]^{-1} \delta_{\ell
  \ell'} \delta_{m m'}
\end{equation} 
thus cancelling the first term of (\ref{oneloop}).

\vskip 1em

\noindent {\it ii. Connected Diagrams:} 
\vskip 1em
They arise when the differentiation on external currents is applied to
different factors in the product. There are $4 \times 3 = 12$ such terms, giving
\begin{eqnarray}  
\langle \phi_{\ell m}\bar{\phi}_{\ell' m'} \rangle^C_1 &=&
 -\frac{\lambda}{4!}\left[8 \frac{\delta_{\ell \ell_4} \delta_{m+
       m_4,0}(-1)^{m_4}}{-\Delta_{\ell}+\mu^2} \frac{\delta_{\ell'
       \ell_3} \delta_{m+ m_3,0}(-1)^{m_3}}{-\Delta_{\ell'}+\mu^2}
   \frac{\delta_{\ell_1 \ell_2}
     \delta_{m_1+m_2,0}(-1)^{m_2}}{-\Delta_{\ell_1}+\mu^2}
   V(1234)+ \right. \nonumber \\
&&  \left.4 \frac{\delta_{\ell \ell_2} \delta_{m+
       m_2,0}(-1)^{m_2}}{-\Delta_{\ell}+\mu^2} \frac{\delta_{\ell'
       \ell_4} \delta_{m+ m_4,0}(-1)^{m_4}}{-\Delta_{\ell'}+\mu^2}
   \frac{\delta_{\ell_1 \ell_3}
     \delta_{m_1+m_3,0}(-1)^{m_3}}{-\Delta_{\ell_1}+\mu^2}
   V(1234)\right] \label{oneloop1}
\end{eqnarray} 
where, keeping in mind the symmetries of the trace, we have decomposed
(\ref{oneloop1}) into planar and nonplanar contributions. In the
planar case, the indices of an adjacent $\hat{T}$'s get
contracted. There are 8 such terms. In the non-planar case, it is the
indices of the alternate $\hat{T}$'s that get contracted, and there
are 4 such terms. 

The planar term can be further simplified, by observing that
$\hat{T}^{\ell_1}_{m_1} \hat{T}^{\ell_1}_{-m_1}(-1)^{m_1} =
\hat{T}^{\ell_1}_{m_1} \hat{T}^{\ell_1 \dagger}_{m_1}$ is rotationally
invariant, and thus proportional to ${\bf 1}$, the constant of
proportionality being $1/(n+1)$ (as seen by taking the trace). Incising
the external legs, the one loop planar contribution is thus
\begin{equation} 
(-\Delta_\ell + \mu^2)^{-1} \langle \phi_{\ell m}\bar{\phi}_{\ell' m'}
  \rangle^{C, planar}_1 (-\Delta_{\ell'}+\mu^2)^{-1} = -\frac{1}{3} 
\delta_{\ell \ell'} \delta_{m+m',0}(-1)^m \sum_{\ell=0}^n
  \frac{2\ell+1}{\ell(\ell+1)+\mu^2}  
\label{oneloopc}
\end{equation}

In the non-planar case, the indices of nonadjacent $\hat{T}$'s get
contracted. To evaluate the non-planar term, we need to make explicit
use of the form (\ref{quarticint}). There are four such terms giving
\begin{eqnarray} 
\lefteqn{(-\Delta_{\ell}+\mu^2)^{-1}\langle \phi_{\ell m}\bar{\phi}_{\ell' m'}
\rangle^{C,nonplanar}_1(-\Delta_{\ell'} + \mu^2)^{-1} = } \\
&&-\frac{1}{6} (n+1)\sum_{\ell_1,m_1,\ell_3,m_3}\prod_{i=1}^4 (2\ell_i
+1)^{1/2}\sum_{l,m}^{l=n} \left\{ \begin{array}{ccc}
                                \ell_1 & \ell_2 & l \\
                                \frac{n}{2}   & \frac{n}{2}   & \frac{n}{2}
                          \end{array} \right\} \left\{ \begin{array}{ccc}
                                   \ell_3 & \ell_4 & l \\
                             \frac{n}{2}   & \frac{n}{2}   & \frac{n}{2}
                                                \end{array} \right\}
                             \times \nonumber \\
&&\times (-1)^m C_{m_1 m_2 m}^{\ell_1\;\; \ell_2\;\; \ell} C_{m_3 m_4
  -m}^{\ell_3\;\; \ell_4\;\; \ell} \frac{\delta_{\ell_1 \ell_3}
	\delta_{m_1+m_3,0}(-1)^{m_3}}{\ell_1(\ell_1+1) + \mu^2}, \\  
&=&-\frac{1}{6}(n+1) \sqrt{(2\ell_2+1)(2\ell_4+1)}
			     \sum_{\ell,m,\ell_1,m_1} (2\ell+1)\left\{
			     \begin{array}{ccc} 
                                \ell_1 & \ell_2 & l \\
                                \frac{n}{2}   & \frac{n}{2}   & \frac{n}{2}
                          \end{array} \right\} \left\{ \begin{array}{ccc}
                                   \ell_1 & \ell_4 & l \\
                             \frac{n}{2}   & \frac{n}{2}   & \frac{n}{2}
                                                \end{array} \right\}
                             \times \nonumber \\
&& \times (-1)^{m-m_1}C_{m_1 m_2 m}^{\ell_1\;\; \ell_2\;\; \ell}
C_{-m_1 m_4 -m}^{\;\ell_1\;\; \ell_4\;\;\ell} 
\end{eqnarray}
We first perform the sum
\begin{equation} 
\sum_{m,m_1}(-1)^{m-m_1}C_{m_1 m_2 m}^{\ell_1\;\; \ell_2\;\; \ell}
C_{-m_1 m_4 -m}^{\;\ell_1\;\; \ell_4\;\;\ell}
\end{equation}
for which we need the identities
\begin{eqnarray} 
C_{m_1 m_2 m}^{\ell_1\;\; \ell_2\;\; \ell} &=& (-1)^{\ell_1-m_1}
\sqrt{\frac{2\ell+1}{2\ell_2+1}} C_{m_1 -m -m_2}^{\ell_1\;\; \ell\;\;
  \ell_2} , \\
C_{-m_1 m_4 -m}^{\;\ell_1\;\; \ell_4\;\;\ell} &=&
(-1)^{\ell-\ell_4+m_1}\sqrt{\frac{2\ell+1}{2\ell_2+1}} C_{m_1 -m
  m_4}^{\;\ell_1\;\; \ell\;\;\ell_4} ,\\
\sum_{m_1,m_2} C_{m_1 m_2 m_3}^{\;\ell_1\;\; \ell_2\;\;\ell_3} C_{m_1 m_2
  m_4}^{\;\ell_1\;\; \ell_2\;\;\ell_4} &=& \delta_{\ell_3 \ell_4}
\delta_{m_3 m_4}. 
\end{eqnarray}   
This simplifies the non-planar contribution to
\begin{multline} 
(-\Delta_{\ell}+\mu^2)^{-1}\langle \phi_{\ell m}\bar{\phi}_{\ell' m'}
\rangle^{C,nonplanar}_1(-\Delta_{\ell'} + \mu^2)^{-1} =
-\frac{1}{6}(n+1)\delta_{\ell_2 \ell_4} \delta_{m_2+
  m_4}(-1)^{m_2-\ell_2} \\
\times \sum_{\ell,\ell_1}(-1)^{\ell_1+\ell}
\frac{(2\ell+1)(2\ell_1+1)}{\ell_1(\ell_1+1)+\mu^2} \left\{ \begin{array}{ccc} 
                                \ell_1 & \ell_2 & l \\
              \frac{n}{2}   & \frac{n}{2}   & \frac{n}{2} \end{array}
\right\} \left\{ \begin{array}{ccc} 
                  \ell_1 & \ell_4 & l \\
                 \frac{n}{2}   & \frac{n}{2}   & \frac{n}{2}
                       \end{array} \right \} \,.
\end{multline}   
This can be simplified even further, using the following identity involving
the $6j$ symbols:
\begin{equation} 
\sum_{\ell} (-1)^{n+\ell} (2\ell+1)\left\{ \begin{array}{ccc} 
                                \ell_1 & \ell_2 & l \\
              \frac{n}{2}   & \frac{n}{2}   & \frac{n}{2} \end{array}
\right\} \left\{ \begin{array}{ccc} 
                  \ell_1 & \ell_4 & l \\
                 \frac{n}{2}   & \frac{n}{2}   & \frac{n}{2}
                       \end{array} \right\} = \left\{
                                \begin{array}{ccc}
			\ell_1 & \frac{n}{2} & \frac{n}{2}\\
			\ell_4 & \frac{n}{2} & \frac{n}{2}
			\end{array} \right\}
\end{equation} 
We finally get
\begin{eqnarray} 
\lefteqn{(-\Delta_{\ell}+\mu^2)^{-1}\langle \phi_{\ell m}\bar{\phi}_{\ell' m'}
\rangle^{C,nonplanar}_1(-\Delta_{\ell'} + \mu^2)^{-1} =} \nonumber \\
&&-\frac{1}{6}(n+1)\delta_{\ell_2 \ell_4} \delta_{m_2+
  m_4}(-1)^{m_2}(-1)^{\ell_4+n} \sum_{\ell_1} (-1)^{\ell_1}
\frac{(n+1)(2\ell_1+1)}{\ell_1(\ell_1+1)+\mu^2} \left\{
                                \begin{array}{ccc}
			\ell_1 & \frac{n}{2} & \frac{n}{2}\\
			\ell_4 & \frac{n}{2} & \frac{n}{2}
			\end{array} \right\}
\end{eqnarray}
 
The surprising fact is that this nonplanar contribution to the
one-loop two-point function does not vanish even in the limit of $n
\rightarrow \infty$ \cite{vaidya1}. In particular the difference
between planar and non-planar contributions remains finite. To see
this, we can use the Racah formula \cite{moskalev}
\begin{equation} 
\left\{ \begin{array}{ccc}
 \ell_1 & \frac{n}{2} & \frac{n}{2}\\
 \ell_4 & \frac{n}{2} & \frac{n}{2}
\end{array} \right\} \simeq \frac{(-1)^{\ell_1 +\ell_4 +n}}{n}
 P_{\ell_1}\left(1-\frac{{2\ell_4}^2}{n^2}\right)
\end{equation} 
where $P_{\ell}$ are the usual Legendre polynomials. Recall that the
planar contribution from each Feynman diagram is
\begin{equation}
\sum_{\ell=0} \frac{2\ell+1}{\ell(\ell+1)+\mu^2}
\end{equation}
which is logarithmically divergent. The {\it difference} 
%between planar and nonplanar terms 
\begin{equation} 
\delta \equiv \sum_{\ell_1=0} \frac{2\ell_1+1}{\ell_1(\ell_1+1)+\mu^2} -
\sum_{\ell_1} (-1)^{\ell_1} \frac{(n+1) (2\ell_1+1)}{\ell_1 (\ell_1+1)
  + \mu^2} \left\{ \begin{array}{ccc}
		\ell_1 & \frac{n}{2} & \frac{n}{2}\\
		\ell_4 & \frac{n}{2} & \frac{n}{2}
		\end{array} \right\}
\end{equation} 
between planar and nonplanar terms then simplifies to 
\begin{equation} 
\delta = \sum_{\ell=0}^n \frac{2\ell+1}{\ell(\ell+1)+\mu^2}
\left[1-P_{\ell_1}\left(1-\frac{2{\ell_4}^2}{n^2}\right)\right]
\end{equation} 
It is easy to see that
\begin{equation} 
\delta \simeq \int \frac{1-P_{\ell_4}(x)}{1-x} =
2\sum_{k=1}^{\ell_4}\big(\frac{1}{k} \big) 
\end{equation} 
This is the the celebrated UV-IR mixing \cite{vaidya1,chu,dolan}:
integrating out high energy (or UV) modes in the loop produces
non-trivial effects even at low (or IR) external momenta.

This mixing has the potential to pose a serious challange to any
lattice program that uses matrix models on $S_F^2$ to discretize
continuum models on the sphere. It is therefore important to ask if
its effect can effectively be restricted to a class of $n$-point
functions. To this end, one can calculate the four-point function
at one-loop. Interestingly in this case, careful analysis shows that
the difference between planar and the non-planar diagrams vanishes in
the limit of large $n$ \cite{dolan}. Since only the quadratic term is
affected by UV-IR mixing (albeit by a complicated momentum
dependence), it suggests that appropriately ``normal-ordered''
vertices may completely eliminate this problem. That this is indeed
the case was shown by Dolan, O'Connor and Presnajder
\cite{dolan}. Working with a modified action
\begin{equation}
S_0 = \frac{1}{n+1}\text{Tr} \left[-\frac{1}{2}[L_i, \hat{\phi}]
  [L_i,\hat{\phi}] + \frac{\mu^2}{2} \hat{\phi}^2 + \frac{\lambda}{4!}
  : \hat{\phi}^4 : \right]
\end{equation} 
where
\begin{equation}
{\rm Tr} : \hat{\phi}^4 : = {\rm Tr} \left[\hat{\phi}^4 - 12 \sum_{\ell,m}
  \frac{\hat{\phi} \hat{T}^\dagger_{\ell m} \hat{T}_{\ell m}
  \hat{\phi}}{\ell(\ell+1) +\mu^2} + 2 \sum_{\ell,m}
  \frac{[\hat{\phi}, \hat{T}_{\ell m}]^\dagger [\hat{\phi},
  \hat{T}_{\ell m}]}{\ell(\ell+1)+\mu^2} \right] \,,
\end{equation}
they showed that one gets the standard action on the sphere in the
continuum limit $n \rightarrow \infty$.

One may ask if normal-ordering can help cure the UV-IR mixing problem
in higher dimensions, say, on $S_F^2 \times S_F^2$. Here the problem
is much more severe, and unfortunately persists \cite{sachin1,sachin2}.

%\begin{chapter}

\chapter{Instantons, Monopoles and Projective Modules}

The two-sphere $S^2$ admits many nontrivial field configurations.

One such configuration is the instanton. It occurs when $S^2$ is
Euclidean space-time. It is of particular importance as a configuration
which tunnels between distinct ``classical vacua'' of a $U(1)$ gauge
theory. An instanton can be regarded as the curvature of a connection
for a $U(1)$-bundle on $S^2$. As there are an infinite number of
$U(1)$-bundles on $S^2$ characterized by an integer $k$ (Chern
number), there are accordingly an infinite number of instantons as
well. 

We can also think of $S^2$ as the spatial slice of space-time $S^2
\times {\mathbb R}$. In that case, the instantons become monopoles
(The monopoles can be visualized as sitting at the center of the
sphere embedded in ${\mathbb R}^3$. If a charged particle moves in its
field, $k$ is the product of its electric charge and monopole charge
\cite{topology1, bal}.).

In algebraic language, what substitutes for bundles are ``projective
modules'' \cite{madore1}. Here we describe what they mean and find
them for monopoles and instantons.

\section{Free Modules, Projective Modules}

Consider $Mat(N+1)=Mat(2L+1)$. It carries the left- and right-regular representations of the fuzzy algebra. Thus for 
each $a \in Mat(2L+1)$ there are two operators $a^L$ and $a^R$ acting on $Mat(2L+1)$ (thought of as a vector space) defined by
\begin{equation} 
a^L b = a b, \quad a^R b = b a, \quad b \in Mat(N+1)
\end{equation} 
with $a^L b^L = (ab)^L$ and $a^R b^R = (ba)^R$.

{{\it Definition}}: A module $V$ for an algebra ${\cal A}$ is a vector space which
carries a representation of ${\cal A}$. 

Thus $V = Mat(N+1)$ is an ${\cal A}$- ($= Mat(N+1)-$) module. As this $V$ carries two actions
of ${\cal A}$, it is a bimodule. (But note that $a^R b^R = (ba)^R$.)

For an ${\cal A}$-module, linear combinations of vectors in $V$ can be
taken with coefficients in ${\cal A}$. Thus if $v_i \in V$ and $a_i
\in {\cal A}$, $a_i v_i \in V$. A vector space over complex
numbers in this language is a ${\mathbb C}$-module.

We consider only ${\cal A}$-modules $V$ whose elements are
finite-dimensional vectors $v_i= (v_{i1} , \cdots v_{iK})$ with
$v_{ij} \in {\cal A}$. The action of $a \in {\cal A}$ on $V$ is then
$v_i \rightarrow a v_i = (a v_{i1}, \cdots, a v_{ik})$.

Consider the identity ${\bf 1}$ belonging to this $V$. Then all
its elements can be got by (left- or right-) ${\cal A}$-action. As an
${\cal A}$-module, it is one-dimensional. It is also ``generated'' by
${\bf 1}$ as an ${\cal A}$-module. It is a ``free'' module as it
has a basis.

Generally, an ${\cal A}$-module $V$ is said to be free if it has a
basis $\{e_i\}$, $e_i \in V$. That means that any $x \in V$ can be uniquely written
as $\sum a_i e_i, a_i \in {\cal A}$. Uniqueness implies linear
independence: $\sum a_i e_i=0 \Leftrightarrow$ all $a_i =0$.

The phrase ``free'' merits comment. It just means that there is no
(additional) condition of the form $b_i e_i=0, b_i \in {\cal A}$, with
at least one $b_j \neq 0$. In other words, $\{e_i\}$ is a basis.

A class of free $Mat(N+1)$-bimodules we can construct from
$V=Mat(N+1)$ are $V \otimes {\mathbb C}^K \equiv V^K$. Elements
of $V^K$ are $v:=(v_1, \ldots v_K), v_i \in V$. The left- and right-
actions of $a \in {\cal A}$ on $V^K$ are the natural ones: $a^L v = (a
v_1, \ldots, a v_K), a^R v = (v_1 a, \ldots, v_K a)$.

$V^K$ is a free module as it has the basis $\langle \{e_i \}: e_i =
(0, \ldots, 0, \underbrace{1}_{i^{th} entry}, 0, \ldots,0)\rangle$.

A projector $P$ on the ${\cal A}$-module $V^K$ is an $N\times N$ matrix
$P= (P_{ij})$ with entries $P_{ij} \in {\cal A}$, fulfilling $P^\dagger =
P, P^2=P$ where $P^\dagger_{ij}=P^*_{ji}$. Consider $PV^K$. (We can also
apply $P$ on the right: $\xi \in V^K P \Rightarrow \xi_i = \xi_j
P_{ji}$). On $PV^K$ we can generally act only on the right with ${\cal 
A}$, so it is only a right- ${\cal A}$-module and not a left one.

Any vector in $PV^K$ is a linear combination of $Pe_i$ with
coefficients in ${\cal A}$ (acting on the right): $\xi \in PV^K
\Rightarrow \xi = \sum_i (Pe_i) a_i, a \in {\cal A}$. But $\{Pe_i =
f_i\}$ cannot be regarded as a basis as $f_i$ are not linearly
independent. There exist $a_i \in {\cal A}$, not all equal to zero,
such that $\sum_i P e_i a_i=0$, that is $\sum e_i a_i$ is in the
kernel of $P$, without $\sum e_i a_i$ being 0. $PV^K$ is an example of
a projective module.

A module projective or otherwise is said to be trivial if it is a
free module. 

Note that $PV^K$ is a summand in the decomposition $V^K =
PV^K \oplus ({\bf 1}-P)V^K$ of the trivial module $V^K$.

These ideas are valid (with possible technical qualifications) for any
algebra ${\cal A}$ and an ${\cal A}$-module $V$. In particular they
are valid if ${\cal A}$ is the commutative algebra $C^\infty (M)$ of
smooth functions on a manifold with point-wise multiplication. We now
show that elements of ${\cal A}$-modules are sections of bundles on
$M$, picking $M=S^2$ for concreteness. In this picture, sections of
twisted bundles on $S^2$, such as twisted $U(1)$-bundles, are elements
of nontrivial projective modules. Such sections have a natural
interpretation as charge-monopole wave functions.

It is a theorem of Serre and Swan \cite{Varilly} that all such sections can
be obtained from projective modules using preceding algebraic constructions.

\section{Projective Modules on ${\cal A}=C^\infty(S^2)$}

Consider the free module ${\cal A}^2={\cal A} \otimes {\mathbb
  C}^2$. If $\hat{x}$ is the coordinate function, $(\hat{x}_i
  a)(x)=x_i a(x), a \in {\cal A}$, we can define the projector
\begin{equation} 
P^{(1)}=\frac{{\bf 1}+ \vec{\tau}\cdot \hat{x}}{2}
\end{equation} 
where $\tau_i$ are the Pauli matrices. $P^{(1)} {\cal A}^2$ is an
example of a projective module. $P^{(1)} {\cal A}^2$ carries an ${\cal
  A}$-action, left- and right- actions being the same. 

The projector $P^{(1)}$ occurs routinely when discussing the
charge-monopole system \cite{jr, Hasenfratz} or the Berry phase \cite{berryphase}. We will
now establish that $P^{(1)} {\cal A}^2$ is a nontrivial projective
module. Its elements are known to be the wave functions for Chern
number $k$ (= product of electric and magnetic charges) $=1$. For
$k=-1$, we can use the projector $P^{(-1)}=\frac{{\bf 1}-\vec{\tau}
  \cdot \hat{x}}{2}$.

At each $x$, $P^{(1)}(x)$ is of rank 1. If $P^{(1)} {\cal A}^2$ has a
basis $e$, then $e(x)$ is an eigenstate of $P^{(1)}(x),
P^{(1)}(x)e(x)=e(x)$, and smooth in $x$. But there is no such
$e$. For suppose that is not so. Let us normalize $e(x): e^\dagger (x)
e(x)=1$. Let $f_a=\epsilon_{ab}e_b (\varepsilon_{ab} = -\varepsilon_{ba} \,, \varepsilon_{12} = +1)$. 
Then $f$ is a smooth normalized vector perpendicular to $e$ and annihilated by
$P^{(1)}: P^{(1)}f=0$. The operator
\begin{equation} 
U= \left(\begin{array}{cc}
           e_1 & f_1 \\
           e_2 & f_2
         \end{array}\right) \,.
\end{equation} 
is unitary at each $x$ $(U^\dagger (x)U(x)={\bf 1})$ and
\begin{equation} 
U^\dagger P^{(1)} U = \frac{{\bf 1}+\tau_3}{2}  
\end{equation} 
So we have rotated the hedgehog (winding number 1) map $\hat{x}: x
\rightarrow \hat{x}(x)$ to the constant map $x \rightarrow
(0,0,1)$. As that is impossible \cite{topology1}, $e$ does not exist.

For higher $k$, we can proceed as follows. Take $k$ copies of
${\mathbb C}^2$ and consider ${\mathbb C}^{2^k}={\mathbb C}^2 \otimes
\cdots \otimes {\mathbb C}^2$. Let $\vec{\tau}^{(i)}$ be the Pauli matrices acting on the $i^{th}$ slot in  
${\mathbb C}^{2^k}$. That is $\vec{\tau}^{(i)} = 1 \otimes \cdots
\otimes \vec{\tau} \otimes \cdots \otimes 1$.
Then the projector for $k$ is
\begin{equation} 
P^{(k)} = \prod_{i=1}^k \frac{{\bf 1}+ \vec{\tau}^{(i)} \cdot
  \hat{x}}{2}
\label{kprojector}
\end{equation} 
and the projective module is 
\begin{equation} 
P^{(k)}[{\cal A} \otimes {\mathbb C}^{2^k}]:=P^{(k)} {\cal A}^{2^k}.
\end{equation} 
For $k=-|k|$, the projector in (\ref{kprojector}) gets replaced by 
\begin{equation} 
P^{(-|k|)}= \prod_{i=1}^{|k|} \frac{{\bf 1}- \vec{\tau}^{(i)} \cdot
  \hat{x}}{2} \,.
\end{equation} 

We can also construct the modules in another way. Let $k>0$. Consider
$z=(z_1,z_2)$ with $\sum_i |z_i|^2=1$. These are the $z$'s of Chapter
2. For $k>0$, let
\begin{equation} 
v_k(z)=\frac{1}{\sqrt{Z_k}}\left(\begin{array}{c}
                                        z_1^k \\
                                        z_2^k
                                 \end{array}\right), \quad Z_k= \sum_i
                                        |z_i|^{2k}.
\end{equation} 
It is legitimate to put $Z_k$ in the denominator: it cannot vanish
without both $z_i=0$, and that is not possible. $v_k(z)$ is normalized:
\begin{equation} 
v^\dagger_k(z) v_k (z) = 1.
\end{equation} 
So $v_k (z) \otimes v^\dagger_k(z)$ is a projector. Under $z_i
\rightarrow z_i e^{i \theta}$, $v_k(z) \rightarrow v_k(z)
e^{ik\theta}$ and the projector is invariant, so it depends only on
$x=z^\dagger \vec{\tau} z \in S^2$. In this way, we get the projector 
$P^{\prime {(k)}}$
\begin{equation} 
{P^\prime}^{(k)}(x) = v_k(z) \otimes v^\dagger_k(\bar{z})
\label{eq:kproj1}
\end{equation} 
For $k=-|k|<0$, such a projector is 
\begin{equation} 
{P^\prime}^{(-|k|)}(x) = \bar{v}_{|k|}(\bar{z}) \otimes
\bar{v}^\dagger_{|k|}(z)
\label{eq:kproj2}
\end{equation} 
The projectors (\ref{eq:kproj1} , \ref{eq:kproj2}) are sometimes refered
to as ``Bott'' projectors.

\section{Equivalence of Projective Modules}

We briefly explain the sense in which the projectors $P^{(k)},
{P^\prime}^{(k)}$ and the modules $P^{(k)} {\cal A}^{2^k}$ and
${P^\prime}^{(k)}{\cal A}^2$ are equivalent. 

Two modules are said to be equivalent if the corresponding projectors
are equivalent. But there are several definitions of equivalence of projectors
\cite{wegge-olsen}. We pick one which appears best for physics. 

The $2^{2^k} \times 2^{2^k}$ matrix $P^{(k)}$ or the $2 \times 2$
matrix ${P'}^{(k)}$ can be embedded in the space of linear operators
on an infinite-dimensional Hilbert space ${\cal H}$. The elements of
${\cal H}$ consist of $a=(a_1,a_2,\ldots), a_i \in C^\infty
(S^2)$. The scalar product for ${\cal H}$ is $(b,a)=\int_{S^2} d
\Omega \sum_l b_l^*(x) a_l (x)$. ${\cal H}$ is clearly an ${\cal A}$-module. 

The embedding is accomplished by putting $P^{(k)}$ and ${P'}^{(k)}$ in the top left-
corner of an ``$\infty \times \infty$'' matrix. The result is 
\begin{equation} 
{\cal P}^{(k)} = \left(\begin{array}{cc}
                         P^{(k)} & 0 \\
                         0 & 0 
                       \end{array}\right), \quad 
{\cal P}^{\prime (k)} = \left(\begin{array}{cc} 
                         {P'}^{(k)} & 0 \\
                         0 & 0 
                       \end{array}\right).   
\end{equation} 
  
A matrix $U$ acting on ${\cal H}$ has ``coefficients'' in ${\cal A}:
U_{ij} \in C^\infty (S^2)$. It is said to be unitary if $U^\dagger
U={\bf 1}$ where each diagonal entry in ${\bf 1}$ is the constant
function on $S^2$ with value $1 \in {\mathbb C}$. 

The projectors $P^{(k)}$ and $P^{\prime (k)}$ are said to be equivalent if
there exists a unitary $U$ such that $U {\cal P}^{(k)} U^\dagger={\cal P}^{\prime (k)}$. 
If there is such a $U$, then $U {\cal P}^{(k)} a =  {\cal P}^{\prime (k)} Ua \,, a \in {\cal H}$.
That means that wave functions given by
${\cal P}^{(k)} {\cal H}$ and ${\cal P'}^{(k)} {\cal H}$ are unitarily
related. It is then reasonable to regard $P^{(k)}{\cal A}^{2^k}$ and
${P'}^{(k)}{\cal A}^2$ as equivalent.

\vskip 2em

{\it Illustration:} 

\vskip 1em

We now illustrate this notion of equivalence using $P^{(k)}$ and
${P'}^{(k)}$. Since $P^{(\pm 1)}={P'}^{(\pm 1)}$, $k=\pm 2$ is the
first nontrivial example.

Let $z_i$ be as above. Then the matrix with components $z_i \bar{z}_j$
is a projector. It is invariant under $z_i \rightarrow z_i e^{i
  \theta}$ and is a function of $x$. In fact
\begin{equation} 
P^{(1)}(x)_{ij} = z_i \bar{z}_j.
\end{equation} 
Similarly, 
\begin{equation} 
P^{(-1)}(x)_{ij}=\bar{z}_i z_j.
\end{equation} 
Inspection shows that $z$ and $\epsilon \bar{z} =(\epsilon_{ij}
\bar{z}_j)$ are eigenvectors of $P^{(1)}(x)$ with eigenvalues 1 and 0,
whereas $\bar{z}$ and $\epsilon z$ are those of $P^{(-1)}(x)$ with the same
eigenvalues.

Previous remarks on the impossibility of diagonalizing $P^{(k)}(x)$
using a unitary $U(x)$ for all $x$ do not contradict the existence of
these eigenvectors: their domain is not $S^2$, but $S^3$.

Just as $P^{(\pm 1)}$, ${P^\prime}^{(k)}$ has eigenvectors $v_k, \epsilon
\bar{v}_k$ for $k>0$, and $v_{|k|}, \epsilon {\bar v}_{-|k|}$ for $k < 0$.

As $P^{(k)}$ is $2^{|k|} \times 2^{|k|}$, let us embed ${P'}^{(k)}$ 
inside a $2^{|k|} \times 2^{|k|}$ matrix ${\cal P}^{\prime (k)}$ in the
manner described above.

Let us first assume that $k>0$.

Let $\xi^{(k)}(j)$be orthonormal eigenvectors of $P^{(k)}$ constructed
as follows: For $\xi^{(k)}(1)$, we set
\begin{equation} 
\xi^{(k)}(1)=
\begin{array}{cccccc}
z& \otimes& z &\cdots& \otimes& z \\
1&        & 2 &      &        & k
\end{array} 
\end{equation} 
The integers $1,2,\cdots,k$ below $z$'s label the vector space ${\mathbb C}^2$
which contains the $z$ above it: the $z$ above $j$ belongs to the
${\mathbb C}^2$ of the $j$-th slot in the tensor product ${\mathbb
C}^2 \otimes {\mathbb C}^2 \otimes \cdots \otimes {\mathbb C}^2 = {\mathbb C}^{2^k}$. 

The next set of vectors $\xi^{(k)}(j)$ $(j=2, \cdots,k+1)$ is obtained
by replacing $z$ above $j$ by $\epsilon \bar{z}$ and not touching the
remaining $z$'s. We say we have ``flipped'' one $z$ at a time to get
these vectors. 

Next we flip 2 $z$'s at a time: there are $_k C_2$ of these.

We proceed in this manner, flipping 3,4, etc $z$'s. When all are flipped, we get the vector
\begin{equation} 
\xi^{(k)}(2^k)=\epsilon \bar{z} \otimes \epsilon \bar{z} \otimes
\cdots \otimes \epsilon \bar{z}.
\end{equation}
 
The following is important: a basis vector after $j$ flips has the
property
\begin{equation} 
\xi^{(k)}(l) \rightarrow e^{i(k-2j)\theta} \xi^{(k)}(l), \quad {\rm
  when} \quad z \rightarrow e^{i \theta} z.
\end{equation} 
Our task is to find an orthonormal basis $\eta^{(k)}(l)$ where 
$\eta^{(k)}(1)$ is the eigenvector of ${\cal P}^{\prime (k)}(x)$ with 
eigenvalue 1,
\begin{gather} 
\eta^{(k)}(1) = (v_k, \vec{0}), \nonumber \\
{\cal P}^{\prime (k)}(x) \eta^{(k)}(1) = \eta^{(k)}(1).
\end{gather} 
Then the rest are in the null space of ${\cal P}^{\prime (k)}(x)$:
\begin{equation} 
{\cal P}^{\prime (k)}(x)\eta^{(k)}(j)=0, \quad j \neq 1.
\end{equation} 
We require in addition that $\eta^{(k)}(l)$ transforms in exactly the
same manner as $\xi^{(k)}(l)$:
\begin{equation} \eta^{(k)}(l) \rightarrow e^{i(k-2j)\theta}
  \eta^{(k)}(l), \quad {\rm when} \quad z \rightarrow e^{i \theta} z.
\end{equation} 
Then the operator 
\begin{equation} 
\hat{U}(z)=\sum_l \xi^{(k)}(l) \otimes \bar{\eta}^{(k)}(l)
\end{equation} 
is unitary,
\begin{equation} 
\hat{U}(z)^\dagger \hat{U}(z) = {\bf 1},
\end{equation} 
and invariant under $z \rightarrow z e^{i \theta}$:
\begin{equation} 
\hat{U}(z e^{i \theta})=\hat{U}(z).
\end{equation} 
Hence we can write 
\begin{equation} 
\hat{U}(z) = U(x)
\end{equation} 
and $U$ provides the equivalence between ${P'}^{(k)}$ and $P^{(k)}$:
\begin{equation} 
U {\cal P}^{\prime (k)} U^\dagger = {\cal P}^{(k)}
\end{equation} 

There are indeed such orthonormal vectors. $\eta^{(k)}(1)$ clearly has
the required property. As for the rest, we show how to find them from
$k=2$ and $3$. The general construction is similar.

If $k=-|k|<0$, the same considerations apply after changing $z$ to
$\epsilon \bar{z}$ in $P^{(k)}$ and $v_k$ to $\bar{v}_{|k|}$ in
${P'}^{(k)}$.

\vskip 2em

\noindent {\bf k=2}

\vskip 2 em

In this case, ${\mathbb C}^{2^k}={\mathbb C}^4$. The basis is
\begin{equation} 
\eta^{(2)}(1) \, \quad \eta^{(2)}(2)=\left(\begin{array}{c}
                                        0 \\ 0 \\ v_2 \end{array}
                                        \right),  \quad \eta^{(2)}(3)=
                                        \left(\begin{array}{c} 
                          0 \\ 0\\ \epsilon {\bar v}_2 \end{array}\right), \quad 
                                        \eta^{(2)}(4)= \left (\begin{array}{c}
                                                              \epsilon \bar{v}_2 \\ 0 \end{array} \right). 
\end{equation}

\vskip 2em 

\noindent {\bf k=3}

Now ${\mathbb C}^{2^k} = {\mathbb C}^8$. The basis is 
\begin{gather}  
\eta^{(3)}(1), \quad \eta^{(3)}(2)=\left(\begin{array}{c}
                                0\\0\\v_3\\0\\0\\0\\0 \end{array}\right), \quad
                                \eta^{(3)}(3)= \left(\begin{array}{c}
                                                     0\\0\\0\\0\\v_3\\0\\0 
                                                    \end{array}\right), \quad  
\eta^{(3)}(4)= \left(\begin{array}{c}
                     0\\0\\0\\0\\0\\0\\v_3 \end{array}\right), \nonumber \\
\eta^{(3)}(5)=\left(\begin{array}{c}
                     0\\0\\ \epsilon \bar{v}_3 \\0\\0\\0\\0
                     \end{array}\right), \quad
\eta^{(3)}(6) =\left(\begin{array}{c}
                     0\\0\\0\\0\\ \epsilon \bar{v}_3 \\0\\0
                     \end{array}\right), \quad 
\eta^{(3)}(7)=\left(\begin{array}{c}
                     0\\0\\0\\0\\0\\0\\ \epsilon \bar{v}_3
                     \end{array}\right), \quad 
\eta^{(3)}(8)=\left(\begin{array}{c}
                     \epsilon \bar{v}_3 \\0\\0\\0\\0\\0\\0
                     \end{array}\right).
\end{gather} 
In this manner, we can always construct $\eta^{(k)}(j)$. 

\section{Projective Modules on Fuzzy Sphere}
%\cite{bbvy}}

We want to construct the analogues of $P^{(k)}$ and ${P'}^{(k)}$ for
the fuzzy sphere. They give us the monopoles and instantons of
$S_F^2$. Let us consider $P^{(k)}$ first, and denote the corresponding
projectors as $P_F^{(k)}$.

\subsection{Fuzzy Monopoles and Projectors $P_F^{(k)}$}

We begin by illustrating the ideas for $k=1$. 

On ${\mathbb C}^2$, the spin 1/2 representation of $SU(2)$ acts with
generators $\tau_i/2$. On $S_F^2$, the spin $\ell$ representation of
$SU(2)$ acts with generators $L_i^L$. Let $P_F^{(1)}$ be the projector
coupling $\ell$ and $1/2$ to $\ell+1/2$. Consider the projective module
$P_F^{(1)}(S_F^2 \otimes {\mathbb C}^2)$. On this module, 
\begin{equation} 
(\vec{L}^L + \vec{\tau}/2)^2 = (\ell+1/2)(\ell+3/2),
\end{equation} 
or
\begin{equation} 
\frac{\vec{L}^L}{\ell} \cdot \vec{\tau}=1.
\end{equation} 
Passing to the limit $\ell \rightarrow \infty$, this becomes $\hat{x}
\cdot \vec{\tau} =1$, so $P_F^{(1)} \rightarrow P^{(1)}$ as $\ell
\rightarrow \infty$.

We can find $P_F^{(1)}$ explicitly.
\begin{equation} 
-2P_F^{(1)} -1 \equiv \Gamma^L = \frac{\vec{\tau} \cdot \vec{L}^L
 +1/2}{\ell + 1/2}.
\end{equation} 
$\Gamma^L$ is an involution,
\begin{equation} 
(\Gamma^L)^2 = {\bf 1}
\end{equation} 
and will turn up in the theory of fuzzy Dirac operators and the
Ginsparg-Wilson system (see chapter 8). It is the chirality operator of the
Watamuras' \cite{watamura}. 

An important feature of $P_F^{(1)}(S_F^2 \otimes {\mathbb C}^2)$ is
that it is still an $SU(2)$-bimodule. On the right, $L_i^R$ act as
before. On the left, $L_i^L$ do not, but $L_i^L + \tau_i/2$ do as they
commute with $P_F^{(1)}$. 

This addition of $\vec{\tau}/2$ to $\vec{L}^L$ stands here for the
phenomenon of ``mixing of spin and isospin'' in the t'Hooft- Polyakov-monopole theory
\cite{jr}. 

But $P_F^{(1)}(S_F^2 \otimes {\mathbb C}^2)$ is not a free
$S_F^2$-module as it does not have a basis $\lbrace e_i = (e_{i1},
e_{i2}): e_{i,j} \in S_F^2 \rbrace$ That is because if $\alpha =(\alpha_1 \,, \alpha_2) 
\in S_F^2 \otimes {\mathbb C}^2 \,, \alpha_i \in S_F^2$ the projector $P_F^{(1)}$ mixes up the
rows of $\alpha_i$.

For $k=-1$, the projector $P_F^{(-1)}$ couples $\ell$ and 1/2 to
$\ell-1/2$. It is just ${\bf 1}-P_F^{(1)}$. 

The construction for any $k$ is similar. For $k=|k|$, we consider
${\mathbb C}^{2^k}={\mathbb C}^2 \otimes {\mathbb C}^2 \cdots \otimes
{\mathbb C}^2$. On this, the $SU(2)$ acts on each ${\mathbb C}^2$, the
generators for the $j$th slot being $\tau_i^{(j)}/2\equiv {\bf
1}\otimes \cdots \otimes \tau_i/2 \otimes \cdots \otimes {\bf 1}$,
the $\tau_i/2$ being in the $j$th slot. Let $P_F^{(k)}$ be the
projector coupling $\ell$ and all the spins 1/2's to the maximum value
$\ell + k/2$. The projective module is $P_F^{(k)}(S_F^2 \otimes
{\mathbb C}^{2^k})$.

For $k=-|k|$, $P_F^{(k)}$ couples $\ell$ and the spins to the least
value $\ell -|k|/2$. 

We can show that $(\tau^{(j)} \cdot L^L)/\ell$ tends to +1 for $k>0$
and -1 for $k<0$ on these modules, so that the $\tau^{(j)} \cdot \hat{x}$
have the correct values in the limit. Thus consider for example $k>0$. As
all angular momenta are coupled to the maximum possible value, every
pair must also be so coupled. So on this module $(\vec{L}^L + \vec{\tau}^{(j)}/2)^2 = (\ell
+1/2)(\ell +3/2)$ and the result follows as for $k=1$. 

Similar considerations apply for $k<0$. 

For higher $k$, we can also proceed in a different manner. If $k=|k|$,
$SU(2)$ acts on ${\mathbb C}^{k+1}$ by angular momentum $k/2$
representation. Hence there is the projector ${P}^{\prime(k)}$ coupling the left $\ell$ and $k/2$ to $\ell
+k/2$. The projective module is then ${P}^{\prime (k)}(S_F^2 \otimes
{\mathbb C}^{k+1})$. 

For $k<0$ we can couple $\ell$ and $|k|$ to $\ell -|k|/2$ instead (we
assume $\ell > |k|/2$).

${P}^{\prime (k)}$ and $P^{(k)}$ are equivalent in the sense discussed
earlier. We can in fact exhibit the two modules so that they look the
same: diagonalize the angular momentum $(\vec{L}^L + \sum_j
\vec{\tau}^{(j)}/2)^2$ and its third component on $P_F^{(k)}(S_F^2
\otimes {\mathbb C}^{2^k})$. Their right angular momenta being both
$\ell$, their equivalence (in any sense!) is clear. 

For reasons indicated above, none of these $S_F^2$-modules are free.

\subsection{Fuzzy Module for Tangent Bundle}

The projectors for $k=2$ are of particular interest as they can be
interpreted as fuzzy sections of the tangent bundle.

To see this, let us begin with the commutative algebra ${\cal 
A}=C^\infty(S^2)$ and the module ${\cal A}^2=C^\infty(S^2) \otimes
{\mathbb C}^3$. In this case, $SU(2)$ acts on ${\mathbb C}^3$ with 
the spin 1 generators $\theta(\alpha)$ where
\begin{equation} 
\theta(\alpha)_{ij} = -i \epsilon_{\alpha ij}.
\end{equation} 

Consider
\begin{equation} 
\theta(\alpha)\hat{x}_\alpha \equiv \theta \cdot \hat{x}.
\end{equation} 
Its eigenvalues at each $x$ are $\pm 1,0$. Let $P^{(T)}$ be the
projector to the subspace $(\theta \cdot \hat{x})^2= {\bf 1}$:
\begin{equation} 
P^{(T)}=(\theta \cdot \hat{x})^2 \,.
\end{equation} 
Any vector in the module $P^{(T)}{\cal A}^3$ can be written as $\xi^+
+ \xi^-$ where $\theta \cdot \hat{x} \xi^{\pm}=\pm \xi^{\pm}$, that is
$-i \epsilon_{\alpha i j} x_\alpha \xi_j^{\pm}(x) = \pm
\xi_i^{\pm}(x)$. It follows from antisymmetry that $x_i \xi_i^\pm
(x)=0$ or that $\xi^\pm (x)$ are tangent to $S^2$ at $x$. The $\xi^\pm$
give sections of the (complexified) tangent bundle $TS^2$.

A smooth split for all $x$ of $TS^2(x)$ into two subspaces $TS^2_\pm
(x)$ gives a complex structure $J$ on $TS^2$. $J(x)$ is $\pm i {\bf 1}$ on $TS^2_\pm
(x)$. Thus a complex structure on $TS^2$ is defined by the decomposition
\begin{eqnarray} 
TS^2 &=& TS^2_+ \oplus TS^2_-, \nonumber \\
J|_{TS^2_\pm} &=& \pm i {\bf 1}.
\end{eqnarray} 
Now $P^{(T)}$ is the sum of projectors which give eigenspaces of
$\theta \cdot \hat{x}$ for eigenvalues $\pm 1$:
\begin{eqnarray}  
P^{(T)} &=& P^{(T)}_+ + P^{(T)}_-, \nonumber \\
P^{(T)}_\pm &=& \frac{\theta \cdot \hat{x} (\theta \cdot \hat{x} \pm
  {\bf 1})}{2}.
\end{eqnarray} 
With 
\begin{equation} 
JP^{(T)}_\pm = \pm i P^{(T)}_\pm
\end{equation} 
we get the required decomposition of $P^{(1)}{\cal A}^3$ for a complex
structure:
\begin{equation} 
P^{(T)}{\cal A}^3 = P^{(T)}_+{\cal A}^3 \oplus P^{(T)}_-{\cal A}^3.
\end{equation}

Fuzzification  of these structures is easy and elegant. 

Instead of working with $S_F^2 \otimes {\mathbb C}^2$
we work with $S_F^2 \otimes {\mathbb C}^3$. The projector $P_F^{(T)}$ we thereby
obtain is the fuzzy version of $P^{T}$. We can show this as follows.

Let $P_F^{(T, \pm)}$ be the projectors coupling $L_\alpha^L$ and
$\theta(\alpha)$ to the values $\ell \pm 1$. Then  
\be
P_F^{(T)} = P_F^{(T, +)} + P_F^{(T, -)} \,.
\ee

On the module $P_F^{(T, +)} (S_F^2 \otimes {\mathbb C}^3)$, 
\be
\lbrack L_\alpha^L + \theta(\alpha) \rbrack^2 = (\ell +1) (\ell +2)
\ee
or 
\be
\frac{L_\alpha^L  \theta(\alpha)}{\ell} = 1 \,.
\ee

On the module $P_F^{(T, -)} (S_F^2 \otimes {\mathbb C}^3)$,
\be
( L_\alpha^L + \theta(\alpha) )^2 = -1 - \frac{1}{\ell} 
\ee
Thus as $\ell \rightarrow \infty$ 
\be
\frac{L_\alpha^L \theta(\alpha)}{\ell} \rightarrow \pm 1 \quad \mbox{on} \quad P_F^{(T, \pm)} (S_F^2 \otimes {\mathbb C}^3) \,.
\ee
As the left hand side tends to $\theta(\alpha) {\hat x}_\alpha$ as $\ell \rightarrow \infty$, we  
have that $P_F^{(T)}(S_F^2 \otimes {\mathbb C}^3)$ defines the fuzzy
tangent bundle and its decomposition $P_F^{(T,+)}(S_F^2 \otimes
{\mathbb C}^3) \oplus P_F^{(T,-)}(S_F^2 \otimes {\mathbb C}^3)$
defines the fuzzy complex structure: the corresponding $J$, call it $J_F$, is $\pm
i$ on $P_F^{(T,\pm)}(S_F^2 \otimes {\mathbb C}^3)$.

%\begin{document}

\chapter{Fuzzy Nonlinear Sigma Models}

\section{Introduction}

In space-time dimensions larger than 2, whenever a global symmetry $G$ is spontaneously broken to a subgroup $H$,
and $G$ and $H$ are Lie groups, there are massless Nambu-Goldstone modes with values in the coset space $G/H$.
Being massless, they dominate low energy physics as is the case with pions in strong
interactions and phonons in crystals. Their theoretical description contains new concepts
because $G/H$ is not a vector space.

Such $G/H$ models have been studied extensively in $2-d$ physics,
even though in that case there is no spontaneous breaking of
continuous symmetries. A reason is that they are often tractable nonperturbatively in the
two-dimensional context, and so can be used to test ideas suspected to be true in
higher dimensions. A certain amount of numerical work has also been done on such $2-d$
models to control conjectures and develop ideas, their discrete versions having been
formulated for this purpose.

This chapter develops discrete fuzzy approximations to $G/H$ models.
We focus on two-dimensional Euclidean quantum field theories with target space
$G/H=SU(N+1)/U(N)={\mathbb C}P^N$. The novelty of this approach is that
it is based on fuzzy physics \cite{madore1} and non-commutative
geometry \cite{connes, Landi, Varilly, Szabo:2001kg, Douglas:2001ba}. Although fuzzy physics has striking
elegance because it preserves the symmetries of the continuum and because techniques of
non-commutative geometry give us powerful tools to describe continuum topological
features, still its numerical efficiency has not been fully tested. 
This chapter approaches $\sigma$-models with this in mind, the idea being to write fuzzy
$G/H$ models in a form adapted to numerical work.

This is not the only approach on fuzzy $G/H$. In \cite{monopole}, a particular
description based on projectors and their orbits was discretized. We shall refine
that work considerably in this paper. Also in the continuum there
is another way to approach $G/H$, namely as gauge theories with gauge invariance under $H$ and
global symmetry under $G$ \cite{baletal}. This approach is extended here to fuzzy
physics. Such a fuzzy gauge theory involves the decomposition of
projectors in terms of partial isometries \cite{wegge-olsen} and brings
new ideas into this field. It is also very pretty. It is equivalent to the projector method
as we shall also see.

Related work on fuzzy $G/H$ model and their solitons is due to 
Govindarajan  and Harikumar \cite{trg}. A different
treatment, based on the Holstein-Primakoff realization of the $SU(2)$
algebra, has been given in \cite{chan}. A more general approach to
these models on noncommutative spaces was proposed in \cite{dabrowski}.

The first two sections describe the standard $\CP^1$-models on $S^2$.
In section 2 we discuss it using projectors, while in section 3 we
reformulate the discussion in such a manner that transition to fuzzy
spaces is simple. Sections 4 and 5 adapt the previous sections to fuzzy spaces.

Long ago, general $G/H$-models on $S^2$ were written as gauge theories
\cite{baletal}. Unfortunately their fuzzification for generic $G$ and
$H$ eludes us. Generalization of the considerations here to the case where $S^2\simeq\CP^1$
is replaced with $\CP^N$, or more generally Grassmannians and flag manifolds
associated with $(N+1)\times(N+1)$ projectors of rank $\le(N+1)/2$,
 is easy as we briefly show in the concluding section 6.
But extension to higher ranks remains a problem.
%%%%%%%%%%%%%%%%%%%%%%%%%%%%%%%%%%%%%%%%%%%%%%%

\section{$CP^1$ Models and Projectors}

Let the unit vector $x=(x_1,x_2,x_3)\in {\mathbb R}^3$ describe a point
of $S^2$. The field $n$ in the ${\mathbb C}P^1$-model is a map from
$S^2$ to $S^2$:
\be
 n=(n_1,n_2,n_3)\,:\ x\rightarrow\ n(x)\;\in\,{\mathbb R}^3,\quad
 n(x)\cdot n(x):=\sum_a n_a(x)^2=1\;.
 \label{ngi}
\ee
These maps $n$ are classified by their winding number ${\kappa}\in{\mathbb Z}$:
\be
 \kappa=\frac{1}{8\pi}\int_{S^2}\epsilon_{abc}\,n_a(x)\,dn_b(x)\,dn_c(x)\;.
 \label{ngii}
\ee
That $\kappa$ is the winding of the map can be seen taking spherical
coordinates $(\Theta,\Phi)$ on the target sphere $ ( n^2=1 ) $ and using the
identity $\sin\Theta d\Theta\,d\Phi=\half\epsilon_{abc}n_a dn_b\,dn_c$. 
We omit wedge symbols in products of forms.

We can think of $n$ as the field at a fixed time $t$ on a (2+1)-dimensional
manifold where the spatial slice is $S^2$. In that case, it can describe a
field of spins, and the fields with $\kappa\ne 0$ describe solitonic
sectors.
We can also think of it as a field on Euclidean space-time $S^2$. In that
case, the fields with $\kappa\ne 0$ describe instantonic sectors.

Let $\tau_a$ be the Pauli matrices. Then each $n(x)$ is associated with the
projector
\be
P(x)=\half (1+\vec\tau\cdot\vec n(x))\;.
\label{ngiii}
\ee
Conversely, given a $2\times 2$ projector $P(x)$ of rank 1, we can write
\be
P(x)=\half (\alpha_0(x)+\vec \tau\cdot\vec \alpha(x))\;.
\label{ngiv}
\ee
Using $\tr P(x)=1,\ P(x)^2=P(x)$ and $P(x)^\dagger =P(x)$, we get
\be
\alpha_0(x)=1,\quad \vec\alpha(x)\cdot\vec\alpha(x)=1,
\quad \alpha^*_a(x)=\alpha_a(x)\;.
\label{ngv}
\ee
Thus $\CP^1$-fields on $S^2$ can be described either by $P$ or
by $ n_a=\tr (\tau_a\,P)$ \cite{wojciech}.

In terms of $P$, $\kappa$ is
\be
\kappa=\frac{1}{2\pi i}\int_{S^2}\tr P\,(dP)\,(dP)\;.
\label{ngvi}
\ee

There is a family of projectors, called Bott projectors \cite{mignaco,landi}
which play a central role in our approach. Let
\be
z=(z_1,z_2),\quad |z|^2:= |z_1|^2+|z_2|^2=1\; .
\label{ngvii}
\ee
The $z$'s are points on $S^3$. We can write $x\in S^2$ in terms of $z$:
\be
x_i(z)=z^\dagger\tau_i z
\label{ngviii}
\ee
The Bott projectors are
\beqa
P_\kappa(x)=v_\kappa(x)v_\kappa^\dagger(z),\quad
&v_\kappa(z)&=\left[\begin{matrix} z_1^\kappa\\z_2^\kappa \\ 
\end{matrix}\right]\frac{1}{\sqrt{Z_\kappa}}\quad 
\hbox{if}\ \kappa\ge 0\ ,\nn\\
&  Z_k&\equiv|z_1|^{2|\kappa|}+|z_2|^{2|\kappa|}\ ,\nn\\
 &v_\kappa(z)&=
 \left[\begin{matrix}z_1^{*|\kappa|}\\z_2^{*|\kappa|} \\ \end{matrix}\right]
\frac{1}{\sqrt{Z_\kappa}}
\quad  \hbox{if}\ 
\kappa<0 \ .
 \label{ngix}
\eeqa
The field $n^{(\kappa)}$ associated with $P_\kappa$ is given by
\be
 n^{(\kappa)}_a(x)=\tr\tau_a P_\kappa(x)=v_\kappa^\dagger(z)\tau_a v_\kappa(z)
 \ .\label{ngx}
\ee
Under the phase change $z\rightarrow z e^{i\theta}$, $v_\kappa(z)$ changes
$v_\kappa(z)\rightarrow v_\kappa(z)e^{i\kappa\theta}$, whereas $x$ is 
invariant.
As this phase cancels in $v_\kappa(z)v_\kappa^\dagger(z)$, $P_\kappa$ is a 
function of
$x$ as written.

The $\kappa$ that appears in eqs.(\ref{ngix})(\ref{ngx}) is the winding 
number as the explicit calculation of section 3 will show. 
But there is also the following argument.

In the map $z\rightarrow v_\kappa(z)$, for $\kappa=0$, all of $S^3$ and 
$S^2$ get mapped to a point, giving zero winding number. 
So, consider $\kappa>0$. Then the points
\be
 \left( z_1e^{i\frac{2\pi}{\kappa}(l+m)},\, z_2e^{i\frac{2\pi}{\kappa}m} 
\right) 
,\quad l,m\in\;\{ 0,1,..,\kappa-1\}
\nn
\ee
have the same image. But the overall phase $e^{i\frac{2\pi}{\kappa}m}$ of
$z$ cancels out in $x$. Thus, generically $\kappa$ points of $S^2$
(labeled by $l$) have the same projector $P_\kappa(x)$, giving winding
number $\kappa$. As for $\kappa<0$, we get $|\kappa|$ points of $S^2$ 
mapped to the same $P_\kappa(x)$. But because of the complex conjugation in 
eq.(\ref{ngix}),
there is an orientation-reversal in the map giving $-|\kappa|=\kappa$ as 
winding numbers. One way to see this is to use 
\be
P_{-|\kappa|}(x)=P_{|\kappa|}(x)^T
\label{ngxi}
\ee
Substituting this in (\ref{ngvi}), we can see that $P_{\pm|\kappa|}$
have opposite winding numbers.

The general projector ${\cal P}_\kappa(x)$ is the gauge transform
of $P_\kappa(x)$:
\be
 {\cal P}_\kappa(x)=U(x)P_\kappa(x)U(x)^\dagger
 \label{ngxii}
\ee
where $U(x)$ is a unitary $2\times 2$ matrix. Its $n^{(\kappa)}$ is also
given by (\ref{ngx}), with $P_\kappa$ replaced by ${\cal P}_\kappa$.
 The winding number is
unaffected by the gauge transformation. That is because $U$ is a map from
$S^2$ to $U(2)$ and all such maps can be deformed to identity since
$\pi_2(U(2))=\{\hbox{identity}\ e\}$.

The identity
\be
 {\cal P}_\kappa(d{\cal P}_\kappa)=(d{\cal P}_\kappa)(\BI-{\cal P}_\kappa)
 \label{ngxiii}
\ee
which follows from ${\cal P}_\kappa^2={\cal P}_\kappa$, is valuable when
working with projectors.

The soliton described by $P_\kappa$ have the action (below) peaked at the north pole $x_3 =1$
or $\frac{x_1 + i x_2}{1 +x_3} = 0$ and a fixed width and shape. The solitons with energy density  
peaked at $\frac{x_1 + i x_2}{1 +x_3} = \eta$ and variable width and shape are given by the projectors
\beqa
P_\kappa (x\,, \eta \,, \lambda) &=& v_\kappa (z, \eta, \lambda) v_\kappa (z, \eta, \lambda)^\dagger \nonumber \\
v_\kappa (z, \eta, \lambda) &=& 
\left (
\begin{array}{c}
\lambda z_1^\kappa \\
z_2^\kappa - \eta z_1^\kappa
\end{array}
\right )
\frac{1}{(|\lambda z_1|^{2 \kappa} + |z_2^\kappa - \eta z_1^\kappa|^2)^\frac{1}{2}}
\eeqa
For $\kappa > 0$, they correspond to the choice
\be
U(x) = v_\kappa (z, \eta, \lambda) v_\kappa(z)^\dagger
\ee
in (\ref{ngxii}). We call the field associated with $P_\kappa (.,
\eta, \lambda)$ as $n^{(\kappa)}(., \eta, \lambda)$:
\be
n^{(\kappa)} (x, \lambda, \eta) = v_\kappa(z,\eta, \lambda)^\dagger v_\kappa(z, \eta, \lambda) \,.
\ee
We can use $v_\kappa (z, \eta, \lambda) = v_{|\kappa|} (\bar{z}, \eta,
\lambda)$ to write the solitons for $\kappa < 0$.

%The solitons described by ${\cal P}_k$ have the action below peaked at
%the north pole $x_3 =1$ or $\frac{x_1 + x_2}{1+ x_3} = 0$ and a fixed
%width and shape. The solitons with energy density peaked at $\frac{x_1
%  + x_2}{1+ x_3} = \eta$ and variable width and shape are given by the
%projectors
%\begin{gather}
%{\cal P}(x, \eta, \lambda) = v_k(z, \eta, \lambda) v_k(z, \eta,
%\lambda)^\dagger \\
%v_k(z, \eta, \lambda) =\left[\begin{matrix} \lambda
%    z_1^\kappa\\z_2^\kappa - \eta z_i^\kappa \\ 
%\end{matrix}\right]\frac{1}{\sqrt{\lambda^2 |z_1|^{2 \kappa}
%    +|z_2^\kappa - \eta z_1^\kappa|^2}} \quad \mbox{if} \quad \kappa
%    \geq 0 \,.
%\label{eq:generalsolitons1}
%\end{gather}
%They correspond to the choice 
%\be
%U(x) = v_\kappa(z, \eta, \lambda) v_\kappa(z)^\dagger
%\ee
%in (\ref{eq:genericpro1}). We can cal the field associated with ${\cal
%  P}_\kappa(\cdot, \eta , \lambda)$ as $n^{(\kappa)}(\cdot, \eta,
%\lambda)$:
%\be
%n^{(\kappa)} (x, \eta, \lambda) = v_\kappa(z, \eta, \lambda)^\dagger
%v_\kappa(z, \eta, \lambda) \,.
%\ee
%
%We can use $v_\kappa(z, \eta, \lambda) = v_{|\kappa|}({\bar z}, \eta,
%\lambda)$ to write the solitons for $\kappa < 0$.
 
\section{ An Action}

Let ${\cal L}_i=-i(x\wedge\nabla)_i$ be the angular momentum operator.
Then a Euclidean action in the $\kappa$-th topological sector
 for $n^{(\kappa)}$ (or a static Hamiltonian in the (2+1)
 picture) is
\be 
 S_\kappa=-\frac{c}{2}\int_{S^2}d\Omega\, ({\cal L}_i n^{(\kappa)}_b)({\cal 
L}_i n^{(\kappa)}_b)
\ ,\quad c=\ \hbox{a positive constant,}
\label{ngxiv}
\ee
where $d\Omega$ is the $S^2$ volume form $d\cos\theta\, d\varphi$. We can also
write
\be 
 S_\kappa=-c\int_{S^2}d\Omega\, \tr\,({\cal L}_i{\cal P}_\kappa)({\cal 
L}_i{\cal P}_\kappa)
\; .
\label{ngxv}       
\ee
The following identities, based on (\ref{ngxiii}), are also useful:
\be
\tr\, {\cal P}_\kappa({\cal L}_i{\cal P}_\kappa)^2=
\tr\,({\cal L}_i{\cal P}_\kappa)
(\BI-{\cal P}_\kappa)({\cal L}_i{\cal P}_\kappa)=\tr
(\BI-{\cal P}_\kappa)({\cal L}_i{\cal P}_\kappa)^2=\half \tr({\cal L}_i{\cal 
P}_\kappa)^2
\label{ngxvb}
\ee
Hence
\be
S_\kappa=-2c\int_{S^2} d\Omega\,\tr{\cal P}_\kappa\; {\cal L}_i{\cal 
P}_\kappa\;{\cal L}_i{\cal P}_\kappa
\label{ngxvi}
\ee

The Euclidean functional integral for the actions $S_\kappa$ is
\be
 Z(\psi)=\sum_\kappa e^{i\kappa\psi}\int{\cal D\,P}_\kappa e^{-S_\kappa}
 \label{ngxviii}
\ee
where the angle $\psi$ is induced by the instanton sectors as in QCD.

Using the identity $dP=-\epsilon_{ijk}\,dx_i\,x_j\,i{\cal L}_k P$, we can
rewrite the definition (\ref{ngii}) or (\ref{ngvi}) of the winding number as
\beqa
\kappa&=&\frac{1}{8\pi}\int_{S^2}d\Omega\, \epsilon_{ijk}x_i\,\epsilon_{abc}
n_a^{(\kappa)}\,i{\cal L}_jn_b^{(\kappa)}\,  i{\cal L}_kn_c^{(\kappa)}
\label{ngxviiib}  \\    
 &=& \frac{1}{2\pi i}\int_{S^2}d\Omega\tr {\cal P}_\kappa\,
\epsilon_{ijk}\,x_i\,i{\cal L}_j{\cal P}_\kappa\,i{\cal L}_k{\cal P}_\kappa\ .
\label{ngxviiia} 
\eeqa
The Belavin-Polyakov bound \cite{yangsigma} 
\be
S_\kappa\ge 4\pi\, c\,|\kappa|
\label{ngxvib}
\ee
 follows from (\ref{ngxviiib}) on integration of
\be
(i{\cal L}_i 
n_a^{(\kappa)}\pm\epsilon_{ijk}x_j\,\epsilon_{abc}\,n_b^{(\kappa)}\, 
i{\cal L}_k n_c^{(\kappa)})^2\ge 0\ ,
\label{nxvii}
\ee
or from (\ref{ngxviiia})  on integration of
\be
\tr\big({\cal P}_\kappa(i{\cal L}_i{\cal P}_\kappa)\pm i\epsilon_{ijk}\,x_j
{\cal P}_\kappa(i{\cal L}_k{\cal P}_\kappa)\big)^\dagger
\big({\cal P}_\kappa(i{\cal L}_i{\cal P}_\kappa)\pm i\epsilon_{ij'k'}\,x_{j'}
{\cal P}_\kappa(i{\cal L}_{k'}{\cal P}_\kappa)\big)
\ge 0\ .
\label{ngxviiic}
\ee

From this last form it is easy to rederive the bound in a way better adapted 
to fuzzification. Using Pauli matrices $\{\sigma_i\}$ 
 we first rewrite (\ref{ngxvi}) and (\ref{ngxviiia}) as
\beqa
S_\kappa&=&c\int_{S^2}d\Omega\tr\,{\cal P}_\kappa 
(i\sigma\cdot {\cal L}\,{\cal P}_\kappa) 
(i\sigma\cdot {\cal L}\,{\cal P}_\kappa)\ ,\nn\\ 
\kappa&=&\frac{-1}{4\pi}\int_{S^2}d\Omega\tr\big(\sigma\cdot x\, {\cal P}_k
(i\sigma\cdot{\cal L}\,{\cal P}_k)(i\sigma\cdot{\cal L}\,{\cal
P}_k)\big)\ .
\label{nglxviiic}
\eeqa
The trace is now over ${\mathbb C}^2\times{\mathbb C}^2={\mathbb C}^4$,
where $\tau_a$ acts on the first ${\mathbb C}^2$ and $\sigma_i$ on
the second ${\mathbb C}^2$ (so they are really
$\tau_a\otimes\BI$ and $\BI\otimes\sigma_i$)
Then, with $\epsilon_1,\epsilon_2=\pm 1$,
\be
\frac{1+\epsilon_2 \tau\cdot n^{(\kappa)}}{2}\sigma_i\big(
(i{\cal L}_i{\cal P}_\kappa)+\epsilon_1 
i\epsilon_{ijk}\,x_j
(i{\cal L}_k{\cal P}_\kappa)\big)
=(1+\epsilon_1\sigma\cdot x)\frac{1+\epsilon_2 \tau\cdot n^{(\kappa)}}{2}
(i\sigma\cdot{\cal L}{\cal P}_\kappa)\ ,
\label{nglxviiid}
\ee
since $x\cdot{\cal L}=0$. The inequality (\ref{ngxviiic}) is equivalent to
\be
\tr\,\left[\frac{1+\epsilon_1\sigma\cdot x}{2}\,
\frac{1+\epsilon_2 \tau\cdot n^{(\kappa)}}{2}\,
(i\sigma\cdot{\cal L}{\cal P}_\kappa)\right]^\dagger
\left[\frac{1+\epsilon_1\sigma\cdot x}{2}\,
\frac{1+\epsilon_2 \tau\cdot n^{(\kappa)}}{2}\,
(i\sigma\cdot {\cal L}{\cal P}_\kappa)\right]   \ge 0\ ,
\label{nglxviiie}
\ee
from which (\ref{ngxvib}) follows by integration.

%%%%%%%%%%%%%%%%%%%%%%%%%%%%%%%%%%%%%%%%%%%%%%%%%%%%%%%%%%%%%%%%%%%%%%
\section{$\CP^1$-Models and Partial Isometries}

If ${\cal P}(x)$ is a rank 1 projector at each $x$, we can find its 
normalized eigenvector $u(z)$:
\be
  {\cal P}(x)u(z)=u(z)\, ,\quad u^\dagger(z)u(z)=1\; .
 \label{ngxix}
\ee
Then
\be
{\cal P}(x)=u(z)u^\dagger(z)\; .
 \label{ngxx}
\ee
If ${\cal P}={\cal P}_\kappa$, an example of $u$ is $v_\kappa$. 
$u$ can be a function of $z$, changing by
a phase under $z\rightarrow z e^{i\theta}$. Still, ${\cal P}$ will depend
only on $x$.

We can regard $u(z)^\dagger $ (or a slight generalization of it) as an
example of a partial isometry \cite{wegge-olsen} in the algebra ${\cal A}=
C^\infty(S^3)\otimes_{\mathbb C}Mat_{2\times 2}({\mathbb C})$
of $2\times 2$ matrices with coefficients in $C^\infty(S^3)$.
A partial isometry in a $*-$algebra $A$ is an element 
${\cal U}^\dagger\in A$
such that ${\cal U\,U}^\dagger$ is a projector; ${\cal U\,U}^\dagger$ 
is the {\it support projector} of ${\cal U}^\dagger$. 
It is an isometry if ${\cal U}^\dagger\,{\cal U}=\BI$. With
\be
 {\cal U}=\left(\begin{matrix} u_1&0\\u_2&0 \end{matrix}\right) 
\in{\cal A} ,
 \label{ngxxi}
\ee
we have 
\be
 {\cal P}={\cal U\,U}^\dagger
 \label{ngxxii}
\ee
so that  ${\cal U}^\dagger$ is a partial isometry.

We will be free with language and  also call $u^\dagger$ as a 
partial isometry.

The partial isometry for $P_\kappa$ is $v^\dagger_\kappa$.

Now consider the one-form
\be
 A_\kappa=v^\dagger_\kappa\, dv_\kappa\;.
 \label{ngxxiii}
\ee
Under $z_i\rightarrow z_ie^{i\theta(x)}$, $A_\kappa$ transforms like a
connection:
\be
A_\kappa\rightarrow A_\kappa+i\kappa\,d\theta
\nn\ee
($A_\kappa$ are connections for $U(1)$ bundles on $S^2$ for Chern numbers
$\kappa$, see later.) Therefore
\be
 D_\kappa=d+A_\kappa
 \label{ngxxiv}
\ee
is a covariant differential, transforming under $z\rightarrow 
z e^{i\theta}$ as
\be
 D_\kappa\rightarrow e^{i\kappa\theta}D_\kappa  e^{-i\kappa\theta}
 \label{ngxxv}
\ee
and
\be
 D_\kappa^2=dA_\kappa
 \label{ngxxvi}
\ee
is its curvature.

At each $z$, there is a unit vector $w_\kappa(z)$ perpendicular to
$v_\kappa(z)$.
An explicit realization of $w_\kappa(z)$ is given by
\be
w_{\kappa,\alpha}=i\tau_{2\,\alpha\beta}\,v^*_{\kappa,\beta}:=
\epsilon_{\alpha\beta}\,v^*_{\kappa,\beta}
 \label{ngxxvii}
\ee
Since $w^\dagger_\kappa v_\kappa=0$,
\be
 B_\kappa=w^\dagger_\kappa\,dv_\kappa\ ,\quad 
B_\kappa^*=(dv^\dagger_\kappa)w_\kappa=-v^\dagger_\kappa\,dw_\kappa
 \label{ngxxviii}
\ee
are {\it gauge covariant},
\be
 B_\kappa(z)\rightarrow e^{i\theta(x)}B_\kappa e^{i\theta(x)}\ ,\quad
 B_\kappa(z)^*\rightarrow e^{-i\theta(x)}B_\kappa^*e^{-i\theta(x)}
 \label{ngxxix}
\ee
under $z\rightarrow z e^{i\theta}$.

We can account for $U(x)$ by considering
\beqa
{\cal V}_\kappa=Uv_\kappa\ &,&\quad {\cal A}_\kappa={\cal
V}_\kappa^\dagger\,d{\cal V}_\kappa\ ,\quad
 {\cal D}_\kappa=d+ {\cal A}_\kappa\ ,\quad{\cal D}_\kappa^2=d{\cal
A}_\kappa \,, \nn\\
 &\,&{\cal W}_\kappa=(\tau_2U^*\tau_2)w_\kappa\ ,\quad
 {\cal B}_\kappa={\cal W}_\kappa^\dagger\,d{\cal V}_\kappa \ .
 \label{ngxxx}
 \eeqa
${\cal A}_\kappa$ is still a connection, and the properties (\ref{ngxxix})
are not affected by $U$. ${\cal P}_\kappa$ is the support projector 
of ${\cal V}_\kappa^\dagger$, and 
\be
 {\cal W}_\kappa{\cal W}_\kappa^\dagger=\BI-{\cal P}_\kappa\ ,\quad
 (\BI-{\cal P}_\kappa){\cal V}_\kappa=0 \ .
 \label{ngxxxiv}
\ee

Gauge invariant quantities being functions on $S^2$, we can contemplate
a formulation of the $\CP^1$-model as a gauge theory. Let ${\cal J}_i$
be the lift of $L_i$ to angular momentum generators appropriate for 
functions of $z$,
\be
 (e^{i\theta_i{\cal J}_i}f)(z)=f(e^{-i\theta_i\tau_i/2}z)\; ,
 \label{ngxxxi}
\ee
and let
\be
 {\cal B}_{\kappa,i}={\cal W}_\kappa^\dagger\,{\cal J}_i {\cal V}_\kappa\; .
 \label{ngxxxii}
\ee
Now, ${\cal W}_\kappa{\cal B}_{\kappa,i}{\cal V}_\kappa^\dagger$ is gauge 
invariant, and should have an expression in terms of ${\cal P}_\kappa$. 
Indeed it is, in view of (\ref{ngxxxiv}),
\be
{\cal W}_\kappa{\cal B}_{\kappa,i}{\cal V}_\kappa^\dagger=
{\cal W}_\kappa{\cal W}_\kappa^\dagger ({\cal J}_i{\cal V}_\kappa){\cal
V}_\kappa^\dagger=
(\BI-{\cal P}_\kappa){\cal J}_i({\cal V}_\kappa{\cal V}_\kappa^\dagger)=
(\BI-{\cal P}_\kappa)({\cal L}_i{\cal P}_\kappa)=({\cal L}_i{\cal
P}_\kappa){\cal P}_\kappa\; .
\label{ngxxxv}
\ee
Therefore we can write the action (\ref{ngxv}, \ref{ngxvi}) in terms of 
the ${\cal B}_{\kappa,i}$:
\beqa
S_\kappa&=&-
2c\int_{S^2}d\Omega\tr\,{\cal P}_\kappa({\cal L}_i{\cal P}_\kappa)
({\cal L}_i{\cal P}_\kappa)=
2c\int_{S^2}d\Omega\tr\,
(({\cal L}_i{\cal P}_\kappa){\cal P}_\kappa )^\dagger
(({\cal L}_i{\cal P}_\kappa){\cal P}_\kappa)=\nn\\
&=&
2c\int_{S^2}d\Omega\tr({\cal W}_\kappa{\cal B}_{\kappa,i}
{\cal V}_\kappa^\dagger)^\dagger
({\cal W}_\kappa{\cal B}_{\kappa,i}{\cal V}_\kappa^\dagger)=
2c\int_{S^2}d\Omega\;{\cal B}_{\kappa,i}^*{\cal B}_{\kappa,i}\ .
\label{ngxxxvi}
\eeqa

It is instructive also to write the gauge invariant $(d{\cal A}_\kappa)$
in terms of ${\cal P}_\kappa$ and relate its integral to the winding number
(\ref{ngvi}). The matrix of forms
\be
 {\cal V}_\kappa(d+{\cal A}_\kappa){\cal V}_\kappa^\dagger
 \label{ngxxxvii}
\ee
is gauge invariant. Here
\be
d{\cal V}_\kappa^\dagger=(d{\cal V}_\kappa^\dagger)+
{\cal V}_\kappa^\dagger\, d 
\nn\ee
where $d$ in the first term differentiates only ${\cal V}_\kappa^\dagger$.
Now
\be
{\cal V}_\kappa(d+{\cal V}_\kappa^\dagger(d{\cal V}_\kappa))
{\cal V}_\kappa^\dagger
\nn\ee
and
\be
{\cal P}_\kappa\,d{\cal P}_\kappa={\cal V}_\kappa{\cal V}_\kappa^\dagger\,d\,
({\cal V}_\kappa{\cal V}_\kappa^\dagger)=
{\cal V}_\kappa{\cal V}_\kappa^\dagger(d{\cal V}_\kappa)
{\cal V}_\kappa^\dagger+{\cal V}_\kappa(d{\cal V}_\kappa^\dagger)
+{\cal V}_\kappa{\cal V}_\kappa^\dagger\, d
\label{ngxxxviib}
\ee
are equal. Hence, squaring
\be
{\cal V}_\kappa(d+{\cal A}_\kappa)^2{\cal V}_\kappa^\dagger=
{\cal V}_\kappa\,(d{\cal A}_\kappa){\cal V}_\kappa^\dagger=
{\cal P}_\kappa\,(d{\cal P}_\kappa) \,(d{\cal P}_\kappa)
\label{ngxxxviic}
\ee
on using $d^2=0$, eq.(\ref{ngxxxviib}) and 
${\cal P}_\kappa(d{\cal P}_\kappa){\cal P}_\kappa=0$ . Thus
\be 
\int_{S^2}(d{\cal A}_\kappa)=\int_{S^2}\tr\, {\cal V}_\kappa(d{\cal A}_\kappa)
{\cal V}_\kappa^\dagger=\int_{S^2}\tr\,{\cal P}_\kappa\,(d{\cal P}_\kappa)
\,(d{\cal P}_\kappa)\ .
\label{ngxxxviii}
\ee 
We can integrate the LHS. For this we write (taking a section
of the bundle $U(1)\rightarrow S^3\rightarrow S^2$ over
$S^2 \backslash \{ \hbox{north pole} (0,0,1)\})$,
\be 
z(x)=e^{-i\tau_3\varphi/2}e^{-i\tau_2\theta/2}e^{-i\tau_3\varphi/2}
\left(\begin{matrix} 1\\0 \end{matrix}\right)=
\left(\begin{matrix} e^{-i\varphi}\cos\frac{\theta}{2}\\
\sin\frac{\theta}{2} \end{matrix}\right)\;.
\label{ngxxxviiia}
\ee
Taking into account the fact that $U(\vec x)$ is independent of 
$\varphi$ at $\theta=0$, we get
\be
 \int_{S^2}(d{\cal A}_\kappa)=-\int e^{i\kappa\varphi}\, d e^{-i\kappa\varphi}
 =2\pi i \kappa \;.
 \label{ngxxxix}
\ee
This and eq.(\ref{ngxxxviii}) reproduce eq.(\ref{ngvi}).

The Belavin-Polyakov bound \cite{yangsigma} for $S_\kappa$ can now be got 
from the inequality
\be
\tr {\cal C}_{\kappa,i}^\dagger{\cal C}_{\kappa,i}\ge 0\ ,\quad
 {\cal C}_{\kappa,i}=
{\cal W}_\kappa{\cal B}_{\kappa,i}{\cal V}_\kappa^\dagger\pm
{\cal W}_\kappa(\epsilon_{ijl}x_j{\cal B}_{\kappa,l}){\cal V}_\kappa^\dagger\;.
\label{ngxl}
\ee

\subsection{Relation Between ${\cal P}^{(\kappa)}$ and ${\cal P}_\kappa$}

The treatment in \cite{monopole}, for $\kappa>0$, the fuzzy $\sigma$-model 
was based on the continuum projector
\be
 P^{(\kappa)}(x)=P_1(x)\otimes...\otimes P_1(x) =\prod_{i=1}^\kappa
 \,\half(1+\tau^{(i)}\cdot x)
 \label{ngxli}
\ee
and its unitary transform
\be
{\cal P}^{(\kappa)}(x)=U^{(\kappa)}(x) P^{(\kappa)}(x)U^{(\kappa)}(x)^{-1}
\ ,\quad
 U^{(\kappa)}(x)=U(x)\otimes...\otimes U(x)\quad (\kappa \hbox{ factors}).
\label{ngxlii}
\ee
At each $x$, the stability group of $P^{(\kappa)}(x)$ is $U(1)$ with
generator $\half\sum_{i=1}^\kappa\tau^{(i)}\cdot x$, and we get a sphere
$S^2$ as $U(x)$ is varied. Thus $U^{(\kappa)}(x)$ gives a section of a
sphere bundle over a sphere, leading us to identify $ {\cal P}^{(\kappa)}$
with a $\CP^1$-field. Furthermore, the R.H.S. of eq.(\ref{ngxxxviii}) 
(with ${\cal P}^{(\kappa)}$ replacing ${\cal P}_\kappa$) gives $\kappa$ as 
the invariant associated with ${\cal P}^{(\kappa)}$, suggesting a 
correspondence between $\kappa$ and winding number. 

We can write
\be  
 {\cal P}^{(\kappa)}= {\cal V}^{(\kappa)} {\cal V}^{(\kappa)\dagger}\ ,\quad
 {\cal V}^{(\kappa)}={\cal V}_1\otimes...\otimes{\cal V}_1
 \quad \kappa \hbox{ factors}),
 \label{ngxliii}
\ee
its connection $ {\cal A}^{(\kappa)}$ and an action as previously.
A computation similar to the one leading to eq.(\ref{ngxxxviii}) shows that
\be
-\frac{i}{2\pi}\int\, d{\cal A}^{(\kappa)}=\kappa\; .
 \label{ngxlix}
\ee
So $\kappa$ is the Chern invariant of the projective module associated with
${\cal P}^{(\kappa)}$. 

For $\kappa<0$, we must change $x$ to $-x$ in (\ref{ngxli}),
and accordingly change other expressions.

We note that $\kappa$ cannot be identified with the winding
number of the map $x\rightarrow{\cal P}_\kappa(x)$. To see this, say for 
$\kappa>0$, we show that there is a winding number $\kappa$ map  from  
${\cal P}^{(\kappa)}$ to ${\cal P}_\kappa(x)$. As that is also the winding
number of the map $x\rightarrow{\cal P}_\kappa(x)$, the map 
$x\rightarrow{\cal P}^{(\kappa)}(x)$ must have winding number 1.

The map ${\cal P}^{(\kappa)}\rightarrow{\cal P}_\kappa(x)$ is induced from the
map
\be  
 {\cal V}^{(\kappa)}\ \rightarrow\  {\cal V}_\kappa=
\left(\begin{matrix} {\cal V}^{(\kappa)}_{11...1}\\
{\cal V}^{(\kappa)}_{22...2}\end{matrix}\right)
 \label{ngl}
\ee
and their expressions in terms of ${\cal V}^{(\kappa)}$ and ${\cal V}_\kappa$.
In (\ref{ngl}) all the points \\
${\cal V}^{(\kappa)}(z_1e^{2\pi i\,j/\kappa},z_2e^{2\pi i\,l/\kappa})$,
$j,l\in\{0,1,...,\kappa-1\}$, have the same image,
but in the passage to ${\cal P}^{(\kappa)}$ and ${\cal P}_\kappa$ the overall
phase
of $z$ is immaterial. However, the projectors for
${\cal V}^{(\kappa)}(z_1e^{2\pi ij/\kappa},z_2)$ and 
${\cal V}_\kappa^\dagger(z_1,z_2e^{2\pi ij/\kappa})$ are distinct and map to the
same
${\cal P}_\kappa$, giving winding number $\kappa$.

We have not understood the relation between the models based on  
${\cal P}^{(\kappa)}$ and ${\cal P}_\kappa$.

%%%%%%%%%%%%%%%%%%%%%%%%%%%%%%%%%%%%%%%%%%%%%%%%%%%%%%%%%%%%%5
\section{Fuzzy $\CP^1$-Models}

The advantage of the preceding formulation using $\{z_\alpha\}$ is that the 
passage to fuzzy models is relatively transparent. Thus let
$\xi=(\xi_1,\xi_2)\in{\mathbb C}^2\backslash\{0\}$. 
We can then identify $z$ and $x$ as
 \be 
 z=\frac{\xi}{|\xi|}\ ,\quad |\xi|=\sqrt{|\xi_1|^2+|\xi_2|^2}\ ,\qquad
 x_i=z^\dagger\tau_i z\ .
 \label{ngli}
\ee

Quantization of the $\xi$'s and $\xi^*$'s consists in replacing 
$\xi_\alpha$ by annihilation operators $a_\alpha$ and $\xi_\alpha^*$ 
by $a_\alpha^\dagger$.
$|\xi|$ is then the square root of the number operator:
\beqa
\hat N&=&\hat N_1+\hat N_2\ ,\quad 
\hat N_1=a^\dagger_1 a_1\ ,\ N_2=a^\dagger_2 a_2\ ,\nn\\ 
\hat z^\dagger_\alpha&=&\frac{1}{\sqrt{\hat N}}
a^\dagger_\alpha=a^\dagger_\alpha\frac{1}{\sqrt{\hat N +1}}\ ,\quad
\hat z_\alpha=\frac{1}{\sqrt{\hat N +1}}a_\alpha=
a_\alpha\frac{1}{\sqrt{\hat N}}\;,\nn\\
\hat x_i&=&\frac{1}{\sqrt{\hat N}}a^\dagger\tau_i a\ .
 \label{nglii}
 \eeqa  
(We have used hats on some symbols to distinguish them as fuzzy operators).

We will apply these operators only on the subspace of the Fock space
with eigenvalue $n \geq 1 $ of $\hat N$, where $\frac{1}{\sqrt{\hat N}}$ is
well-defined. This restriction is natural and reflects the fact that $\xi$
cannot be zero.

%%%%%%%%%%%%%%%%%%%%%%%%%%%%%%%%%%%%%%%%%%%%%
\subsection{The Fuzzy Projectors for $\kappa>0$}

On referring to (\ref{ngix}), we see that if $\kappa>0$, for the
quantized versions $\hat v_\kappa,\;\hat v^\dagger_\kappa$ of 
$v_\kappa,\; v^*_\kappa$, we have
\beqa
\hat v_\kappa&=&\left[ \begin{matrix}a_1^\kappa\\a_2^\kappa\end{matrix}\right]
\frac{1}{\sqrt{\hat Z_\kappa}}\ ,\qquad
\hat v_\kappa^\dagger=\frac{1}{\sqrt{\hat Z_\kappa}}
\left[ \begin{matrix}
(a_1^\dagger)^\kappa&(a_2^\dagger)^\kappa\end{matrix}\right]
\ ,\qquad \hat v_\kappa^\dagger\hat v_\kappa=\BI\ ,\nn\\
\hat Z_\kappa&=&\hat Z_\kappa^{(1)}+\hat Z_\kappa^{(2)}\quad ,\quad
\hat Z_\kappa^{(\alpha)}=
\hat N_\alpha(\hat N_\alpha-1)...(\hat N_\alpha-\kappa+1)\ ..
\label{ngliii}
\eeqa

The fuzzy analogue of $U$ is a $2\times 2$ unitary matrix $\hat U$
whose entries $\hat U_{ij}$ are polynomials in $a^\dagger_a a_b$. 
As for $\hat{\cal V}_\kappa$, the quantized version of ${\cal V}_\kappa$, it is
just
\be
 \hat{\cal V}_\kappa=\hat U\,\hat v_\kappa
 \label{ngliv}
\ee
and fulfills
\be
 \hat{\cal V}_\kappa^\dagger\,\hat{\cal V}_\kappa=\BI\; ,
 \label{nglv}
\ee
$ \hat{\cal V}_\kappa^\dagger$ being the quantized version of 
${\cal V}_\kappa^\dagger$.
We thus have the fuzzy projectors
\be
\hat P_\kappa=\hat v_\kappa\hat\, v_\kappa^\dagger\ ,\qquad
 \hat {\cal P}_\kappa=\hat{\cal V}_\kappa\,\hat{\cal V}_\kappa^\dagger\;.
 \label{nglvi}
\ee 

Unlike $\hat v_\kappa,\hat{\cal V}_\kappa $ and their adjoints, 
$\hat P_\kappa$ and $\hat {\cal P}_\kappa$ commute with the number operator
$\hat N$. So we can formulate a finite-dimensional matrix model
for these projectors as follows. 
Let ${\cal F}_n$ be the subspace of the Fock 
space where $\hat N=n$. It is of dimension $n+1$, and carries the $SU(2)$
representation with angular momentum $n/2$, the $SU(2)$ generators being
\be
L_i=\frac{1}{2}a^\dagger\tau_i a \ .
\label{nglvii}
\ee
Its standard orthonormal basis is $|\frac{n}{2},m>\;,\ m=-\frac{n}{2},
-\frac{n}{2}+1,...,\frac{n}{2}$.
Now consider ${\cal F}_n\otimes_{\mathbb C}{\mathbb C}^2:={\cal F}_n^{(2)}$,
with elements $f=(f_1,f_2),\;f_a\in{\cal F}_n$. Then 
$\hat P_\kappa,\; \hat {\cal P}_\kappa$ act on ${\cal F}_n^{(2)}$ in the
natural way.
For example
\be
 f\rightarrow\hat {\cal P}_\kappa f, \quad
(\hat{\cal P}_\kappa f)_a=(\hat{\cal P}_\kappa)_{ab}f_b=
 (\hat{\cal V}_{\kappa,a}\hat{\cal V}_{\kappa,b}^\dagger)f_b \ .
 \label{nglviii}
\ee

We can now write explicit matrices for $\hat P_\kappa$ and 
$\hat {\cal P}_\kappa$. We have:
\beqa
\hat P_\kappa&=&\left( \begin{matrix}
a_1^\kappa\frac{1}{\hat Z_\kappa}a_1^{\dagger\,\kappa}
&a_1^\kappa\frac{1}{\hat Z_\kappa}a_2^{\dagger\,\kappa}\\
a_2^\kappa\frac{1}{\hat Z_\kappa}a_1^{\dagger\,\kappa}&
a_2^\kappa\frac{1}{\hat Z_\kappa}a_2^{\dagger\,\kappa}\end{matrix}\right)
\ ,\label{nglix}\\
a_1^\kappa\frac{1}{\hat Z_\kappa}&=&\frac{1}{(\hat N_1+\kappa)...(\hat N_1+1)
+\hat Z^{(2)}_\kappa}a_1^\kappa\ ,\quad
a_1^\kappa a_1^{\dagger\,\kappa}=(\hat N_1+\kappa)...(\hat N_1+1)\ ,\nn
\eeqa 
from which its matrix $\hat P_\kappa(n)$ for $\hat N=n$ can be obtained.

The matrix $ \hat {\cal P}_\kappa$ is the unitary transform 
$\hat U\hat P_\kappa(n)\hat U^\dagger$ where $\hat U$ is a 
$2\times 2$ matrix and 
$\hat U_{ab}$ is itself an $(n+1)\times (n+1)$ matrix.
As for the fuzzy analogue of ${\cal L}_i$, we define it by
\be
{\cal L}_i\hat {\cal P}_\kappa=[L_i,\hat {\cal P}_\kappa]\ .
\label{nglx}
\ee

The fuzzy action 
\be
 S_{F,\kappa}(n)=\frac{c}{2(n+1)}\tr_{\hat N=n}\,({\cal L}_i
\hat {\cal P}_\kappa)^\dagger
 ({\cal L}_i\hat {\cal P}_\kappa)\ ,\quad c=\hbox{constant}\ ,
 \label{nglxi}
\ee
follows, the trace being over the space ${\cal F}_n^{(2)}$.

\subsection{The Fuzzy Projector for $\kappa<0$\;.}

For $\kappa<0$, following an early indication, we must exchange the roles of
$a_a$ and $a^\dagger_a$.

\subsection{Fuzzy Winding Number}

In the literature \cite{GKP2}, there are suggestions on how to 
extend (\ref{ngvi}) to
the fuzzy case. They do not lead to an integer value for this number
except in the limit $n\rightarrow\infty$.

There is also an approach to topological invariants using Dirac 
operator and 
cyclic cohomology. Elsewhere this approach was applied to the fuzzy case
\cite{monopole, instanton1} and gave integer values, and even a fuzzy
analogue of the Belavin-Polyakov bound.  However they were not for 
the action $S_{F,\kappa}$, 
but for an action which approaches it as $n\rightarrow\infty$.
In the subsection below, we present an alternative
approach to this bound which works for $S_{F,\kappa}$. It looks like 
 (\ref{ngxvib}), except that $\kappa$ becomes an integer only in the limit
$n\rightarrow\infty$.

There is also a very simple way to associate an integer to 
$\hat{\cal V}_\kappa$ \cite{GKP2, sachin, instanton1}. It
is equivalent to the Dirac operator approach. We can assume that the 
domain of $\hat{\cal V}_\kappa$ are vectors with a fixed value $n$ of $\hat N$. 
Then after applying $\hat{\cal V}_\kappa$, $n$ becomes $n-\kappa$ if $\kappa>0$ and 
$n+|\kappa|$ is $\kappa<0$.Thus $\kappa$ is just the difference in the
value of $\hat N$, or equivalently twice the difference in the value
of the angular momentum, between its domain and its range.

We conclude this section by deriving the bound for $S_{F,\kappa}(n)$.

\subsection{The Generalized Fuzzy Projector : Duality or BPS States}

We introduced the projectors ${\cal P}_\kappa(\cdot, \eta, \lambda)$
and their fields $n^{(\kappa)}(\cdot, \eta, \lambda)$ earlier.
They describe solitons localized at $\frac{x_1 + ix_2}{1 +x_3} = \eta$
and a shape and width controlled by $\lambda$.
As inspection shows, they are very easy to quantize by replacing
$\xi_i$ by $a_i$ and $\bar{\xi}_j$ by $a_j^\dagger$.

The fields $n^{(\kappa)}(\cdot, \eta, \lambda)$ and their projectors
${\cal P}_\kappa(\cdot, \eta, \lambda)$ have a particular significance.
$P_{|\kappa|}(\cdot, \eta, \lambda)$ saturates the bounds (\ref{ngxviiic}) with
the plus sign, $P_{-\kappa} (\cdot, \eta, \lambda)$ saturates it
with the minus sign. This result is due to their holomorphicity 
(anti-holomorphicity) properties as has been explained elsewhere \cite{bal}. 

It is very natural to identify their fuzzy versions as fuzzy BPS
states. But as we note below, they do not saturate the bound on the fuzzy action.
 
\subsection{The Fuzzy Bound.}

A proper generalization of the Belavin-Polyakov bound to its fuzzy version
involves a slightly more elaborate approach. This is because the 
straightforward fuzzification
of $\vec\sigma\cdot\vec x$ and $\vec\tau\cdot\vec n^{(\kappa)}$
and their corresponding projectors do not commute, and the product of such
fuzzy projectors is not a projector. {\it We use this elaborated approach
only in this section.} It is not needed elsewhere. In any case, what is 
there in other sections is trivially adapted to this formalism.

The operators $a^\dagger_\alpha a_\beta$ acting on the vector space with
$\hat N=n$ generate the algebra $Mat(n+1)$ of $(n+1)\times(n+1)$ matrices.
The extra structure comes from regarding them not as observables, but
as a Hilbert space of matrices $m,\ m',...$ with scalar product
$(m',m)=\frac{1}{n+1}\tr_{{\mathbb C}^{n+1}}{m'}^\dagger\;m$, with the 
observables acting thereon.

To each $\alpha\in Mat(n+1)$, we can associate two linear operators
 $\alpha^{L, R}$ on  $Mat(n+1)$  according to
\be
\alpha^Lm=\alpha m\ ,\quad \alpha^R m=m\alpha\ ,\quad m\in\;Mat(n+1)\ .
\label{ngci}
\ee
$\alpha^L-\alpha^R$ has a smooth commutative limit for operators
of interest. It actually vanishes, and $\alpha^{L, R}\to 0$ if $\alpha$
remains bounded during this limit.

Consider the angular momentum operators $L_i\in Mat(n+1)$. The associated
`left' and `right' angular momenta $L_i^{L,R}$ fulfil
\be
(L_i^L)^2=(L^R_i)^2=\frac{n}{2}(\frac{n}{2}+1)\ .
\label{ngcii}
\ee

We now regard $a_\alpha,\ a_\alpha^\dagger$ of section 6.5.1 as left operators
$a_\alpha^L$ and $ a_\alpha^{\dagger L}$. $\hat P_\kappa^L$ thus becomes 
a $2\times 2$ matrix with each entry being a left multiplication
operator. It is the linear operator $\hat{\cal P}^L_\kappa$
on $Mat(n+1)\otimes{\mathbb C}^2$.
We tensor this vector space with another ${\mathbb C}^2$ as before to get
${\cal H}=Mat(n+1)\otimes{\mathbb C}^2\otimes{\mathbb C}^2$, with
$\sigma_i$ acting on the last ${\mathbb C}^2$, and 
 $\sigma\cdot{\cal L}\hat{\cal P}^L_\kappa$ denoting the operator
 $\sigma_i({\cal L}_i\hat{\cal P}_\kappa)^L$.

We can repeat the previous steps if there are fuzzy analogues $\gamma$
and $\Gamma$ of continuum `world volume' and `target space' chiralities
$\vec\sigma\cdot\vec x$ and $\vec\tau\cdot\vec n^{(\kappa)}$ which
mutually commute. Then $\frac{1}{2}(1\pm\gamma)$, $\frac{1}{2}(1\pm\Gamma)$
are commuting projectors and the expressions derived at the end of Section 3
 generalize, as we shall see.

There is such 
a $\gamma$, due to Watamuras\cite{watamura}, and  discussed further
by \cite{monopole}. Following \cite{monopole}, we take
\be
\gamma\equiv\gamma^L=\frac{2\sigma\cdot L^L+1}{n+1}\ .
\label{ngciv}
\ee
The index $L$ has been put to emphasize its left action on $Mat(n+1)$.

As for $\Gamma$, we can do the following. $\hat{\cal P}_\kappa$ acts on the
left on $Mat(n+1)$, let us call it $\hat{\cal P}_\kappa^L$. It has a
$\hat {\cal P}_\kappa^R$ acting on the right and an associated
\be
\Gamma\equiv\Gamma^R_\kappa=2\hat{\cal P}_\kappa^R-1\quad ,
\quad(\Gamma^R_\kappa)^2
=1\ .
\label{ngcv}
\ee
As it acts on the right and involves $\tau$'s while $\gamma$ acts on the left
and involves $\sigma$'s,
\be
\gamma^L\Gamma^R_\kappa=\Gamma^R_\kappa\gamma^L\ .
\label{ngcvi}
\ee

The bound for (\ref{nglxi}) now follows from
\be
\tr_{\cal H}\left(
\frac{1+\epsilon_1\gamma^L}{2}\frac{1+\epsilon_2\Gamma^R_\kappa}{2}
\sigma\cdot{\cal L}\hat{\cal P}_\kappa^L
\right)^\dagger
\left(
\frac{1+\epsilon_1\gamma^L}{2}\frac{1+\epsilon_2\Gamma^R_\kappa}{2}
\sigma\cdot{\cal L}\hat{\cal P}_\kappa^L
\right)\ge 0
\label{ngcvii}
\ee
($\epsilon_1,\epsilon_2=\pm 1$), and reads
\beqa
S_{F,\kappa}&=&
\frac{c}{4(n+1)}\tr_{\cal H}(\sigma\cdot{\cal L}\hat{\cal P}_\kappa^L)^\dagger
(\sigma\cdot{\cal L}\hat{\cal P}_\kappa^L)\nn\\
&\ge&\frac{c}{4(n+1)}\tr_{\cal H}\left((\epsilon_1\gamma^L+
\epsilon_2\Gamma^R_\kappa)(\sigma\cdot{\cal L}\hat{\cal P}_\kappa^L)
(\sigma\cdot{\cal L}\hat{\cal P}_\kappa^L)\right)\nn\\
&&+\,\frac{c}{4(n+1)}\tr_{\cal H}\left(\epsilon_1\epsilon_2\gamma^L\Gamma^R
(\sigma\cdot{\cal L}\hat{\cal P}_\kappa^L)
(\sigma\cdot{\cal L}\hat{\cal P}_\kappa^L)
\right)
\label{ngcviii}
\eeqa
The analogue of the first term on the R.H.S. is zero in the continuum,
being absent in (\ref{ngxvib}), but not so now. As $n\to\infty$,
 (\ref{ngcviii}) reproduces (\ref{ngxvib}) to leading order $n$, 
but has corrections 
which vanish in the large $n$ limit.

A minor clarification: if $\tau$'s are substituted by $\sigma$'s in
$2\hat{\cal P}^L_1-1$, then it is $\gamma^L$. The different projectors
are thus being constructed using the same principles.

%%%%%%%%%%%%%%%%%%%%%%%%%%%%%%%%%%%%%%%%%%%%%%%%%%%%%%%%%%%
\section{$\CP^N$-Models}

We need a generalization of the Bott projectors to adapt the previous approach to all $\CP^N$.

Fortunately this can be easily done. The space $\CP^N$ is the space of $(N+1)\times(N+1)$ {\it rank 1} projectors.  
The important point is the rank.  So we can write
\be
\CP^N= \langle U^{(N+1)}P_0U^{(N+1)\dagger}:\  P_0=
\hbox{diag.}\underbrace{(0,....,0,1)}_{N+1 \;entries}\, \ U^{(N+1)}\in U(N+1) \rangle \;.
\label{nglxii}
\ee

As before, let $z=(z_1,z_2),\ |z_1|^2+|z_2|^2=1$, and $x_i=z^\dagger\tau_iz$.
Then we define
\be
v_\kappa^{(N)}(z)=\left(
\begin{matrix}z_1^\kappa\\z_2^\kappa\\0\\.\\.\\0\end{matrix}
 \right)\frac{1}{\sqrt{Z_\kappa}}\;,\ \kappa>0\,;\quad v_\kappa^{(N)}(z)
 =\left(\begin{matrix}z_1^{*\kappa}\\z_2^{*\kappa}\\0\\.\\.\\0\end{matrix}
 \right)\frac{1}{\sqrt{Z_\kappa}}\;,\ \kappa<0\;. 
 \label{nglxiii}
\ee
Since
\begin{gather}
v_\kappa^{(N)}(z)^\dagger v_\kappa^{(N)}(z) = 1\,,\nn \\
P^{(N)}_\kappa(x) = v_\kappa^{(N)}(z)v_\kappa^{(N)}(z)^\dagger\,\in\,\CP^N\; .
\label{nglxiv}
\end{gather}

We can now easily generalize the previous discussion, using 
$P^{(N)}_\kappa$ for $P_\kappa$ and $U^{(N+1)}$ 
for $U$, and subsequently quantizing $z_\alpha,\,z_\alpha^*$. 
In that way we get fuzzy $\CP^N$-models.

$\CP^N$-models can be generalized by replacing the target space
 by a general Grassmannian or a flag manifold. They
can also be elegantly formulated as gauge theories \cite{baletal}. 
But we are able to formulate only a limited class of such manifolds
in such a way that they can be made fuzzy.
The natural idea would be to look for several vectors
\be
 v_{k_i}^{(N)(i)}(z)\ ,\quad i=1,..,N
 \label{nglxv}
\ee 
in $(N+1)$-dimensions which are normalized and orthogonal,
\be
v_{k_i}^{(N)(i)\dagger}(z) v_{k_j}^{(N)(j)}(z) =\delta_{ij}
\label{nglxvi}
\ee
and have the equivariance property
\be
v_{k_i}^{(N)(i)}(ze^{i\theta})= v_{k_i}^{(N)(i)}(z)e^{i\,k_i\theta} \,.
\label{nglxvii}
\ee
The orbit of the projector $\sum_{i=1}^M v_{k_i}^{(N)(i)}(z) v_{k_i}^{(N)(i)\dagger}(z)$
under $U^{(N+1)}$ will then be a Grassmannian for each $M\le N$, while the 
orbit of $\sum_i \lambda_iv_{k_i}^{(N)(i)}(z) v_{k_i}^{(N)(i)\dagger}(z)$ 
with possibly unequal $\lambda_i$ under $U^{(N+1)}$ will be a flag manifold.

But we can find such $v_{k_i}^{(N)(i)}$ only for $i=1,2,...,M\le\frac{N+1}{2}$.

For instance in an $(N+1)=2L$-dimensional vector space, for integer $L$, 
we can form the vectors
\be
v_{k_1}^{(N)(1)}(z)=\left(\begin{matrix}z_1^{k_1}\\z_2^{k_1}\\0\\ \cdot\\0
\end{matrix}\right)\frac{1}{\sqrt{Z_{k_1}}}\ ,\ 
v_{k_2}^{(N)(2)}(z)=
\left(\begin{matrix}0\\0\\z_1^{k_2}\\z_2^{k_2}\\0\\ \cdot\\0
\end{matrix}\right)\frac{1}{\sqrt{Z_{k_2}}}\ ,\ ...\ ,\ 
v_{k_L}^{(N)(L)}(z)=\left(\begin{matrix}0\\ \cdot\\0\\z_1^{k_L}\\z_2^{k_L}
\end{matrix}\right)\frac{1}{\sqrt{Z_{k_L}}}
\label{nglxviii}
\ee
for $k_i>0$. For those $k_i$ which are negative, we replace 
$v^{(N)(i)}_{k_i}(z)$ here by $v^{(N)(i)}_{|k_i|}(z)^*$:
\be
v^{(N)(i)}_{k_i}(z)=v^{(N)(i)}_{|k_i|}(z)^*\ ,\ k_i<0\ .
\label{nglxix}
\ee
These $v^{(N)(i)}_{k_i}$ are orthonormal for all $z$ with 
$\sum_\alpha |z_\alpha|^2=1$, so that we can handle Grassmannians and
flag manifolds involving projectors up to rank $L$.

If $N$ instead is $2L$, we can write
\be
v_{k_1}^{(N)(1)}(z)=\left(\begin{matrix}z_1^{k_1}\\z_2^{k_1}\\0\\ \cdot\\0
\end{matrix}\right)\frac{1}{\sqrt{Z_{k_1}}}\ ,\ 
v_{k_2}^{(N)(2)}(z)=
\left(\begin{matrix}0\\0\\z_1^{k_2}\\z_2^{k_2}\\0\\ \cdot\\0
\end{matrix}\right)\frac{1}{\sqrt{Z_{k_2}}}\ ,\ ...\ ,\ 
v_{k_L}^{(N)(L)}(z)=\left(\begin{matrix}0\\ \cdot\\0\\z_1^{k_L}\\z_2^{k_L}\\
0\end{matrix}\right)\frac{1}{\sqrt{Z_{k_L}}}
\label{nglxviiib}
\ee
for $k_i>0$, and use (\ref{nglxix}) for $k_i<0$. 

But we can find no vector $v_{k_{L+1}}^{(N)(L+1)}(z)$ fulfilling
\be
v_{k_i}^{(N)(i)}(z)^\dagger v_{k_{L+1}}^{(N)(L+1)}(z)=\delta_{i,L+1},\ 
i=1,2,..,L+1 \ ,\quad
v_{k_{L+1}}^{(N)(L+1)}(ze^{i\theta})=v_{k_{L+1}}^{(N)(L+1)}(z)
e^{ik_{L+1}\theta}\ . 
\label{nglxixa}
\ee

The quantization or fuzzification of these models can be done as before.
But lacking suitable $v^{(i)}_{k_i}$ for $i>L$, the method fails if the 
target flag manifold involves projectors of rank $>\frac{N+1}{2}$.

Note that we cannot consider vectors like
\be
v'(z)=\left(\begin{matrix}0\\ \cdot\\0\\z_i^{k}\\0\\ \cdot\\0
\end{matrix}\right)\frac{1}{|z_i|^k}\ ,\quad k>0\ ,\ i=1\;or\;2
\label{nglxxx}
\ee
and $v'(z)^*$. That is because $z_i$ can vanish compatibly with the 
constraint
$|z_1|^2+|z_2|^2=1$, and $v'(z),\, v'(z)^*$ are ill-defined when $z_i=0$.
 
As mentioned before, the flag manifolds are coset spaces
${\cal M}=SU(K)/SU(k_1) \otimes U(k_2)\otimes..\otimes U(k_\sigma),
\ \sum k_i=K$. Since $\pi_2({\cal M})=
\underbrace{{\mathbb Z}\oplus...\oplus{\mathbb Z}}_{\sigma\; terms}$,
a soliton on ${\cal M}$ is now characterized by $\sigma$ winding numbers,
with each number allowed to take either sign. The two possible signs
for $k_i$ in $v^{(i)}_{k_i}$ reflect this freedom.

%\end{document}

\chapter{Fuzzy Gauge Theories}

Gauge transformations on commutative spaces are based on transformations which depend on space-time points $P$. Thus if $G$ is a 
conventional global group, the associated gauge group is the group of maps ${\cal G}$ from space-time to ${\cal G}$, the group 
multiplication being point-wise multiplication. For each irreducible representation (IRR) $\sigma$ of $G$, there is an IRR $\Sigma$ 
of ${\cal G}$ given by 
$\Sigma (g \in {\cal G}) (p) = \sigma (g(p))$. The construction works for any connected Lie group $G$. There is no problem in 
composing representations of $G$ either: if $\Sigma_i$ are representations of ${\cal G}$ associated with representations of $\sigma_i$
of $G$, then we can define the representations $\Sigma_1 {\hat \otimes} \Sigma_2$ which has the same relation to $\sigma_1 \otimes
\sigma_2$ that $\Sigma_i$ have to $\sigma_i$:  $\Sigma_1 {\hat \otimes} \Sigma_2 (g) (p) = \lbrack \sigma_1 \otimes \sigma_2 \rbrack
(g (p) ) = \sigma_1 (g(p)) \otimes \sigma_2 (g(p))$. Thus such products of $\Sigma$ are defined using those of $G$ at each $p$.
Existence of these products is essential to describe gauge theories of particles and fields transforming by different representations
of $G$.

An additional point of significance is that there is no condition on $G$, except that it is a compact connected Lie group.

For general noncommutative manifolds, several of these essential features of ${\cal G}$ are absent. Thus in particular
\begin{itemize}
\item Noncommutative manifolds require $G$ to be a $U(N)$ group,
\item Only a very limited and quite inadequate number of representations of the gauge group can be defined.
\end{itemize}

We shall illustrate these points below for the fuzzy gauge groups ${\cal G}_F$ based on $S^2_F$, but one can see the generalities
of the considerations.

There is an important map, the Seiberg-Witten(SW), map for a noncommutative deformation of ${\mathbb R}^N$.
In that case the deformed algebra ${\mathbb R}_\theta^N$ depends continously on a parameter $\theta$, becoming
the commutative algebra for $\theta =0$. If a certain gauge group on ${\mathbb R}_\theta^N$ is ${\cal G}_\theta$, it becomes a 
standard gauge group ${\cal G}_0$ on ${\mathbb R}_0^N = {\mathbb R}^N$. The SW map is based on a homomorphism from 
${\mathbb R}_\theta^N$ to ${\mathbb R}_0^N$ and connects gauge theories for different $\theta$. The aforementioned problems can 
be more or less overcome on ${\mathbb R}_\theta^N$ using this map.

But fuzzy spheres have no continuous parameter like $\theta$. What plays the role of $\theta$ is $\frac{1}{L}$ where $2 L$ is the
cut-off angular momentum, and $\frac{1}{L}$ assumes discrete values. Fuzzy spheres have no SW map as originally conceived, and we can
not circumvent its gauge-theoretic problems along the lines for ${\mathbb R}_\theta^N$.

There is however a complementary positive feature of fuzzy spaces. While $S_F^2$ for example presents problems in describing particles
of charge $\frac{1}{3}$ and $\frac{2}{3}$ at the same time (because we can not ``tensor'' representations of the fuzzy $U(1)$ gauge 
group ${\cal G}_F(U(1))$), we can describe particles with differing magnetic charges. The projective modules for all magnetic
charges were already explained in Chapter 5 and 6. There is no symmetry (``duality'') here between electric and magnetic charges.

\section{Limits on Gauge Groups}

The conditions on gauge groups on the fuzzy sphere arise algebraically. They can be understood at the Lie algebraic level.

If $\lbrace \lambda_a \rbrace$ are the basis for the Lie algebra of $G$ in a representation $\sigma$, the Lie algebra 
of ${\cal G}_F$, the fuzzy gauge group of $G$ are generated by 
\be
\lambda_a \xi_a 
\ee
where $\xi_a$ are $(2L+1) \times (2L +1)$ matrices. $\xi_a$ become functions on $S^2$ in the large $L$-limit.

Now consider the commutator
\be 
\lbrack \lambda_a \xi_a \,, \lambda_b \eta_b \rbrack \,, \quad \quad  \eta_b = (2L+1) \times (2L+1) \, \, \mbox{matrix}
\ee
of two such Lie algebra elements. We get 
\begin{gather}
\lbrack \lambda_a \,, \lambda_b \rbrack \xi_a \eta_b + \lambda_a \lambda_b \lbrack \xi_a \,, \eta_b \rbrack 
= i C_{ab}^c \xi_a \eta_b \lambda_c + \lambda_a \lambda_b \lbrack \xi_a \,, \eta_b \rbrack \,, \nonumber \\
C_{ab}^c = \mbox{structure constants of the Lie algebra of} \, \, {\cal G} \,.
\end{gather}
Since $ C_{ab}^c \xi_a \eta_b \in S_F^2$, the first term is of the appropriate form for a fuzzy gauge group of $G$. But the last 
term is not, it involves $\lambda_a \lambda_b$ which is a product of two generators. By taking repeated commutators, we will generate 
products of all orders and their commutators. If $\sigma$ is irreducible and of dimension $d$, we will get all the $d \times d$  
hermitian matrices this way and not just the $\lambda_a$. That means that the fuzzy gauge group is that of $U(d)$.

In the commutative limit, $\lbrack \xi_a \,, \eta_b \rbrack$ is zero and this problem does not occur.

This escalation of  the gauge group to $U(d)$ is difficult to
control. No convincing proposal to minimize its effect exists. [But
  see \cite{chaichian1}].

In any case, $U(d)$ gauge theories without matter fields can be consistently formulated on fuzzy spheres.

For applications, there is one mitigating circumstance: In the
standard model, if we gauge just $SU(3)_C$ and $U(1)_{EM}$, namely
the $SU(3)$ of colour and $U(1)$ of electromagnetism, the group is
actually $U(3)$ \cite{nonabelianmonopole}. Likewise, the weak group 
is not $SU(2) \times U(1)$, but $U(2)$. Thus gauge fields without
matter in these sectors can be studied on fuzzy spheres.

Unfortunately, this does not mean that these gauge theories can be 
formulated satisfactorily on $S_F^2$ or (for a four-dimensional
continuum limit) on $S_F^2 \times S_F^2$ say, when quarks and leptons
are included. For example with different flavours, different 
charges like $2/3$ and $- 1 / 3$ occur, and there is no good way to
treat arbitrary representations of gauge groups 
in noncommutative geometry \cite{chaichian1}. We explain this problem now.

\section{Limits on Representations of Gauge Groups}
 
For the fuzzy $U(d)$ gauge group on fuzzy sphere $S_F^2 (2L+1)$, we
consider $S_F^2 (2L+1) \otimes {\mathbb C}^d$. The fuzzy $U(d)$ gauge
group $U(d)_F$ consists of $d \times d$ matrices $U$ with coefficients 
in $S_F^2(2L+1): U_{ij} \in S_F^2 (2L+1)$.  The  $U(d)_F$ can
act in three different ways on  $S_F^2 (2L+1) \otimes {\mathbb C}^d$ : 
on left, right and both:
\begin{itemize}
\item[ {\it i.}] Left action : $U \rightarrow U^L$ where $U^L X = U X$ for $X \in S_F^2 (2L+1) \otimes {\mathbb C}^d$ \,,
\item[{\it ii.}] Right action : $U \rightarrow (U^\dagger)^R : (U^\dagger)^R X = x U^\dagger \,,$    
\item[{\it iii.}] Adjoint action : $U \rightarrow AdU : AdU \, X = U x U^\dagger$ \,.
\end{itemize}
If ${\it i.}$ gives representation $\Lambda$, then ${\it ii.}$ is its complex conjugate $\lambda^*$ and ${\it iii.}$ is its 
adjoint representation $Ad \, \lambda$. We are guaranteed that these representations can always be constructed.

But can we construct other representations such as the one
corresponding to $\Sigma_1 {\widehat \otimes} \Sigma_2$? The answer 
appears to be no.

The reason is as follows ${\widehat \otimes}$ is not the tensor product $\otimes$. In $\Sigma \otimes \Sigma$, we get functions of
two variables $p$ and $q$: $(\Sigma(g) \otimes \Sigma(g) ) (p,q) = \sigma (g(p)) \otimes \sigma (g(q))$. We must restrict  
$(\Sigma(g) \otimes \Sigma(g))$ to the diagonal points $(p,p)$ to get ${\widehat \otimes}$.

In noncommutative geometry, the tensor product $\Lambda_1  \otimes \Lambda_2$ exists of course since $\Lambda_1(U) \otimes 
\Lambda_2(U)$ is defined, and gives a representation of $U(d)_F$.
But noncommutative geometry has no sharp points. That obstructs the construction of an analogue of diagonal points, or the 
restriction of $\otimes$ to an analogue of ${\widehat \otimes}$.

There exist proposals \cite{chaichian1} to get around this problem using Higgs fields.

\section{Connection and Curvature}

As a convention we choose the gauge potential to act on the left of $S_F^2 (2L+1) \otimes {\mathbb C}^d$. So the components of the
gauge potentials are 
\be
A_i^L = (A_i^L)^a \lambda_a \,, \quad (A_i^L)^a \in S_F^2(2l+1) \,.
\ee
where $\lambda_a \,, (a= 1 \,, \cdots \,, d^2)$ are the $d \times d$ basis matrices for the Lie algebra of $U(d)$. They can be the 
Gell-Mann matrices.

The covariant derivative $\nabla$ is then the usual one:
\be
\nabla_i = {\cal L}_i + A_i^L 
\ee
The curvature is 
\beqa
F_{ij} &=& \lbrack \nabla_i \,, \nabla_j \rbrack - i \varepsilon_{ijk} \nabla_k \nonumber \\
&=& \lbrack {\cal L}_i \,, {\cal L}_j \rbrack + {\cal L}_i A_j^L  - {\cal L}_j A_i^L + \lbrack A_i^L \,, A_j^L \rbrack -i 
\varepsilon_{ijk} ({\cal L}_k + A_k^L) \nonumber \\
&=& {\cal L}_i A_j^L - {\cal L}_j A_i^L + \lbrack A_i^L \,, A_j^L \rbrack - i \varepsilon_{ijk} A_k^L \,.
\label{eq:curvature}
\eeqa
The subtraction of $i \varepsilon_{ijk} \nabla_k$ is needed to cancel the $\lbrack {\cal L}_i \,, {\cal L}_j \rbrack$ term in
$ \lbrack \nabla_i \,, \nabla_j \rbrack$.

There is one important condition on $\nabla_i$. On $S^2$, $A^L$ becomes a commutative gauge field $a$ and its components $a_i$ have
to be tangent to $S^2$:
\be
x_i a_i = 0  \,.
\label{eq:cvpn}
\ee
We need a condition on $\nabla_i$ which becomes this condition for large $L$. 

A simple condition of such a nature is due to Nair and Polychronakos \cite{Nair-Poly} and reads
\be
(L_i^L + A_i^L )^2 = L( L +1 ) \,.
\label{eq:vpn1}
\ee
This is compatible with gauge invariance. Its expansion is 
\be
L_i^L A_i^L + A_i^L L_i^L + A_i^L A_i^L = 0 \,.
\label{eq:vpn}   
\ee
We have that $\frac{A_i^L}{L} \rightarrow 0$ as $L \rightarrow \infty$. Dividing (\ref{eq:vpn}) by $L$ and passing to the limit, we
thus get (\ref{eq:cvpn}).

The fuzzy Yang-Mills action is 
\be
{\mathcal S}_F = \frac{1}{4 e^2} Tr F_{ij}^2 + \lambda ( \nabla^2_i- L(L+1)) \,, \quad \lambda \geq 0 \,,
\label{eq:gaugeac}
\ee
where the second term is a Lagrange multiplier: it enforces the constraint (\ref{eq:vpn1}) as $\lambda \rightarrow \infty$.

\section{Instanton Sectors}

The above action is good in the sector with no instantons. But $U(d)$ gauge theories on $S^2$ have instantons, or equivalently, 
twisted $U(1)$-bundles on $S^2$. We outline how to incorporate instantons on the fuzzy sphere, taking $d =1$ for simplicity.

The projective modules for instanton sectors were constructed previously. We review it briefly constructing the modules in a different
(but Morita equivalent) manner.

The instanton sectors on $S^2$ correspond to $U(1)$ bundles thereon. To build the corresponding projective module for Chern number
$2 T \in {\mathbb Z}^+$, introduce ${\mathbb C}^{2T +1}$ carrying the angular momentum $T$ representation of $SU(2)$. Let
$T_i$ be the angular momentum operators in this representation with standard commutation relations. Let $Mat (2L+1) \otimes
{\mathbb C}^{(2T+1)} \equiv Mat(2L+1)^{(2T+1)}$. We let $P^{L+T}$ be the projector coupling left angular momentum operators $L^L$ 
and $T$ to produce maximum angular momentum $L + T$. Then the projective module $P^{L+T} Mat(2L+1)^{(2T+1)}$ is a
fuzzy analogue of sections of $U(1)$ bundles on $S^2$ with Chern number $2 T > 0$ \cite{monopole}. If instead we couple $L^L$ 
and $T$ to produce the least angular momentum $L - T$ using the projector $P^{L-T}$, then the projective module
$P^{L-T} Mat(2L+1)^{(2T+1)}$ corresponds to Chern number $-2 T$. (We
assume that $L \geq T$). 
    
The derivation ${\cal L}_i$ does not commute with $P^{L \pm T}$ and has no action on these modules. But, ${\cal J}_i = {\cal L}_i + 
T_i$ does commute with $P^{L \pm T}$. Thus ${\cal L}_i$ must be replaced with ${\cal J}_i$ in further considerations. ${\cal J}_i$ is
to be considered the total angular momentum . The addition of $T_i$ to ${\cal L}_i$ here is the algebraic analogue of 
``mixing of spin and isospin''. \cite{Hasenfratz, jr}.
It is interesting that the mixing of `spin and isospin' occurs already in our finite-dimensional matrix
model and does not need noncompact spatial slices and spontaneous
symmetry breaking.

We must next gauge ${\cal J}_i$. In the zero instanton sector, the fuzzy gauge fields $A_i^L$ were functions of $L_i^L$. But that is 
not possible now since $A_i^L$ does not commute with $P^{L \pm T}$. Instead we require $A_i^L$ to be a function of  
$\vec{L}^L + \vec{T}$ and write for the covariant derivative
\be
{\bf \nabla}_i = {\cal J}_i + A_i^L \,.
\ee
When $L\rightarrow \infty$, $\vec{T}$ can be ignored, and then $A_i^L$ becomes a function of just $x$ as we want.

The transversality condition must be modified. It is now
\be
(L_i^L + T_i + A_i^L )^2 = (L_i^L + T_i)^2  
\ee
where
\be
(L_i^L + T_i)^2 = (L \pm T) ( L \pm T +1)
\ee  
on $P^{L \pm T} Mat(2L+1)^{(2T+1)}$.

The curvature $F_{ij}$ and the action ${\mathcal S}_F$ are as in (\ref{eq:curvature}) and (\ref{eq:gaugeac}).

\section{The Partition Function and the $\theta$-parameter}  

Existence of instanton bundles on a commutative manifold brings in a new parameter, generally called $\theta$ as, in $QCD$. The
partition function $Z_\theta$ depends on $\theta$.

Let us denote the action in the instanton number $K \in {\mathbb Z}$ sector by ${\mathcal S}_F^K$. Then
\be
Z_\theta = \sum_k \int D A_i^L e^{-{\mathcal S}_F^K + i K \theta } \,.
\ee

We thus have a matrix model for $U(d)$ gluons.

In the continuum, $K$ can be written as the integral of curvature $tr F$ (where trace $tr$ (with lower case $t$) is over the internal 
indices). In four dimensions it is the integral of $tr F \wedge F$. But on $S_F^2$, $Tr \varepsilon^{ij} F_{ij}$ is not an
integer. A similar difficulty arises for $S_F^2 \times S_F^2$ or ${\mathbb C}P^2$.

In continuum gauge theory, $F$ and $F \wedge F$ play a role in discussions of chiral symmetry
breaking. They arise as the local anomaly term in the continuity equation for chiral current. Therefore although $Z_\theta$ defines
the theory, it is still helpful to have fuzzy analogues of the topological densities $tr F$ and 
$tr F \wedge F$.

It is possible to construct fuzzy topological densities using cyclic cohomology \cite{connes}. 
We will not review cyclic cohomology here.

%\end{document}

%\begin{document}

\chapter{The Dirac Operator and Axial Anomaly}

\section{Introduction}

The Dirac operator is central for fundemental physics. It is also
central in noncommutative geometry. In Connes' approach 
\cite{connes}, it is possible to formulate metrical, differential
geometric and bundle-theoretic ideas using the Dirac operator
in a form generalisable to noncommutative manifolds.

In this chapter, we explain the theory of the fuzzy Dirac operator
basing it on the Ginsparg-Wilson (GW) algebra \cite{ginsparg}. 
This algebra appeared first in the context of lattice gauge theories
as a device to write the Dirac operator overcoming the 
well-known fermion-doubling problem. The same algebra appears
naturally for the fuzzy sphere. The theory of the fuzzy Dirac operator 
can be based on this algebra. It has no fermion doubling and correctly 
and elegantly reproduces the integrated $U(1)_A$-(axial) anomaly.

Incidentally the association of the GW-algebra with the fuzzy sphere
is surprising as the latter is not designed with this algebra in mind.

Below we review the GW-algebra in its generality. We then adapt it to
$S_F^2$. Our discussion here closely follows \cite{giorgio}.

\section{A Review of the Ginsparg-Wilson Algebra.}

In its generality, the Ginsparg-Wilson algebra $\cal A$ can be defined
as the unital $*$-algebra over $\mathbb C$ generated by two
$*$-invariant involutions $\Gamma$ and $\Gamma'$:
\be
{\cal A}= \big \langle\Gamma,\Gamma':\quad \Gamma^2={\Gamma'}^2=\BI,\quad 
\Gamma^*=\Gamma,\quad {\Gamma'}^*=\Gamma' \big \rangle\ ,
\label{gwfi}
\ee
$*$ denoting the adjoint. The unity of $\cal A$ has been indicated by $\BI$.

In any such algebra, we can define a Dirac operator
\be 
D'=\frac{1}{a}\Gamma(\Gamma+\Gamma')\ ,
\label{gwfii}
\ee
where $a$ is the ``lattice spacing''. It fulfills
\be
{D'}^*=\Gamma\, D'\,\Gamma,\quad \lbrace \Gamma,D' \rbrace = a \,D'\,\Gamma\,D'\ .
\label{gwfiii}
\ee
(\ref{gwfii}) and (\ref{gwfiii}) give the original formulation \cite{ginsparg}.  But they are equivalent to (\ref{gwfi}), since 
(\ref{gwfii}) and (\ref{gwfiii}) imply that
\be
\Gamma'=\Gamma(aD')-\Gamma
\label{gwfiv}
\ee
is a $*$-invariant involution \cite{luscher} \cite{fujikawa}.

Each representation of (\ref{gwfi}) is a particular realization of the Ginsparg-Wilson algebra. Representations of physical 
interest are reducible.

Here we choose
\be
D=\frac{1}{a}(\Gamma+\Gamma')\ ,
\label{gwfv}
\ee
instead of $D'$ as our Dirac operator, as it is self-adjoint and has the desired continuum limit.

From $\Gamma$ and $\Gamma'$, we can construct the following elements of
$\cal A$:
\beqa
\Gamma_0&=&\frac{1}{2}\lbrace \Gamma,\Gamma' \rbrace \ ,   \label{gwfvi}\\
\Gamma_1&=&\frac{1}{2}(\Gamma+\Gamma')\ ,   \label{gwfvii}\\
\Gamma_2&=&\frac{1}{2}(\Gamma-\Gamma')\ ,   \label{gwfviii}\\
\Gamma_3&=&\frac{1}{2i}[\Gamma,\Gamma']\ .   \label{gwfix}
\eeqa

Let us first look at the centre ${\cal C}({\cal A})$ of $\cal A$ in terms of these operators. It is generated by $\Gamma_0$ which 
commutes with $\Gamma$ and $\Gamma'$ and hence with every element of $\cal A$\,.
$\Gamma_i^2,\ i=1,2,3$ also commute with every element of  $\cal A$, but they are not independent of $\Gamma_0$. Rather,
\beqa
\Gamma_1^2=\frac{1}{2}(\BI+\Gamma_0) &,&\label{gwfx}\\
\Gamma_2^2=\frac{1}{2}(\BI-\Gamma_0) &,&\label{gwfxi}\\ 
\rightarrow\quad \Gamma_1^2+\Gamma_2^2=\BI\ &,&\label{gwfxii}\\
\Gamma_0^2+\Gamma_3^2=\BI\ &.&\label{gwfxiii}
\eeqa
Notice also that
\be
\lbrace \Gamma_i,\Gamma_j \rbrace = 0 \ ,\ i,j=1,2,3,\ i\ne j\ .
\label{gfwxiv}
\ee

From now on by $\cal A$ we will mean a representation of $\cal A$.

The relations (\ref{gwfx})-(\ref{gwfxiii}) contain spectral information. From (\ref{gwfxiii}) we see that
\be
-1\le\Gamma_0\le 1\ ,
\label{gwfxvii}
\ee
where the inequalities mean that the eigenvalues of $\Gamma_0$ are accordingly bounded. By (\ref{gwfx}), this implies that the 
eigenvalues of $\Gamma_1$ are similarly bounded.

We now discuss three cases associated with (\ref{gwfxvii}).

\vskip 1em

\noindent{\it Case 1} : 

\vskip 1em

$\Gamma_0= \BI$.\ Call the subspace where $\Gamma_0= \BI$ as $V_{+1}$. On $V_{+1}$, $\Gamma_1^2=\BI$ and
$\Gamma_2=\Gamma_3=0$ by (\ref{gwfx}-\ref{gwfxiii}). This is subspace of the top modes of the operator $|D|$.

\vskip 1em

\noindent{\it Case 2} :

\vskip 1em

$\Gamma_0= -\BI$.\ Call the subspace where $\Gamma_0= -\BI$ as $V_{-1}$. On $V_{-1}$, 
$\Gamma_2^2=\BI$ and $\Gamma_1=\Gamma_3=0$ by (\ref{gwfx}-\ref{gwfxiii}). This is the subspace of zero modes of the Dirac operator 
$D$.

\vskip 1em

\noindent{\it Case 3} :

\vskip 1em

$\Gamma_0^2\ne \BI$. \ Call the subspace where $\Gamma_0^2\ne \BI$ as $V$. On this subspace,  $\Gamma_i^2\ne 0$ 
for $i=1,2,3$ by (\ref{gwfix}-\ref{gwfxii}), and therefore  
\be
sign\,\Gamma_i=\frac{\Gamma_i}{|\Gamma_i|}\ ,\quad |\Gamma_i|=\hbox{positive square root of}\ \Gamma_i^2
\label{gwfxv}
\ee
are well defined and by (\ref{gfwxiv}) generate a Clifford algebra on $V$: 
\be
\lbrace sign\,\Gamma_i,sign\,\Gamma_j \rbrace = 2\delta_{ij}\ .
\label{gwfxvi}
\ee

Consider $\Gamma_2$. It anticommutes with $\Gamma_1$ and $D$. Also
\be
\tr\,\Gamma_2=(\tr_V+\tr_{V_{+1}}+\tr_{V_{-1}})\Gamma_2\ ,
\label{gwfxviii}
\ee
where the subscripts refer to the subspaces over which the trace is taken. These traces can be calculated:
\beqa
\tr_V\Gamma_2&=&\tr_V(sign\,\Gamma_i)\Gamma_2(sign\,\Gamma_i)
\quad (i\ \hbox{fixed,}\ \ne 2) \nn\\
&=&-\tr_V\Gamma_2\quad \hbox{by} (\ref{gwfxvi})\nn\\
&=&0,     \label{gwfxix}\\           
\tr_{V_{+1}}\Gamma_2&=&0,\quad \hbox{as}\ \Gamma_2=0\ \ \hbox{on}\ 
V_{+1}\ .
\label{gwfxx}
\eeqa
So
\be
\tr\Gamma_2=\tr_{V_{-1}}\Gamma_2=\tr_{V_{-1}}(\frac{1+\Gamma_2}{2}-
\frac{1-\Gamma_2}{2})=\ \hbox{index of}\ \Gamma_1\ .
\label{gwfxxi}
\ee

Following Fujikawa \cite{fujikawa}, we can use $\Gamma_2$ as the generator of chiral transformations. It is not involutive on 
$V\oplus V_{+1}$
\be 
\Gamma_2^2=\BI-\frac{\BI+\Gamma_0}{2}\ .
\label{gwfxxii}
\ee
But this is not a problem for fuzzy physics. In the fuzzy model below, in the continuum limit, $\Gamma_0\to -\BI$ on all states with 
$|D|\le $ a fixed `energy' $E_0$ independent of $a$ (and is $-\BI$ on $V_{-1}$ where $D=0$). We can see this as follows. 
$\Gamma_1=aD$, so that if $|D|\le E_0, \ \Gamma_1\to 0$ as $a\to 0$. Hence by (\ref{gwfx},\ref{gwfxii}), $\Gamma_0\to -\BI$ and 
$\Gamma_2^2\to\;\BI$ on these levels.

There are of course states, such as those of $V_{+1}$, on which $\Gamma_2^2$ does not go to $\BI$ as $a\to 0$. But their (Euclidean)
energy diverges and their contribution to functional integrals vanishes in the continuum limit.

We can interpret (\ref{gwfxxii}) as follows. The chiral charge of levels with $D\ne 0$ gets renormalized in fuzzy physics. For levels
with $|D|\le E_0$, this renormalization vanishes in the naive continuum limit.

We note that the last feature is positive: it resolves a problem in
faced in \cite{fermion}, where all the top modes had to be 
projected out because of insistence that chirality squares to $\BI$ on $V_{+1}$ (see below).

For Dirac operators of maximum symmetry, $\Gamma_0$ is a function of the conserved total angular momentum $\vec J$ as we shall show. 
It increases with $\vec{J}^2$ so that $V_{+1}$ consists of states of maximum $\vec{J}^2$. This maximum value diverges as $a\to 0$ as 
the general argument above shows.

%%%%%%%%%%%%%%%%%%%%%%%%%%%%%%%%%%%%%%%%%%%%%%%%%%
\section{Fuzzy Models}

\subsection{Review of the Basic Algebra}

Let us briefly recollect the basic algebraic details.

The algebra for the fuzzy sphere characterized by cut-off $2L$ is the full matrix algebra $Mat(2L+1)\equiv M_{2L+1}$ of 
$(2L+1)\times (2L+1)$ matrices. On $M_{2L+1}$, the $SU(2)$ Lie algebra acts either on the left or on the
right. Call the operators for left action as $L^L_i$ and for right action as $L^R_i$. We have
\be
L^L_ia=L_i a \ ,\  L^R_ia=aL_i\  ,\  a\in M_{2L+1}\ ,\nn
\ee
\be
[L_i^L,L_j^L]=i\epsilon_{ijk}L^L_k\  , \quad
[L_i^R,L_j^R]=-i\epsilon_{ijk}L^R_k\ ,\quad 
(L_i^L)^2=(L_i^R)^2=L(L+1)\BI\ ,
\label{gwfxxiii}
\ee
where $L_i$ is the standard matrix for the $i$-th component of the angular momentum in the the $(2L+1)$-dimensional irreducible
representation (IRR). The orbital angular momentum which becomes $-i(\vec r\wedge\vec\nabla)_i$ as $L\to\infty$ is
\be
{\cal L}_i=L_i^L-L_i^R\ ,\quad {\cal L}_ia=[L_i,a]\ .
\label{gwfxxiv}
\ee

As $L\to\infty$, both $\vec L^L/L$ and  $\vec L^R/L$ approach the unit vector $\hat x$ with commuting components:
\be
\frac{\vec L^{L,R}}{L}\  
{\lower .7ex\hbox{$\;\stackrel{\longrightarrow}
{\scriptstyle L\to\infty }\;$}}
\ \hat x\ ,\qquad \hat x\cdot\hat x=1\ ,\quad 
[\hat x_i,\hat x_j]=0\ .
\label{gwfxxiva}
\ee
$\hat x$ labels a point on the sphere $S^2$ in the continuum limit.

\subsection{The Fuzzy Dirac Operator (No Instantons or Gauge Fields)}

Consider $M_{2L+1}\otimes{\mathbb C}^2$. ${\mathbb C}^2$ is the carrier of the spin $1/2$ representation of $SU(2)$ with generators
$\frac{1}{2}\sigma_i,\ \sigma_i=$ Pauli matrices. We can couple its spin $1/2$ and the angular momentum $L$ of $L^L_i$ to the value
$L+1/2$. If $(1+\Gamma)/2$ is the corresponding projector, then
\cite{fermion} \cite{watamura} \cite{monopole}
\be
\Gamma=\frac{\vec\sigma\cdot\vec L^L+1/2}{L+1/2} \ .
\label{gwfxxv}
\ee
$\Gamma$ is a self-adjoint involution,
\be
\Gamma^*=\Gamma\quad,\quad\Gamma^2=\BI\  .
\label{gwfxxvi}
\ee

There is likewise the projector $(\BI+\Gamma')/2$ coupling the spin $1/2$ of ${\mathbb C}^2$ and the right angular momentum $-L^R_i$ 
to $L+1/2$, where
\be
\Gamma'=\frac{-\vec\sigma\cdot\vec L^R+1/2}{L+1/2}={\Gamma'}^*\, \quad{\Gamma'}^2=\BI\ . 
\label{gwfxxvii}
\ee
The algebra $\cal A$ is generated by $\Gamma$ and $\Gamma'$.

The fuzzy Dirac operator of Grosse et al.\cite{GKP1} is
\be
D=\frac{1}{a}(\Gamma+\Gamma')=\frac{2}{a}\Gamma_1=\vec\sigma\cdot(\vec L^L-\vec L^R)+1\ ,\quad a=\frac{1}{L+1/2}\ .
\label{gwfxxviii}
\ee
Thus the Dirac operator is in this case  an element of the Ginsparg-Wilson algebra $\cal A$.

We can calculate $\Gamma_0$ in terms of $\vec J=\vec{\cal L}+\vec\sigma/2$:
\be
\Gamma_0=\frac{a^2}{2}[\vec J^2-2L(L+1)-\frac{1}{4}]\ .
\label{gwfxxix}
\ee
Thus the eigenvalues of $\Gamma_0$ increase monotonically with the eigenvalues $j(j+1)$ of $\vec J^2$ starting with a minimum 
for $j=1/2$ and attaining a maximum of $1$ for $j=2L+1/2$.

$\Gamma_2$ is the chirality. It anticommutes with $D$. For fixed $j$, as $L\to\infty$, $\Gamma_0\to -\BI$ and $\Gamma_2^2=\BI$ as 
expected. In fact, $\Gamma_2$ in the naive continuum limit is the standard chirality for fixed $j$. As $L\to\infty, \ \Gamma_2\to
\sigma\cdot\hat x$. As mentioned earlier, use of $\Gamma_2$ as chirality resolves a difficulty addressed elsewhere
\cite{fermion}, where $sign\,(\Gamma_2)$ was used as chirality. That necessitates projecting out $V_{+1}$ and creates
a very inelegant situation.

Finally we note that there is a simple reconstruction of $\Gamma$ and
$\Gamma'$ from their continuum limits 
\cite{ydrithesis}. If $\vec x$ 
is not normalized, $\vec\sigma\cdot\hat x= \frac{\vec\sigma\cdot \vec x}{|\vec\sigma\cdot \vec x|},\ |\vec\sigma\cdot 
\vec x|\equiv|\big((\vec\sigma\cdot \vec x)^2\big)^{1/2}|$. As $\vec x$ can be represented by $\vec L^L$ or $\vec L^R$ in fuzzy 
physics, natural choices for $\Gamma$ and $\Gamma'$ are $sign\,(\vec\sigma\cdot L^L)$ and $-sign\,(\vec\sigma\cdot L^R)$. The first 
operator is $+1$ on vectors having $\vec\sigma\cdot \vec L^L>0$ and $-1$ if instead $\vec\sigma\cdot \vec L^L<0$. 
But if $(\vec L^L+\vec\sigma/2)^2=(L+1/2)(L+3/2)$, then $\vec\sigma\cdot \vec L^L=L>0$, while if $(\vec L^L+\vec\sigma/2)^2=(L-1/2)
(L+1/2)$, $\vec\sigma\cdot \vec L^L= -(L+1)<0$. $\Gamma$ is $+1$ on former states and $-1$ on latter states.
Thus
\be sign\,(\vec\sigma\cdot \vec L^L)=\Gamma\ ,
\label{gwfxxx}
\ee
and similarly 
\be sign\,(\vec\sigma\cdot \vec L^R)=-\Gamma'\ .
\label{gwfxxxi}
\ee

It is easy to calculate the spectrum of $D$. We can write
\be
a D= \vec{{\cal J}}^2 -\vec{{\cal L}}^2 - \frac{3}{4} +1 
\ee
We observe that $\lbrack  \vec{{\cal J}}^2  \,, \vec{{\cal L}}^2 \rbrack =0$. The spectrum of $\vec{{\cal L}}^2$ is
\be
spec \, \vec{{\cal L}}^2 = \lbrace \ell(\ell +1) : \ell = 0,1, \cdots, 2 L \rbrace \,,
\label{eq:specL}
\ee
whereas that of $\vec{{\cal J}}^2$ is 
\be
spec \, \vec{{\cal J}}^2 = \left \lbrace j(j+1) : j = \frac{1}{2} ,
\frac{3}{2}, \cdots, 2L +\frac{1}{2}  \right \rbrace \,.
\ee
Here each $j$ can come from $\ell = j \pm \frac{1}{2}$ by adding spin, except $j = 2L +\frac{1}{2}$ which comes only from$\ell =2L$.
It follows that the eigenvalue of $D$ for $\ell = j - \frac{1}{2}$ is $j + \frac{1}{2} = \ell +1 \,, \ell \leq 2L$ and for
$\ell = j + \frac{1}{2}$ is $-(j + \frac{1}{2}) = - \ell  \,, \ell \leq 2L$.

The spectrum found here agrees {\it exactly} with what is found in the continuum for $j \leq 2L - \frac{3}{2}$. For 
$j = 2L + \frac{1}{2}$ we get the positive eigenvalue correctly, but the negative one is missing. That is an edge effect caused 
by cutting off the angular momentum at $2L$.   

%We emphasize that this spectrum agrees completely with the spectrum of
%the continuum Dirac operator, except at the $j=(2L+1/2)$ level.

\subsection{The Fuzzy Gauged Dirac Operator (No Instanton Fields)}

We adopt the convention that gauge fields are built from operators on $Mat(2L+1)$ which act by left multiplication. For $U(k)$ gauge 
theory, we start from $Mat(2L+1)\otimes{\mathbb C}^k$. The fuzzy gauge fields $A_i^L$ are $k\times k$ matrices $[(A^L_i)_{mn}]$ where 
each entry is the operator of left-multiplication by $(A_i)_{mn}\in\;Mat(2L+1)$ on $Mat(2L+1)$. $A^L_i$
thus acts on $\xi=(\xi_1,\ldots,\xi_k), \ \xi_i\in\;Mat(2L+1)$ according to
\be
(A_i^L\xi)_m=(A_i)_{mn}\xi_n\ .
\label{gwfxxxii}
\ee
The gauge-covariant derivative is then
\be
\nabla_i(A^L)={\cal L}_i+A^L_i=L_i^L-L_i^R+A_i^L\ .
\label{gwfxxxiv}
\ee
Note how only the left angular momentum is augmented by a gauge field.

The hermiticity condition on $A^L_i$ is 
\be
(A^L_i)^*=A^L_i\ ,
\label{gwfxxxv}
\ee
where
\be
((A^L_i)^*\xi)_m=(A^*_i)_{nm}\xi_n\ ,
\label{gwfxxxvi}
\ee
$(A^*_i)_{nm}$ being hermitean conjugate of $(A_i)_{nm}$. The corresponding field strength $F_{ij}$ is defined by
\be
[(L+A)^L_i,(L+A)^L_j]=i\epsilon_{ijk}(L+A)^L_k+iF_{ij}\ .
\label{gwfxxxvib}
\ee

There is a further point to attend to. We need a gauge-invariant condition which in the continuum limit eliminates the component 
of $A_i$ normal to $S^2$. There are different such conditions, the following simple one was disccussed in chapter 7, 
(cf. \ref{eq:vpn1}): 
\be
(L^L_i+A^L_i)^2=(L^L_i)^2=L(L+1)\ .
\label{gwfxxxvii}
\ee

The Ginsparg-Wilson system can be introduced as follows. As $\Gamma$ squares to $\BI$, there are no zero modes for $\Gamma$ and hence 
for $\vec\sigma\cdot\vec L^L+1/2$. By continuity, for generic $\vec A^L$, its gauged version $\vec\sigma\cdot(\vec L^L+\vec A^L)+1/2$ 
also has no zero modes. Hence we can set 
\be
\Gamma(A^L)=\frac{\vec\sigma\cdot(\vec L^L+\vec A^L)+1/2}{
|\vec\sigma\cdot(\vec L^L+\vec A^L)+1/2|}\ ,\quad
\Gamma(A^L)^*=\Gamma(A^L)\ ,\quad \Gamma(A^L)^2=\BI\ .
\label{gwfxxxix}
\ee
It is the gauged involution which reduces to $\Gamma=\Gamma(0)$ for zero $\vec A^L$.

As for the second involution $\Gamma'(A^L)$, we can set
\be
\Gamma'(A^L)=\Gamma'(0)\equiv\Gamma'
\label{gwfxxxixa}
\ee

On following (\ref{gwfvi}-\ref{gwfix}), these idempotents generate the Ginsparg-Wilson algebra with operators $\Gamma_\lambda(A^L)$, 
where $\Gamma_\lambda(0)=\Gamma_\lambda$. 

The operators $\vec L^{L,R}$ do not individually have continuum limits as their squares $L(L+1)$ diverge as $L\to\infty$. In contrast 
$\vec{\cal L}$ and $\vec A^L$ do have continuum limits. This was remarked earlier on for the latter, while $\vec{\cal L}$ just becomes
orbital angular momentum. 

To see more precisely how  $D(A^L)$, the Dirac operator for gauge field $A^L$, ($D(0)$ being $D$ of (\ref{gwfxxviii})), and  
$\Gamma_2(A^L)$, behave in the continuum limit, we note that from (\ref{gwfxxxvib}),(\ref{gwfxxxvii}) 
\be
\big(\vec\sigma\cdot(\vec L^L+\vec A^L)+\frac{1}{2}\big)^2=
(L+\frac{1}{2})^2 - \frac{1}{2}\epsilon_{ijk}\sigma_iF_{ij}\ ,
\label{gwfxla}
\ee
and therefore we have the expansions 
\be
\frac{1}{|\vec\sigma\cdot(\vec L^L+\vec A^L)+\frac{1}{2}|}
=\frac{2}{\sqrt\pi}\int_0^\infty ds\,e^{-s^2
(\vec\sigma\cdot(\vec L^L+\vec A^L)+\frac{1}{2})^2}=\frac{1}{L+\frac{1}{2}}
%\big(1+\frac{1}{(2L+1)^2}\epsilon_{ijk}\sigma_iF_{jk}+..\big).
+\frac{1}{4(L+\frac{1}{2})^3}\epsilon_{ijk}\sigma_iF_{jk}+...,
\label{gwfxlb}
\ee
\beqa
D(A^L)&=&(2L+1)\Gamma_1(A^L)=
\vec\sigma\cdot(\vec L^L-\vec L^R+\vec A^L)+1 \;+
\frac{\vec\sigma\cdot(\vec L^L+\vec A^L)+\frac{1}{2}}{4(L+\frac{1}{2})^2}
\epsilon_{ijk}\sigma_kF_{ij}+..   \nn\\
\Gamma_2(A^L)&=&\frac{\vec\sigma\cdot(\vec L^L+\vec A^L)+\frac{1}{2}}
{2(L+\frac{1}{2} )}-
\frac{-\vec\sigma\cdot\vec L^R+\frac{1}{2}}{2(L+\frac{1}{2})}
\;+\;\frac{\vec\sigma\cdot(\vec L^L+\vec A^L)+\frac{1}{2}}{8(L+\frac{1}{2})^3}
\epsilon_{ijk}\sigma_kF_{ij}+...\ .\nn\\
\label{gwfxlc}
\eeqa
So in the continuum limit, $D(A^L)\to\vec\sigma\cdot(\vec{\cal L}+\vec A)+1\,$, and $\Gamma_2(A)\to\vec\sigma\cdot\hat x$, exactly 
as we want.

It is remarkable that even in the presence of gauge field, there is the operator
\be
\Gamma_0(\vec A^L)=\frac{1}{2}[\Gamma(\vec A^L),\Gamma'(\vec A^L)]_+
\label{gwfxl}
\ee
which is in the centre of $\cal A$. It assumes the role of $\vec J^2$ in the presence of $\vec A^L$. In the continuum limit, it has 
the following meaning. With $D(A^L)$ denoting the Dirac operator for gauge field $A^L$, ($D(0)$ being $D$ of (\ref{gwfxxviii})), 
$sign\,(D(A^L))$ and $\Gamma_2(A^L)$ generate a Clifford algebra in that limit and the Hilbert space splits into a direct sum
of subspaces, each carrying its IRR. $\Gamma_0(A^L)$ is a label for these subspaces.

%%%%%%%%%%%%%%%%%%%%%%%%%%%%%%%%%%%%%%%%%%
\section{The Basic Instanton Coupling}

The instanton sectors on $S^2$ correspond to $U(1)$ bundles thereon. The connection on these bundles is not unique. Those with maximum
symmetry have a particular simplicity and are therefore important for analysis.

In a similar way, on $S^2_F$, there are projective modules which in the algebraic approach substitute for sections of bundles 
\cite{connes} \cite{landi} \cite{monopole}(see chapter 5 and 6). There are particular connections on these modules
with maximum symmetry and simplicity. In this section we build the Ginsparg-Wilson system for such connections. The Dirac operator
then is also simple. It has zero modes which are responsible for the axial anomaly. Their presence will also be shown by simple 
reasoning.

To build the projective module for Chern number $2T$, $T>0$, we follow chapters 6 and 7 and introduce ${\mathbb C}^{2T+1}$ carrying 
the angular momentum $T$ representation of $SU(2)$. Let $T_\alpha,\ \alpha=1,2,3$ be the angular momentum operators in
this representation with standard commutation relations. Let $Mat(2L+1)^{2T+1}\equiv Mat(2L+1)\otimes{\mathbb C}^{2T+1}$. We let 
$P^{(L+T)}$ be the projector coupling left angular momentum operators $\vec L^L$ with $\vec T$ to produce maximum angular momentum 
$L+T$. Then the projective module $P^{(L+T)}Mat(2L+1)^{2T+1}$ is the fuzzy analogue of sections of $U(1)$ bundles on $S^2$ with 
Chern number $2T>0$ \cite{monopole}. If instead we couple $\vec L^L$ and $\vec T$ to produce the least angular momentum 
$(L-T)$ using the projector $P^{(L-T)}$, $P^{(L-T)}Mat(2L+1)^{2T+1}$ corresponds to Chern number $-2T$ (we assume that $L\ge T$).
 
We go about as follows to set up the Ginsparg-Wilson system. For $\Gamma$ we now choose
\be
\Gamma^\pm=\frac{\vec\sigma\cdot(\vec L^L+\vec T)+1/2}{L\pm T+1/2} \,,
\label{gwfxli} 
\ee
The domain of $\Gamma^\pm$ is $P^{(L\pm T)}Mat(2L+1)^{2T+1}\otimes{\mathbb C}^2$ with $\sigma$ acting on 
${\mathbb C}^2$. On this module $(\vec L^L+\vec T)^2=(L\pm T)(L\pm T+1)$  and $(\Gamma^\pm)^2=\BI$. 

As for $\Gamma'$, we choose it to be the same as in eq.(\ref{gwfxxvii}).

$\Gamma^\pm$ and $\Gamma'$ generate the new Ginsparg-Wilson
system. The operators $\Gamma_\lambda$ are defined as before as also the 
new Dirac operator $D^{(L\pm T)} = \frac{2}{a}\Gamma_1$. For $T>0$ it 
is convenient to choose
\be
a=\frac{1}{\sqrt{(L+\half)(L\pm T+\half)}}\ .
\label{gwfxlia}
\ee
 
\subsection{Mixing of Spin and Isospin}
%{\bf Line missing noted from a previous reading of Bal's}
The total angular momentum $\vec J$ which commutes with $P^{(L\pm T)}$
and hence acts on\\ $P^{(L\pm T)}Mat(2L+1)\otimes{\mathbb C}^2$
is not $\vec L^L-\vec L^R+\vec\sigma/2$, but $\vec L^L+\vec T-\vec
L^R+\vec\sigma/2$. The addition of $\vec T$ here is the algebraic 
analogue of the `mixing of spin and isospin' \cite{jr} as remarked
in chapter 7. Such a term is essential in $\vec J$ since 
$\vec L^L-\vec L^R+\vec\sigma/2$, not commuting with $P^{(L\pm T)}$, 
would not preserve the modules. 

%It is interesting that a mixing of `spin and isospin' occurs already in our finite-dimensional matrix model
%and does not need noncompact spatial slices and spontaneous symmetry breaking.

\subsection{The Spectrum of the Dirac operator}

The spectrum of $\Gamma_1$ and $D^{(L\pm T)}$ can be derived simply by angular momentum addition, confirming the results of section 2.
On the   $P^{(L\pm T)}Mat(2L+1)^{2T+1}$ modules, $(\vec L^L+\vec T)^2$ has the fixed values $(L\pm T)(L\pm T+1)$, and 
\beqa
(\Gamma_1)^2&=&\frac{1}{(2(L\pm T)+1)(2L+1)}\big((\vec L^L+\vec T-\vec L^R+
\frac{1}{2}\vec\sigma)^2+\frac{1}{4}-T^2\big)\ ,
\label{ai}
\\
\Gamma^\pm&=&\frac{(\vec L^L+\vec T+\frac{1}{2}\vec\sigma)^2-(L\pm T)(L\pm T+1)
-\frac{1}{4}}{(L\pm T)+\frac{1}{2}}\ ,
\label{aii}\\
\Gamma'&=&\frac{(-\vec L^R+\frac{1}{2}\vec\sigma)^2-L(L+1)-\frac{1}{4}}{L+
\frac{1}{2}}\ .
\label{aiii}
\eeqa
Comparing (\ref{ai}) with (\ref{gwfx}) we see that the `total angular momentum' $(\vec J)^2=(\vec L^L+\vec T-\vec L^R+\frac{1}{2}
\vec\sigma)^2$ is linearly related to $\Gamma_0=\frac{1}{2}[\Gamma^\pm,\Gamma']_+$. 
The eigenvalues $(\gamma_1)^2$ of $(\Gamma_1)^2$ are determined by those of  $(\vec J)^2$, call them $j(j+1)$.

For $j=j_{max}=L\pm T+L+ \frac{1}{2}$ we have $(\Gamma_1)^2=1$, so this is $V_{+1}$, and the degeneracy is $2j_{max}+1=2(2L\pm T+1)$. 
The maximum value of $j$ can be achieved only if
\be
(\vec L^L+\vec T+\frac{1}{2}\vec\sigma)^2=(L\pm T+\frac{1}{2})
(L\pm T+\frac{3}{2})\ ,\quad (-\vec L^R+
\frac{1}{2}\vec\sigma)^2=(L+\frac{1}{2})(L+\frac{3}{2})\ .
\label{aiiia}
\ee 
Replacing these values in (\ref{aii},\ref{aiii}) we see that on $V_{+1}$ we have $\gamma_1=1$, and $\Gamma_2=0$.

The case $T=0$ has been treated before \cite{GKP1}\cite{monopole}\cite{fermion}.So we here assume that $T>0$. In that case,
for either module $j_{min}=T-\frac{1}{2}$, which gives an eigenvalue $(\gamma_1)^2=0$ with degeneracy $2T$; we are in $V_{-1}$, 
the space of the zero modes. To realize this minimum value of $j$ we must have
\be
(\vec L^L+\vec T+\frac{1}{2}\vec\sigma)^2=(L\pm T\mp\frac{1}{2})
(L\pm T\mp\frac{1}{2}+1)\ ,\quad (-\vec L^R+
\frac{1}{2}\vec\sigma)^2=(L\pm\frac{1}{2})(L\pm\frac{1}{2}+1)\ .
\label{aiiib}
\ee
Replacing these values in (\ref{aii}, \ref{aiii}) we find that on the corresponding eigenstates $\Gamma_2=\mp 1$: they are all 
either chiral left or chiral right. These are the results needed by continuum index theory and axial anomaly.

For $j_{min}<j<j_{max}$, that is on $V$, we have $0<(\gamma_1)^2<1$, and by (\ref{gwfxii}), $\Gamma_2\ne 0$. Since 
$[\Gamma_1,\Gamma_2]_+=0$, to each state $\psi$ such that $\Gamma_1\psi= \gamma_1\psi$ corresponds a state $\psi'=\Gamma_2\psi$ 
such that $\Gamma_1\psi'=-\gamma_1\psi'$.

For any value of $j$ we can write $j=n+T-\frac{1}{2}$ with $n=0,1,..., 2L+1$ when the projector is $P^{(L+T)}$, and  
$n=0,1,...,2(L-T)+1$ when the projector is $P^{(L-T)}$, while correspondingly,
\be
(\gamma_1)^2=\frac{n(n+2T)}{(2(L\pm T)+1)(2L+1)}\ .
\label{aiv}
\ee
With the choice (\ref{gwfxlia}) for $a$ this gives for the squared Dirac operator the eigenvalues $\rho^2=n(n+2T)$. This spectrum 
agrees {\it exactly} with what one finds in the continuum \cite{bassetto}, except at the top value of $n$. Such a result is true also 
for $T=0$ \cite{fermion}\cite{monopole}. For the top value of $n$, $\Gamma_2=0$, and we get only the eigenvalue $\gamma_1=1$, whereas in 
the continuum, $\Gamma_2\ne 0$ and both eigenvalues $\gamma_1=\pm 1$ occur. This result \cite{fermion}\cite{monopole}, valid also for $T=0$,
 has been known for a long time. 

Finally, we can check that summing the degeneracies of the eigenvalues we have found, we get exactly the dimension of the 
corresponding module. In fact:
\beqa
2T+2\sum_{n=1}^{2L}\big(2(n+T-\half)+1\big)+2(2L+T+1)&=&2(2L+1)(2(L+T)+1)\ ,
\nn\\
2T+2\sum_{n=1}^{2(L-T)}\big(2(n+T-\half)+1\big)+2(2L-T+1)&=&2(2L+1)(2(L-T)+1)
\ .\nn\\
\label{avi}
\eeqa

We show below that the axial anomaly on $S^2_F$ is stable against perturbations compatible with the chiral properties of the
Dirac operator, and is hence a `topological' invariant.

%%%%%%%%%%%%%%%%%%%%%%%%%%%%%%%%%%%%%%%%%%%%%%%%%%%%%%
\section{ Gauging the Dirac Operator in Instanton Sectors}
%{\bf In a previous reading of Bal's it was noted that a page 8.13-2 is
% missing. I will check to see if I have a copy of it. }
The operator $\vec{\cal L}+\vec T$ commutes with $P^{(L\pm T)}$ and hence preserves the projective modules. It is important to
preserve this feature on gauging as well. So the gauge field $\vec A^L$ is taken to be a function of $\vec L^L+\vec T$ (which remains
bounded as $L\to\infty$). For $L\to\infty$, it becomes a function of $x$. The limiting transversality of $\vec T+\vec A^L$ can be
guaranteed by imposing the condition 
\be 
(\vec L^L+\vec T+\vec A^L)^2=(\vec L^L+\vec T)^2=(L\pm T)(L\pm T+1)\ ,
\label{gwfil}
\ee   
which generalizes (\ref{gwfxxxvii}).

We can now construct the Ginsparg-Wilson system using
\be
\Gamma(A^L)=\frac{\sigma\cdot(\vec L^L+\vec T+\vec A^L)+1/2}
{|\sigma\cdot(\vec L^L+\vec T+\vec A^L)+1/2|}
\label{gwfl}
\ee
and the $\Gamma'$ of (\ref{gwfxxvii}), $\Gamma(0)$ being $\Gamma$ of(\ref{gwfxli}). $\sigma\cdot(\vec L^L+\vec T)+1/2$ has no zero 
modes, and therefore (\ref{gwfl}) is well-defined for generic $\vec A^L$. We can now use section 2 to construct the Dirac theory. 

We have a continuous number of Ginsparg-Wilson algebras labeled by $\vec A^L$. For each, (\ref{gwfxxi}) holds:
\be
\tr\Gamma_2(A^L)=n(A^L)\ .
\label{gwfli}
\ee
Here as $n(A^L)\in\mathbb Z$, it is in fact a constant by
continuity. The index of the Dirac operator and 
the global $U(1)_A$ axial anomaly implied by 
(\ref{gwfli}) are thus independent of $\vec A^L$ as previously indicated.
[See Fujikawa \cite{fujikawa} and \cite{BalNair} for the connection of
  (\ref{gwfli}) to the global axial anomaly.]
 
The expansions (\ref{gwfxla}-\ref{gwfxlc}) are easily extended to the instanton sectors, and imply the desired continuum limit of 
$D^{(L\pm T)}(\vec A^L)$ and chirality $\Gamma_2(\vec A^L)$
\beqa 
D^{(L\pm T)}(\vec A^L)\ &\to &\ \vec\sigma\cdot(\vec{\cal L}+\vec T+\vec
A)+ 1\ ,\nn\\
\Gamma_2(A^L)\ &\to& \ \vec\sigma\cdot\hat x\ .
\label{gwflii}
\eeqa
Chirality is thus independent of the gauge field  in the limiting case, but not otherwise.

\section{Further Remarks on the Axial Anomaly}

The local form of $U(1)_A$-anomaly has not been treated in the present
approach. (See however \cite{chiral}\cite{bassetto}\cite{aoki}.)
As for gauge anomalies, the central and familiar problem is that 
noncommutative algebras allow gauging only by the particular groups 
$U(N)$, and that too by their particular representations (see chapter 7). 
This is so in a naive approach. There are clever methods to overcome
this problem on the Moyal planes \cite{wess} using the Seiberg-Witten map 
\cite{seiberg}, but they fail for the fuzzy spaces. Thus gauge
anomalies can be studied for fuzzy spaces only in a very limited
manner, but even this is yet to be done. More elaborate issues like
anomaly cancellation in a fuzzy version of the standard model have to 
wait till the above mentioned problems are solved.

\chapter{Fuzzy Supersymmetry} 

Another important feature we encounter in studying fuzzy
discretizations is their ability to preserve supersymmetry 
(SUSY) exactly: They allow the formulation of regularized and exactly
supersymmetric field theories. It is very difficult 
to formulate models with exact SUSY in conventional lattice
discretizations. At least for this reason, fuzzy 
supersymmetric spaces merit careful study. 

The original idea of a fuzzy supersphere is due to Grosse et al.\cite{GKP1, fuzzyS}. A slightly different approach for 
its construction, which is closer to ours is given in \cite {klimcik1}. 

We start this chapter describing the supersphere $S^{(2,2)}$ and
its fuzzy version $S^{(2,2)}_F$. Although, the mathematical structure underlying the 
formulation of the supersphere is a generalization of that of the $2$-sphere, it is not widely known. 
Therefore, we here collect the necessary information on representation theory and basic properties of Lie 
superalgebras $osp(2,1)$ and $osp(2,2)$ and their corresponding supergroups $OSp(2,1)$ and $OSp(2,2)$: they underlie 
the construction of $S^{(2,2)}$ and consequently that of $S_F^{(2,2)}$. 

In section \ref{sec-SCS} construction of generalized coherent states is
extended to the supergroup $OSp(2,1)$. 

In section \ref{sec-SUSYac} we outline the SUSY action of Grosse et
al. \cite{GKP1} on $S^{(2,2)}$. It is a quadratic action in scalar and spinor fields.
It is the simplest SUSY action one can formulate and is closest to the
quadratic scalar field action on $S^2$. We then 
discuss its fuzzy version. The latter has exact SUSY.

Following three sections discuss the construction and differential
geometric properties of an associative $*$-product of 
functions on $S_F ^{(2,2)} $ and on ``sections of bundles'' on $S_F
^{(2,2)}$. 

We conclude the chapter by a brief discussion on construction of
non-linear sigma models on  $S^{(2,2)}_F$.    

Our discussion in this chapter follows and expands upon \cite{seckin1}.

\section{$osp(2,1)$ and $osp(2,2)$ Superalgebras and their Representations}

Here we review some of the basic features regarding the Lie
superalgebras $osp(2,1)$ and $osp(2,2)$. For detailed discussions, 
the reader is refered to the references \cite{nahm1, nahm2, pais, Dewitt, cornwell}.

The Lie superalgebras $osp(2,1)$ and $osp(2,2)$ can be defined in
  terms of $3 \times 3$ matrices acting on ${\mathbb C}^3$. The vector
  space ${\mathbb C}^3$ is graded: it is to be regarded as ${\mathbb
  C}^2 \oplus {\mathbb C}^1$ where ${\mathbb C}^2$ is the even- and
  ${\mathbb C}^1$ is the odd-subspace. As ${\mathbb C}^3$ is so graded,
  it is denoted by ${\mathbb C}(2,1)$ while linear operators on
  ${\mathbb C}^{(2,1)}$ are denoted by $Mat(2,1)$. (${\mathbb C}(2,1)$ is to be
  distinguished from the superspace ${\cal C}^{(2,1)}$ which will
  appear in section \ref{sec-susyspace}) By convention the above ${\mathbb C}^2$ and
  ${\mathbb C}^1$ are embedded in ${\mathbb C}^3$ as follows:
\beqa
{\mathbb C}^2 &=& \left \lbrace (\xi_1, \xi_2, 0) \, : \, \xi_i \in
  {\mathbb C} \right \rbrace \subset {\mathbb C}^{(2,1)} \,, \nonumber \\
{\mathbb C}^1 &=& \left \lbrace (0, 0, \eta) \, : \, \eta \in
  {\mathbb C} \right \rbrace \subset {\mathbb C}^{(2,1)} \,.
\eeqa

The grade of ${\mathbb C}^2$ is $0$ (mod $2$) and that of ${\mathbb
  C}^1$ is $1$ (mod $2$). The grading of ${\mathbb C}(2,1)$ induces a
  grading  of $Mat(2,1)$. A linear operator $L \in \, Mat(2,1)$ has
  grade $|L| = 0$ (mod $2$) or is ``even''. If it does not change the
  grade of underlying vectors of definite grade. Such an $L$ is
  block-diagonal:
\beqa
L = \left ( 
\begin{array}{ccc}
\ell_1 & \ell_2 & 0 \\
\ell_3 & \ell_4 & 0 \\
0 & 0 & \ell
\end{array} 
\right )
\, \quad \ell_i \,, \ell \in {\mathbb C} \, \, {\mbox if} \, \, |L| = 0
\quad (\mbox{mod} \, 2) \,.
\eeqa

If $L$ instead changes the grade of an underlying vector of definite
grade by $1$ (mod $2$) unit, its grade is $|L| =1$ 
(mod $2$) or it is ``odd''. Such an $L$ is off-diagonal:
\beqa
L = \left ( 
\begin{array}{ccc}
0 & 0 & s_1 \\
0 & 0 & s_2 \\
t_1 & t_2 & 0
\end{array} 
\right )
\, \quad s_i \,, t_i \in {\mathbb C} \, \quad \mbox{if} \quad  |L| = 1
\quad (\mbox{mod} \, 2) \,.
\eeqa     
 
A generic element of ${\mathbb C}(2,1)$ and $Mat(2,1)$ will be a sum of
 elements of both grades and will have no definite grade.

If $M, N \in Mat(2,1)$ have definite grades $|M|$, $|N|$ their graded
Lie bracket $\lbrack M , N \rbrace$ is defined by 
\be 
\lbrack M \,, N \rbrace = M N - (-1)^{|M| |N|} N M \,.
\ee

The even part of $osp(2,1)$ is the Lie algebra $su(2)$ for which
${\mathbb C}^2$ has spin $\frac{1}{2}$ and ${\mathbb C}^1$ has spin
$0$. $su(2)$ has the usual basis $\Lambda_i ^{(\frac{1}{2})}$
\be
\Lambda_i ^{(\frac{1}{2})} = \frac{1}{2}\left(
\begin{array}{cc}
 \sigma_i & \,0 \\
\,0 & 0
\end{array}
\right ) \,, \sigma_i = \mbox{Pauli matrices} \,.
\ee
The superscript $\frac{1}{2}$ here denotes this representation:
irreducible representations of $osp(2,1)$ are labelled by the highest
angular momentum.

$osp(2,1)$ has two more generators $\Lambda_\alpha ^{(\frac{1}{2})}$
($\alpha=4,5$) in its basis: 
\be
\Lambda_4^{(\frac{1}{2})} = \frac{1}{2} \left(
\begin{array}{ccc}
0 & 0 & -1\\
0 & 0 & 0 \\
0 & -1 & 0
\end{array} 
\right) \,, \quad
\Lambda_5^{(\frac{1}{2})} =  \frac{1}{2} \left(
\begin{array}{ccc}
0 & 0 & 0 \\
0 & 0 & -1 \\
1 & 0 & 0
\end{array} 
\right) \,.
\ee

The full $osp(2,1)$ superalgebra is defined by the graded commutators
\be
\lbrack \Lambda_i^{(\frac{1}{2})} ,\Lambda_j^{(\frac{1}{2})} \rbrack =
i \epsilon_{ijk} \Lambda_k^{(\frac{1}{2})} \,, \quad
\lbrack\Lambda_i^{(\frac{1}{2})} , \Lambda_\alpha^{(\frac{1}{2})}
\rbrack = \frac{1}{2} (\sigma_i)_{ \beta \alpha} 
\Lambda_\beta^{(\frac{1}{2})}  \,, \quad
\{ \Lambda_\alpha^{(\frac{1}{2})} , \Lambda_\beta^{(\frac{1}{2})} \} =
\frac{1}{2} (C \sigma_i)_{\alpha \beta} \Lambda_i^{(\frac{1}{2})} \,, 
\label{eq:gradedcom1}
\ee
where $C_{\alpha \beta} = - C_{\beta \alpha}$ is the Levi-Civita symbol with $C_{45} = 1$.
(Here the rows and columns of $\sigma_i$ and $C$ are being labeled by
$4,5$).

The abstract $osp(2,1)$ Lie superalgebra has basis $\Lambda_i,
\Lambda_\alpha \, (i =1,2,3 \,, \alpha = 4, 5)$ with graded commutators
obtained from (\ref{eq:gradedcom1}) by dropping the superscript
$\frac{1}{2}$:
\be
\lbrack \Lambda_i ,\Lambda_j \rbrack = i \epsilon_{ijk} \Lambda_k \,, \quad
\lbrack\Lambda_i , \Lambda_\alpha \rbrack = \frac{1}{2} (\sigma_i)_{ \beta \alpha} \Lambda_\beta  \,, \quad
\{ \Lambda_\alpha , \Lambda_\beta \} = \frac{1}{2} (C \sigma_i)_{\alpha \beta} \Lambda_i \,. 
\label{eq:gradedcom2}
\ee   

Thus $\Lambda_\alpha$ transforms like an $su(2)$ spinor.

The Lie algebra $su(2)$ is isomorphic to the Lie algebra $osp(2)$ 
of the ortho-symplectic group $OSp(2)$. 
%{\bf These osp notations I have not encountered in literature!!!!, still to check.} 
The above graded Lie algebra 
has in addition one spinor in its basis. For this reason, it is
denoted by $osp(2,1)$.

In customary Lie algebra theory, compactness of the underlying group
is reflected in the adjointness properties of its Lie algebra elements.
Thus these Lie algebras allow a star $*$ or adjoint operation $\dagger$
and their elements are invariant under $\dagger$ (in the convention of
physicists) if the underlying group is compact. As $\dagger$ complex
conjugates complex numbers, the Lie algebras of compact Lie groups are
real as vector spaces: they are real Lie algebras.

In graded Lie algebras, the operation $\dagger$ is replaced by the
grade adjoint (or grade star) operation $\ddagger$.
Its relation to the properties of the underlying supergroup will be
indicated later. The properties and definition of $\ddagger$ are as
follows.

First, we note that the grade adjoint of an even (odd) element is 
even (odd). Next, one has $(A ^{\ddagger})^\ddagger = (-1)^{|A|} A$ 
for an even or odd (that is homogeneous) element $A$ of degree 
$|A|$ $(\mbox{mod} \,2)$, or equally well, integer $(\mbox{mod} \,2)$. (So,
depending on $|A|$, $|A|$ itself can be taken $0$ or $1$.) Thus, it is the usual
$\dagger$ on the even part, while on an odd element $A$, it squares to $-1$.
Further $(AB)^\ddagger= (-1)^{|A||B|} B^\ddagger A^\ddagger$ so that, ${\lbrack A, B \}}^\ddagger
= (-1)^{|A||B|} \lbrack B^\ddagger, A^\ddagger \}$ for homogeneous elements
$A, B$.

Henceforth, we will denote the degree of $a$ (which may be a Lie superalgebra element, a linear operator
or an index) by $|a| (\mbox{mod} \,2)$, $|a|$ denoting any integer in its equivalence class
$\langle |a| + 2n \,: n \in {\mathbb Z} \rangle $.

The basis elements of the $osp(2,1)$ (and $osp(2,2)$, see later) graded Lie
algebras are taken to fulfill certain ``reality'' properties implemented by
$\ddagger$. For the generators of $osp(2,1)$, these are given by
\begin{equation}
\Lambda_i^{\ddagger} = \Lambda_i^{\dagger} = \Lambda_i, \quad 
\Lambda_\alpha ^{\ddagger} = - \sum_{\beta = 4,5}C_{\alpha \beta} \Lambda_\beta \, \quad \alpha = 4,5 \,.
\label{eq:six}
\end{equation}

Let $V$ be a graded vector space $V$ so that $V = V_0 \oplus V_1$ where $V_0$
and $V_1$ are even and odd subspaces \cite{cornwell}.
In a (grade star) representation of a graded Lie algebra on $V$, 
$V_0$ and $V_1$ are invariant under the even elements of the graded
Lie algebra while its odd elements map one to the other. 

This representation becomes a grade-$*$ representation if the
following is also true. Let us
assume that $V$ is endowed with the inner product $\langle u|v \rangle $ for all
$u, v \in V$. Now if $L$ is a linear operator acting on $V$, then the grade
adjoint of $L$ is defined by
\begin{equation}
\langle L^\ddagger \,u | v \rangle  = (-1)^{|u| \,|L|}\, \langle u|L \,v \rangle
\label{eq:gradeadj}
\end{equation}
for homogenous elements $u$, $L$.
In a basis adapted to the above decomposition of $V$, a generic $L$ has the matrix representation
\begin{equation}
M_L = \left(
\begin{array}{cc}
\alpha_1 & \alpha_2 \\
\alpha_3 & \alpha_4
\end{array}
\right) = M_0 + M_1 \,, \quad 
M_0 = \left(
\begin{array}{cc}
\alpha_1 & 0 \\
0 & \alpha_4
\end{array}
\right) \,, \quad 
M_1 = \left(
\begin{array}{cc}
0 & \alpha_2 \\
\alpha_3 & 0
\end{array}
\right)
\end{equation}
where $M_0$ and $M_1$ are the even and odd parts of $M_L$. The formula for $\ddagger$ is then
\begin{equation}
M_L^{\ddagger} = \left(
\begin{array}{cc}
\alpha_1^{\dagger} & - \alpha_3^{\dagger} \\
\alpha_2^{\dagger} & \alpha_4^{\dagger}
\end{array}
\right) \,,
\end{equation}
$\alpha_i^{\dagger}$ being matrix adjoint of $\alpha_i$. 

Then in a grade-$*$ representation, the image of $L^\ddagger$ is $M_L^\ddagger$.

We note that the supertrace $str$ of $M_L$ is by definition
\begin{equation}
strM_L = Tr \alpha_1 - Tr\alpha_4 \,.
\end{equation} 
   
The irreducible representations of $osp(2,1)$ are characterized by an integer or half-integer 
non-negative quantum  number $J_{osp(2,1)}$ called superspin. From the point of view of the irreducible 
representations of $su(2)$, the superspin $J_{osp(2,1)}$ representation has the decomposition
\begin{equation}
J_{osp(2,1)} = J_{su(2)} \oplus \left(J - \frac{1}{2}
\right)_{su(2)}\,,
\end{equation}
where $J_{su(2)}$ is the $su(2)$ representation for angular momentum $J_{su(2)}$.
All these are grade-$*$ representations : the relations
(\ref{eq:six}) are preserved in the representation.
 
The fundamental and adjoint representations of
$osp(2,1)$ correspond to \mbox{$J_{osp(2,1)}=\frac{1}{2}$} and 
$J_{osp(2,1)}=1$ respectively, being 3 and 5 dimensional. The 
quadratic Casimir operator is
\begin{equation}
K_2^{osp(2,1)} = \Lambda_i \Lambda_i + C_{\alpha \beta} \Lambda_\alpha 
\Lambda_\beta.
\end{equation}
It has eigenvalues $J_{osp(2,1)}(J_{osp(2,1)} + \frac{1}{2})$.

It is also worthwhile to make the following technical remark. The
superspin multiplets in $J_{osp(2,1)}$ representation 
may be denoted by $|J_{osp(2,1)} \,,$ $J_{su(2)} \,, J_3 \rangle$, 
and $|J_{osp(2,1)} \,, \big(J -\frac{1}{2} \big)_{su(2)} \,, 
J_3 \rangle$. One of the multiplets generates the even and the other 
generates the odd subspace of the representation space. 
Although, this can be arbitrarily assigned, the choice consistent 
with the reality conditions we have chosen in 
(\ref{eq:six}) and the definition of grade adjoint operation in 
(\ref{eq:gradeadj}) fixes the multiplet $|J_{osp(2,1)} \,, J_{su(2)}
\,, J_3 \rangle$ to be of even degree while $|J_{osp(2,1)} \,, 
\big(J -\frac{1}{2} \big)_{su(2)} \,, J_3 \rangle$ is odd.      

The $osp(2,2)$ superalgebra can be defined by introducing 
an even generator $\Lambda_8$ commuting with the $\Lambda_i$
and odd generators $\Lambda_\alpha$  with $\alpha = 6,7$ in
addition to the already existing ones for $osp(2,1)$. The graded 
commutation relations for $osp(2,2)$ are then
\begin{gather}
\lbrack \Lambda_i ,\Lambda_j \rbrack = i \epsilon_{ijk} \Lambda_k \,, \quad 
\lbrack\Lambda_i , \Lambda_\alpha \rbrack = \frac{1}{2} (\tilde{\sigma}_i)_{ \beta \alpha} 
\Lambda_\beta \,, \quad \lbrack \Lambda_i ,\Lambda_8 \rbrack = 0 \,, \nonumber \\
\lbrack \Lambda_8 ,\Lambda_\alpha \rbrack = \tilde{\varepsilon}_{\alpha \beta} \Lambda_\beta \,, \quad
\{ \Lambda_\alpha ,\Lambda_\beta \} = \frac{1}{2}(\tilde{C} \tilde{\sigma_i})_{\alpha \beta} 
\Lambda_i + \frac{1}{4} (\tilde{\varepsilon}\tilde{C})_{\alpha \beta} \Lambda_8 \,,
\label{eq:osp22algebra}
\end{gather}
where $i,j= 1,2,3$ and $\alpha, \beta= 4,5,6,7$. In above we have used the matrices
\begin{equation}
\tilde{\sigma}_i = \left(
\begin{array}{cc}
\sigma_i & 0 \\
0 & \sigma_i
\end{array}
\right) 
\,, \quad  
\tilde{C} = \left(
\begin{array}{cc}
C & 0 \\
0 & -C
\end{array} 
\right)
\,, \quad 
\tilde{\varepsilon} = \left(
\begin{array}{cc}
0 & I_{2\times 2} \\
 I_{2\times 2}& 0
\end{array} 
\right)\,.
\label{eq:matrices}
\end{equation}
Their matrix elements are indexed by $4,\ldots ,7$.

In addition to (\ref{eq:six}), the new generators satisfy the ``reality" conditions
\begin{equation}
\Lambda_\alpha ^{\ddagger} =  - \sum_{\beta = 6,7} \tilde{C}_{\alpha \beta}
\Lambda_\beta \,,\quad \alpha = 6,7 \,, \quad \quad
\Lambda_8^{\ddagger} = \Lambda_8^{\dagger} = \Lambda_8 \,.
\label{eq:reality1}
\end{equation}
So we can write the $osp(2,2)$ reality conditions for all $\alpha$ as 
$\Lambda_\alpha ^\ddagger = - \tilde{C}_{\alpha \beta} \Lambda_\beta$.
 
Irreducible representations of $osp(2,2)$ fall into two categories,
namely the typical and non-typical ones.
Both are grade $*$-representations which preserve the reality
conditions (\ref{eq:six}) and (\ref{eq:reality1}).
Typical ones are reducible with respect to the $osp(2,1)$ superalgebra 
(except for the trivial representation) whereas non-typical 
ones are irreducible. Typical representations are labeled by an integer or half integer non-negative number 
$J_{osp(2,2)}$, called $osp(2,2)$ superspin and the maximum eigenvalue
$k$ of $\Lambda_8$ in that IRR. They can be denoted by $(J_{osp(2,2)},
k)$. Independently of $k$, these have the $osp(2,1)$ content $J_{osp(2,2)}= J_{osp(2,1)} \oplus 
(J - \frac{1}{2})_{osp(2,1)}$ for $J_{osp(2,2)} \geq \frac{1}{2}$ while $(0)_{osp(2,2)}= (0)_{osp(2,1)}$. Hence
\begin{equation}
\left (J_{osp(2,2)}, k \right ) = \left\{
\begin{array}{ll}
J_{su(2)} \oplus \left( J - \frac{1}{2}\right)_{su(2)} \oplus \left( J
- \frac{1}{2}\right)_{su(2)} \oplus (J - 1)_{su(2)} \,,
& J_{osp(2,2)} \geq 1  \,; \\
(\frac{1}{2})_{su(2)} + (0)_{su(2)} + (0)_{su(2)} \,, & J_{osp(2,2)} = \frac{1}{2} \,.
\end{array} 
\right.
\end{equation}

$osp(2,2)$ has the quadratic Casimir operator
\beqa
K_2^{osp(2,2)} &=& \Lambda_i \Lambda_i + \tilde{C}_{\alpha \beta}
\Lambda_\alpha \Lambda_\beta - \frac{1}{4} \Lambda_8 ^2 \nonumber \\
&=& K_2^{osp(2,1)} - \left( \sum_{\alpha, \beta = 6,7} -
\tilde{C}_{\alpha \beta} \Lambda_\alpha \Lambda_\beta + \frac{1}{4} 
\Lambda_8 ^2 \right) \,.
\eeqa 
It has also a cubic Casimir operator \cite{nahm1, sorba}. We do not
show it here , as we will not use it.

Note that since all the generators of $osp(2,1)$ commute with 
$K_2^{osp(2,2)}$ and $K_2^{osp(2,1)}$, they also commute with
\begin{equation}
K_2^{osp(2,1)} - K_2^{osp(2,2)} = -\sum_{\alpha, \beta = 6,7} \tilde{C}_{\alpha \beta} \Lambda_\alpha \Lambda_\beta + 
\frac{1}{4} \Lambda_8 ^2 \,.
\label{eq:casdif}
\end{equation}
The $osp(2,2)$ Casimir $K_2^{osp(2,2)}$ vanishes on non-typical
representations:
\be
K_2^{osp(2,2)} \Big |_{nontypical} = 0 \,.
\ee

The substitutions
\begin{equation}
\Lambda_i \rightarrow \Lambda_i, \quad \Lambda_\alpha \rightarrow \Lambda_\alpha, \quad 
\alpha = 4,5; \quad \Lambda_\alpha \rightarrow - \Lambda_\alpha, \quad \alpha = 6,7; \quad
\Lambda_8 \rightarrow - \Lambda_8 \,
\end{equation}
define an automorphism of $osp(2,2)$. This automorphism changes the
irreducible representation $(J_{osp(2,2)}, k)$ into an inequivalent one
$(J_{osp(2,2)}, -k)$ (except for the trivial representation with $J = 0$),
while preserving the reality conditions given in (\ref{eq:six}) and (\ref{eq:reality1}) \cite{nahm2}.
In the nontypical case, we discriminate between these two
representations  associated with $J_{osp(2,1)}$ 
as follows: For $J>0$, $J_{osp(2,2) + }$ will denote the representation in which the eigenvalue of
the representative of $\Lambda_8$ on vectors with angular momentum $J$ is positive
and $J_{osp(2,2) - }$ will denote its partner where this eigenvalue is negative.
(This eigenvalue is zero only in the trivial representation with $J=0$.) 
Here while considering nontypical IRR's we concentrate on $J_{osp(2,2)
  +}$. 
The results for $J_{osp(2,2)-}$ are similar and will be occasionally indicated.

Another important result in this regard is that every non-typical representation $J_{osp(2,2) \pm}$ of $osp(2,2)$, is 
at the same time an irreducible representation of $osp(2,1)$ with superspin $J_{osp(2,1)}$. For this reason the $osp(2,2)$ generators 
$\Lambda_{6,7,8}$ can be nonlinearly realized in terms of the $osp(2,1)$ generators. Repercussions of this result will be seen
later on.  
     
Below we list some of the well-known results and standard notations that are used throughout the text.
The fundamental representation of $osp(2,2)$ is non-typical and we concentrate on the one given by
$J_{osp(2,2) + } = (\frac{1}{2})_{osp(2,2) + }$.
It is generated by the $(3 \times 3)$ supertraceless matrices $\Lambda_a ^{(\frac{1}{2})}$
satisfying the ``reality'' conditions of (\ref{eq:six}) and (\ref{eq:reality1}):
\begin{align}
\Lambda_i ^{(\frac{1}{2})} &= \frac{1}{2}\left(
\begin{array}{cc}
 \sigma_i & \,0 \\
\,0 & 0
\end{array}
\right) \,, & 
\Lambda_4  ^{(\frac{1}{2})} &= \frac{1}{2} \left(
\begin{array}{cc}
0 & \xi \\
\eta^T & 0
\end{array} 
\right) \,, &
\Lambda_5^{(\frac{1}{2})} &=  \frac{1}{2} \left(
\begin{array}{cc}
0 & \eta \\
-\xi^T & 0
\end{array} 
\right) \,, \nonumber \\
\Lambda_6 ^{(\frac{1}{2})} &=  \frac{1}{2} \left(
\begin{array}{cc}
0 & -\xi \\
\eta^T & 0
\end{array} 
\right) \,, & 
\Lambda_7 ^{(\frac{1}{2})} & =  \frac{1}{2} \left(
\begin{array}{cc}
0 & -\eta \\
-\xi^T & 0
\end{array}
\right) \,, &
\Lambda_8 ^{(\frac{1}{2})} & = \left(
\begin{array}{cc}
I_{2\times 2} & 0 \\
\,0 & 2
\end{array}
\right) \,,
\label{eq:nontyp}
\end{align}
where
\begin{equation}
\xi = \left(
\begin{array}{c}
-1 \\
0
\end{array}
\right)
\quad {\mbox{and}} \quad 
\eta = \left(
\begin{array}{c}
0 \\
-1
\end{array}
\right) \,.
\end{equation}
These generators satisfy 
\begin{equation}
\Lambda_a ^{(\frac{1}{2})} \Lambda_b ^{(\frac{1}{2})} = S_{ab}{\bf 1} + \frac{1}{2}\, \big(d_{abc} + 
i f_{abc} \big) \,\Lambda_c  ^{(\frac{1}{2})} \quad (a,b,c = 1,2,\dots 8) \,.
\end{equation}
It is possible to write 
\be
S_{ab} = str \Big(\Lambda_a ^{(\frac{1}{2})} \Lambda_b ^{(\frac{1}{2})} \Big) \,, \quad 
f_{abc} = str \Big(- i \lbrack \Lambda_a ^{(\frac{1}{2})} , \Lambda_b ^{(\frac{1}{2})} \} \Lambda_c ^{(\frac{1}{2})}\Big) 
\,, \quad 
d_{abc} = str\Big( \{ \Lambda_a ^{(\frac{1}{2})} , \Lambda_b ^{(\frac{1}{2})} \rbrack \Lambda_c ^{(\frac{1}{2})}\Big) \,. 
\ee
Here $a = i = 1,2,3$, and $a =8$ label the even generators whereas $a = \alpha = 4,5,6,7$ label the odd generators.
%Also here and in what follows, 
In above $[A,B\} , \{A,B]$ denote the graded commutator and the graded
anticommutator respectively. The former is already defined, while the
latter is given by $\{A,B] = AB + (-1)^{|A||B|} BA$ for homogenous
elements $A$ and $B$.

$S_{ab}$ defines the invariant metric of the Lie superalgebra $osp(2,2)$. In their block diagonal form, $S$ and its inverse read
\begin{equation}
S = \left(
\begin{array}{ccc}
\frac{1}{2}\,I & & \\
 & -\frac{1}{2}\,\tilde{C} & \\
 & & -2
 \end{array}
\right) _{8 \times 8}
%\quad {\mbox{and}}
\,, \quad \quad
S ^{-1} = \left(
\begin{array}{ccc}
2\,I & & \\
 & 2\,\tilde{C} & \\
 & & -\frac{1}{2}
\end{array}
\right) _{8 \times 8} \,.
\end{equation} 
The explicit values of the structure constants $f_{abc}$ can be read from (\ref{eq:matrices}), since $\lbrack \Lambda_a, \Lambda_b \} 
= i f_{abc} \Lambda_c$. Those of $d_{abc}$ are as follows\footnote{The tensor $d_{abc}$ given explicitly in (\ref{eq:dtensor})
for $J_{osp(2,2) + }$ becomes $- d_{abc}$ for $J_{osp(2,2) - }$.}:
\begin{gather}
d_{ij8} = - \frac{1}{2} \delta_{ij} \,, \quad d_{\alpha \beta 8} = \frac{3}{4} \tilde{C}_{\alpha \beta} \,, \quad
d_{\alpha 8 \beta} = 3 \delta_{\alpha \beta} \,, \quad d_{i8j} = 2 \delta_{ij} \,, \nonumber \\
d_{\alpha \beta i} = - \frac{1}{2}(\tilde{\varepsilon} \tilde{C} \tilde{\sigma}_i)_{\alpha\beta} \,, \quad  
d_{i \alpha \beta} = - \frac{1}{2}(\tilde{\varepsilon} \tilde{\sigma}_i)_{\beta \alpha} \,, \quad  d_{888} = 6 \,.
\label{eq:dtensor}
\end{gather}

We close this subsection with a final remark. Discussion in the subsequent
sections will involve the use of linear operators acting on the adjoint representation of $osp(2,2)$.
These are linear operators ${\widehat{\cal Q}}$ acting on $\Lambda_a$ according to
${\widehat{\cal Q}} \Lambda_a = \Lambda_b {\cal Q}_{ba}$, ${\cal Q}$ being the matrix representation
of ${\widehat{\cal Q}}$. They are graded because $\Lambda_a$'s are, and hence the linear operators on the
adjoint representation are graded.
The degree (or grade) of a matrix ${\cal Q}$ with only the nonzero entry ${\cal Q}_{ab}$ is
$(|\Lambda_a| + |\Lambda_b|) \, (mod \,2) \equiv (|a| + |b|) \,(mod \,2)$.
The grade star operation on ${\widehat{\cal Q}}$ now follows from the sesquilinear form
\begin{equation}
\big(\alpha = \alpha_a \Lambda_A, \,\beta= \beta_b \Lambda_b \big) = \bar{\alpha}_a S^{-1}_{ab} \beta_b,
\quad \quad \alpha_a, \beta_b \in {\mathbb C} \,
\end{equation}
and is given by
\begin{equation}
({\widehat{\cal Q}}^\ddagger \alpha, \beta) = (-1)^{|\alpha|
\,|{\widehat{\cal Q}}|} (\alpha, {\widehat{\cal Q}} \beta) \,. 
\end{equation}
  
\section{Passage to Supergroups}

We recollect here the passage from these superalgebras to their corresponding supergroups \cite{cornwell, chaichian}.
Let $\xi \equiv (\xi_1 \,, \cdots \,, \xi_8)$ be the elements of the superspace ${\mathbb R}^{(4,4)}$. Here $\xi_a$ for
$a = i = 1,2,3$ and $a =8$ label the even and for $a = \alpha =
4,5,6,7$ label the odd elements of a real Grassmann algebra ${\cal G}$.
$ \xi_a$'s satisfy the graded commutation relations mutually and with the algebra elements:
\begin{equation}
\lbrack \xi_a,\xi_b \} = 0 \,, \quad \quad \lbrack \xi_a,\Lambda_b \} = 0 \,. 
\end{equation}
We assume that $\xi_i^{\ddagger} = \xi_i,  \xi_8^{\ddagger} = \xi_8$ and $\xi_\alpha^{\ddagger} = 
- \tilde{C}_{\alpha \beta} \xi_\beta$. Then $\xi_a \Lambda_a$ is
grade-$*$ even:
\begin{equation}
(\xi_a \Lambda_a)^{\ddagger} = \xi_a \Lambda_a.
\label{eq:hermit}
\end{equation}
An element of $OSp(2,2)$ is given by $g = e^{i \xi_a \Lambda_a}$, 
while for $a$ restricted to $a \leq 5$, $g$ gives an element of 
$OSp(2,1)$. (\ref{eq:hermit}) corresponds to the usual hermiticity 
property of Lie algebras which yields unitary representations of the group. 

\section{On the Superspaces}
\label{sec-susyspace}

\subsection{The Superspace ${\cal C}^{2,1}$ and the Noncommutative ${\cal C}_F^{2,1}$}

${\cal C}^{2,1}$ is the $(2 \,,1)$-dimensional superspace specified by
two even and one odd element of a complex Grassmann algebra ${\cal
G}$. Let ${\cal G}_0$ and ${\cal G}_1$ denote the even and odd 
subspaces of ${\cal G}$.  We write 
\be
{\cal C}^{2,1} \equiv \lbrace \psi \equiv (z_1 \,, z_2 \,, \theta) \rbrace\,,
\ee
where $z_1 \,, z_2 \in  {\cal G}_0$ and $\theta \in {\cal G}_1$ satisfy
\be  
\lbrace \theta \,, {\bar {\theta}} \rbrace \equiv \theta {\bar {\theta}} + {\bar {\theta}} \theta = 0 \,, \quad  
\theta \theta = {\bar {\theta}} {\bar {\theta}} = 0 \,.
\ee
We note that under $\ddagger$ operation
\be
z_i^\ddagger = z_i^\dagger = {\bar z}_i \,, \quad \theta^\ddagger = {\bar \theta} \,,
\quad {\bar \theta}^\ddagger = - \theta \,.
\label{eq:realityonzt}
\ee

The noncommutative ${\cal C}^{2,1}$, denoted by ${\cal C}_F^{2,1}$ hereafter, is obtained by replacing $\psi \in 
{\cal C}^{2,1}$, by $\Psi \equiv (a_1 \,, a_2 \,, b)$, where the operators $a_i$ and $b$ obey the commutation and 
anticommutation relations
\begin{gather}
\lbrack a_i \, a_j \rbrack = \lbrack a_i^\dagger \, a_j^\dagger \rbrack = 0 \,, \quad \lbrack a_i \, a_j^\dagger  
\rbrack = \delta_{ij} \,, \quad \lbrack a_i \,, b \rbrack = \lbrack a_i \,, b^\dagger \rbrack = 0 \nonumber \\
\lbrace b \,, b \rbrace = \lbrace b^\dagger \,, b^\dagger \rbrace = 0 \,, \quad \lbrace b \,, b^\dagger \rbrace = 1 \,.  
\end{gather}
Under $\dagger$ they fulfill $a_i^\ddagger = a_i$, $(a_i^\dagger)^\ddagger = a_i$,
$b^\ddagger = b^\dagger$, $(b^\dagger)^\ddagger = -b$. 

Using the notation 
\be
( \Psi_1 \,, \Psi_2 \,,  \Psi_0 ) \equiv  (a_1 \,, a_2 \,, \, b ) \,,
\label{eq:gosterim1}
\ee 
the commutation relations can be more compactly expressed as 
\be
\lbrack \Psi_{\mu} \,, \Psi_{\nu} \rbrace = \lbrack \Psi_{\mu}^\dagger \,, \Psi_{\nu}^\dagger \rbrace = 0 \,, \quad
\lbrack \Psi_{\mu} \,, \Psi_{\nu}^\dagger \rbrace = \delta_{\mu \nu} \,,
\ee
where $\mu = 1,2,0$. $\Psi_\mu$, $\Psi_\mu^\dagger$ and the identity 
operator ${\bf 1}$ span the graded Heisenberg-Weyl 
algebra, with ${\bf 1}$ being its center.

\subsection{The Supersphere $S^{(3,2)}$ and the Noncommutative $S^{(3,2)}$}

Dividing $\psi$ by its modulus $|\psi| \equiv |z_1|^2 + |z_2|^2 + \bar{\theta} \theta$, we define $\psi^\prime = 
\frac{\psi}{|\psi|} \in {\cal C}^{2,1} \setminus \{0 \}$ with $|\psi^\prime| = 1$. The $(3,2)$ dimensional supersphere 
$S^{(3,2)}$ can then be defined as
\be
S^{(3,2)} \equiv \big \langle \psi^\prime = \frac{\psi}{|\psi|} \in {\mathbb C}^{2,1} \setminus \{0 \} \big \rangle \,.
\ee
Obviously $S^{(3,2)}$ has the $3$-sphere $S^3$ as its even part. 

The noncommutative $S^{(3,2)}$ is obtained by replacing $\psi^\prime$
by $\Psi \frac{1}{\sqrt{{\widehat N}}}$ where ${\widehat N} =
a_i^\dagger a_i + b^\dagger b$ is the number operator. We have
\beqa
\psi^\prime_\mu \quad \longrightarrow \quad S_\mu  &:=& \Psi_\mu \,\frac{1}{\sqrt{{\widehat N}}} = 
\frac{1}{\sqrt{{\widehat N} + 1}}\, \Psi_\mu \,, \nonumber \\
\psi^{\prime \dagger}_\mu \quad \longrightarrow \quad S_\mu ^\dagger  &:=& \frac{1}{\sqrt{{\widehat N}}} \, \Psi_\mu ^\dagger = 
\Psi_\mu ^\dagger \, \frac{1}{\sqrt{{\widehat N} + 1}} \,,
\eeqa
where ${\widehat N} \neq 0$. Furthermore, we have that $\lbrack S_\mu , S_\nu \} = \lbrack S_\mu ^\dagger , S_\nu ^\dagger \} =0$, 
while
after a small calculation we get
\begin{equation}
\lbrack S_\mu , S_\nu ^\dagger \} = \frac{1}{{\widehat N} + 1}\, \left( \delta_{\mu \nu}- (-1)^{|S_\mu| |S_\nu|}\, S_\nu ^\dagger
S_\mu \right) \,.
\label{eq:gcommutator}
\end{equation}
We note that as the eigenvalue of ${\widehat N}$ approaches to infinity we recover $S^{(3,2)}$ back.
              
Noncommutative $S^{(3,2)}$ suffers from the same problem as
noncommutative $S^3$ does: $S_\mu$ an $S_\mu^\dagger$ 
act on an infinite-dimensional Hilbert space so that we do not 
obtain finite-dimensional models for noncommutative $S^{(3,2)}$ either. 
Nevertheless, the structure of the non-commutative $S^{(3,2)}$
described above is quite useful in the construction of $S^{(2,2)}_F$ as 
well as for obtaining $*$-products on the ``sections of bundles'' over 
$S^{(2,2)}_F$ as we will discuss later in this chapter.

\subsection{The Commutative Supersphere $S^{(2,2)}$}

There is a supersymmetric generalization of the Hopf fibration. 
In this subsection we construct this (super)-Hopf fibration through studying the 
actions of $OSp(2,1)$ and $OSp(2,2)$ on $S^{(3,2)}$. We also establish
that $S^{(2,2)}$ is the adjoint orbit of $OSp(2,1)$, while it is a
closely related (but not the adjoint) orbit of $OSp(2,2)$. We
elaborate on the subtle features of the latter, which are important for 
future developments in this chapter.

We first note that the group manifold of $OSp(2,1)$ is nothing but
$S^{(3,2)}$. Also note that $|\psi|^2$ is preserved under the 
group action $\psi \longrightarrow g \psi$ for $g \in OSp(2,1)$. 
Let us then consider the following map $\Pi$ from the functions on 
$(3,2)$-dimensional supersphere $S^{(3,2)}$ to functions on $S^{(2,2)}$:
\begin{equation}
\Pi \quad : \quad \psi^\prime \longrightarrow \quad w_a (\psi \,,
\bar{\psi}) := {\bar \psi}^\prime \Lambda_a^{(\frac{1}{2})} 
\psi^\prime \, =  \frac{2}{|\psi|^2} \bar{\psi} \Lambda_a^{(\frac{1}{2})} \psi \,.
\label{eq:promap1}
\end{equation}
The fibres in this map are $U(1)$ as the overall phase in $\psi
\rightarrow \psi e^{i \gamma}$ cancels out while no other degree of 
freedom is lost on r.h.s. Quotienting  $S^{(3,2)} \equiv OSp(2,1)$ by 
the $U(1)$ fibres we get the $(2,2)$ dimensional base space 
\footnote{In what follows we do not show the $\bar{\psi}$ dependence 
of $w_a$ to abbreviate the notation a little bit.} 
\begin{equation}
S^{(2,2)} \,: = \, S^{(3,2)}/ \,U(1) \, \equiv \Big \lbrace w(\psi) 
= \big ( w_1(\psi), \cdots, \,w_5(\psi) \big ) \Big \rbrace \,.
\end{equation}  
$\Pi$ is thus the projection map of the ``super-Hopf fibration'' over 
$S^{(2,2)}$ \cite{Berezin:jz, marmo, landi}, and $S^{(2,2)}$ can be 
thought as the supersphere generalizing $S^2$.

We now characterize $S^{(2,2)}$ as an adjoint orbit of $OSp(2,1)$. 
First observe that $w(\psi)$ is a (super)-vector in the adjoint 
representation of $OSp(2,1)$. Under the action
\begin{equation}
w \,\rightarrow gw, \quad (gw)(\psi) = w(g^{-1}\psi), \quad g \in OSp(2,1) \,,
\end{equation}
it transforms by the adjoint representation $g \rightarrow \, Ad \, g \,$ :
\begin{equation}
w_a(g^{-1} \psi) = w_b \,(\psi) \,(Ad \,g)_{ba} \,.
\end{equation}       

The generators of $osp(2,1)$ in the adjoint representation are $ad \Lambda_a$ where
\begin{equation}
(ad \,\Lambda_a)_{cb} = i f_{abc} \,.
\end{equation}
From this and the infinitesimal variations $\delta w(\psi) = \varepsilon_a\, ad \, \Lambda_a \,
w(\psi)$ of $w(\psi)$
under the adjoint action, where $\varepsilon_i$'s are even and
$\varepsilon_\alpha$'s are odd Grassmann variables, we can verify that 
\begin{equation}
\delta ( w_i (\psi)^2 + C_{\alpha \beta} w_\alpha (\psi) w_\beta(\psi)) = 0 \,.
\end{equation}
Hence, $S^{(2,2)}$ is an $OSp(2,1)$ orbit with the invariant
\begin{equation}
\frac{1}{2} (w_a (S^{-1})_{ab} w_b) = w_i(\psi)^2 + C_{\alpha \beta} w_\alpha (\psi) w_\beta(\psi) \,.
\label{eq:invariant1}
\end{equation}  
The value of the invariant can of course be changed by scaling. 
Now the even components of $w_a(\psi)$ are real while its odd 
entries depend on both $\theta$ and $\bar{\theta}$:
\begin{equation}
w_i(\psi) = \frac{1}{|\psi|^2} \bar{z}\sigma_i z \,,\quad
w_4(\psi) = - \frac{1}{|\psi|^2}
(\bar{z}_1 \theta + z_2 \bar{\theta}) \,,\quad w_5(\psi) 
= \frac{1}{|\psi|^2}(- \bar{z}_2 \theta + z_1 \bar{\theta})\,.
\label{eq:promap2}
\end{equation}   

From (\ref{eq:realityonzt}) and (\ref{eq:promap2}), one deduces the reality conditions
\begin{equation}
w_i(\psi)^\ddagger = w_i(\psi) \quad \quad \quad
w_\alpha(\psi)^\ddagger = - C_{\alpha \beta} \,w_\beta (\psi) \,.
\label{eq:real1}
\end{equation}

The $OSp(2,1)$ orbit is preserved under this operation as can be checked directly using (\ref{eq:real1}) in (\ref{eq:invariant1}). 
The reality condition (\ref{eq:real1}) reduces the degrees of freedom in $w_\alpha (\psi)$ to two. The $(3,2)$ number of variables 
$w_a (\psi)$ are further reduced to $(2,2)$ on fixing the value of the invariant (\ref{eq:invariant1}). As $(2,2)$ is the 
dimension of $S^{(2,2)}$, there remains no further invariant in this orbit. Thus
\begin{equation}
S^{(2,2)} = \Big \langle \eta \in {\mathbb R}^{(3,2)} \Big | \, \eta_i^2 + C_{\alpha \beta} \, \eta_\alpha ^{(-)} \, 
\eta_\beta ^{(-)} = 1 \,, (\eta_i)^\ddagger = \eta_i \,, (\eta_\alpha ^{(-)})^\ddagger = 
- C_{\alpha \beta} \eta_\beta ^{(+)} \Big \rangle \,,
\label{eq:adjointorbit}
\end{equation}
where we have chosen $\frac{1}{4}$ for the value of the invariant. It is important to note that the superspace ${\mathbb R}^{(3,2)}$ 
in (\ref{eq:adjointorbit}) is defined as the algebra of polynomials in
generators $\eta_i$ and $\eta^{(-)}_\alpha$ satisfying the 
reality conditions $\eta_i^\ddagger = \eta_i \,, \eta^{(-) \ddagger}=
- C_{\alpha \beta} \eta^{(+)}_\beta$. Thus $S^{(2,2)}$ is embedded
in ${\mathbb R}^{(3,2)}$ as described by (\ref{eq:adjointorbit}). 
%
%The (super)adjoint action of $osp(2,1)$ generators on the algebra of
%functions over $S^{(2,2)}$ can be written as graded linear
%differential operators in the embedding coordinates 
%$(\eta_i \,, \eta^{(+)}_\alpha) \in {\mathbb R}^{(3,2)}$. We have
%\begin{eqnarray} 
%\zeta_i &=& -i \varepsilon_{ijk} \eta_j \partial_k - 
%\frac{1}{2}(\sigma_i)_{\beta \alpha} \eta_\beta \partial_{\eta^\alpha} \nonumber 
%\,, \\
%\zeta_\alpha &=& - \frac{1}{2}(\sigma_i)_{\beta \alpha} \eta_\beta
%\partial_i 
%+ \frac{1}{2}(C \sigma_i)_{\alpha \beta} \eta_i
%\partial_{\eta^\beta} \,,
%\label{eq:diffop1}
%\end{eqnarray}
%corresponding to the $osp(2,1)$ generators $\Lambda_i$ and $\Lambda_\alpha$, respectively.

As $OSp(2,2)$ acts on $\psi$, that is on $S^{(3,2)}$, preserving the
$U(1)$ fibres in the map $S^{(3,2)} \rightarrow S^{(2,2)}$, it 
has an action on the latter. It is not the adjoint action, but closely 
related to it, as we now explain.

The nature of the $OSp(2,2)$ action on $S^{(2,2)}$ has elements of
subtlety. If $g \in OSp(2,2)$ and $\psi \in S^{(3,2)}$ then
$g \,\psi \in S^{(3,2)}$ and hence $w(g \,\psi) \in S^{(2,2)}$ :
\begin{gather}
w_i(g \,\psi)^2 + C_{\alpha \beta} \, w_\alpha (g \,\psi) \, w_\beta(g \,\psi) = 1 \,, \nonumber \\
w_i (g \,\psi)^\ddagger = w_i (g \,\psi) \,, \quad  w_\alpha ^\ddagger
(g \,\psi) = - C_{\alpha \beta} \,
w_\beta (g \,\psi) \,.
\end{gather}
But the expansion of $w_\alpha (g\,\psi)$ for infinitesimal $g$ 
contains not only the odd Majorana spinors
$\eta_\alpha ^{(-)}$, but also the even ones $ \eta_\alpha ^{(+)}$,
where $(\eta_\alpha ^{(+)}) ^\ddagger = - \sum_{\beta =6,7} 
\tilde{C}_{\alpha \beta} \, \eta_\beta ^{(+)}$ $(\alpha = 6,7)$. 
We cannot thus think of the $OSp(2,2)$ action 
as an adjoint action on the adjoint space of $OSp(2,1)$. The reason of
course is that the Lie superalgebra $osp(2,1)$ is not 
invariant under graded commutation with the generators $\Lambda_{6,7,8}$ of $osp(2,2)$.

Now consider the generalization of the map (\ref{eq:promap1}) 
to the $osp(2,2)$ Lie algebra, 
\begin{equation}
\psi^\prime \quad \longrightarrow \quad {\cal W}_a (\psi) 
:= {\bar \psi}^\prime \Lambda_a ^{(\frac{1}{2})} \psi^\prime \, =  
\frac{2}{|\psi|^2} \bar{\psi} \Lambda_a^{(\frac{1}{2})} \psi \,, \quad a = ( 1, \ldots , 8 ) \,,
\end{equation}
where the $\bar{\psi}$ dependence of ${\cal W}_a$ has been suppressed
for notational brevity. Just as for $OSp(2,1)$, we find,
\begin{equation}
{\cal W}_a(g^{-1} \psi) = {\cal W}_b(\psi) (Ad \,g)_{ba} \,, 
\quad a,b = \,1,\ldots ,8 \,, \quad g \in OSp(2,2)\,. 
\end{equation}
Thus this extended vector ${\cal W}(\psi) = ({\cal W}_1(\psi) 
\,,{\cal W}_2(\psi), \ldots ,{\cal W}_8(\psi))$ transforms as 
an adjoint (super)-vector of $osp(2,2)$ under $OSp(2,2)$ action. 
The formula given in (\ref{eq:promap2}) extends to this case 
when index $a$ there also takes the values $(6,7,8)$. Explicitly we have 
\begin{gather}
{\cal W}_6(\psi) = \frac{1}{|\psi|^2} (\bar{z}_1 \theta -
z_2 \bar{\theta}) \,,\quad {\cal W}_7(\psi) =  
\frac{1}{|\psi|^2} (\bar{z}_2 \theta + z_1 \bar{\theta}) \,, \nonumber \\
{\cal W}_8(\psi) = 2 \frac{1}{|\psi|^2}(\bar{z}_i z_i + 2 \bar{\theta}
\theta) = 2 \big(2 - \frac{1}{|\psi|^2} \bar{z}_i z_i \big) \,.
\end{gather} 
The reality conditions for ${\cal W}_6(\psi) \,, {\cal W}_7(\psi) \,, 
{\cal W}_8(\psi)$ are  
\begin{equation}
{\cal W}_8(\psi)^\ddagger = {\cal W}_8(\psi) \,, \quad 
{\cal W}_\alpha(\psi)^\ddagger = - \sum_{\beta = 6,7} \tilde{C}_{\alpha \beta} \,
{\cal W}_\beta (\psi) \,, \quad \alpha = 6,7 \,,
\end{equation}
showing that the new spinor ${\cal W}_\alpha (\psi)$, $(\alpha = 6,7)$
is an even Majorana spinor as previous remarks suggested.

%The differential operators for the additional $osp(2,2)$ generators have the form
%\begin{eqnarray}
%{\tilde \zeta}_\alpha &=& -r \Big ( 1 + \frac{2}{r^2} \Big ) C_{\alpha \beta} \partial_{\eta_\beta} + \frac{1}{2r} 
%(\sigma_i)_{\beta \alpha} 
%\eta_\beta {\cal L}_i - \frac{\eta_\alpha}{2r} \eta_i \partial_i \,, \nonumber \\ 
%\zeta_8 &=& \Big ( \frac{\eta_+ x_3}{r} + \frac{\eta_- x_+}{r} \Big) \partial_+ +
%\Big ( \frac{\eta_+ x_-}{r} - \frac{\eta_- x_3}{r} \Big) \partial_- \equiv 2(\eta_- v_+ -\eta_+ v_-) \,,
%\label{eq:diffop2}
%\end{eqnarray}
%where in our case $r = \frac{1}{2}$ and $\partial_\pm$ stand for
%derivatives with respect to $\eta_\pm = \eta_1 \pm i \eta_2$.

As ${\cal W}(\psi)$ transforms as an adjoint vector under $OSp(2,2)$, 
the $OSp(2,2)$ Casimir function evaluated at ${\cal W}(\psi)$ 
is a constant on this orbit:
\begin{equation}
\frac{1}{2} ({\cal W}_a (S^{-1})_{ab} {\cal W}_b) = {\cal W}_i^2
(\psi) + \tilde{C}_{\alpha \beta} \,{\cal W}_\alpha (\psi) {\cal W}_\beta(\psi) - 
\frac{1}{4} {\cal W}_8 ^2 (\psi) = {\mbox{constant}} \,. 
\end{equation}
But we saw that the sum of the first term, and the second term with 
$\alpha \,,\beta = 4,5$ only, is invariant under $OSp(2,1)$.
Hence so are the remaining terms:
\begin{equation}
\sum_{\alpha, \beta = 6,7} \tilde{C}_{\alpha \beta} {\cal W}_\alpha
(\psi) {\cal W}_\beta(\psi) - \frac{1}{4} {\cal W}_8(\psi)^2 = {\mbox{constant}} \,.
\end{equation}
Its value is $-1$ as can be calculated by setting $\psi=(1,0,0)$.

In fact, since the $OSp(2,1)$ orbit has the dimension of
$S^{(3,2)}/U(1)$ and ${\cal W}_a(\psi) = {\cal W}_a(\psi \,e^{i \gamma})$ 
are functions of this orbit, we can completely express the latter 
in terms of $w(\psi)$. We find\footnote{${\cal W}_{6,7,8}$
become  $-{\cal W}_{6,7,8}$ for $J_{osp(2,2) -}$.} 
\begin{gather}
{\cal W}_\alpha(\psi) = -w_\beta \left ( \frac{\sigma \cdot
  w(\psi)}{r} \right)_{\beta \,,\alpha-2} \nonumber \\
 {\cal W}_8(\psi) = \frac{2}{r} (r^2 + C_{\alpha \beta} w_\alpha
  w_\beta) \,, \quad r^2 = w_iw_i \,.
\label{eq:nonl}
\end{gather}
%\begin{gather} 
%{\cal W}_6(\psi) = -2 \Big ( w_3(\psi) w_4(\psi) + \big 
%(w_1(\psi) + i w_2(\psi) \big ) w_5(\psi) \Big ) \,, \nonumber \\ 
%{\cal W}_7(\psi) = 2 \Big (w_3(\psi) w_5(\psi) - \big ( w_1(\psi) -i 
%w_2(\psi) \big) w_4(\psi) \Big) \,, \nonumber \\
%{\cal W}_8(\psi) =  2 \Big ( 1 - \sqrt{w_i(\psi)^2} \Big ) \,.  
%\label{eq:nonlre}
%\end{gather}

\subsection{Fuzzy Supersphere $S_F^{(2,2)}$}

We are now ready to construct the fuzzy supersphere $S_F^{(2,2)}$. We do so by 
replacing the coordinates $w_a$ of $S^{(2,2)}$ by ${\hat w}_a$: 
\be
w_a \, \longrightarrow \, {\hat w}_a = S^\dagger
\Lambda_a^{(\frac{1}{2})} S = \frac{1}{\sqrt{{\widehat N}}}\, \Psi^\dagger
\Lambda_a^{(\frac{1}{2})} \Psi \,\frac{1}{\sqrt{{\widehat N}}} 
= \frac{1}{{\widehat N}} \Psi^\dagger \Lambda_a^{(\frac{1}{2})} \Psi \,.
\ee
Obviously, we have ${\hat w}_a$ commuting with the number operator ${\widehat N}$:
\be
\lbrack {\hat w}_a \,, {\widehat N} \rbrack = 0 \,.
\ee
Consequently, we can confine ${\hat w}_a$ to the subspace ${\tilde
{\cal H}}_n$ of the Fock space of dimension $(2n+1)$ spanned by the kets
\be
|n_1 \,, n_2 \,, n_3 \rangle \equiv \frac{(a_1^\dagger)^{n_1}}{\sqrt{n_1!}} 
\frac{(a_2^\dagger)^{n_2}}{\sqrt{n_2!}} (b^\dagger)^{n_3}
|0 \rangle \,, \quad n_1 + n_2 + n_3 = n \,,
\label{eq:superstatevec}
\ee
where $n_3$ takes on the values $0$ and $1$ only. The Hilbert space  
${\tilde {\cal H}}_n$ splits into the even subspace 
${\tilde {\cal H}}_n^e$ and the odd subspace ${\tilde {\cal H}}_n^o$ 
of dimensions $n+1$ and $n$, respectively.
    
Linear operators, and hence $w_a$, acting on ${\tilde {\cal H}}_n$ 
generate the algebra of supermatrices $Mat(n+1, n)$ of dimension 
$(2n+1)^2$ which is customarily identified with the fuzzy
supersphere. Similar to the fuzzy sphere, $S_F^{(2,2)}$ also has a 
``quantum'' structure: $Mat(n+1, n)$ is its inner product space with 
the inner product
\be
(m_1 \,, m_2) = Str \,m_1^\ddagger m_2 \,, \quad \quad m_i \in Mat(n+1,n) \,,
\ee      
where the identity matrix is already normalized to have the unit norm in this form.

In order to be more explicit, we first note that the $osp(2,1)$ 
(and hence $osp(2,2)$) Lie superalgebras can be realized 
as a supersymmetric generalization of the Schwinger construction by
\be
\lambda_a = \Psi^\dagger (\Lambda_a ^{(\frac{1}{2})}) \Psi \,, 
\quad \lbrack \lambda_a \,, \lambda_b \rbrace = i f_{abc} \lambda_c \,.
\label{eq:lLambda} 
\ee
The vector states in (\ref{eq:superstatevec}) for $n=1$ give the 
superspin $J = \frac{1}{2}$ representation of $osp(2,1)$, while for
generic $n$ they correspond to the $n$-fold graded symmetric tensor
product of $J = \frac{1}{2}$ superspins that span the superspin 
$J = \frac{n}{2}$ representation of $osp(2,1)$. Therefore, on the
Hilbert space ${\tilde {\cal H}}_n$, we have 
\be 
\big (\lambda_i \lambda_i + C_{\alpha \beta} \lambda_\alpha
\lambda_\beta \big ) {\tilde {\cal H}}_n = \frac{n}{2} \Big 
( \frac{n}{2} + \frac{1}{2} \Big) {\tilde {\cal H}}_n \,.
\ee
Using the relation
\be
{\hat w}_a {\tilde {\cal H}}_n = \frac{2}{n} \lambda_a {\tilde {\cal H}}_n \,, 
\ee
we obtain    
\begin{gather}
\lbrack {\hat w}_a \,, {\hat w}_b \rbrace {\tilde {\cal H}}_n = \frac{2}{n} i f_{abc} {\hat w}_c {\tilde {\cal H}}_n \\
\big( {\hat w}_i {\hat w}_i +  C_{\alpha \beta} {\hat w}_\alpha {\hat w}_\beta \big) 
{\tilde {\cal H}}_n = \Big ( 1 +{\frac{1}{n}} \Big ) {\tilde {\cal H}}_n\,.   
\end{gather}
The radius $\sqrt{\Big ( 1 +{\frac{1}{n}} \Big )}$ of $S_F^{(2,2)}$ goes to $1$ as $n$ 
tends to infinity. The graded commutative limit is recovered when $J \rightarrow \infty \Rightarrow \lbrack 
{\hat w}_a \,, {\hat w}_b \rbrace \rightarrow 0$.

The Schwinger construction above naturally extends to the generators of $osp(2,2)$ as well. In general we can write
\be
{\widehat {\cal W}}_a := \frac{2}{n} \lambda_a \,, \quad a = (1 \,, \cdots \,, 8) \,. 
\label{eq:shwosp22}
\ee
(\ref{eq:shwosp22}) generate the $osp(2,2)$ algebra where
\be
{\widehat {\cal W}}_a \rightarrow  {\cal W}_a \quad \mbox{as} \quad n \rightarrow \infty.
\ee 
The generators ${\widehat {\cal W}}_{6,7,8}$ can be realized in terms of the $osp(2,1)$ generators.
This fact becomes important for field theories on both $S^{(2,2)}$ and $S_F^{(2,2)}$; Even though, 
these field theories have the $OSp(2,1)$ invariance, $osp(2,2)$
structure is needed to uncover it as we will see later in the chapter.
 
The observables of $S_F^{(2,2)}$ are defined as the linear operators 
$\alpha \in Mat(n+1,n)$ acting on $Mat(n+1, n)$. They have the 
graded right- and left- action on the Hilbert space $Mat(n+1, n)$ given by
\be
\alpha^L m = \alpha m \,, \quad \alpha^R m = (-1)^{|\alpha||m|} m
\alpha \,, \quad \forall \, m \in Mat(n+1, n) \,.
\ee  
They satisfy
\be
(\alpha \beta)^L = \alpha^L \beta^L \,, \quad (\alpha \beta)^R = 
(-1)^{|\alpha||\beta|} \beta \alpha \,,
\ee
and commute in the graded sense:
\be
\lbrack \alpha^L \,, \beta^R \rbrace = 0 \,, \quad \forall \, 
\alpha \,, \beta \in Mat(n+1, n) \,.
\ee

In particular $osp(2,1)$ and $osp(2,2)$ act on $Mat(n+1,n)$ by the (super)-adjoint action:
\be
ad \, \Lambda_a \, m = \big( \Lambda_a^L - \Lambda_a^R \big ) \, m = \lbrack \Lambda_a \,, m \rbrace \,,
\label{eq:gradeadac}
\ee
which is a graded derivation on the algebra $Mat(n+1,n)$.

%Just as $S_F^2$ preserves the $SU(2)$ symmetry of the $2$-sphere, 
%$S_F^{(2,2)}$ preserves the $OSp(2,1)$ symmetry of $S^{(2,2)}$
%since under the $OSp(2,1)$ action $Ad g {\widehat {\cal W}}_a = 
%g {\widehat {\cal W}}_a g^{-1}$ we have  
%\be
%\lbrack  g {\widehat {\cal W}}_a g^{-1} \,,  g {\widehat {\cal W}}_b
%g^{-1} \rbrace 
%= \frac{1}{n} i f_{abc} g {\widehat {\cal W}}_c g^{-1} \,. 
%\ee
%Thus we see that $OSp(2,1)$ supersymmetry encoded in $S_F^{(2,2)}$ is
%exact, which enables us to develop field theories on 
%$S_F^{(2,2)}$ that possess exact supersymmetry.

Before closing this section we note that left- and right-action 
of $\Psi_\mu$ and $\Psi_\mu^\dagger$ can also be defined on 
$Mat(n+1, n)$. They shift the dimension of the Hilbert space by an
increment of $1$ and will naturally arise in 
discussions of ``fuzzy sections of bundles'' in section \ref{sec-starpro}.      

\section{More on Coherent States}
\label{sec-SCS} 

In this section we construct the $OSp(2,1)$ supercoherent states (SCS)
by projecting them from the coherent states associated to 
${\mathbb C}^{2,1}$ \cite{seckin1}. 
In the literature the construction of $OSp(2,1)$ coherent states has
been discussed \cite{chaichian, gradechi}. 
Here we explicitly show that our SCS is equivalent to the one obtained
using the Perelomov's construction of 
the generalized coherent states, considered in chapter 3. 
 
We start our discussion by introducing the coherent state including 
the bosonic and fermionic degrees of freedom 
\cite{perelomov, Klauder}:
\begin{equation}
|\psi \rangle \equiv |z,\theta \rangle  = e^{-1/2 \,|\psi|^2}\, 
e^{a_\alpha ^\dagger z_\alpha + b^\dagger \theta}\,|0 \rangle \,.
\label{eq:scoherentstate}
\end{equation}
We can see from section \ref{sec-susyspace} that the labels $\psi$ of
the states $|\psi \rangle$ are in one to one correspondence with 
points of the superspace ${\cal C}^{(2,1)}$. We recall that 
$|\psi|^2 \equiv |z_1|^2 + |z_2|^2 + \bar{\theta} \theta$. 
%{\bf It unnecessary to set $|\psi|^2$ to $1$ here, the coherent state is
%  normalized as written in 9.75, without imposing any other condition}
Hence $|\psi \rangle$'s are normalized to $1$ as written.

The projection operator to the subspace ${\tilde {\cal H}}_n$ of the 
Fock space can be written as 
\begin{equation}
P_n = \sum_{n = n_1+n_2+n_3} 
\frac{1}{n_1 ! \,n_2 !}\,(a_1 ^\dagger)^{n_1}(a_2 ^\dagger)^{n_2}
(b ^\dagger)^{n_3}|0 \rangle \langle 0|(b)^{n_3}(a_2)^{n_2}(a_1)^{n_1}  \,,
\end{equation}
where $n_3 =0 \,{\mbox{or}}\,1$. Clearly $P_n^2 = P_n \,, P_n ^\dagger = P_n$.

Projecting $|\psi \rangle$ with $P_n$ and renormalizing the result 
by the factor $(\langle \psi|P_n |\psi \rangle)^{-1/2}$, we get
\begin{equation}
|\psi^\prime , n \rangle = \frac{1}{\sqrt{n!}} \, \frac{(a_\alpha ^\dagger z_\alpha + b^\dagger \theta)^n}{(|\psi|)^n}
\, |0 \rangle = \frac{(\Psi_\mu^\dagger \psi_\mu^\prime)^n}{\sqrt{n!}} | 0 \rangle \,.
\label{eq:5}
\end{equation}
This is the supercoherent state associated to $OSp(2,1)$. It is normalized to unity :
\be
\langle \psi^\prime , n | \psi^\prime , n \rangle = 1 \,.
\ee

We first establish the relation of (\ref{eq:5}) to the Perelomov's
construction of coherent states. To this end consider the following highest weight 
state in the $J_{osp(2,1)} = \frac{1}{2}$ representation of $osp(2,1)$ for which ${\widehat N}=1$: 
\begin{equation}
|J_{osp(2,1)} \, J_{su(2)} \,, J_3 \rangle = |\frac{1}{2}, \frac{1}{2}, \frac{1}{2} \rangle \,.
\label{eq:state}
\end{equation}
This is also the highest weight state in the associated non-typical representation 
$J_{osp(2,2) +} = (\frac{1}{2})_{osp(2,2) + }$ of $osp(2,2)$. Consider now the action of the $OSp(2,1)$ 
on (\ref{eq:state}). This can be realized by taking $g \in OSp(2,1)$ and ${\cal U}(g)$ as the corresponding element in the 
$3 \times 3$ fundamental representation. Thus let
\begin{equation}
|g \rangle  = {\cal U}(g) |\frac{1}{2}, \frac{1}{2}, \frac{1}{2} \rangle \,,
\label{eq:gcoherent}
\end{equation}
where $|g \rangle$ is the super-analogue of the Perelomov coherent
state \cite{perelomov}. We can write
\begin{equation}
|\frac{1}{2}, \frac{1}{2}, \frac{1}{2} \rangle = \Psi_1 ^\dagger |0 \rangle \,
\end{equation}
where $\Psi^\dagger = \left(\Psi_1 ^\dagger \,, \, \Psi_2^\dagger \,, \, \Psi_0 ^\dagger
\right) \equiv \left(a_1 ^\dagger \,, \, a_2 ^\dagger \,, \, b^\dagger \right)$ as given in (\ref{eq:gosterim1}).
In the basis spanned by the $\{\Psi_\mu ^\dagger|0 \rangle \}$, $(\mu = 1,2,0)$ the matrix of ${\cal U}(g)$ can be expressed as
\cite{chaichian}
\begin{equation}
{\cal D}(g) = \left(
\begin{array}{ccc}
z_1 ' & - \bar{z}_2 ' & - \theta^\prime \\
z_2 ' &  \bar{z}_1 ' & -{\bar \theta}^\prime \\
\chi  & -\bar{\chi} & \lambda
\end{array}
\right) \,, \quad \sum_{i} |z_i '|^2 + \bar{\theta} ' \theta ' = 1 \,.
\end{equation}
Then
\beqa
|g \rangle &=& \big({\cal D}(g) \big)_{1 \mu} \Psi_\mu ^\dagger |0 \rangle \nonumber \\
&=& \big( a_\alpha ^\dagger z_\alpha ' + b^\dagger \theta' \big ) |0
\rangle = \Psi_\mu ^\dagger \,\psi_\mu ' |0 \rangle \,.
\label{eq:D} 
\eeqa
Clearly (\ref{eq:D}) is exactly equal to $|\psi^\prime \,, 1 \rangle$ in (\ref{eq:5}).

For the case of general $n$, we start from the highest weight 
state $|\frac{n}{2}, \frac{n}{2}, \frac{n}{2} \rangle$ in the 
$n$-fold graded symmetric tensor product $\otimes_G ^n$ of the
$J_{osp(2,1)} = \frac{1}{2}$ representation and the 
corresponding representative ${\cal U}^{\otimes_G ^n} (g)$ of $g$: 
\begin{eqnarray}
|\frac{n}{2}, \frac{n}{2}, \frac{n}{2} \rangle &:=& | \frac{1}{2},
\frac{1}{2}, \frac{1}{2} \rangle \otimes_G \cdots \cdots 
\otimes_G |\frac{1}{2}, \frac{1}{2}, \frac{1}{2} \rangle \,, \nonumber \\
{\cal U}^{\otimes_G ^n} \,(g) &:=& {\cal U} \,(g) \otimes_G \cdots \cdots \otimes_G \,{\cal U} \,(g) \,.
\label{eq:tensorp1}
\end{eqnarray}
Note that, since ${\cal U} \,(g)$ is an element of $OSp(2,1)$, it is even. The corresponding coherent state is
\begin{equation}
|g; \frac{n}{2} \rangle = {\cal U}^{\otimes_G^n} |\frac{n}{2}, \frac{n}{2}, \frac{n}{2} \rangle = 
{\cal U} \,(g) \,|\frac{1}{2}, \frac{1}{2}, \frac{1}{2} \rangle \otimes_G \cdots \cdots 
\otimes_G \, {\cal U} \,(g) \,|\frac{1}{2}, \frac{1}{2}, \frac{1}{2} \rangle \,.
\end{equation}
Upon using (\ref{eq:D}) this becomes equal to (\ref{eq:5}) as we intended to show.

The coherent state in (\ref{eq:scoherentstate}) can be written as a sum of its even and odd 
components by expanding it in powers of $b^\dagger$:
\beqa
|\psi \rangle \equiv |z,\theta \rangle  &=& e^{-1/2 \,|\psi|^2} \, e^{a_\alpha ^\dagger z_\alpha} \, \big(|0 \,, 0 
\rangle - \theta \, | 0 \,, 1 \rangle \big ) \nonumber \\
&=& |z \,, 0 \rangle - \theta \, |z \,, 1 \rangle \,.
\label{eq:evenoddex}
\eeqa

We proved in chapter 2 that the diagonal matrix elements of an
operator $K$ in the coherent states $|z \rangle$ completely determine
$K$. That proof can be adapted to $| \psi \rangle$ as can be
infered from (\ref{eq:evenoddex}). It can next be adapted to
$|\psi^\prime, n \rangle$ for operators leaving the the subspace $N =n$
invariant. The line of reasoning is similar to the one used for
$SU(2)$ coherent states in chapter 2. 

\section{The Action on Supersphere $S^{(2,2)}$}
\label{sec-SUSYac}

The simplest $Osp(2,1)$-invariant Lagrangian density ${\cal L}$ can be
written as $\Phi^\ddagger V \Phi$, where $\Phi$ is the scalar
superfield and $V$ an appropriate differential operator. 
%To this can be added a mass term like $m^2 \Phi^\ddagger \Phi$. We can also impose
%the reality condition $\Phi^\dagger = \Phi$. We ignore these possibilities
We focus on ${\cal L}$ in what follows.

The superfield $\Phi$ is a function on $S^{(2,2)}$, that is , it is a
function of $w_a \,, (a= 1,2, \cdots,5)$ fulfilling the
constraint in (\ref{eq:real1}).

For functional integrals, what is important is not ${\cal L}$, but the
action $S$. Thus we need a method to integrate ${\cal L}$ over
$S^{(2,2)}$ maintaining SUSY.

We also need a choice of $V$ to find $S$. The appropriate choice is not
obvious, and was discovered by Fronsdal \cite{Fronsdal}. It was
adapted to $Osp(2,1)$ by Grosse et al. \cite{GKP1}.

We now describe these two aspects of $S$ and indicate also the
calculation of $S$.

\vskip 2em
{\it i. Integration on $S^{(2,2)}$}
\vskip 2em

Let $K$ be a scalar superfield on  $S^{(2,2)}$. It is a function of
$w_i$ and $w_\alpha$. We can write it as 
\be
K = k_0 + C_{\alpha \beta} k_\alpha w_\beta + k_1 C_{\alpha \beta}
w_\alpha w_\beta
\label{eq:kfield}
\ee
where $k_0$ and $k_1$ are even, $k_\alpha (\alpha= 4,5)$ is odd and $k's$ do not
depend on $w_\alpha$'s, but can depend on $w_i$'s.       

There is no need to include $w_{6, 7}$ in (\ref{eq:kfield}) as they
are nonlinearly related to $w_{4, 5}$.

The integral of $K$ over $S^{(2,2)}$ (of radius
$R$) can be defined as 
\be
I(K) = \int d \Omega r^2 \, dr \,  d w_4 \, d w_5 \, \delta(r^2 + C_{\alpha \beta}
w_\alpha w_\beta - R^2) \, K
\label{eq:invint1}
\ee
where $R > 0$ and $d \Omega = d \cos (\theta) d \psi$ is the volume
form on $S^2$. 

In the coefficients of $K$ in the integrand of $I(K)$, we do not
constrain $w_i \,, w_\alpha$ to fulfil $w_i^2 + C_{\alpha \beta}
w_\alpha w_\beta = R^2$.
  
The grade-adjoint representation of $osp(2,1)$ is $5$-dimensional. It
acts on ${\mathbb R}^{3,2} := {\mathbb R}^3 \oplus {\mathbb R}^2$ with
an even subspace ${\mathbb R}^3$ (spanned by $w_i$) and an odd
subspace ${\mathbb R}^2$ (spanned by $w_\alpha$).  Integration in      
(\ref{eq:invint1}) uses the $OSp(2,1)$-invariant volume form on
${\mathbb R}^{3,2}$ and the $OSp(2,1)$-invariant $\delta$-function to
restrict the integral to $S^{(2,2)}$. Thus $I(K)$ is invariant under
the action of SUSY on $K$.

$I(K)$ is in fact $OSp(2,2)$ invariant. That is because
$OSp(2,2)$ leaves the argument of the $\delta$-function invariant as we already
saw. The volume form as well is invariant because of the nonlinear
realization of $W_{6,7,8}$ as is easily checked.

We can write
\beqa
\delta(r^2 + C_{\alpha \beta} w_\alpha w_\beta - R^2) &=& \delta(r^2-
R^2) + 2 w_4 w_5 \frac{d}{d r^2} \delta(r^2- R^2) \nonumber \\   
&=& \frac{1}{2R} \delta(r - R) + \frac{1}{2 R r} w_4 w_5 \frac{d}{d r}
  \delta(r - R) \,,
\eeqa
where we have dropped terms involving $\delta(r + R)$ and $\frac{d}{d
  r } \delta (r + R)$ as they do not contribute to the $\lbrack 0 \,,
\infty)$, $dr$-integral. Thus using also
\be 
\int d w_4 \, d w_5 \, w_4 \, w_5 = -1 \,,
\ee
we get
\be
I(K) = \int d \Omega \left \lbrack \frac{d}{dr} (r k_0) - R k_1 \right
\rbrack_{r = R} \,.
\label{eq:Iaction}
\ee
This is a basic formula.

\vskip 2em

{\it ii. The $OSp(2,1)$-invariant operator $V$}   

\vskip 2em

The first guess would be the Casimir $K_2$ of $OSp(2,1)$, written in
terms of differential and superdifferential operators \cite{Fronsdal,GKP1}.
But this choice is not satisfactory. The simplest $OSp(2,1)$-invariant
action is that of the Wess-Zumino model \cite{Wess} and contains just
the standard quadratic (``kinetic energy'') terms of the scalar and
spinor fields. But $K_2$ gives a different action with nonstandard
spinor field terms \cite{Fronsdal, GKP1}.

But the $OSp(2,1)$ representation is also the nontypical
representation of $OSp(2,2)$ and its $OSp(2,2)$ Casimir $K_2^\prime$
is certainly $OSp(2,1)$ invariant. Thus so is $V$:
\be
V := K_2^\prime - K_2 = \Lambda_6  \Lambda_7 -  \Lambda_7  \Lambda_6 +
\frac{1}{4} \Lambda_8^2 \,.
\label{eq:opV}
\ee
It happens that this $V$ correctly reproduces the needed simple
action.

\vskip 2em

{\it iii. How to calculate : A sketch}

\vskip 2em

SUSY calculations are typically a bit tedious. For that reason, we
just sketch the details and give the final answer. 

We first expand the superfield $\Phi$ in the standard manner:
\be
\Phi(w_i, w_\alpha) = \varphi_0(w_i) + C_{\alpha \beta} \psi_\alpha
w_\beta + \chi(w_i) C_{\alpha \beta} w_\alpha w_\beta \,.
\label{eq:susfi1}
\ee
Here $(\alpha, \beta = 4, 5)$, $\varphi_0$ and $\chi$ are even fields
(commuting with $w_\alpha$) and $\psi_\alpha$ are odd fields
(anticommuting with $w_\alpha$). 

The aim is to calculate
\be
S= I( \Phi^\ddagger V \Phi) \,.
\ee

For $V$ we take (\ref{eq:opV}) where $\Lambda_{6,7, 8}$ represent the
$OSp(2,2)$ generators acting on $w_i, w_\alpha$. Thus we need to know
how they act on the constituents of $\Phi$ in (\ref{eq:susfi1}).

The action of $\Lambda_\alpha$ on $w_\beta$ follows from
(\ref{eq:osp22algebra}) since $w_\beta$ transform  like $osp(2,2)$ generators:
\be 
\Lambda_\alpha w_i = \frac{1}{2} w_\beta ({\tilde \sigma}_i)_{\beta
  \alpha} \,, \quad \Lambda_\alpha w_{\beta-2} = \frac{1}{2} C_{\alpha \beta}
w_8 \,, \quad \alpha \,, \beta = 6,7 \,.
\ee
We now write $w_{6, 7}$ in terms of  $w_{4, 5}$ using the relation
(\ref{eq:nonl}) to find 
\beqa
\Lambda_\alpha w_i &=& -\frac{1}{2} w_{\gamma - 2} (\sigma \cdot {\hat
  w})_{\gamma \beta} ({\tilde \sigma}_i)_{\beta \alpha} \,,  \nonumber \\
\Lambda_\alpha w_{\beta - 2} &=& \frac{1}{2} C_{\alpha \beta} \frac{2}{r}
(r^2 + 2 w_4 w_5)  \,, \quad \alpha, \beta, \gamma = 6, 7 \,.
\eeqa

The action of of $\Lambda_\alpha$ on the fields of (\ref{eq:susfi1}) follows from
the chain rule. For example, 
\be
\Lambda_\alpha \varphi_0 (w_i) = (\Lambda_\alpha w_i) \frac{\partial}{\partial w_i} \varphi_0(w_i) \,.
\ee

The ingredients for working out the action are now at hand. The
calculation can be conveniently done for a real superfield:
\be
\Phi^\ddagger = \Phi \,.
\ee
$\Phi$ can be decomposed in component fields as follows:
\be
\Phi = \psi_0 + C_{\alpha \beta} \psi_\alpha \theta_\beta + \frac{1}{2} \chi C_{\alpha \beta} \theta_{\alpha \beta}
\ee
Then with $\theta_\alpha$ an odd Majorana spinor,
\be
\theta_\alpha^\ddagger = - C_{\alpha \beta} \theta_{\beta} \,,
\ee
we find that so is $\psi$:
\be
\psi_\alpha^\ddagger = - C_{\alpha \beta} \psi_{\beta} \,.
\ee

We give the answer for the action 
\be
S(\Phi)= \int d \Omega \, r^2 \, dr \, \delta(r^2+ C_{\alpha \beta} w_\alpha w_\beta -1) \Phi V \Phi \,.
\ee
We have set $R=1$  whereas in previous sections we had $R
=\frac{1}{2}$. We have 
\beqa
&&S(\Phi) = \int d \Omega \left \lbrace - \frac{1}{4} ({\cal L}
  \varphi_0)^2 + \frac{1}{4} (\chi - \varphi_0^\prime)^2 - \frac{1}{4}
(C \psi)_\alpha (D \psi)_\alpha \right \rbrace \nonumber \\  
&& \varphi^\prime_0 = \frac{1}{r} \frac{d}{d r} \psi_0 \,, \quad D = - {\tilde \sigma}
\cdot {\cal L} +1 \,, \quad {\cal L}_i = i (\vec{r} \times
\vec{\nabla})_i \,.
\label{eq:susyactioncomp1}
\eeqa
The Dirac operator $D$ here is unitarily equivalent to the Dirac
operator in chapter 8. 

%For details of deriving (\ref{eq:susyactioncomp1}) see \cite{SUSY1}.

$(\chi_0 - \varphi_0^\prime)$ is the auxiliary field $F$. Having no
kinetic energy term, it can be eliminated. SUSY transformations mix
all the fields. 

A complex superfield $\Phi$ can be decomposed into two real
superfields: 
\begin{gather}
\Phi = \Phi^{(1)} + i \Phi^{(2)} \\
\Phi = \frac{\Phi + \Phi^\ddagger}{2} \quad \,, \Phi^{(2)} =
\frac{\Phi - \Phi^\ddagger}{2i} 
\end{gather}         
The action for $\Phi$ is the sum of actions for $\Phi^{(i)}$. We can
use (\ref{eq:susyactioncomp1}) to write it. No separate calculation is
needed.

%For inclusion of mass term and interactions, see \cite{SUSY1}.

\section{The Action on the Fuzzy Supersphere $S_F^{(2,2)}$}

Finding the action on  $S_F^{(2,2)}$ is the crucial step for
regularizing supersymmetric field theories using finite-dimensional
matrix models, preserving $OSp(2,1)$-invariance.

We have seen that $S^2$ and $S_F^2$ allow instanton sectors. They
affect chiral symmetry and are important for physics.

There are SUSY generalizations of these instantons. They are discussed
in \cite{susybreaking}.

\subsection{The Integral and Supertrace}

In fuzzy physics with no SUSY, trace substitutes for $SU(2)$-invariant
integration. The trace $tr M$ of an $(n+1) \times (n+1)$ matrix $M$ is
invariant under the $SU(2)$ action $M \rightarrow U(g) M U(g)^{-1}$ by
its angular momentum $\frac{n}{2}$ representation $SU(2) :g
\rightarrow U(g)$. It becomes the invariant integration in the large
$n$ -limit. 

In fuzzy SUSY physics, the corresponding $OSp(2,2)$ invariant trace
is supertrace $str$.

But (\ref{eq:susyactioncomp1}) gives invariant integration in the
(graded) commutative limit. We now establish that str goes over to the
invariant integration as the cut-off $n \rightarrow \infty$.

A simple way to establish this is to use the supercoherent states. We
have already defined them in (\ref{eq:5}). Here we drop the $\prime$
on $\psi$ and write  
\be
|\psi, n \rangle = \frac{(a_\alpha ^\dagger z_\alpha + b^\dagger
  \theta)^n}{\sqrt{n !}}
\, |0 \rangle \,.
\ee      

Then as we saw, to every operator ${\hat K}$ commuting with $N=
a^\dagger_i a_i + b^\dagger b$, we can define its symbol $K$, a function
of $w's$, by 
\be
K(w) = \langle \psi, N | {\hat K} | \psi, N \rangle \,. 
%w_i = {\bar z} \tau_i z = {\bar \psi} \Lambda_i^{(\frac{1}{2})} \psi
%\,,\quad  w_\alpha = {\bar \psi} \lambda_\alpha^{(\frac{1}{2})} \psi
%\,, \quad \quad \alpha= 4, 5, 
%\end{gather}
%$\tau_i$ being the Pauli matrices.
\ee

An invariant ``integral'' ${\hat I}$ on ${\hat K}$ can then be defined as
\be
{\hat I}({\hat K}) = I(K) \,.
\ee
With the normalization
\be
\int d \Omega =1 \quad \mbox{or} \quad d \Omega = \frac{d cos \theta \wedge d \phi}{ 4
  \pi} \,,
\ee
we can show that 
\be
{\hat I}({\hat K}) = \frac{1}{2} str K \,.
\ee
It is then clear that $str$ becomes $2I$ as $n \rightarrow \infty$. 

The proof is easy. First note that for the non-SUSY coherent state
\be
|z \,, n \rangle =  \frac{(a^\dagger \cdot {\hat
    z}^n)}{\sqrt{n!}} |0 \rangle \,, \quad {\hat z} \cdot {\hat z} = 1 \,.
\ee
\be
\int d \omega \langle {\hat z}, n | {\hat A} | {\hat z}, n \rangle =
\frac{1}{n+1} Tr {\hat A} \, 
\ee
if ${\hat A}$ is an operator on the subspace spanned by $|{\hat z}, n
\rangle$ for fixed $n$. 

Terms linear in $b$ and $b^\dagger$ have zero str. Hence we can assume
that 
\be
{\hat K} = M_0 + M_1 b^\dagger b
\ee
where $M_j$ are polynomials in $a_i^\dagger a_j$.

It can be easily checked that $str {\hat K}$ is $OSp(2,2)$-invariant
as well.

In the $OSp(2,1)$ IRR $\left \lbrack \frac{N}{2} \right \rbrack_{osp(2,1)}$, the even subspace of
its carrier space has angular momentum $\frac{N}{2}$ and the odd
subspace has angular momentum $\frac{N-1}{2}$. Hence 
\be
str {\hat K} = tr_{N+1} M_0 - tr_N M_0 - tr_N M_1
\ee
where $tr_m$ indicates trace over an $m$-dimensional space. 

As for ${\hat I}({\hat K})$, we note that 
\be
|\psi, N \rangle = |z, N \rangle + \sqrt{N} b^\dagger \theta |z, N-1 \rangle \,.
\ee
   
Hence
\be
K(w) = \langle z \,, N| M_0 | z \,, N \rangle + N {\bar \theta}
\theta \langle z \,, N-1 | M_0 | z, N-1 \rangle + N {\bar \theta} 
\theta \langle z \,, N-1 | M_1 | z \,, N-1 \rangle \,.
\ee 

But by (\ref{eq:promap2}), ${\bar \theta} \theta = w_4 w_5$. So on using (\ref{eq:Iaction}), we get 
\beqa
&&I(K) = - \frac{1}{2} \int d \Omega \Big \lbrace N \, \langle z  \,, N-1 | M_0 | z \,, N-1
\rangle + N \, \langle z \,, N-1 | M_1 | z \,, N-1 \rangle \nonumber \\ 
&& \quad \quad \quad \quad \quad - (N+1) \, \rangle z \,, N | M_0 | z \,, N \Big \rbrace = \frac{1}{2} str {\hat K} \,.
\eeqa

\subsection{$OSp(2,1)$ IRR's with Cut-Off $N$}

The Clebsh-Gordan series for $OSp(2,1)$ is 

\be
[J]_{osp(2,1)} \otimes [K]_{osp(2,1)} = [J+K]_{osp(2,1)} \oplus 
\left \lbrack J+K -\frac{1}{2} \right \rbrack_{osp(2,1)} \oplus 
\cdots \oplus \left \lbrack |J-K|\right \rbrack_{osp(2,1)} \,.
\ee

The series on R.H.S thus descends in steps of $\frac{1}{2}$ (and not
in steps of $1$ as for $su(2)$) from $J+K$ to $|J-K|$.

Under the (graded) adjoint action of $osp(2,1)$, the linear operators in the 
representation space of $\left \lbrack \frac{N+1}{2} 
\right \rbrack_{osp(2,1)}$ transform as $\left \lbrack \frac{N+1}{2} 
\right \rbrack_{osp(2,1)} \otimes \left \lbrack \frac{N+1}{2} 
\right \rbrack_{osp(2,1)}$. Hence the $osp(2,1)$ content of the fuzzy
supersphere is
\begin{multline}
\left \lbrack \frac{N+1}{2} \right \rbrack_{osp(2,1)} \otimes 
\left \lbrack \frac{N+1}{2} \right \rbrack_{osp(2,1)} = \\
\lbrack N+1 \rbrack_{osp(2,1)} \oplus \left \lbrack \frac{N+1}{2} 
\right \rbrack_{osp(2,1)} \oplus \left \lbrack N +\frac{1}{2}
\right \rbrack_{osp(2,1)} \oplus \cdots \oplus [0]_{osp(2,1)} \,.
\end{multline}

We now discuss
\begin{itemize}
\item{The highest weight angular momentum states in each of these IRR's
and the realization of $osp(2,2)$ on these $osp(2,1)$ multiplets, and}
\item{The spectrum of $V$ and the free supersymmetric scalar field action
on the fuzzy supersphere.}
\end{itemize}

\subsection{The Highest Weight States and the $osp(2,2)$ Action}

The graded Lie algebra $osp(2,1)$ is of rank $1$. We can diagonalize
(a multiple of) one operator in $osp(2,1)$ in each IRR. We choose it
to be $\Lambda_3$, the third component of angular momentum. 

$\Lambda_4$ is a raising operator for $\Lambda_3$, raising its
eigenvalues by $\frac{1}{2}$. The vector state annihilated by
$\Lambda_4$ in an IRR of $osp(2,1)$ is it highest weight state.

$\Lambda_+ = \Lambda_1 + i \Lambda_2$ is also a raising operator for
$\Lambda_3$, raising its eigenvalue by $+1$. Vector states annihilated
by $\Lambda_4$ are the highest weight states for the $su(2)$ IRR's contained
in an $osp(2,1)$ IRR. A vector state in an IRR annihilated by $\Lambda_4$
is also annihilated by $\Lambda_+$.

The matrices of the fuzzy supersphere are polynomials in $a_i^\dagger
a_j$, $a_i^\dagger b$, $b^\dagger a_i$ restricted to the subspace with
$N = a^\dagger_i a_i + b^\dagger b$ fixed. Supersymmetry acts on them
by adjoint action. The expression for $\Lambda_4$ is given in (\ref{eq:lLambda}) 
while
\be
\Lambda_+ = a_1^\dagger a_2 \,.
\ee
It follows that for $J$ integral, 
\begin{gather}
\mbox{The highest weight state for} \quad [J]_{osp(2,1)} =
(a_1^\dagger a_2)^J \,, \nonumber \\ 
\mbox{the highest weight state for} \quad [J-\frac{1}{2}]_{osp(2,1)} =
(a_1^\dagger a_2)^{J -1} \Lambda_6 \,.
\label{eq:hwstates11}
\end{gather}

The fact that $(a_1^\dagger a_2)^{J -1} \Lambda_6$ anticommutes 
with $\Lambda_4$ follows from $\lbrace \Lambda_4 \,, \Lambda_6 \rbrace = 0$.

The states with angular momentum $J-\frac{1}{2}$ in $[J]_{osp(2,1)}$ and $J-1$ in
$[J- \frac{1}{2}]_{osp(2,1)}$ which are $su(2)$-highest weight states can be
got acting with ad$\Lambda_5$ on heigest weight states in
(\ref{eq:hwstates11}).
\beqa
\begin{array}{cccc}
[J]_{osp(2,1)}: & (a_1^\dagger a_2)^J &
\stackrel{\mbox{ad}\Lambda_5}{\longrightarrow} 
& (a_1^\dagger a_2)^{J-1} \Lambda_4 \\
& & & \\
& ad \Lambda_7 \downarrow & \swarrow ad \Lambda_8 & ad \Lambda_7 \downarrow \\                 
& & & \\
\lbrack J - \frac{1}{2} \rbrack_{osp(2,1)}: & (a_1^\dagger a_2)^{J-1} \Lambda_6 & 
\stackrel{\mbox{ad}\Lambda_5} {\longrightarrow} & X 
\end{array}
\eeqa
where
\be
X = \frac{1+ N- J}{4} (a_1^\dagger a_2)^{J-1} + \frac{2J - 1}{4}(a_1^\dagger a_2)^{J-1} b^\dagger b
\ee
As usual, $ad$ denotes graded adjoint action as in
{\ref{eq:gradeadac}}. The vectors are not normalized. The arrows
indicate the adjoint actions of $\Lambda_{5,7,8}$.
They establish that $osp(2,2)$ acts irreducibly on $[J]_{osp(2,1)}
\oplus \lbrack \frac{J-1}{2} \rbrack_{osp(2,1)}$.

%Using (ref(eq:...)) 
We also see that
\be
J = \left (0, \frac{1}{2} \,, \cdots , \frac{N+1}{2} \right ) \,.
\ee

\subsection{The Spectrum of $V$}

We show that for $J$ integer
\begin{subequations}
\begin{align}
V \big |_{[J]_{osp(2,1)}} &= \frac{J}{2} {\bf 1} \,, 
\label{eq:spectrumofVa} 
\\
V \big |_{[J-\frac{1}{2}]_{osp(2,1)}} &= - \frac{J}{2} {\bf 1} \,.
\label{eq:spectrumofVb}
\end{align}
\end{subequations}

\vskip 2em

{\it Proof of} (\ref{eq:spectrumofVa})

\vskip 2em

It is enough to evaluate $V$ on the highest weight state
$(a_1^\dagger a_2)^J$. Since
\be
ad \Lambda_8 \, (a_1^\dagger a_2)^J = ad \Lambda_6 \, (a_1^\dagger a_2)^J = 0
\,,
\ee
we have
\beqa
V (a_1^\dagger a_2)^J &=& (ad \Lambda_6 ad \Lambda_7 - ad \Lambda_7 ad
\Lambda_6)
(a_1^\dagger a_2)^J   \nonumber \\
 &=& (ad \Lambda_6 ad \Lambda_7 + ad \Lambda_7 ad \Lambda_6) (a_1^\dagger a_2)^J
\nonumber \\
&=& ad \lbrace \Lambda_6 \,, \Lambda_7 \rbrace \, (a_1^\dagger a_2)^J
\nonumber \\
&=& -\frac{1}{2} (\varepsilon \sigma_i)_{67} \, ad \Lambda_i
(a_1^\dagger a_2)^J \nonumber \\
&=& \frac{1}{2} ad \Lambda_3 \, (a_1^\dagger a_2)^J \nonumber \\
&=& \frac{J}{2} (a_1^\dagger a_2)^J \,.
\eeqa

\vskip 2em

{\it Proof of} (\ref{eq:spectrumofVb})

\vskip 2em

We evaluate $V$ on $(a_1^\dagger a_2)^{J-1} \Lambda_6$. We have
\be
ad \Lambda_8 \,(a_1^\dagger a_2)^{J-1} \Lambda_6 = (a_1^\dagger
a_2)^{J-1} \Lambda_4 \,.
\ee
Thus
\be
\frac{1}{4} (ad \Lambda_8)^2 (a_1^\dagger a_2)^{J-1} \Lambda_6 =
\frac{1}{4} (a_1^\dagger a_2)^{J-1} \Lambda_6 \,.
\ee
Now the $osp(2,1)$ Casimir $K_2$ has value $J(J + \frac{1}{2}) {\bf
1}$ in the IRR $[J]_{osp(2,1)}$ while
\be
(ad \Lambda_i)^2 (a_1^\dagger a_2)^{J-1} \Lambda_4 = (J-
\frac{1}{2})(J+ \frac{1}{2}) \, (a_1^\dagger a_2)^{J-1} \Lambda_4
\label{eq:Lambda8kare}
\ee
Hence with $\alpha \,, \beta \in [4, 5]$,
\be
(\varepsilon_{\alpha \beta} \, ad \Lambda_\alpha \, ad \Lambda_\beta)
\, (a_1^\dagger a_2)^{J-1} \Lambda_4 = \frac{2J+1}{4} (a_1^\dagger
a_2)^{J-1} \Lambda_4 \,.
\ee
But
\be
e^{i \frac{\pi}{2} \Lambda_8} \, \Lambda_{4,5} \, e^{-i \frac{\pi}{2}
\Lambda_8} = i \,\Lambda_{6,7} \,.
\ee
Hence
\beqa
e^{i \frac{\pi}{2} \Lambda_8} \, (\varepsilon_{\alpha \beta} \,ad
\Lambda_\alpha \, ad \Lambda_\beta \,
(a_1^\dagger a_2)^{J-1} \Lambda_4) \, e^{-i \frac{\pi}{2} \Lambda_8} &=&
- (\varepsilon_{\alpha \beta} \, ad \Lambda_{\alpha^\prime} \, ad
\, \Lambda_{\beta^\prime}) \, ( i (a_1^\dagger a_2)^{J-1}
\Lambda_6 ) \nonumber \\
&=& \frac{2J+1}{4} \,i  \, (a_1^\dagger a_2)^{J-1} \Lambda_6 \,.
\eeqa
(\ref{eq:spectrumofVb}) follows upon using this and (\ref{eq:Lambda8kare}).

\subsection{The Fuzzy SUSY Action}

Let $J$ be integral. We can write the highest weight component in
angular momentum $J$ of the superfield in the IRR $[J]_{osp(2,1)}$ as 
\be
\Phi_J = c_j (a_1^\dagger a_2)^J + (a_1^\dagger a_2)^{J-1} \xi_{J-
  \frac{1}{2}} \Lambda_4 \,
\ee
where $c_j$ is a (commuting) complex number and $\xi_{J-\frac{1}{2}}$ is a Grasmmann
number. The $osp(2,2)$ transformations map $[J]_{osp(2,1)}$ to  
$[J-\frac{1}{2}]_{osp(2,1)}$. The highest weight component in the
latter can be written as
\be
\Phi_{J -\frac{1}{2}} = \eta_{J- \frac{1}{2}} (a_1^\dagger a_2)^{J-1}
\Lambda_5 + d_{J-1} X \,,
\ee
where $\eta_{J -\frac{1}{2}}$ is a Grassmann and $d_{J-1}$ a complex
number.

The fuzzy action for the heighest weight state in $[J]_{osp(2,1)}$ is
\begin{multline}
S_F^J(M=J) = 
%S_F^J(M) \quad \mbox{for} \quad M=J \longrightarrow
 \frac{J}{2} str \Phi_J^\ddagger \Phi_J \\
= \frac{J}{2} \left \lbrack |c_J|^2 str \lbrack(a_2^\dagger a_1)^J (a_1^\dagger a_2)^J
\rbrack + \xi^\ddagger_{J-\frac{1}{2}} \xi_{J-\frac{1}{2}} \, str \lbrack \Lambda_4^\ddagger \, (a_2^\dagger
a_1)^{J-1} \, (a_1^\dagger a_2)^{J-1} \rbrack \right \rbrack \,, \quad
\Lambda_4^\ddagger = - \Lambda_5 \,,
\end{multline}
since the two terms in $\Phi_J$ are str-orthogonal. For $[J -
\frac{1}{2}]_{osp(2,1)}$, instead,
\begin{multline}
S_F^{J- \frac{1}{2}}(M = J -\frac{1}{2}) =  
%S_F^{J- \frac{1}{2}}(M) \quad \mbox{for} \quad M = J -\frac{1}{2} \longrightarrow 
\frac{J}{2} str \Phi_{J-\frac{1}{2}}^\ddagger \Phi_{J- \frac{1}{2}} \\
= -\frac{J}{2} \left \lbrack \eta_{J-\frac{1}{2}}^\ddagger
\eta_{J-\frac{1}{2}} \, str \lbrack \Lambda_5^\ddagger (a_2^\dagger a_1)^{J-1}
(a_1^\dagger a_2)^{J-1} \Lambda_5 \rbrack + |d_{J-1}|^2 \, str X^\ddagger X
\right \rbrack \,, \quad \Lambda_5^\ddagger = \Lambda_4 \,,
\end{multline}
since the two terms in $\Phi_{J- \frac{1}{2}}$ are also
str-orthogonal. The second term here is the integral spin term. It is
positive since
\be
str \, X^\ddagger X < 0 
\ee
as can be verified.

Str-orthogonality extends also to $\Phi_J$ and $\Phi_{J
-\frac{1}{2}}$:
\be
str \Phi^\ddagger \Phi_{J -\frac{1}{2}} = 0 \,.
\ee

Hence for the heighest weight states of $[J]_{osp(2,2)} =
[J]_{osp(2,1)} \oplus [J-\frac{1}{2}]_{osp(2,1)}$, the
actions add up:
\be
S_F^{J \oplus (J -\frac{1}{2})} \quad \mbox{for heighest weight states}
\quad = S_F^J(J) + S_F^{J - \frac{1}{2}} (J - \frac{1}{2}) \,.
\ee

The superfield $\Phi$ is a superposition of such terms. We must first
include all angular momentum desecendents of $\Phi_J$ and $\Phi_{J -
\frac{1}{2}}$. We must also sum on $J$ from $0$ to $N$ in steps of
$\frac{1}{2}$.

For the fuzzy sphere $S_F^2$, such calculations are best performed using
spherical tensors ${\widehat T}_{LM}(N)$ and their properties. Similarly,
perhaps such calculations are best performed on the fuzzy supersphere
using supersymmetric spherical tensors. But as yet only certain basic
results about these tensor are available \cite{fuzzyS}.

Reality conditions like $\Phi^\ddagger = \Phi$ constrain the Fourier
coefficents $c_j\,, \xi_{J- \frac{1}{2}}\,, \eta_{J-\frac{1}{2}}\,,
d_{J-1}$.

%$OSp(2,1)$-invariant interactions can also be included. \cite{SUSY1}

\section{The $*$-Products}
\label{sec-starpro}

\subsection{The $*$-Product on $S_F^{(2,2)}$}

The diagonal matrix elements of operators in the supercoherent state 
$|\psi^\prime \,, n\rangle$ define functions on $S^{(2,2)}_F$. 
The $*$-product of functions on $S^{(2,2)}_F$ is induced by this map 
of operators to functions. To determine this map explicitly it is 
sufficient to compute the matrix elements of the operators ${\widehat
{\cal W}}_a$. Generalization to arbitrary operators can
then be made easily as we will see. 

The diagonal coherent state matrix element for ${\widehat {\cal W}}_a$'s are
\be
{\cal W}_a \,(\psi^\prime \,, \bar{\psi}^\prime \,, n) = \langle
\psi^\prime, n| {\widehat {\cal W}}_a | \psi^\prime, n \rangle = 
\frac{2}{|\psi|^2} \, \bar{\psi}
\Lambda_a ^{(\frac{1}{2})} \psi = \bar{\psi^\prime} \,\Lambda_a^{(\frac{1}{2})} \,\psi^\prime \,. 
\label{eq:w12}
\ee
This defines the map
\be
{\widehat {\cal W}}_a \longrightarrow {\cal W}_a
\ee
of the operator ${\widehat {\cal W}}_a$ to functions ${\cal W}_a$.
${\cal W}_a$ is a superfunction on $S^{(2,2)}_F$ since it is
invariant under the $U(1)$ phase $\psi^\prime \rightarrow \psi^\prime e^{i \gamma}$.

We are now ready to define and compute the $*$-product of two functions of the form ${\cal W}_a$ and ${\cal W}_b$. 
It depends on $n$, and to emphasise this we include it in the argument of the product. It is given by 
\begin{equation}
{\cal W}_a * {\cal W}_b \,(\psi^\prime ,\bar{\psi}^\prime \,, n) = \langle \psi^\prime \,, n|{\widehat {\cal W}}_a \,
{\widehat {\cal W}}_b | \psi^\prime \,, n \rangle
\label{eq:starproduct}
\end{equation}
which becomes, after a little manipulation
\begin{equation}
{\cal W}_a *_n {\cal W}_b \, \big (\psi^\prime, \bar{\psi}^\prime , n \big ) = \frac{1}{n} \,\bar{\psi}^\prime\,
\Big (\Lambda_a ^{(\frac{1}{2})}  \, \Lambda_b ^{(\frac{1}{2})} \Big) \,\psi^\prime + \frac{n-1}{n} \, \Big( \bar{\psi}^\prime \,
\Lambda_a^{(\frac{1}{2})} \,\psi^\prime \Big ) \Big(\bar{\psi}^\prime \, \Lambda_b^{(\frac{1}{2})} \,\psi^\prime \Big ) \,.
\label{eq:manip}
\end{equation}
Furthermore, since $\psi^\prime \Lambda_a^{(\frac{1}{2})} \Lambda_b^{(\frac{1}{2})} \psi^\prime$ is ${\cal W}_a * 
{\cal W}_b (\psi^\prime \,, {\bar \psi}^\prime \,, 1)$, (\ref{eq:manip}) can be rewritten as
\begin{equation}
{\cal W}_a *_n {\cal W}_b \, \big ( \psi^\prime ,\bar{\psi}^\prime ,n \big ) = \frac{1}{n} {\cal W}_a *_1 {\cal W}_b \,
(\psi^\prime , \bar{\psi}^\prime, 1) + \frac{n-1}{n} \, {\cal W}_a \, (\psi^\prime, \bar{\psi}^\prime) \,{\cal W}_b \, 
(\psi^\prime , \bar{\psi}^\prime) \,.
\label{eq:starproduct1}
\end{equation}
Introducing the matrix $K$ with
\begin{equation}
K_{ab}:= {\cal W}_a *_1 {\cal W}_b - {\cal W}_a \, {\cal W}_b\,,
\label{eq:Kab}
\end{equation}
we can express (\ref{eq:starproduct1}) as
\begin{equation}
{\cal W}_a *_n \,{\cal W}_b = \frac{1}{n} K_{ab} + {\cal W}_a\,{\cal W}_b\,.
\label{eq:simpleproduct}
\end{equation} 
In this form it is apparent that in the graded commutative limit $n \to \infty$, we recover the graded commutative product of 
functions ${\cal W}_a$ and ${\cal W}_b$.

The $*$-product of arbitrary functions on $S^{(2,2)}_F$ can also be obtained via a similar procedure used to derive that on $S_F^2$.
In this case, one also needs to pay attention to the graded structure of the operators. Thus we can start from the generic operators 
$F$ and $G$ in the representation $(\frac{n}{2})_{osp(2,2) + }$ expressed as   
\begin{eqnarray}
{\widehat F} &=& F^{a_1 a_2 \cdots a_n} \, {\widehat {\cal W}}_{a_1} \otimes_G \cdots \otimes_G  {\widehat {\cal W}}_{a_n} \,, 
\nonumber \\
{\widehat G} &=& G^{b_1 b_2 \cdots b_n} \, {\widehat {\cal W}}_{b_1} \otimes_G \cdots \otimes_G  {\widehat {\cal W}}_{b_n} \,,
\end{eqnarray}
where for example $F^{a_1 \cdots a_i a_j \cdots a_n} = (-1)^ {|a_i| |a_j|} F^{a_1 \cdots a_j a_i \cdots a_n}$, $|a_i| 
\,(mod \,2)$ being the degree of the index $a_i$.  After a long but a
straightforward calculation, the following finite-series 
formula is obtained (details can be found in \cite{seckin1}):
\begin{equation}
{\cal F}_n *_n {\cal G}_n ({\cal W}) = {\cal F}_n \,{\cal G}_n ({\cal W}) +
\sum_{m = 1}^n \frac{(n-m)!}{n! \,m!} \,{\cal F}_n  ({\cal W}) \,\,\vdots
\underbrace{\big(\partial^{^{\!\!\!\!\!\leftarrow}}
\,K \,\partial^{^{\!\!\!\!\rightarrow}} \big) \cdots \big( \partial^{^{\!\!\!\!\!\leftarrow}}
\,K \,\partial^{^{\!\!\!\!\rightarrow}})}_{m \,factors} \,\vdots \, {\cal G}_n ({\cal W} \big) \,.
\label{eq:finalstar}
\end{equation}  
Here we have introduced the ordering $\vdots \cdots \vdots$, in which 
$\partial^{^{\!\!\!\!\!\leftarrow}}_{{\cal W}_{a_i}}$ 
$\big(\partial^{^{\!\!\!\!\!\rightarrow}}_{{\cal W}_{b_i}} \big)$ are moved to the left (right) extreme and 
$\partial^{^{\!\!\!\!\!\leftarrow}}_{{\cal W}_{a_i}}$'s ($\partial^{^{\!\!\!\!\!\rightarrow}}_{{\cal W}_{b_i}}$)'s
act on everything to their left (right). In doing so one always has to
remember to include the overall factor coming from graded 
commutations. Thus for example, $\vdots \big(\partial^{^{\!\!\!\!\!\leftarrow}}
\,K \,\partial^{^{\!\!\!\!\rightarrow}} \big) \big( \partial^{^{\!\!\!\!\!\leftarrow}}
\,K \,\partial^{^{\!\!\!\!\rightarrow}}) \,\vdots = (-1)^{|a||c|+ |b|(|c| +|d|)} \partial^{^{\!\!\!\!\!\leftarrow}}_{{\cal W}_{a}} 
\partial^{^{\!\!\!\!\!\leftarrow}}_{{\cal W}_{c}} K_{ab} K_{cd} \partial^{^{\!\!\!\!\!\rightarrow}}_{{\cal W}_{b}}
\partial^{^{\!\!\!\!\!\rightarrow}}_{{\cal W}_{d}}$. 
From (\ref{eq:finalstar}) it is apparent that, in the graded commutative limit \mbox{$(n \to \infty)$}, we get back the 
ordinary point-wise multiplication ${\cal F}_n \,{\cal G}_n ({\cal
  W})$. This formula was first derived in \cite{seckin1}.

A consequence of (\ref{eq:starproduct1}) is the graded commutator of the $*$-product
\begin{equation}
\lbrack {\cal W}_a, {\cal W}_b \}_{*_n} = \frac{i}{n} f_{abc} {\cal W}_c
\end{equation}
which generalizes a familiar result for the usual $*$-products.

A special case of our result for the $*$-product follows if we restrict ourselves to the even subspace $S_F^2$ of $S_F^{(2,2)}$, 
namely the fuzzy sphere. In this case, ${\cal F}_n({\cal W})$ and ${\cal G}_n({\cal W})$ become ${\cal F}_n(\vec{x})$ and 
${\cal G}_n(\vec{x})$ and we get from 
%(\ref{eq:starproduct1}) 
(\ref{eq:finalstar}):
\begin{multline}
{\cal F}_n *_n {\cal G}_n (\vec{x}) = {\cal F}_n \,{\cal G}_n (\vec{x}) +  
\sum_{m=1}^n \frac{(n - m)!}{n! \,m!}\, 2^m \partial_{i_1} \cdots \partial_{i_m} {\cal F}_n \,(\vec{x}) \\ 
\times \Big(\frac{1}{2} \Big)^m {\cal K}^+_{i_1 j_1} \cdots {\cal K}^+_{i_m j_m}\,\, 2^m \partial_{j_1} \cdots \partial_{j_m} 
{\cal G}_n \,(\vec{x}) \,,    
\end{multline} 
%$$
%\partial_{a_i} \equiv \partial_{{\cal W}_{a_i}}\,, \quad \partial_{b_j} \equiv \partial_{{\cal W}_{b_j}} \,,
%$$
which is the formula given in (3.99).

\subsection{$*$-Product on Fuzzy ``Sections of Bundles''}

Let us first remark that the left- and right-action of $\Psi_\mu^{L,R}$ and $(\Psi_\mu^\dagger)^{L,R}$ on $Mat(n+1, n)$ 
are defined and changes $n$ by an increment of $1$:
\beqa
\Psi_\mu^{L,R} Mat(n+1, n) : \quad \quad {\tilde {\cal H}}_n \rightarrow {\tilde {\cal H}}_{n-1} \,,\nonumber \\
(\Psi_\mu^{L,R})^\dagger Mat(n+1, n) : \quad \quad {\tilde {\cal H}}_n \rightarrow {\tilde {\cal H}}_{n+1} \,.
\eeqa

On $| \psi^\prime \,, n \rangle$ we find
\be
S_\mu |\psi^\prime, n \rangle = \psi_\mu^\prime |\psi^\prime, n - 1 \rangle \,, \quad 
\langle \psi^\prime, n| S_\mu ^\dagger = \langle \psi^\prime, n - 1| \bar{\psi_\mu}^\prime \,.
\label{eq:vectors}
\ee
Thus we get the matrix elements
\be
\langle \psi^\prime, n - 1|S_\mu|\psi^\prime, n \rangle = \psi_\mu^\prime \,, \quad 
\langle \psi^\prime, n |S_\mu ^\dagger |\psi^\prime ,n - 1 \rangle = \bar{\psi_\mu}^\prime\,.
\label{eq:secofb}
\ee
We observe that the r.h.s. of the equations in (\ref{eq:secofb}) defines functions on $S^{(3 , 2)}$.
Thus these matrix elements correspond to fuzzy sections of bundles on
$S^{(2,2)}$. It is possible to obtain the $*$-product for these fuzzy sections of bundles.
The results below also provide an alternative way to compute the $*$-products in (\ref{eq:starproduct1}) and (\ref{eq:finalstar}).
 
For the $*$-product of $\psi^\prime$ with $\bar{\psi}^\prime$ we find  
\begin{eqnarray}
\psi_\mu^\prime * \bar{\psi}_\nu^\prime &=&  \langle \psi^\prime, n |S_\mu S_\nu ^\dagger |\psi^{\dagger \prime}, 
n \rangle \nonumber \\
&=& \langle \psi ^\prime, n | (-1)^{|S_\mu| |S_\nu|}\, \frac{n}{n + 1}\,S_\nu ^\dagger S_\mu + \frac{1}{n + 1}\,\delta_{\mu \nu} 
|\psi ', n \rangle \nonumber \\
&=& \frac{n}{n + 1} \, \psi_\mu^\prime \bar{\psi}_\nu^\prime + \frac{1}{n + 1}\, \delta_{\mu \nu} \,.
\end{eqnarray}
Here we have used (\ref{eq:gcommutator}) and the fact that $\psi_\mu^\prime \bar{\psi}_\nu^\prime = (-1)^{|S_\mu| |S_\nu|} 
\,\bar{\psi}_\nu^\prime \,\psi_\mu^\prime$ to get rid of $(-1)^{|S_\mu| |S_\nu|}$. Rearranging the last result we can write
\begin{eqnarray}
\psi_\mu^\prime * \bar{\psi}_\nu^\prime &=& \frac{1}{n + 1}\, \Omega_{\mu \nu}
+ \psi_\mu ^\prime \bar{\psi}_\nu^\prime \,, \nonumber \\
\Omega_{\mu \nu} &\equiv& \delta_{\mu \nu} - \psi_\mu^\prime \, \bar{\psi}_\nu^\prime \,. 
\label{eq:63}
\end{eqnarray} 
The significance of $\Omega_{\mu \nu}$ will be be discussed shortly. Before that, as a check of our results of the previous section, 
we can compute ${\cal W} _a *_n {\cal W}_b$, using the method above. First note that 
\begin{equation}
{\cal W}_a = \bar{\psi}^\prime \, \Lambda_a^{(\frac{1}{2})} \, \psi^\prime = \, \langle \psi^\prime, n |S ^\dagger \, 
\Lambda_a^{(\frac{1}{2})} \,S|\psi^\prime, n \rangle \,.
\end{equation}
Hence
\begin{eqnarray}
{\cal W}_a *_n {\cal W}_b &=& \langle \psi^\prime, n |S_\mu ^\dagger \,(\Lambda_a^{(\frac{1}{2})})_{\mu \nu} \,S_\nu
S_\alpha ^\dagger \,(\Lambda_b^{(\frac{1}{2})})_{\alpha \beta} \,S_\beta |\psi^\prime, n \rangle \nonumber \\
&=& \bar{\psi}_\mu^\prime \,(\Lambda_a^{(\frac{1}{2})})_{\mu \nu}\, \left( \frac{1}{n}\,\Omega_{\nu \alpha} + \psi_\nu^\prime 
\bar{\psi}_\alpha^\prime \right) \,(\Lambda_b^{(\frac{1}{2})})_{\alpha \beta} \, \psi_\beta^\prime \nonumber \\
&=& \bar{\psi}_\mu^\prime \,(\Lambda_a^{(\frac{1}{2})})_{\mu \nu}\, \left( \frac{1}{n}\,\delta_{\nu \alpha} + \frac{n - 1}{n}\,
\psi_\nu^\prime \bar{\psi}_\alpha^\prime \right) \,(\Lambda_b^{(\frac{1}{2})})_{\alpha \beta} \, \psi_\beta^\prime \nonumber \\
&=& \frac{1}{n}\, {\cal W}_a *_1 {\cal W}_b + \frac{n - 1}{n}\, {\cal W}_a \,{\cal W}_b \,,
\label{eq:conclusion2}
\end{eqnarray}
which is (\ref{eq:starproduct1}).

Comparing the second line of the last equation with (\ref{eq:simpleproduct}) we get the important 
result
\begin{eqnarray}
K_{ab} & = & ({\cal W}_a \,\partial^{^{\!\!\!\!\!\leftarrow}}_\mu) \,\Omega_{\mu \nu} \,(\vec{\bar{\partial}}_\nu \,{\cal W}_b) 
\nonumber \\
&\equiv &{\cal W}_a \,\partial^{^{\!\!\!\!\!\leftarrow}} \,\Omega \,\vec{\bar{\partial}} \,{\cal W}_b \,, 
\label{eq:Kprojector}
\end{eqnarray}
where $\partial^{^{\!\!\!\!\!\leftarrow}} \,\Omega \,\vec{\bar{\partial}} \equiv  \partial^{^{\!\!\!\!\!\leftarrow}}_\mu \,
\Omega_{\mu \nu} \,\vec{\bar{\partial}}$ and $\partial_\mu = \frac{\partial}{\partial \,\psi_\mu^\prime}$.

We would like to note that this result can be used to write (\ref{eq:finalstar}) in terms
of $\partial^{^{\!\!\!\!\!\leftarrow}} \,\Omega \,\vec{\bar{\partial}}$. To this end we write
\begin{equation}
{\cal F}_n *_n {\cal G}_n ({\cal W}) = 
(-1)^{\sum_{j>i} |a_j| |b_i|} \, F^{a_1 a_2 \cdots a_n} \, \prod_i ({\cal W}_{a_i}(1 + \partial^{^{\!\!\!\!\!\leftarrow}}
\,\Omega \,\vec{\bar{\partial}}){\cal W}_{b_i}) 
\, G^{b_1 b_2 \cdots b_n} \,. 
\label{eq:omegastar1}
\end{equation}
Carrying out a similar calculation that lead to (\ref{eq:finalstar}), one finally finds
\begin{equation}
{\cal F}_n *_n {\cal G}_n ({\cal W}) = {\cal F}_n \,{\cal G}_n ({\cal W}) +
\sum_{m = 1}^n \frac{(n-m)!}{n! m!} \,{\cal F}_n  ({\cal W}) \,\,\vdots
\underbrace{(\partial^{^{\!\!\!\!\!\leftarrow}}
\,\Omega \,\vec{\bar{\partial}}) \cdots (\partial^{^{\!\!\!\!\!\leftarrow}}
\,\Omega \,\vec{\bar{\partial}})}_{m factors} \, \vdots \, {\cal G}_n ({\cal W}) \,,
\label{eq:omegastar2}
\end{equation}
where now $\vdots \cdots \vdots$ takes $\partial^{^{\!\!\!\!\!\leftarrow}}$ and $\vec{\bar{\partial}}$ to the left and right extreme
respectively. (When $\partial^{^{\!\!\!\!\!\leftarrow}}$'s and $\vec{\bar{\partial}}$'s are moved in this fashion, the phases coming
from the graded commutators should be included just as for (\ref{eq:finalstar})). 
  
It can be explicitly shown that $\Omega = (\Omega_{\mu\nu})$ is a projector, i.e.,
\begin{equation}
\Omega^2 = \Omega \quad {\mbox{and}} \quad 
\Omega^\ddagger = \Omega \,.
\end{equation}
Due to (\ref{eq:Kprojector}), the last equation implies similar properties for \footnote{
We consider all the indices down through out this chapter. In the following section the relevant object under investigation 
is ${\cal K}_{ab}$ corresponding to $K_a {}^b$ in a notation where indices are raised and lowered by the metric.} 

\begin{equation}
{\cal K}_{ab} \equiv (K \,S^{-1})_{ab} \,.
\label{eq:convofk}
\end{equation}
which we discuss next.

\section{More on the Properties of ${\cal K}_{ab}$}

A closer look at the properties of ${\cal K}_{ab} \equiv (K \,S^{-1})_{ab}$, where
\begin{align}
K_{ab}(\psi) &= {\cal W}_a *_1 {\cal W}_b (\psi) - {\cal W}_a (\psi) \,{\cal W}_b (\psi) \nonumber \\ 
&= \langle \psi^\prime, 1|{\widehat {\cal W}}_a {\widehat {\cal W}}_b |\psi^\prime, 1 \rangle - \langle \psi^\prime, 1|
{\widehat {\cal W}}_a |\psi^\prime, 1 \rangle \langle \psi^\prime, 1| {\widehat {\cal W}}_b | \psi^\prime, 1 \rangle \,,
\label{eq:conv2}
\end{align}
will give us  more insight on the structure of the $*$-product found in the previous section. First note that ${\cal K}_{ab}$ 
depends on both $\psi$ and $\bar{\psi}$. We denote this dependence by ${\cal K}_{ab}(\psi)$ for short, omitting to write the
$\bar{\psi}$ dependence. Now we would like to show that the matrix ${\cal K}(\psi) \,= ({\cal K}_{ab}(\psi))$ is a projector.

We first recall that the $(\frac{1}{2})_{osp(2,2) + }$, representation of $osp(2,2)$ is at the same time the $J_{osp(2,1)} = 
\frac{1}{2}$ irreducible representation of $osp(2,1)$. Their highest and lowest weight states are given by
\begin{equation}
|J_{osp(2,1)}, J_{su(2)}, J_3 \rangle  = \left\{
\begin{array}{ll}
|\frac{1}{2}, \frac{1}{2}, \frac{1}{2} \rangle  & \equiv {\mbox{highest weight state}}, \\
|\frac{1}{2}, \frac{1}{2}, - \frac{1}{2} \rangle   & \equiv  {\mbox{lowest weight state}}
\end{array}
\right.
\label{eq:weigths}
\end{equation}

We note that, starting from the lowest weight state $|1/2, 1/2, -1/2 \rangle = \Psi_2 ^\dagger|0 \rangle$, one can construct 
another supercoherent state, expressed by a formula similar to (\ref{eq:D}). Now consider the following fiducial point for 
${\cal W}(\psi)$ at $\psi = \psi^0 = (1,0,0)$ obtained from computing ${\cal W}_a (\psi ^0)$ in the supercoherent states induced 
from the states given in (\ref{eq:weigths}):  
\begin{equation}
{\cal W}^{\pm} (\psi^0) = ({\cal W}_1 (\psi^0) \cdots {\cal W}_8 (\psi^0)) = \big (0,0,\pm \frac{1}{2},0,0,0,0,1 \big )\,.
\label{eq:fudicial}
\end{equation} 
In (\ref{eq:fudicial}) $+(-)$ corresponds to upper(lower) entries in (\ref{eq:weigths}) and the calculation is done using 
(\ref{eq:promap2}) and (\ref{eq:nonl}).
 
Although not essential in what follows, we remark that ${\cal W}^- (\psi = (1,0,0)) = {\cal W}^+ (\psi = (0,1,0))$, that is,
\begin{equation}
{\cal W}_a ^- (\psi ^0) = {\cal W}_b ^+(\psi^0) \,(Ad \,e^{i \pi \Lambda_2 ^{(\frac{1}{2})}})_{ba} \,.
\end{equation}
 
Note that all other points in $S_F^{(2,2)}$ can be obtained from ${\cal W}^{\pm} (\psi^0)$ by the adjoint action of the group, i.e., 
\begin{equation}
{\cal W}_a ^{\pm} (\psi) = {\cal W}_b ^{\pm} (\psi^0) (Ad\,g^{-1})_{ba} \,
\end{equation}
where $\psi = g \psi^0$.

We define ${\cal K}^\pm (\psi^0)$ using ${\cal W}^\pm (\psi^0)$ for ${\cal W}$, and the equations (\ref{eq:convofk}), 
(\ref{eq:conv2}). The matrices ${\cal K}^{\pm}(\psi^0)$ when computed at the fiducial points (using for instance 
(\ref{eq:nontyp}), (\ref{eq:manip}), (\ref{eq:Kab})) have the block diagonal forms
\begin{equation}
{\cal K} ^\pm \,(\psi^0) = ({\cal K}_{ab} ^\pm \,(\psi^0)) = \left(
\begin{array}{ccc}
\left(\frac{1}{2}\,\delta_{ij} \pm \frac{i}{2}\, \epsilon_{ij3} -2 \,{\cal W}_i ^\pm (\psi^0) \,({\cal W}_j ^\pm \,(\psi^0)) 
\right)_{3 \times 3} & 0 & 0 \\
0 & \left( \Sigma^\pm _{\alpha \beta} \right)_{4 \times 4} & 0\\
0& 0 & 0
\end{array}
\right)
\label{eq:matrixK}
\end{equation}
with
\begin{equation}
\Sigma^\pm = ( \Sigma^\pm _{\alpha \beta}) = \frac{1}{4}\left(
\begin{array}{cc}
1 \pm \sigma_3 & -(1 \pm \sigma_3) \\
-(1 \pm \sigma_3) & 1 \pm \sigma_3
\end{array}
\right)
\end{equation}
where the upper (lower) sign stands for the upper (lower) sign in ${\cal W} ^\pm \,(\psi^0)$. The supermatrices ${\cal K}^\pm 
\,(\psi^0)$ are even and consequently do not mix the $1,2,3,8$ and $4,5,6,7$ entries of a (super)vector. Its grade adjoint is its 
ordinary adjoint $\dagger$.
Now from (\ref{eq:matrixK}), it is straightforward to check that the relations
\begin{gather}
({\cal K}^{\pm} \,(\psi^0))^2 = {\cal K}^{\pm} \,(\psi^0) \,, \nonumber \\
({\cal K}^{\pm} \,(\psi^0))^\ddagger = {\cal K}^{\pm} \,(\psi^0)\,, \nonumber \\
{\cal K}^+ \,(\psi^0) \,{\cal K}^- \,(\psi^0) = 0 
\label{eq:compros}
\end{gather}
are fulfilled. (\ref{eq:compros}) establishes that ${\cal K}^{\pm} \,(\psi^0)$ are orthogonal projectors. By the adjoint action of 
the group, we have
\begin{equation}
{\cal K}_{ab} ^{\pm} \,(\psi) = ((Ad \,g)^T)^{-1}_{ad} \,{\cal K}_{de} ^{\pm} \,(\psi^0) \,(Ad \, g)^T _{eb} \,,
\label{eq:Adj}
\end{equation}
with $T$ denoting the transpose. (\ref{eq:Adj}) implies that ${\cal K} ^{\pm} \,(\psi)$ are projectors for all $g \in \,OSp(2,2)$.

We further observe that a super-analogue ${\cal J}$ of the complex structure can be defined over the supersphere. To show this, 
we first observe that the projective module for ``sections of the supertangent bundle'' $TS^{(2,2)}$ over
$S^{(2,2)}$ is ${\cal P}{\cal A}^8$, where ${\cal A}$ is the algebra of superfunctions over $S^{(2,2)}$, ${\cal A}^8 = {\cal A} 
\otimes_{{\mathbb C}} \,{\mathbb C}^8$ and 
\be
{\cal P}{(\psi)} = {\cal K}^+ \,(\psi) + {\cal K}^- \,(\psi)
\ee
is a projector. The super-complex structure is the operator with eigenvalues 
$\pm i$ on the subspaces $TS_\pm^{(2,2)}$ of $TS^{(2,2)}$ with  
$TS^{(2,2)} = TS_+^{(2,2)} \oplus TS_-^{(2,2)}$. It is given by the matrix 
${\cal J}$ with elements 
\begin{equation}
{\cal J}_{ab} (\psi) = - i \,({\cal K}^+ - {\cal K}^-)_{ab} (\psi) \,,
\label{eq:complexstruc}
\end{equation}
and acts on  ${\cal P}{\cal A}^8$. Since
\begin{equation}
{\cal J}^2 \,(\psi) \Big{\vert}_{{\cal P}{\cal A}^8} = - {\cal P} (\psi) \,\Big{\vert}_{{\cal P}{\cal A}^8}
= -{\bf 1} \,\Big{\vert}_{{\cal P}{\cal A}^8}
\end{equation}
($\delta \Big{\vert}_\varepsilon$ denoting the restriction of $\delta$ to $\varepsilon$), it indeed defines a super complex 
structure. Furthermore, due to the relation
\begin{equation}
{\cal J} \Big{\vert}_{{\cal K} ^{\pm} {\cal A}^8} = \mp i \,\Big{\vert}_{{\cal K} ^{\pm} {\cal A}^8} \,,
\end{equation}
${\cal K} ^{\pm} {\cal A}^8$ give the ``holomorphic'' and ``anti-holomorphic'' parts of ${\cal P}{\cal A}^8$.
Finally, we can also write
\begin{equation} 
{\cal K} ^{\pm} \,(\psi) = \frac{1}{2}\,(-{\cal J}^2 \pm i{\cal J})(\psi) \,.
\label{eq:projectorK}
\end{equation}

\section{The $O(3)$ Nonlinear Sigma Model on $S^{(2,2)}$}

As a final topic in this chapter, we describe the ``$O(3)$ nonlinear
SUSY sigma model'' on $S^{(2,2)}$ and $S^{(2,2)}_F$. We follow the
discussion in \cite{seckin2}.

\subsection{The Model on $S^{(2,2)}$}

On $S^{(2,2)}$ it is defined by the action
\be
{\cal S}^{SUSY} = - \frac{1}{4 \pi} \int d \mu \Big ( C_{\alpha \beta}
d_\alpha \Phi^a d_\beta \Phi^a + \frac{1}{4} \gamma \Phi^a \gamma \Phi^a \Big ) \,,
\label{eq:susyaction2}
\ee
where $\Phi^a = \Phi^a(x_i \,, \theta_\alpha)$, $(a= 1,2,3)$  is a real triplet superfield fulfilling the constraint 
\be 
\Phi^a \Phi^a = 1 \,, \quad (a =1,2,3) \,.
\label{eq:constraint}
\ee 
Obviously, the world sheet for this theory is $S^{(2,2)}$ while the target manifold is a $2$-sphere.

A closely related model, is the one formulated on the standard $(2,1)$-dimensional superspace ${\cal C}^{(2,1)}$, 
first studied by Witten, and Di Vecchia et al.\cite{witten, divecchia}. 
 
The triplet superfield $\Phi^a$ can be expanded in powers of $\theta_\alpha$ as
\begin{equation}
\Phi^a(x_i \,, \theta_\alpha) = n^a(x_i) + C_{\alpha \beta} \theta_\beta \psi^a_\alpha (x_i) + \frac{1}{2} F^a(x_i) 
C_{\alpha \beta} \theta_\alpha \theta_\beta
\label{eq:fieldexp}
\end{equation}
where $\psi^a(x_i)$ are two component Majorana spinors : $\psi_\alpha^{a \ddagger} = C_{\alpha \beta} \psi^a_\beta$, and $F^a(x_i)$
are auxiliary scalar fields. In terms of the component fields the constraint equation (\ref{eq:constraint}) splits to
\begin{subequations}\label{eq:compcons1}
\begin{align}
n^a n^a = 1 \,, \label{eq:compcons1a} \\
n^a F^a = \frac{1}{2} \psi^{a \ddagger} \psi^a \,, \label{eq:compcons1b} \\
n^a \psi_\alpha^a = 0 \,. \label{eq:compcons1c}
\end{align}
\end{subequations}
(\ref{eq:compcons1a}) is the usual constraint of $O(3)$ non-linear
sigma model defined earlier in chapter 6 
by the action \cite{bal}
\begin{equation}
{\cal S} = - \frac{1}{8 \pi} \int_{S^2} d \Omega ({\cal L}_i n_a) ({\cal L}_i n_a) \,.
\label{eq:action11}
\end{equation}
Thus, we see that bosonic sector of the $S^{SUSY}$ coincides with the ${\mathbb C}P^1$ sigma model. The other two constraints are
additional. We note that (\ref{eq:compcons1b}) can be used along with the equations of motion for $F^a$ to eliminate $F^a$'s from 
the action. The techniques for performing such calculations can be found for
instance in \cite{divecchia}.

%{\bf This can be done in the same way as di vecchia et
%al. does for example. One can put 9.157a,b,c as lagrange multipliers to the
%Lagrangian and find equation of motion for $F^a$, from which it is
%possible to solve for $F^a$ and the related lagrange multiplier, using
%9.157a,b again. See (22) through (25) of di Vecchia. Also note that
%the $\partial n^a$ bit inside $F$ is immaterial, since it $n^a
%\partial n^a=0$ is zero  due to $n^a n^a =1$}

\subsection{The Model on $S_F^{(2,2)}$}

The fuzzy action approaching the (\ref{eq:susyaction2}) for large $n$ is \cite{seckin2} 
\be
{\cal S}^{SUSY} = str \Big ( C_{\alpha \beta} \, \lbrack D_\alpha \,, {\hat \Phi^a} \rbrace \, \lbrack D_\beta \,, {\hat \Phi^a} 
\rbrace + \frac{1}{4} \lbrack \Gamma \,, {\hat \Phi^a} \rbrack \, \lbrack \Gamma \,, {\hat \Phi^a} \rbrack \Big ) \,,
\label{eq:zupac}
\ee
where
\be
{\hat \Phi}^a {\hat \Phi}^a = {\bf 1}_{2n+1} \,, \quad {\bf 1}_{2n+1} \in Mat (n+1, n) \,.
\label{eq:cocon}
\ee
(\ref{eq:cocon}) can be expressed in terms of the $*$-product on $S_F^{(2,2)}$ as 
\be
\Phi^a * \Phi^a (\psi^\prime, {\bar \psi}^\prime, n) = 1 \,.
\ee
This expression involves the product of derivatives of $\Phi^a$ up to
$n^{th}$ order, and not easy to work with. Alternatively
we can construct supersymmetric extensions of ``Bott Projectors''
introduced in chapter 6 to study this model, as we indicate below. 

\subsection{Supersymmetric Extensions of Bott Projectors}

A possible supersymmetric extension of the projector ${\cal P}_\kappa(x)$  can be obtained in the following manner.
Let ${\cal U}(x_i \,, \theta_\alpha)$ be a graded unitary operator :
\be
{\cal U} {\cal U}^\ddagger = {\cal U}^\ddagger {\cal U} =1 \,.
\ee
${\cal U}(x_i \,, \theta_\alpha)$ can be thought as a $2 \times 2$ supermatrix whose entries are functions on $S^{(2,2)}$.
${\cal U}(x_i \,, \theta_\alpha)$ acts on ${\cal P}_\kappa$ by conjugation and generates a set of supersymmetric projectors 
${\cal Q}_\kappa (x_i \,, \theta_\alpha)$:
\begin{equation}
{\cal Q}_\kappa(x_i \,, \theta_\alpha) = {\cal U}^\ddagger \, {\cal P}_\kappa(x) \, {\cal U} \,.
\label{eq:supertr1} 
\end{equation}
It is easy to see that ${\cal Q}_\kappa(x_i \,, \theta_\alpha)$ satisfies 
\be
{\cal Q}_\kappa^2(x_i \,, \theta_\alpha) = Q_\kappa(x_i \,, \theta_\alpha) \,, \quad \mbox{and} \quad {\cal Q}_\kappa^\ddagger(x_i\,,
\theta_\alpha) = {\cal Q}_\kappa(x_i \,, \theta_\alpha) \,.
\ee
Thus ${\cal Q}_\kappa(x_i \,, \theta_\alpha)$ is a (super)projector. The real superfields on $S^{(2,2)}$ associated to 
${\cal Q}_\kappa(x_i \,, \theta_\alpha)$ are given by
\begin{equation}
\Phi_a^\prime (x_i \,, \theta_\alpha) = Tr \, \tau_a {\cal Q}_\kappa \,.
\label{eq:4sf1}
\end{equation}

In order to check that ${\cal Q}_\kappa(x_i \,, \theta_\alpha)$ reproduces the superfields on $S^{(2,2)}$ subject to 
\begin{equation}
\Phi^\prime_a \Phi^\prime_a = 1 \,,
\label{eq:31}
\end{equation}
we proceed as follows. First we expand ${\cal U}(x_i \,, \theta_\alpha)$ in powers of Grassmann variables as 
\begin{equation}
{\cal U}(x_i \,, \theta_\alpha) = {\cal U}_0(x_i) + C_{\alpha \beta} \theta_\beta {\cal U}_\alpha(x_i) + \frac{1}{2} {\cal U}_2(x_i) 
C_{\alpha \beta} \theta_\alpha \theta_\beta 
\label{eq:expansion1}
\end{equation}
where ${\cal U}_0 \,, {\cal U}_\alpha (\alpha= \pm)$ and ${\cal U}_2$ are all $2 \times 2$ graded unitary matrices.
The requirement of graded unitarity for ${\cal U}(x_i \,,
\theta_\alpha)$ implies the following for the component 
matrices:
\begin{itemize}
\item [{\it i.}]  ${\cal U}_0(x_i)$ is unitary,
\item [{\it ii.}] ${\cal U}_\alpha(x_i)$ are uniquely determined by 
\be
{\cal U}_\alpha(x_i) = H_\alpha(x_i) {\cal U}_0(x_i) \,,
\ee
where $H_\alpha$ are $2 \times 2$ odd supermatrices satisfying the reality condition $H_\alpha^\ddagger = - C_{\alpha \beta} 
H_\beta$,
\item [{\it iii.}] ${\cal U}_2$ is of the form ${\cal U}_2 = A {\cal U}_0$ with $A$ being an $2 \times 2$ even supermatrix, whose 
symmetric part satisfies 
\begin{equation}
A + A^\dagger = - C_{\alpha \beta} H_\alpha H_\beta \,.
\label{eq:constraint1}
\end{equation}    
\end{itemize}
Using (\ref{eq:expansion1}) in (\ref{eq:supertr1}) and the conditions listed above, we can extract the 
component fields of the superfield $\Phi_a^\prime(x_i \,, \theta_\alpha)$. We find 
\begin{eqnarray}
n_a^{\kappa \prime} &:=& Tr \, \tau_a U_0^\dagger {\cal P}_\kappa U_0 \,, \\   
\psi_\alpha^{a \prime} &:=& Tr \, \tau_a U_0^\dagger \lbrack H_\alpha \,, {\cal P}_\kappa \rbrack U_0 = 
- 2 i (\vec{n}^{\kappa \prime} \times \vec{H}_\alpha^\prime)^a \,,
\end{eqnarray}
and, after using (\ref{eq:constraint1}),
\beqa
F_a^\prime &:=& Tr \, \tau_a U_0^\dagger ( {\cal P}_\kappa A +A^\dagger {\cal P}_\kappa - C_{\alpha \beta} H_\beta {\cal P}_\kappa 
H_\alpha ) U_0 \\
&=& 4 ( \vec{H}_+^\prime \cdot \vec{H}_-^\prime) n_a^{\kappa \prime} - 2\vec{H}_+^{a \prime} (\vec{n}^{\kappa \prime} 
\cdot \vec{H}_-^\prime) - (\vec{n}^{\kappa \prime} \cdot \vec{H}_+^\prime) 2\vec{H}_-^{a \prime} +i( \vec{n}^{\kappa \prime} 
\times (\vec{A}^\prime- \vec{A}^{\dagger \prime}))^a \nonumber \,,  
\label{eq:fterm}
\eeqa  
where  $\vec{H}_\alpha^{\prime}= H_\alpha^{1 \prime} \tau^1 +H_\alpha^{2 \prime} \tau^2 $ and $\vec{A}^{ \prime} = A^{3 \prime} 
\tau^3$. By direct computation from above it follows that
\begin{equation}
n_a^{\kappa \prime} n_a^{\kappa \prime} = 1 \,, \quad \quad n_a^{\kappa \prime} F_a^{\prime} 
= \frac{1}{2} \psi^{\ddagger \prime}_a \psi_a^\prime \,, \quad \quad
n_a^{\kappa \prime} \psi_\pm^{a \prime} = 0 \,.
\label{eq:constraints2}
\end{equation}
Comparing (\ref{eq:constraints2}) with (\ref{eq:compcons1}) we observe that they are identical. Therefore, we conclude that the 
superfield associated to the super-projector ${\cal Q}_\kappa$ is the same as the superfield of the supersymetric non-linear sigma 
model discussed previously.

\subsection{SUSY Action Revisited}

We now extend (9.129) by including winding number sectors.

Equipped with the supersymmetric projector ${\cal Q}_\kappa$ we can write, in close analogy with the ${\mathbb C}P^1$ model, the 
action for the supersymmetric nonlinear $O(3)$ sigma model for winding number $\kappa$ as 
\begin{equation}
{\cal S}_\kappa^{SUSY} = - \frac{1}{2 \pi} \int d \mu \, Tr \Big \lbrack C_{\alpha \beta} \, (d_\alpha {\cal Q}_\kappa) 
(d_\beta {\cal Q}_\kappa) + \frac{1}{4} (\gamma {\cal Q}_\kappa) (\gamma {\cal Q}_\kappa) \Big \rbrack \,.
\label{eq:superaction2}
\end{equation} 
The even part of this action, as well as the one given in (\ref{eq:susyaction2}) is nothing but the action $S_\kappa$ of the 
${\mathbb C}P^1$ theory given in (\ref{ngxv}) and  (\ref{eq:action11}), respectively. In other words, the action 
$S_\kappa^{SUSY}$ is the supersymmetric extension of $S_\kappa$ on $S^2$ to $S^{(2,2)}$. Consequently, in the supersymmetric theory,
it is possible to interpret the index $\kappa$ carried by the action as the winding number of the corresponding ${\mathbb C}P^1$
theory. For $\kappa = 0$ we get back (9.129).
    
We recall that $d_\alpha$ and $\gamma$ are both graded derivations in the superalgebra $osp(2,2)$. Therefore, they obey a 
graded Leibnitz rule. From ${\cal Q}_\kappa^2 = {\cal Q}_\kappa$, we find
\begin{equation}
{\cal Q}_\kappa d_\alpha {\cal Q}_\kappa = d_\alpha {\cal Q}_\kappa ({\bf 1} -{\cal Q}_\kappa) \,.
\label{eq:property1}
\end{equation}
This enables us to write
\begin{equation}
Tr d_\alpha {\cal Q}_\kappa ({\bf 1} - {\cal Q}_\kappa) d_\alpha {\cal Q}_\kappa = Tr ({\bf 1} -{\cal Q}_\kappa)
(d_\alpha {\cal Q}_\kappa)^2 = \frac{1}{2} Tr (d_\alpha {\cal Q}_\kappa)^2 \,.
\label{eq:property2}
\end{equation}
Equations (\ref{eq:property1}) and (\ref{eq:property2}) continue to hold when $d_\alpha$ is replaced by $\gamma$ as well. The action
can also be written as
\begin{equation}
{\cal S}_\kappa^{SUSY} = - \frac{1}{\pi} \int d \mu\,  Tr \Big \lbrack C_{\alpha \beta} {\cal Q}_\kappa  (d_\alpha {\cal Q}_\kappa) 
(d_\beta {\cal Q}_\kappa) + \frac{1}{4} {\cal Q}_\kappa (\gamma {\cal Q}_\kappa) (\gamma {\cal Q}_\kappa) \Big \rbrack \,.
\label{eq:action3}
\end{equation}  

\subsection{Fuzzy Projectors and Sigma Models}

In much the same way that the supersymmetric projectors ${\cal Q}_\kappa$ have been constructed from ${\cal P}_\kappa$ 
in the previous section, we can construct the supersymmetric extensions of ${\widehat {\cal P}}_\kappa$ by the graded 
unitary transformation
\begin{equation}
{\widehat {\cal Q}}_\kappa = {\widehat {\cal U}}^\ddagger {\widehat {\cal P}}_\kappa {\widehat {\cal U}}
\label{eq:fuzzyp}
\end{equation}
where now ${\widehat {\cal U}}$ is a $2 \times 2$ supermatrix whose entries are polynomials in not only 
$a_\alpha^\dagger a_\beta$ but also in $b^\dagger b$. The domain of ${\cal U}_{ij}$ is ${\tilde {\cal H}}_n$.

${\widehat {\cal Q}}_\kappa$  acts on the finite-dimensional space ${\tilde {\cal H}}^2_n = 
{\tilde {\cal H}}_n \otimes {\mathbb C}^2$. We can check that
\be 
\lbrack {\widehat {\cal Q}}_\kappa \,, {\widehat N} \rbrace = 0 \,,
\ee
where ${\widehat N} = a_\alpha^\dagger a_\alpha + b^\dagger b$ is the number operator on ${\tilde {\cal H}}_n$.  
In close analogy with the fuzzy ${\mathbb C}P^1$ model, it is
now possible to write down a finite-dimensional (super)matrix model for the (super)projectors ${\widehat {\cal Q}}_\kappa$.

The action for the fuzzy supersymmetric model becomes
\begin{equation}
S_{F \,, \kappa}^{SUSY} = \frac{1}{2\pi} \, Str_{{\widehat N} =n}  \, \Big (  C_{\alpha \beta} 
\lbrack D_\alpha \,, {\widehat {\cal Q}}_\kappa \rbrace \, \lbrack D_\beta \,, {\widehat {\cal Q}}_\kappa \rbrace  
+ \frac{1}{4} \lbrack \Gamma \,, {\widehat {\cal Q}}_\kappa \rbrack \, \lbrack \Gamma \,, {\widehat {\cal Q}}_\kappa \rbrack \Big )
\,,
\label{eq:superaction2f}
\end{equation}
$Str$ in the above expression is the supertrace over ${\tilde {\cal H}}^2_n$. In the large 
${\widehat N} = n$ limit (\ref{eq:superaction2f}) approximates the action given in (\ref{eq:superaction2}). 

This concludes our discussion of the non-linear sigma model on $S_F^{(2,2)}$.

\chapter{Fuzzy Spaces as Hopf Algebras}

\section{Overview}

So far we have studied the formal structure of fuzzy supersymmetric spaces, as well as the structure of field theories on such 
spaces, focusing our attention to the fuzzy supersphere, $S_F^{(2,2)}$. In this chapter we will explore
yet another intriguing aspect of fuzzy spaces, namely their potential use as quantum symmetry algebras. To be more precise we will
establish, through studying fuzzy sphere as an example, that fuzzy spaces possess a Hopf algebra structure.

%Fuzzy spaces provide finite-dimensional approximations to certain symplectic manifolds $M$ such as $S^2 \simeq {\mathbb C}P^1$, $S^2 
%\times S^2$ and ${\mathbb C}P^2$. They are typically full matrix algebras \mbox{$Mat(N+1)$} of dimension \mbox{$(N+1) \times (N+1)$}.
%The fuzzy sphere $S_F^2(J)$ for angular momentum $J= \frac{N}{2}$ for example is \mbox{$Mat(N+1)$}. As $N \rightarrow \infty$, a fuzzy
%space provides an increasingly better approximation to the affiliated commutative algebra $C^\infty(M)$. Quantum field theories
%(QFT's) on fuzzy spaces being finite, they are thus new regularizations of QFT's in the continuum.
%
%Fuzzy spaces are obtained by quantizing adjoint orbits of compact Lie groups $G$. $S^2_F(J)$ is associated in this manner with 
%$SU(2)$.

It is a fact that for an algebra ${\cal A}$, it is not always possible to compose two of its representations $\rho$ and $
\sigma$ to obtain a third one. For groups we can do so and obtain the tensor product $\rho \otimes \sigma$. Such
a composition of representations is also possible for coalgebras ${\cal C}$ \cite{sweedler}. A coalgebra ${\cal C}$ has a coproduct 
$\Delta$ which is a homomorphism from ${\cal C}$ to ${\cal C} \otimes {\cal C}$ and the composition of its representations 
$\rho$ and $\sigma$ is the map $(\rho \otimes \sigma)\Delta$. If ${\cal C}$ has a more refined structure and is a Hopf algebra, then 
it closely resembles a group, in fact sufficiently so that it can be used as a ``quantum symmetry group'' 
\cite{Mach-Schomerus}.

We follow the reference \cite{seckin3} in this chapter.
In order to make our discussin self contained we review some of the basic definitions about coalgebras, bialgebras and 
Hopf algebras in terms of the language of commutative diagrams and set our notations and conventions, which are the
standard ones used in the literature. A well known example of a Hopf algebra is the group algebra $G^*$ associated to a group $G$.
Our interest mainly lies on the compact Lie groups $G$, as they are the ones whose adjoint orbits once quantized yield fuzzy spaces. 
The group algebra $G^*$ of such $G$ consists of elements $\int_G d \mu(g) \alpha(g) g$ where $\alpha(g)$ is a smooth complex function
and $d \mu(g)$ is the $G$-invariant measure. It is isomorphic to the convolution algebra of functions on $G$. Basic definitions and
properties related to $G^*$ will be given in section \ref{sec-conv1}.

In section \ref{sec-starhom} and \ref{sec-hopf1}, we establish that fuzzy spaces are irreducible representations 
$\rho$ of $G^*$ and inherit its Hopf algebra structure. For fixed $G$, their direct sum is homomorphic to $G^*$. For example 
both $S_F^2(J)$ and \mbox{$\oplus_J S_F^2(J) \simeq SU(2)^*$} are Hopf algebras. This means that we can define a coproduct on 
$S_F^2(J)$ and \mbox{$\oplus_J S_F^2(J)$} and compose two fuzzy spheres preserving algebraic properties intact.

A group algebra $G^*$ and a fuzzy space from a group $G$ carry several actions of $G$. $G$ acts on $G$ and $G^*$
by left and right multiplications and by conjugation. Also for example, the fuzzy space $S_F^2(J)$ consists of $(2J+1) \times (2J+1)$
matrices and the spin $J$ representation of $SU(2)$ acts on these matrices by left and right multiplication and by conjugation.
The map $\rho$ of $G^*$ to a fuzzy space and the coproduct $\Delta$ are compatible with all these actions: they are
$G$-equivariant.

Elements $m$ of fuzzy spaces being matrices, we can take their hermitian conjugates. They are $*$-algebras if $*$ is hermitian
conjugation. $G^*$ also is a $*$-algebra. $\rho$ and $\Delta$ are $*$-homomorphisms as well: $\rho(\alpha^*)= 
\rho(\alpha)^\dagger$, $\Delta(m^*) = \Delta(m)^*$.

The last two properties of $\Delta$ on fuzzy spaces also derive from the same properties of $\Delta$ for $G^*$.

All this means that fuzzy spaces can be used as symmetry algebras. In that context however, $G$-invariance implies $G^*$-
invariance and we can substitute the familiar group invariance for fuzzy space invariance.

The remarkable significance of the Hopf structure seems to lie elsewhere. Fuzzy spaces approximate space-time algebras. $S_F^2(J)$ is 
an approximation to the Euclidean version of (causal) de Sitter space homeomorphic to $S^1 \times {\mathbb R}$, 
or for large radii of $S^1$, of Minkowski space \cite{Figari}. 
The Hopf structure then gives orderly rules for splitting and joining fuzzy spaces. The decomposition of $(\rho \otimes \sigma) 
\Delta$ into irreducible $*$-representations (IRR's) $\tau$ gives fusion rules for states in $\rho$ and $\sigma$ combining to become 
$\tau$, while $\Delta$ on an IRR such as $\tau$ gives amplitudes for $\tau$ becoming $\rho$ and $\sigma$. In other words, $\Delta$ 
gives Clebsch-Gordan coefficients for space-times joining and splitting.
Equivariance means that these processes occur compatibly with $G$-invariance: $G$ gives selection rules for these processes in the
ordinary sense. The Hopf structure has a further remarkable consequence: An observable on a state in $\tau$ can be split into
observables on its decay products in $\rho$ and $\sigma$. 

There are similar results for field theories on $\tau$, $\rho$ and $\sigma$, indicating the possibility of many orderly calculations.

These mathematical results are very suggestive, but their physical consequences are yet to be explored.
  
The coproduct $\Delta$ on the matrix algebra Mat$(N+1)$ is not unique. Its choice depends on the group actions we care to preserve,
that of $SU(2)$ for $S_F^2$, $SU(N+1)$ for the fuzzy ${\mathbb C}P^N$ algebra 
${\mathbb C}P_F^N$ and so forth. It is thus the particular equivariance that determines the choice of $\Delta$.

We focus attention on the fuzzy sphere for specificity in what follows, but one can see that the arguments are valid for any fuzzy
space. Proofs for the fuzzy sphere are thus often assumed to be valid for any fuzzy space without comment.
%{\bf we also give a summary of results for the case of $S_F^{(2,2)}$, which should be automatic after the discussions of the previous
%two chapters}.

Fuzzy algebras such as ${\mathbb C}P^N_F$ can be further ``$q$-deformed'' into certain quantum group algebras relevant for the study 
of $D$-branes. This theory has been developed in detail by Pawelczyk and Steinacker \cite{steinacker}. 

\section{Basics}
Here we collect some of the basic formulae related to the group $SU(2)$ and its representations which will be
used later in the chapter.
 
The canonical angular momentum generators of $SU(2)$ are $J_i \,(i=1,2,3)$. The unitary irreducible representations (UIRR's)
of $SU(2)$ act for any half-integer or integer $J$ on Hilbert spaces ${\cal H}^J$ of dimension $2J+1$. They have orthonormal basis 
$|J, M \rangle$, with $J_3 |J, M \rangle = M |J, M \rangle$ and obeying conventional phase conventions. The unitary matrix 
$D^J(g)$ of $g \in SU(2)$ acting on ${\cal H}^J$ has matrix elements \mbox{$\langle J ,M| D^J(g) |J, N \rangle = D^J(g)_{MN}$} 
in this basis.

Let 
\begin{equation}
V= \int_{SU(2)} d \mu (g)
\label{eq:volume}
\end{equation}
be the volume of $SU(2)$ with respect to the Haar measure $d \mu$. It is then well-known that \cite{Bal-Trahern} 
\begin{subequations}
\begin{align}
\int_{SU(2)} d \mu (g) D^J(g)_{ij} \, D^K(g)_{kl}^\dagger &= \frac{V}{2J+1} \, \delta_{JK} \, \delta_{il} \, \delta_{jk}\,, 
\label{eq:iki} \\
\frac{2J+1}{V} \sum_{J \,, i j} D^J_{ij}(g) \, {\bar {D}}^J_{ij}(g^\prime) &= \delta_g(g^\prime) \,, \label{eq:comp}
\end{align}
\end{subequations}
where bar stands for complex conjugation and $\delta_g$ is the $\delta$-function on $SU(2)$ supported at $g$:
\begin{equation}
\int_{SU(2)} d \mu (g^\prime) \,\delta_g(g^\prime) \alpha(g^\prime) = \alpha(g)
\label{eq:deltaf}
\end{equation}
for smooth functions $\alpha$ on $G$.

We have also the Clebsch-Gordan series
\begin{equation}
D^K_{\mu_1 m_1} \, D^L_{\mu_2 m_2} = \sum_J C(K\,, L \,, J \,; \,\mu_1 \,, \mu_2) \, C(K \,, L \,, J\,; m_1 \,, m_2) \, 
D^J_{\mu_1 + \mu_2 \,, m_1+m_2}
\label{eq:clebschg}
\end{equation}
where $C$'s are the Clebsch-Gordan coefficients.

\section{The Group and the Convolution Algebras}
\label{sec-conv1}

The group algebra consists of the linear combinations
\begin{equation}
\int_{G} d \mu(g) \, \alpha(g) \,g \,, \quad \quad d \mu(g)= \mbox{Haar \, measure \, on} \,\, G
\label{eq:groupalg}
\end{equation}
of elements $g$ of $G$, $\alpha$ being any smooth ${\mathbb C}$-valued function on $G$. The algebra product is induced from the
group product:
\begin{equation}
\int_G d \mu(g) \, \alpha(g) \,g \int_G d \mu(g^\prime) \, \beta(g^\prime) \,g^\prime :=
\int_G d \mu(g) \int_G d \mu(g^\prime) \alpha(g) \, \beta(g^\prime) (g g^\prime) \,.
\label{eq:groupproduct}
\end{equation}

We will henceforth omit the symbol $G$ under integrals. 

The right hand side of (\ref{eq:groupproduct}) is
\begin{equation}
\int d\mu(s) \, (\alpha *_c \beta)(s) \, s
\label{eq:conv}
\end{equation}
where $*_c$ is the convolution product:
\begin{equation}
(\alpha *_c \beta)(s) = \int d \mu(g) \alpha(g) \, \beta(g^{-1} s) \,.
\label{eq:convproduct}
\end{equation}
The convolution algebra consists of smooth functions $\alpha$ on $G$ with $*_c$ as their product. Under the map
\begin{equation}
\int d \mu (g) \alpha(g) g \rightarrow \alpha \,,
\label{eq:map1}
\end{equation}
(\ref{eq:groupproduct}) goes over to  $\alpha *_c \beta$ so that the group algebra and convolution algebra are isomorphic.
We call either as $G^*$. 

Using invariance properties of $d \mu$, (\ref{eq:map1}) shows that under the action
\begin{equation}
\int d \mu(g) \, \alpha(g) \,g \rightarrow
h_1  \, \left ( \int d \mu(g) \alpha(g) g  \, \right ) h_2^{-1} = \int d \mu(g) \alpha(g) h_1 g h_2^{-1} \,, \quad \quad h_i \in G \,,  
\label{eq:invariance1}
\end{equation}
$\alpha \rightarrow \alpha^\prime$ where 
\begin{equation}
\alpha^\prime(g) =\alpha (h_1^{-1} g h_2). 
\end{equation}
Thus the map (\ref{eq:map1}) is compatible with left- and right- $G$-actions. 

The group algebra is a $*$-algebra \cite{sweedler}, the $*$-operation being 
\begin{equation}
\left \lbrack  \int d \mu(g) \, \alpha (g) \, g \right \rbrack ^* = \int d \mu (g) \, {\bar \alpha (g)} g^{-1} \,.
\label{eq:starop}
\end{equation}
The $*$-operation in $G^*$ is 
\begin{gather}
* : \alpha \rightarrow \alpha^* \,, \nonumber \\
\alpha^*(g) = {\bar \alpha}(g^{-1}) \,.
\label{eq:qstarop1}
\end{gather}
Under the map (\ref{eq:map1}),
\begin{equation}
\left \lbrack \int d \mu (g) \alpha(g) g \right \rbrack ^* \rightarrow \alpha^*
\label{eq:map2}
\end{equation}
since
\begin{equation}
d \mu (g) = i \, Tr(g^{-1} \, d g) \wedge g^{-1} \, d g \wedge g^{-1} \, d g  = - d \mu (g^{-1}) \,.
\label{eq:measure1}
\end{equation}
The minus sign in (\ref{eq:measure1}) is compensated by flips in ``limits of integration'', thus $\int d \mu (g) =
\int d \mu (g^{-1}) = V$. Hence the map (\ref{eq:map1}) is a $*$-morphism, that is, it preserves ``hermitian conjugation''.

\section{A Prelude to Hopf Algebras}
\label{sec-hopf}

This section reviews the basic ingredients that go into the definition of Hopf algebras. It also sets some notations
and conventions, which are standard in the literature. Our approach here will be illustrative and will closely follow the exposition
of \cite{dascalescu}. Unless, stated otherwise we always work over the complex number field ${\mathbb C}$, but definitions given below
extend to any number field $k$ without any further remarks.

In the language of commutative diagrams an algebra ${\cal A}$ is defined as the triple 
${\cal A} \equiv ( {\cal A} \,, M \,, u)$ where ${\cal A}$ is a vector space, $M: {\cal A} \otimes {\cal A} \rightarrow {\cal A} $ 
and $u: {\mathbb C} \rightarrow {\cal A}$ are morphisms (linear maps) of vector spaces such that the following diagrams are
commutative.
\begin{diagram}[width=6em]
{\cal A} \otimes {\cal A} \otimes {\cal A} & \rTo^{id \otimes M} & {\cal A} \otimes {\cal A} \\
\dTo^{M \otimes id} & & \dTo_M \\
{\cal A} \otimes {\cal A} & \rTo^M & {\cal A} \\
\end{diagram}
\vskip 1em
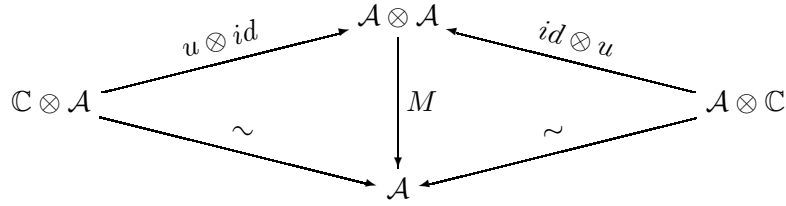
\begin{diagram}[width=6em]
& &{\cal A} \otimes {\cal A}& &\\
{\mathbb C} \otimes {\cal A}&\ruTo(2,1)^{u \otimes id} &\dTo_M &\luTo(2,1)^{id \otimes u} & {\cal A} \otimes {\mathbb C} \\
& \rdTo(2,1)^{\sim} &{\cal A}&\ldTo(2,1)^{\sim} & \\
\end{diagram}
In this definition $M$ is called the product and $u$ is called the unit. The commutativity of the first diagram simply implies the
associativity of the product $M$, whereas for the latter it expresses the fact that $u$ is the unit of the algebra. The unlabeled 
arrows are the canonical isomorphisms of the algebra onto itself. Also in above and what follows $id$ denotes the identity map.

A coalgebra  ${\cal C}$ is the triple ${\cal C} \equiv ({\cal C} \,, \Delta \,, \varepsilon)$, where ${\cal C}$ is a vector space, 
$\Delta : {\cal C} \rightarrow {\cal C} \otimes {\cal C} $ and $\varepsilon: {\cal C} \rightarrow {\mathbb C}$ are morphisms 
of vector spaces such that the following diagrams are commutative.
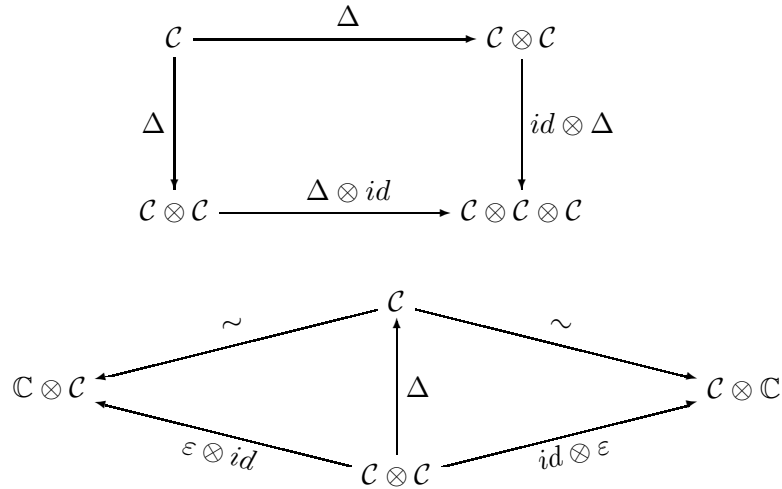
\begin{diagram}[width=6em]
{\cal C} & \rTo^{\Delta} & {\cal C} \otimes {\cal C} \\
\dTo^{\Delta} & & \dTo_{id \otimes \Delta} \\
{\cal C} \otimes {\cal C} & \rTo^{\Delta \otimes id} & {\cal C} \otimes {\cal C} \otimes {\cal C}  \\
\end{diagram}

\begin{diagram}[width=6em]
& &{\cal C}& &\\
{\mathbb C} \otimes {\cal C}&\ldTo(2,1)^{\sim} &\uTo_{\Delta} &\rdTo(2,1)^{\sim} & {\cal C} \otimes {\mathbb C} \\
& \luTo(2,1)_{\varepsilon \otimes id} &{\cal C} \otimes {\cal C} &\ruTo(2,1)_{id \otimes \varepsilon} & \\
\end{diagram}
In this definition $\Delta$ is called the coproduct and $\varepsilon$ is called the counit. The commutativity of the first diagram 
implies the coassociativity of the coproduct $\Delta$, whereas for the latter it expresses the fact that $\varepsilon$ is the
counit of the coalgebra.

An immediate example of a coalgebra is the vector space of $n \times n$ matrices $Mat(n)$, with the coproduct and the counit
\be
\Delta(e^{ij}) = \sum_{1 \leq p \leq n} e^{ip} \otimes e^{pj} \,, \quad \varepsilon(e^{ij}) = \delta^{ij} \,, 
\label{eq:matrixcop}
\ee
where $e^{ij} \,, {1 \leq i \,, j \leq n}$ is a basis for $Mat(n)$\footnote{One might be tempted to call (\ref{eq:matrixcop}) as
the coproduct of $S_F^2$, since elements of $S_F^2$ are described by matrices in $Mat(n+1)$. But, (\ref{eq:matrixcop}) is not
equivariant under $SU(2)$ actions and therefore has no chance of being the appropriate coproduct for  $S_F^2$.}. 

In what follows we adopt the sigma notation which is standard in literature and write for $c \in {\cal C}$
\be
\Delta(c) = \sum c_1 \otimes c_2 \,,
\ee
which with the usual summation convention should have been
\be
\Delta(c) = \sum_{i = 1}^n c_{i1} \otimes c_{i2} \,. 
\ee

One by one we are exhausting the steps leading to the definition of a Hopf algebra. The next step is to define the bialgebra 
structure. A bialgebra is a vector space $H$ endowed with both an algebra and a coalgebra structure such that the 
following diagrams are commutative.
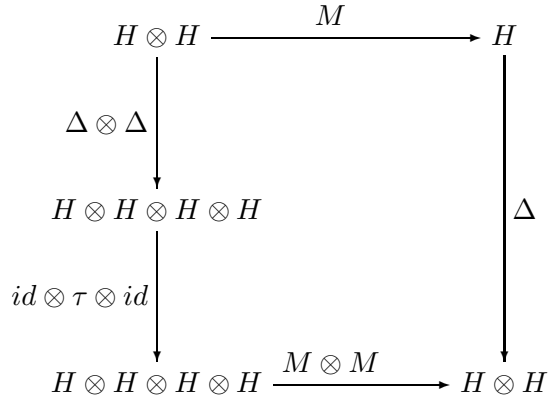
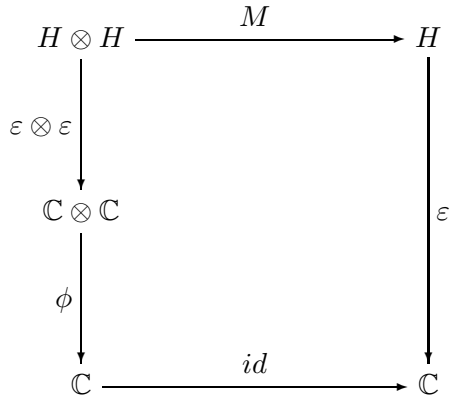
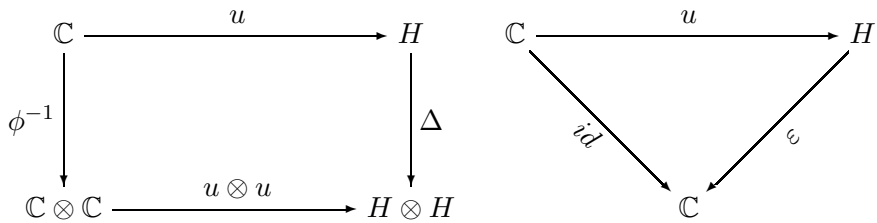
\begin{diagram}[width=6em]
H \otimes H & \rTo^M & H \\
\dTo^{\Delta \otimes \Delta} & &  \\
H \otimes H \otimes H \otimes H & & \dTo_{\Delta} \\
\dTo^{id \otimes \tau \otimes id} & & \\
H \otimes H \otimes H \otimes H & \rTo^{M \otimes M} & H \otimes H \\
\end{diagram}
\vskip 1em
\begin{diagram}[width=6em]
H \otimes H & \rTo^M & H \\
\dTo^{\varepsilon \otimes \varepsilon} & & \\
{\mathbb C} \otimes {\mathbb C}  & & \dTo_{\varepsilon} \\
\dTo^{\phi}& & \\
{\mathbb C} & \rTo^{id} & {\mathbb C} \\
\end{diagram}
\vskip 1em
\begin{displaymath}
\begin{diagram}[width=6em]
{\mathbb C}  & \rTo^u & H \\
\dTo^{\phi^{-1}} & & \dTo_{\Delta} \\
{\mathbb C} \otimes {\mathbb C} & \rTo^{u \otimes u} & H \otimes H \\
\end{diagram}
\qquad
\begin{diagram}[width=6em]
{\mathbb C}  & \rTo^u & H \\
& \rdTo(1,2)_{id} \ldTo(1,2)_{\varepsilon} & \\
& {\mathbb C} & \\
\end{diagram}
\end{displaymath}
In above $\tau : H \otimes H \rightarrow H \otimes H$ is the twist map defined by $\tau (h_1 \otimes h_2) = h_2 \otimes h_1 \,, 
\forall \, h_{1,2} \in H$. In terms of the sigma notation the above four diagrams read
\begin{gather}
\Delta (h g) = \sum h_1 g_1 \otimes h_2 g_2 \,, \quad \varepsilon(hg) = \varepsilon(h) \varepsilon(g) \nonumber \\  
\Delta (1) =1 \otimes 1 \,, \quad \varepsilon(1) = 1 \,.
\end{gather} 

Now, let $S$ be a map from a bialgebra $H$ onto itself. Then $S$ is called an antipode if the following diagram is commutative.
\begin{diagram}[width=6em]
H & \rTo^{\varepsilon} & {\mathbb C} & \rTo^u & H \\
\dTo^{\Delta} & & & & \uTo^M \\
H \otimes H & & \rTo^{id \otimes S \,, S \otimes id} & & H \otimes H \\
\end{diagram}
In terms of the sigma notation this means
\be
\sum S(h_1) h_2 = \sum h_1 S(h_2) = \varepsilon(h) {\bf 1} \,, \quad \quad {\bf 1} \in H \,.
\ee

By definition a {\bf Hopf algebra} is a bialgebra with an antipode. Perhaps, the simplest example for a Hopf algebra is the group 
algebra, and it also happens to be the one of our interest. The group algebra $G^*$ can be made into a Hopf algebra by defining
the coproduct $\Delta$, the counit $\varepsilon$ and antipode $S$ as follows:   
\begin{subequations}\label{eq:hopfcop1}
\begin{align}
\Delta (g) &= g \otimes g \,, \label{eq:co1} \\ 
\varepsilon (g) &= 1 \in {\mathbb C} \,, \label{eq:counit} \\  
S(g) &= g^{-1} \,.
\end{align}
\end{subequations}
Here $\varepsilon$ is the one-dimensional trivial representation of $G$ and $S$ maps $g$ to its inverse. $\Delta$, $\varepsilon$ and
$S$ fulfill all the consistency conditions implied by the commutativity of the diagrams defining the Hopf algebra structure 
as can easily be verified. For instance we have
\be
\sum S(g_1) g_2 = S(g) g = g^{-1} g = {\bf 1} = \varepsilon(g) {\bf 1} \,,
\ee
and similarly $\sum g_1 S(g_2) = \varepsilon(g) {\bf 1}$ for any $g \in G$.   

\section{The $*$-Homomorphism $G^* \rightarrow S_F^2$}
\label{sec-starhom}

As mentioned earlier, henceforth we identify the group and convolution algebras and denote either by $G^*$.
We specialize to $SU(2)$ for simplicity.
We work with group algebra and and group elements, but one may prefer the convolution algebra instead for reasons of rigor.
(The image of $g$ is the Dirac distribution $\delta_g$ and not a smooth function.)

The fuzzy sphere algebra is not unique, but depends on the angular momentum $J$ as shown by the 
notation $S_F^2(J)$, which is $Mat(2J+1)$. Let
\begin{equation}
{\cal S}_F^2 = \oplus_{J} S_F^2(J) = \oplus_{J} Mat(2J+1) \,.
\label{eq:fuzzys}
\end{equation}

Let $\rho (J)$ be the unitary irreducible representation of angular momentum $J$ for $SU(2)$:
\begin{equation}
\rho (J) : \quad g \rightarrow \langle \rho (J) , g \rangle := D^J(g) \,.
\label{eq:uirr1}
\end{equation}
We have
\begin{equation}
\langle \rho (J) , g \rangle \, \langle \rho (J) , h \rangle = \langle \rho (J) , g h \rangle \,.
\label{eq:uirr2}
\end{equation}
Choosing the $*$-operation on $D^J(g)$ as hermitian conjugation, $\rho (J)$ extends by linearity to a $*$-homomorphism 
on $G^*$:
\begin{gather}
\Big \langle \rho (J) , \int d \mu (g) \alpha (g) g \Big \rangle = \int d \mu (g) \alpha (g) D^J(g) \nonumber \\
\Big \langle \rho (J) , \Big(\int d \mu (g) \alpha (g) g \Big)^* \Big \rangle = \int d \mu (g) \bar {\alpha}(g) D^J(g)^\dagger \,. 
\label{eq:PCSU}
\end{gather}
$\rho (J)$ is also compatible with group actions on $G^*$ (that is, it is equivariant with respect to these actions):
\begin{equation}
\Big \langle \rho (J) , \int d \mu (g) \alpha (g) h_1 g h_2^{-1} \Big \rangle= \int d \mu (g) \alpha (g) D^J(h_1)  D^J(g)  
D^J(h_2^{-1}) \quad h_i \in SU(2) \,.  
\label{eq:equivariance1}
\end{equation}

As by (\ref{eq:iki}),
\begin{gather}
\Big \langle \rho (J), \frac{2K+1}{V} \int d \mu (g) (D^K_{ij})^\dagger (g) g \Big \rangle = e^{j i} (J) \delta_{KJ} \,, \nonumber \\
e^{j i} (J)_{rs} = \delta_{jr} \delta_{is} \,, \quad i,j,r,s \in [-J \,, \cdots 0 \,, \cdots \,, J] \,,
\label{eq:PCSU2}
\end{gather}
we see by (\ref{eq:uirr2}) and (\ref{eq:PCSU}) that $\rho(J)$ is a $*$-homomorphism from $G^*$
to \mbox{$S_F^2(J) \oplus \lbrace 0 \rbrace$}, where $\lbrace 0 \rbrace$ denotes the zero elements of $\oplus_{K \neq J} S_F^2(K)$,
the $*$-operation on $S_F^2(J)$ being hermitian conjugation. Identifying $S_F^2(J) \oplus \lbrace 0 \rbrace$ with $S_F^2(J)$, 
we thus get a $*$-homomorphism $ \rho (J) : G^* \rightarrow S_F^2(J)$. It is also seen to be equivariant with respect 
to $SU(2)$ actions, they are given on the basis $e^{ji}(J)$ by $D^J(h_1) e^{ji}(J) D^J(h_2)^{-1}$.

We can think of (\ref{eq:PCSU}) as giving a map
\begin{equation}
\rho : g \quad \rightarrow  \quad \langle \rho (.) \,, g \rangle := g(.)
\label{eq:Pmap} 
\end{equation}
to a matrix valued function $g(.)$ on the space of UIRR's of $SU(2)$ where
\begin{equation}
g(J) = \langle \rho (J) \,, g \rangle \,.
\label{eq:Pmap2}
\end{equation}
The homomorphism property (\ref{eq:PCSU}) is expressed as the product $g(.) h(.)$ of these functions where
\begin{equation}
g(.) h(.) (J) = g(J) h(J)
\label{eq:propoint}
\end{equation}
is the point-wise product of matrices. This point of view is helpful for later discussions.

As emphasized earlier, this discussion works for any group $G$, its UIRR's, and its fuzzy spaces barring technical problems. 
Thus $G^*$ is $*$-isomorphic to the $*$-algebra of functions $g(.)$ on the space of its UIRR's $\tau$, with $g(\tau) 
= D^{\tau}(g)$, the linear operator of $g$ in the UIRR $\tau$ and $g^*(\tau) = D^{\tau}(g)^\dagger$.

A fuzzy space is obtained by quantizing an adjoint orbit $G/H$, $H \subset G$ and approximates $G/H$. It is a full matrix algebra
associated with a particular UIRR $\tau$ of $G$. There is thus a $G$-equivariant $*$-homomorphism from $G^*$ to the fuzzy 
space.

At this point we encounter a difference with $S_F^2(J)$. For a given $G/H$ we generally get only a subset of UIRR's $\tau$. For
example ${\mathbb C}P^2 = SU(3)/U(2)$ is associated with just the symmetric products of just $3$'s (or just $3^*$'s) of $SU(3)$.
Thus the direct sum of matrix algebras from a given $G/H$ is only homomorphic to $G^*$. 

Henceforth we call the space of UIRR's of $G$ as ${\hat G}$. For a compact group, ${\hat G}$ can be identified with the set of 
discrete parameters specifying all UIRR's.

The properties of a group $G$ are captured by the algebra of matrix-valued functions $g(.)$ on ${\hat G}$ with point-wise 
multiplication, this algebra being isomorphic to $G^*$. In terms of $g(.)$, (\ref{eq:hopfcop1}) translate to 
\begin{subequations}\label{eq:hopfcop2}
\begin{align}
\Delta \big(g(.)\big) &= g(.) \otimes g(.) \,, \\
\varepsilon (g(.)) &= {\bf 1} \in {\mathbb C} \,, \\  
S\big( g(.) \big) &= g^{-1}(.) \,.
\end{align}
\end{subequations}
Note that $ g(.) \otimes g(.)$ is a function on ${\hat G} \otimes {\hat G}$.

\section{Hopf Algebra for the Fuzzy Spaces}
\label{sec-hopf1}

Any fuzzy space has a Hopf algebra, we show it here for the fuzzy sphere.

Let $\delta_J$ be the $\delta$-function on ${\widehat {SU(2)}}$: 
\begin{equation}
\delta_J(K) := \delta_{JK} \,.
\end{equation}
(Since the sets of $J$ and $K$ are discrete we have Kronecker delta and not a delta function).

Then 
\begin{equation}
e^{ji}(J) \, \delta_J = \frac{2J+1}{V} \int d \mu (g) D_{ij}^J(g)^\dagger g(.)
%\label{eq:
\end{equation}
Hence 
\begin{equation}
\Delta ( e^{ji}(J) \delta_J) = \frac{2J+1}{V} \int d \mu (g) D_{ij}^J(g)^\dagger g(.) \otimes g(.) \,.
\label{eq:edeltaj}
\end{equation}
At $(K, L) \in {\widehat {SU(2)}} \otimes {\widehat {SU(2)}}$, this is
\begin{equation}
\Delta \big(e^{ji}(J) \big)(K,L) = \frac{2J+1}{V} \int d \mu (g) D_{ij}^J(g)^\dagger \, D^K(g) \otimes D^L(g) \,.
\label{eq:KL}
\end{equation}

As $\delta_J^2 = \delta_J$ and $\delta_J e^{ji}(J) = e^{ji}(J) \delta_J$, we can identify $e^{ji}(J) \delta_J$ with
$e^{ji}(J)$:
\begin{equation}
e^{ji}(J) \delta_J \simeq e^{ji}(J) \,.
\end{equation}
Then (\ref{eq:edeltaj}) or (\ref{eq:KL}) show that there are many coproducts $\Delta = \Delta_{KL}$ we can define and they are 
controlled by the choice of $K$ and $L$:
\begin{equation}
\Delta \big (e^{ji}(J) \delta_J \big)(K,L) := \Delta_{KL} \big( e^{ji}(J) \big).
\label{eq:copKL}
\end{equation}

From section (\ref{sec-hopf}) we know that technically a coproduct $\Delta$ is a homomorphism from ${\cal C}$ to ${\cal C} 
\otimes {\cal C}$ so that only $\Delta_{JJ}$ is a coproduct. But, we will be free of language and call all $\Delta_{KL}$ as 
coproducts. Indeed, it is the very fact that $K \neq L$ in general in (\ref{eq:copKL}) that gives $S_F^2$ its ``generalized'' Hopf
alegbra structure.

Let us now simplify the RHS of (\ref{eq:KL}). Using (\ref{eq:clebschg}), (\ref{eq:KL}) can be written as
\begin{eqnarray}
\Delta \big (e^{ji}(J) \delta_J \big)_{\mu_1 \mu_2 \,, m_1 m_2} =& \frac{2J+1}{V} \int d \mu (g) D_{ij}^J(g)^\dagger \,  
\sum\limits_{J^\prime} C(K,L,J^{\prime}; \mu_1 \,, \mu_2) \times \nonumber \\ 
&\quad \quad \quad \quad C(K, L, J^{\prime}; m_1 \,, m_2) \, D^{J^{\prime}}_{\mu_1 + \mu_2 \,, m_1 + m_2} \,,
\label{eq:copexp}
\end{eqnarray}
with $\mu_1 \,, \mu_2$ and $m_1 \,, m_2$ being row and column indices. The RHS of (\ref{eq:copexp}) is
\begin{multline}
C(K,L,J; \mu_1 \,, \mu_2) \, C(K, L, J; m_1 \,, m_2) \delta_{j \,, \mu_1+\mu_2} \delta_{i, m_1+m_2} \\
= \sum_{\substack{\mu^{\prime}_1+\mu^{\prime}_2 = j \\ m^{\prime}_1+m^{\prime}_2= i}} 
C(K,L,J; \mu^{\prime}_1 \,, \mu^{\prime}_2) \, C(K, L, J; m^{\prime}_1 \,, m^{\prime}_2) \,
\big(e^{\mu^{\prime}_1 m^{\prime}_1} (K) \big)_{\mu_1 m_1} \otimes \big(e^{\mu^{\prime}_2 m^{\prime}_2} (L) \big)_{\mu_2 m_2} \,.
\label{eq:copexp2}  
\end{multline}
Hence we have the coproduct
\begin{equation}
\Delta_{KL} \big( e^{ji}(J) \big) = \sum_{\substack{\mu_1+\mu_2 = j \\ m_1+m_2= i}}
C(K,L,J; \mu_1 \,, \mu_2) \, C(K, L, J; m_1 \,, m_2) \, e^{\mu_1 m_1}(K) \otimes e^{\mu_2 m_2}(L) \,.
\label{eq:copKL2}
\end{equation}

Writing $C(K,L,J; \mu_1 \,, \mu_2\,, j)$ $=C(K,L,J; \mu_1 \,, \mu_2) \delta_{\mu_1+\mu_2 \,, j}$ for the first Clebsch-Gordan 
coefficient, we can delete the constraint $j=\mu_1+\mu_2$ in summation. $C(K,L,J$; $\mu_1 \,, \mu_2 \,, j)$ is an invariant tensor 
when $\mu_1 \,, \mu_2$ and $j$ are transformed appropriately by $SU(2)$. Hence (\ref{eq:copKL2}) is preserved by $SU(2)$ action
on $j, \mu_1, \mu_2$. The same is the case for $SU(2)$ action on $i, m_1, m_2$. In other words, the coproduct in (\ref{eq:copKL2}) 
is equivariant with respect to both $SU(2)$ actions. 

Since any $M \in Mat(2J+1)$ is $\sum_{i,j} M_{ji} e^{ji}(J)$, (\ref{eq:copKL2}) gives
\begin{multline}
\Delta_{KL}(M)= \sum_{\substack{\mu_1 \,, \mu_2  \\ m_1 \,, m_2}} C(K,L,J; \mu_1 \,, \mu_2) \, C(K, L, J; m_1 \,, m_2) \\
\times M_{\mu_1+\mu_2 \,, m_1+m_2} e^{\mu_1 m_1}(K) \otimes e^{\mu_2 m_2}(L) \,.
\label{eq:cobasic}
\end{multline}
This is the basic formula. It preserves conjugation $*$ (induced by hermitian conjugation of matrices):
\begin{equation}
\Delta(M^\dagger) = \Delta(M)^\dagger \,.
\label{eq:cobasic2}
\end{equation}

It is instructive to check directly that $\Delta_{KL}$ is a homomorphism, that is that $\Delta_{KL}(M N) 
= \Delta_{KL}(M) \Delta_{KL}(N)$. Starting from (\ref{eq:copKL2}) we have
\begin{multline}
\Delta_{KL} \big( e^{ji}(J) \big) \Delta_{KL} \big( e^{j^\prime i^\prime}(J) \big)
=\sum_{\substack{\mu_1 \,, \mu_2  \\ m_1 \,, m_2}} \sum_{\substack{\mu_1^\prime \mu_2^\prime \\ m_1^\prime \,, m_2^\prime}} 
C(K,L,J; \mu_1 \,, \mu_2 \,, j) \, C(K,L,J; m_1 \,, m_2 \,, i) \\
\times C(K,L,J; \mu_1^\prime \,, \mu_2^\prime \,, j^\prime) \, C(K,L,J; m_1^\prime \,, m_2^\prime \,, i^\prime) 
\Big(e^{\mu_1 m_1}(K) \otimes e^{\mu_2 m_2}(L) \Big) \\
\times \Big(e^{\mu_1^\prime  m_1^\prime}(K) \otimes e^{\mu_2^\prime m_2^\prime}(L) \Big) \,.
\label{eq:mull1}
\end{multline}
Using $(A \otimes B) (C \otimes D) = AC \otimes BD$, we have
\begin{multline}
\Big( e^{\mu_1 m_1}(K) \otimes e^{\mu_2 m_2}(L) \Big) \Big(e^{\mu_1^\prime  m_1^\prime}(K) 
\otimes e^{\mu_2^\prime m_2^\prime}(L) \Big) \\
= e^{\mu_1 m_1}(K) e^{\mu_1^\prime m_1^\prime}(K) \otimes e^{\mu_2 m_2}(L) e^{\mu_2^\prime m_2^\prime} (L) 
= \delta_{m_1 \mu_1^\prime} \delta_{m_2 \mu_2^\prime}  e^{\mu_1 m_1^\prime}(K) \otimes e^{\mu_2 m_2^\prime}(L) \,.
\label{eq:mull2}
\end{multline}
To get the second line in (\ref{eq:mull2}) we have made use of
\begin{equation}
\Big( e^{\mu_1 m_1}(K) e^{\mu_1^\prime m_1^\prime}(K) \Big)_{\alpha \beta} =
e^{\mu_1 m_1}(K)_{\alpha \gamma}  e^{\mu_1^\prime m_1^\prime}(K)_{\gamma \beta} =
%&=&\delta_{\mu_1 \alpha} \delta_{m_1 \gamma} \delta_{\mu_1^\prime \gamma} \delta_{m_1^\prime \beta} 
%\nonumber \\ 
%&=&\delta_{m_1 \mu_1^\prime} \delta_{\mu_1 \alpha} \delta_{m_1^\prime \beta} \nonumber \\
\delta_{m_1 \mu_1^\prime} e^{\mu_1 m_1^\prime}(K)_{\alpha \beta} \,.
\end{equation}
Inserting (\ref{eq:mull2}) in (\ref{eq:mull1}) we get
\begin{multline}
\Delta_{KL} \big( e^{ji}(J) \big) \Delta_{KL} \big( e^{j^\prime i^\prime}(J) \big)
=\sum_{\substack{\mu_1 \,, \mu_2  \\ m_1 \,, m_2}} \sum_{\substack{\mu_1^\prime \mu_2^\prime \\ m_1^\prime \,, m_2^\prime}} 
C(K,L,J;\mu_1 \,, \mu_2 \,, j) \, C(K,L,J;m_1 \,, m_2 \,, i) \\
\times C(K,L,J;\mu_1^\prime \,, \mu_2^\prime \,, j^\prime) \, C(K,L,J;m_1^\prime \,, m_2^\prime \,, i^\prime)
\delta_{m_1 \mu_1^\prime} \delta_{m_2 \mu_2^\prime}  e^{\mu_1 m_1^\prime}(K) \otimes  e^{\mu_2 m_2^\prime}(L) \\ 
=\sum_{\mu_1 \,, \mu_2} \sum_{m_1^\prime \,, m_2^\prime} C(K,L,J; \mu_1 \,, \mu_2 \,, j) C(K,L,J; m_1^\prime \,, m_2^\prime \,, 
i^\prime) \\ 
\times \underbrace{\Big(\sum_{m_1 \,, m_2} C(K,L,J; m_1 \,, m_2 \,, i) C(K,L,J; m_1 \,, m_2 \,, j^\prime) \Big)}
_{= \delta_{i j^\prime}} e^{\mu_1 m_1^\prime}(K) \otimes  e^{\mu_2 m_2^\prime}(L)  \,,
\end{multline}
where the orthogonality of Clebsch-Gordan coefficients is used to obtain $\delta_{i j^\prime}$ for the factor with the under brace. 
Thus,
\begin{multline}
\Delta_{KL} \big( e^{ji}(J) \big) \Delta_{KL} \big( e^{j^\prime i^\prime}(J) \big) \\
=\sum_{\substack{\mu_1 \,, \mu_2  \\ m_1 \,, m_2}}
%=\sum_{\mu_1 \,, \mu_2 \,, m_1^\prime \,, m_2^\prime} 
C(K,L,J; \mu_1 \,, \mu_2 \,, j) C(K,L,J; m_1^\prime \,, m_2^\prime \,, i^\prime) \delta_{i j^\prime}
e^{\mu_1 m_1^\prime}(K) \otimes  e^{\mu_2 m_2^\prime}(L) \\
= \delta_{i j^\prime} \Delta_{KL}(e^{j i^\prime}) \,.
\label{eq:pro2}
\end{multline}
Upon multiplying both sides of (\ref{eq:pro2}) by the coefficients $M_{ji} N_{j^\prime i^\prime}$ we finally get 
\begin{multline}
\Delta_{KL}\Big(\sum_{ji} M_{ji}e^{ji}(J)\Big) \Delta_{KL} \Big(\sum_{j^\prime i^\prime} 
N_{j^\prime i^\prime} e^{j^\prime i^\prime}(J)\Big) = \Delta_{KL}(M) \Delta_{KL}(N) \\
= (MN)_{ji^\prime} \Delta_{KL} (e^{j i^\prime}) = \Delta_{KL}(MN) \,,
\end{multline}
as we intended to demonstrate.

It remains to record the fuzzy analogues of counit $\varepsilon$ and antipode $S$. For the counit we have
\begin{multline}
\varepsilon \big(e^{ji}(J) \delta_J\big) = \frac{2J+1}{V} \int d \mu (g) D_{ij}^J(g)^\dagger \varepsilon \big( g(.) \big) 
= \frac{2J+1}{V} \int d \mu (g) D_{ij}^J(g)^\dagger 1 \\
= \frac{2J+1}{V} \int d \mu (g) D_{ij}^J(g)^\dagger D^0(g) \,.
\end{multline} 
Using equation (\ref{eq:iki}) and the fact that $D^0(g)$ is a unit matrix with only one entry which we denote by
$00$, we have 
%\begin{equation}
%\varepsilon \big(e^{ji}(J) \delta_J \big)_{00} = \frac{2J+1}{V} \int d \mu (g) D_{ij}^J(g)^\dagger D^0(g)_{00} 
%= \delta_{0J} \delta_{j0} \delta_{i0} \,.
%\end{equation}
%Thus
\begin{equation}
\varepsilon \big(e^{ji}(J) \delta_J \big)_{00}(K) = \delta_{0J} \delta_{j0} \delta_{i0}  \,, \quad \forall \, K \in 
{\widehat {SU(2)}} \,.
\end{equation}

For the antipode, we have
\begin{equation}
S \big(e^{ji}(J) \delta_J\big) = \frac{2J+1}{V} \int d \mu (g) D_{ij}^J(g)^\dagger S \big( g(.) \big) 
=\frac{2J+1}{V} \int d \mu (g) D_{ij}^J(g)^\dagger g^{-1}(.) 
\label{eq:antipode1}
\end{equation}
or
\begin{equation}
S \big(e^{ji}(J) \delta_J\big)(K) = \frac{2J+1}{V} \int d \mu (g) D_{ij}^{J}(g)^\dagger D^K(g^{-1}) \,.
\label{eq:antipode2}
\end{equation} 

In an UIRR $K$ we have $C=e^{-i \pi J_2}$ as the charge conjugation matrix. 
It fulfills $C D^K(g) C^{-1} = {\bar D}^K(g)$. Then since $D^K(g^{-1}) = D^K(g)^\dagger$,
\begin{equation}
D^K(g^{-1}) = C D^K(g)^T C^{-1} \,,
\end{equation}
where $T$ denotes transposition. We insert this in (\ref{eq:antipode2}) and use (\ref{eq:iki}) to find  
\begin{eqnarray}
S \big(e^{ji}(J) \delta_J\big)_{k \ell}(K) &=& \frac{2J+1}{V} \int d \mu (g) D_{ij}^J(g)^\dagger 
\big ( C_{ku} D^K(g)^T_{u \upsilon} C^{-1}_{\upsilon \ell} \big) \nonumber \\
&=& \frac{2J+1}{V} \int d \mu (g) D_{ij}^J(g)^\dagger C_{ku} D^K(g)_{\upsilon u} C^{-1}_{\upsilon \ell} \nonumber \\ 
&=& \delta _{JK} C_{ku} \delta_{ui} \delta_{\upsilon j} C^{-1}_{\upsilon \ell} \nonumber \\
&=& \delta_{JK} C_{ki} C^{-1}_{j \ell} \,.
\end{eqnarray}
This can be simplified further. Since in the UIRR $K$,
\begin{equation}
\big(e^{- i \pi J_2}\big)_{ki} = \delta_{-ki}(-1)^{K+k} = \delta_{-ki}(-1)^{K-i}  \,,
\end{equation}
and $C^{-1}=C^T$, we find
\begin{eqnarray}
S \big(e^{ji}(J) \delta_J\big)_{k \ell}(K) &=& \delta_{JK} \delta_{-ki} \delta_{-\ell j} (-1)^{2K-i-j} \nonumber \\
&=& \delta_{JK} (-1)^{2J-i-j} e^{-i\,,-j}(J)_{k \ell} \,.
\end{eqnarray}
Thus
\begin{equation}
S \big(e^{ji}(J) \delta_J\big)(K) = \delta_{JK} (-1)^{2J-i-j} e^{-i\,,-j}(J) \,.
\end{equation}

\section{Interpretation}

We recall from chapter 2 that the matrix $M \in Mat(2J+1)$ can be interpreted as the wave function of a particle on the 
spatial slice $S_F^2(J)$. The Hilbert space for these wave functions is $Mat(2J+1)$ with the scalar product given by 
$(M, N ) = Tr M^\dagger N$, $M, N \in S_F^2(J)$.

We can also regard $M$ as a fuzzy two-dimensional Euclidean scalar field as we did earlier 
or even as a field on a spatial slice $S_F^2(J)$ of a three dimensional space-time $S_F^2(J) \times {\mathbb R}$.

Let us look at the particle interpretation. Then (\ref{eq:cobasic}) gives the amplitude, up to an overall factor, for $M \in
S_F^2(J)$ splitting into a superposition of wave functions on $S_F^2(K) \otimes S_F^2(L)$. It models the process where a fuzzy 
sphere splits into two others \cite{BalPaulo}. The overall factor is the reduced matrix element much like the reduced 
matrix elements in angular momentum selection rules. It is unaffected by algebraic operations on    
$S_F^2(J), S_F^2(K)$ or $S_F^2(L)$ and is determined by dynamics. 

Now (\ref{eq:cobasic}) preserves trace and scalar product:
\begin{gather}
Tr \Delta_{KL}(M) = Tr M \,, \nonumber \\
\big( \Delta_{KL}(M), \Delta_{KL}(N) \big) = (M, N) \,. 
\label{eq:traces}
\end{gather}
So (\ref{eq:cobasic}) is a unitary branching process. This means that the overall factor is a phase.

$\Delta_{KL}(S_F^2(J))$ has all the properties of $S_F^2(J)$. So (\ref{eq:cobasic}) is also a precise rule on how $S_F^2(J)$ sits in
$S_F^2(K) \otimes S_F^2(L)$. We can understand ``how $\Delta_{KL}(M)$ sits'' as follows. A basis for 
$S_F^2(K) \otimes S_F^2(L)$ is $e^{\mu_1 m_1} (K) \otimes e^{\mu_2 m_2} (L)$. We can choose another basis where 
left- and right- angular momenta are separately diagonal by 
coupling $\mu_1$ and  $\mu_2$ to give angular momentum $\sigma \in \lbrack 0, \frac{1}{2}, 1, \hdots , K+L \rbrack$, 
and $m_1$ and $m_2$ to give 
angular momentum $\tau \in \lbrack 0, \frac{1}{2}, 1, \hdots , K+L \rbrack$. In this basis, $\Delta_{KL}(M)$ is zero 
except in the block with $\sigma = \tau =J$.    

So the probability amplitude for $M \in S_F^2(J)$ splitting into $P \otimes Q \in S_F^2(K) \otimes S_F^2(L)$ for normalized wave
functions is 
\begin{equation}
phase \times Tr(P \otimes Q)^\dagger \Delta_{KL}(M) \,.
\label{eq:probamp1}
\end{equation}  
Branching rules for different choices of $M, P$ and $Q$ are independent of the constant phase and can be determined.

Written in full, (\ref{eq:probamp1}) is seen to be just the coupling conserving left- and right- angular momenta of 
$P^\dagger, Q^\dagger$ and $M$. That alone determines (\ref{eq:probamp1}).

An observable $A$ is a self-adjoint operator on a wave function $M \in S_F^2(J)$. Any linear operator on $S_F^2(J)$ can be written as
$\sum B_\alpha^L C_\alpha^R$ where $B_\alpha \,, C_\alpha \in S_F^2(J)$ and $B_\alpha^L$ and $C_\alpha^R$ act by left- and right-
multiplication: $B_\alpha^L M = B_\alpha M \,, C_\alpha^R M = M C_\alpha$. Any observable on $S_F^2(J)$ has an action on its branched
image $\Delta_{KL} (S_F^2(J))$:
\begin{equation}
\Delta_{KL} (A) \Delta_{KL}(M) :=\Delta_{KL} (AM) \,.    
\label{eq:obaction}
\end{equation}
By construction, (\ref{eq:obaction}) preserves algebraic properties of operators. $\Delta_{KL} (A)$ can actually act on all of 
$S^2_F(K) \otimes S_F^2(L)$, but in the basis described above it is zero on vectors with $\sigma \neq J$ and/or $\tau \neq J$.

This equation is helpful to address several physical questions. For example if $M$ is a wave function with a
definite eigenvalue for $A$, then $\Delta_{KL}(M)$ is a wave function with the same eigenvalue for $\Delta_{KL}(A)$. This follows from
$\Delta_{KL}(BM) = \Delta_{KL}(B) \Delta_{KL}(M)$ and $\Delta_{KL}(MB) = \Delta_{KL}(M) \Delta_{KL}(B)$. Combining this with 
(\ref{eq:traces}) and the other observations, we see that mean value of $\Delta_{KL}(A)$ in $\Delta_{KL}(M)$ and of $A$ in $M$ are
equal.

In summary all this means that every operator on $S_F^2(J)$ is a constant of motion for the branching process (\ref{eq:cobasic}).

Now suppose $R \in S_F^2(K) \otimes S_F^2(L)$ is a wave function which is not necessarily of the form $P \otimes Q$. 
Then we can also give a formula for the probability amplitude
for finding $R$ in the state described by $M$. Note that $R$ and $M$ live in different fuzzy spaces. The answer is 
\begin{equation}
constant \times Tr R^\dagger \Delta_{KL}(M) \,.
\end{equation}

If $M, P, Q$ are fields with $S_F^2(I) \, (I= J,K,L)$ a spatial slice or space-time, (\ref{eq:probamp1}) is an interaction of 
fields on different fuzzy manifolds. It can give dynamics to the branching process of fuzzy topologies discussed above.

\section{The Pre\v{s}najder Map}

This section is somewhat disconnected from the material in the rest of the chapter.

We recall that $S_F^2(J)$ can be realized as an algebra generated by the spherical harmonics $Y_{lm} \, (l \leq 2J)$ which are 
functions on the two-sphere $S^2$. Their product can be the coherent state $*_c$ or Moyal $*_M$ product.

But we saw that $S_F^2(J)$ is isomorphic to the convolution algebra of functions $D_{MN}^J$ on $SU(2) \simeq S^3$. 

It is reasonable to wonder how functions on $S^2$ and $S^3$ get related preserving the respective algebraic properties.

The map connecting these spaces is described by a function on $SU(2) \times S^2 \approx S^3 \times S^2$ and was first introduced 
by Pre\v{s}najder \cite{GP1, chiral}. We give its definition and introduce its properties here. It generalizes to any group $G$.

Let $a_i \,, a_j^\dagger$ $(i = 1,2)$ be Schwinger oscillators for $SU(2)$ and let us also recall that for $J = \frac{n}{2}$ 
\begin{equation}
|z \,, 2J \rangle = \frac{(z_i a_i^\dagger)^{2J}}{\sqrt{2J!}} \,|0 \rangle \,, \quad \sum|z_i|^2 =1 \,
\label{eq:coherentstate1}
\end{equation}
are the normalized Perelomov coherent states. If $U(g)$ is the unitary operator implementing $g \in SU(2)$ in the 
spin $J$ UIRR, the Pre\v{s}najder function \cite{GP1, chiral} $P_J$ is given by 
\begin{gather}
P_J(g, \vec{n})= \langle z \,, 2J | U(g) |z \,, 2J \rangle = D_{JJ}^J(h^{-1} g h)\,, \nonumber \\      
{\vec n} = z^\dagger {\vec \tau} z \,, \quad  {\vec n} \cdot {\vec n} =1 \,, 
\nonumber \\
h = \left (
\begin{array}{cc}
z_1 & -{\bar z}_2 \\
z_2 & {\bar z}_1 \\
\end{array}
\right ) \,.
\label{eq:presnajdermap}
\end{gather}
Now ${\vec n} \in S^2$. As the phase change $z_i \rightarrow z_i e^{i \theta}$ does not effect $P_J$, besides $g$, it 
depends only on ${\vec n}$. It is a function on
\begin{equation}
\big( SU(2) \simeq S^3 \big) \times \big \lbrack SU(2)/U(1) \big \rbrack \simeq S^3 \times S^2 \,.
\label{eq:domain1}
\end{equation}

A basis of $SU(2)$ functions for spin $J$ is $D_{ij}^J$. A basis of $S^2$ functions for spin $J$ is $E^{ij}(J \,, .)$ where 
\begin{equation}
E^{ij} (J \,, {\vec n}) = \langle z \,, 2J | e^{ij}(J) |z \,, 2 J \rangle = D^J(h^{-1})_{Ji} D^J(h)_{jJ} \,, \quad  
\mbox{no} \, \, \mbox{sum} \, \, \mbox{on} \, \, J \,.
\label{eq:basisonS2}
\end{equation}
The transform of $D_{ij}^J$ to $E^{ij}(J,.)$ is given by
\begin{equation}
E^{ij}(J \,, {\vec n}) = \frac{(2J+1)}{V} \int d \mu(g) {\bar P_J}(g, {\vec n}) D_{ij}^J(g) \,.
%D_{ij}^J(g) &=& \int d \Omega ({\overrightarrow n}) P_J(g, {\overrightarrow n}) E^{ij}({\overrightarrow n}) \,.
\label{eq:transforms}
\end{equation}
This can be inverted by constructing a function $Q_J$ on $SU(2) \times S^2$ such that 
\begin{equation}
\int_{S^2} d \Omega ({\vec n}) Q_J(g^\prime \,, {\vec n}) {\bar P}_J (g \,, {\vec n})
= \sum_{ij} D_{ij}^J(g^\prime) {\bar D}_{ij}^J(g) \,, \quad \quad d \Omega ({\vec n}) 
= \frac{d \cos \theta d \varphi}{4 \pi} \,,
\label{eq:qfunction}
\end{equation}
$\theta$ and $\varphi$ being the polar and azimuthal angles on $S^2$. Then using (2), we get
\begin{equation}
D_{ij}^J (g^\prime) = \int_{S^2} d \Omega ({\vec n}) Q_J(g^\prime \,, {\vec n}) 
E^{ij}(J \,, {\vec n})\,.  
\label{eq:transform1}
\end{equation}

Consider first $J = \frac{1}{2}$. In that case 
\begin{equation}
{\bar P}_{\frac{1}{2}} (g \,, {\vec n}) = {\bar g}_{kl} {\bar z}_k z_l
={\bar g}_{kl} \Big( \frac{1 + {\vec \sigma} \cdot {\vec n}}{2} \Big)_{lk}
\label{eq:presnajdermap1/2}
\end{equation}
where $g$ is a $2 \times 2$ $SU(2)$ matrix and $\sigma_i$ are Pauli matrices. Since
\begin{equation}
\int_{S^2} d \Omega ({\vec n}) n_i n_j = \frac{1}{3} \delta_{ij} \,,   
\label{eq:intid1}
\end{equation}
we find 
\begin{gather}
Q_{\frac{1}{2}} (g^\prime \,, {\vec n}) = Tr {\tilde g}^\prime (1+ 3  {\vec \sigma} \cdot {\vec n}) \,, \\
g^\prime = 2 \times 2 \, \, SU(2) \, \, \mbox{matrix} \,, \nonumber \\
{\tilde g}^\prime = \mbox{transpose \, of} \, \, g^\prime \,. \nonumber   
\label{eq:q1/2}
\end{gather}

For $J=\frac{n}{2}$, $D^J(g)$ acts on the symmetric product of $n$ ${\mathbb C}^2$'s and can be written as 
$\underbrace{g \otimes g \otimes \cdots \otimes g}_{N \, factors}$ and (\ref{eq:presnajdermap1/2}) gets replaced by
\begin{equation}
{\bar P}_J (g \,, {\vec n}) = \Big \lbrack Tr {\bar g} \Big( \frac{1 + {\vec \sigma} 
\cdot {\vec n}}{2} \Big) \Big \rbrack^N \,.  
\end{equation}
Then $Q_J (g^\prime \,, {\vec n})$ is defined by (\ref{eq:qfunction}). It exists, but we have not found a neat formula
for it. 

As the relation between $E^{ij}$ and $Y_{lm}$ can be worked out, it is possible to suitably substitute $Y_{lm}$ for $E^{ij}$ in 
these formulae.

These equations establish an isomorphism (with all the nice properties like preserving $*$ and $SU(2)$-actions) between the 
convolution algebra $\rho(J)$ ($G^*$) at spin $J$ and the $*$-product algebra of $S_F^2(J)$. That is because we saw that 
$\rho (J)(G^*)$ and $S_F^2(J) \simeq Mat(2J+1)$ are isomorphic, while it is known that $Mat(2J+1)$ and the $*$-product algebra
of $S^2$ at level $J$ are isomorphic.

There are evident generalizations of $P_J$ for other groups and their orbits.

%%%%%%%%%%%%%%%%%%%%%%%%%%%%%%%%%%%%%%%%%%%%%%%%%%%%%%%%%%%%%%%%%%%%%%%%%%%%%%%%%%


\begin{thebibliography}{99}

\bibitem{courseweb}
A. P. Balachandran, video conference course on ``Fuzzy Physics'', at
http://www.phy.syr.edu/courses/Fuzzy Physics and
http://bach.if.usp.br/FUZZY/. 

\bibitem{madore2} 
J.~Madore,
The Fuzzy Sphere,
Class.\ Quant.\ Grav.\  {\bf 9}, 69 (1992).

\bibitem{madore1} J. Madore, {\it An Introduction to Non-commutative
Differential Geometry and its Physical Applications}, Cambridge University Press,
Cambridge (1995); 

\bibitem{kostant} A.A.~Kirillov, {\it Encyclopedia of Mathematical Sciences},vol 4, p.230; 
B.~Kostant, {\it Lecture Notes in Mathematics}, vol.170,
Springer-Verlag(1970), p.87.

\bibitem{berezin} F.~A.~Berezin, 
General Concept of Quantization,
Commun.\ Math.\ Phys.\  {\bf 40}, 153 (1975).

\bibitem{GKP1} H. Grosse, C. Klimcik, P. Pre\v{s}najder, 
Field Theory on a Supersymmetric Lattice, 
Commun. Math. Phys., {\bf 185} (1997) 155-175 and hep-th/9507074;

H. Grosse, C. Klim\v{c}ik, P. Pre\v{s}najder, 
N=2 Superalgebra and Non-Commutative Geometry, 
hep-th/9603071;

\bibitem{fuzzyS} H. Grosse, G. Reiter, 
The Fuzzy Supersphere, 
J. Geom. and Phys., {\bf 28} (1998) 349-383 and math-ph/9804013.

\bibitem{seckin1} 
A.~P.~Balachandran, S.~Kurkcuoglu and E.~Rojas,
The Star Product on the Fuzzy Supersphere,
JHEP {\bf 0207}, 056 (2002)
[arXiv:hep-th/0204170];

\bibitem{seckinthesis}
S. Kurkcuoglu, Ph.D. Thesis, Syracuse University, Syracuse NY, 2004. 

\bibitem{caterall}  J.~Ambjorn and S.~Catterall, Stripes from (Noncommutative) Stars,
  Phys.\ Lett.\ B {\bf 549}, 253 (2002)
  [arXiv:hep-lat/0209106].

\bibitem{ref1} See for example, P. H. Frampton {\it Gauge Field
  Theories}, The Benjamin Cummings Publishing Company, Menlo Park CA, (1987). 

\bibitem{ref2}R.D. Sorkin Int.J. Theory. Phys. {\bf 30} (1991) 923; 

A.~P.~Balachandran, G.~Bimonte, E.~Ercolessi, G.~Landi, F.~Lizzi, 
G.~Sparano and P.~Teotonio-Sobrinho,
Finite Quantum Physics and Noncommutative Geometry,
Nucl.\ Phys.\ Proc.\ Suppl.\  {\bf 37C}, 20 (1995)
[arXiv:hep-th/9403067].

\bibitem{Raamsdonk} S.~Minwalla, M.~Van Raamsdonk and N.~Seiberg,
Noncommutative Perturbative Dynamics, JHEP {\bf 0002}, 020 (2000)
[arXiv:hep-th/9912072].

\bibitem{ydri1} 
B.~Ydri,
Non-commutative geometry as a regulator,
Phys.\ Rev.\ D {\bf 63}, 025004 (2001)
[arXiv:hep-th/0003232].

\bibitem{vaidya1}
S.~Vaidya,
Perturbative dynamics on fuzzy S(2) and RP(2),
Phys.\ Lett.\ B {\bf 512}, 403 (2001)
[arXiv:hep-th/0102212];

\bibitem{chu}
C.~S.~Chu, J.~Madore and H.~Steinacker,
Scaling limits of the fuzzy sphere at one loop,
JHEP {\bf 0108}, 038 (2001) [arXiv:hep-th/0106205].

\bibitem{dolan}
B.~P.~Dolan, D.~O'Connor and P.~Pre\v{s}najder,
Matrix $\phi^4$ models on the fuzzy sphere and their continuum limits'',
JHEP {\bf 0203}, 013 (2002)
[arXiv:hep-th/0109084];

\bibitem{sachin1}
S.~Vaidya and B.~Ydri,
On the origin of the UV-IR mixing in non-commutative matrix geometry,
Nucl.\ Phys.\ B {\bf 671}, 401 (2003)
[arXiv:hep-th/0305201].

\bibitem{sachin2} S.~Vaidya and B.~Ydri,
New scaling limit for fuzzy spheres, arXiv:hep-th/0209131.

\bibitem{GrosseCP2}  H.~Grosse and A.~Strohmaier,
Noncommutative geometry and the regularization problem of 4D quantum  field
theory,
Lett.\ Math.\ Phys.\  {\bf 48}, 163 (1999)
[arXiv:hep-th/9902138].

\bibitem{fuzzycp2} 
G.~Alexanian, A.~P.~Balachandran, G.~Immirzi and B.~Ydri,
Fuzzy CP(2),
J.\ Geom.\ Phys.\  {\bf 42}, 28 (2002)
[arXiv:hep-th/0103023].

\bibitem{BDBJ}
A.~P.~Balachandran, B.~P.~Dolan, J.~H.~Lee, X.~Martin and D.~O'Connor,
Fuzzy complex projective spaces and their star-products,
J.\ Geom.\ Phys.\  {\bf 43}, 184 (2002)
[arXiv:hep-th/0107099].

\bibitem{Nair}  D.~Karabali and V.~P.~Nair,
Quantum Hall effect in higher dimensions,
  Nucl.\ Phys.\ B {\bf 641}, 533 (2002)
  [arXiv:hep-th/0203264];

D.~Karabali and V.~P.~Nair,
The effective action for edge states in higher dimensional quantum
Hall systems,
  Nucl.\ Phys.\ B {\bf 679}, 427 (2004)
  [arXiv:hep-th/0307281];

D.~Karabali and V.~P.~Nair,
Edge states for quantum Hall droplets in higher dimensions and a
generalized WZW model, Nucl.\ Phys.\ B {\bf 697}, 513 (2004)
  [arXiv:hep-th/0403111];

D.~Karabali, V.~P.~Nair and S.~Randjbar-Daemi,
  Fuzzy spaces, the M(atrix) model and the quantum Hall effect,
  arXiv:hep-th/0407007.

\bibitem{Julieta} J.~Medina and D.~O'Connor, Scalar field theory on fuzzy S(4),
JHEP {\bf 0311}, 051 (2003), [arXiv:hep-th/0212170].

\bibitem{connes} A. Connes, {\it Non-commutative Geometry} San Diego, Academic Press, 1994.

\bibitem{Landi} G. Landi, {\it An Introduction to Non-commutative Spaces and their
Geometries} (Springer-Verlag, 1997); 
  
\bibitem{Varilly} J.M. Gracia-Bond\'{\i}a, J.C. V\'arilly and H. Figueroa, {\it 
Elements of Non-commutative Geometry} (Birkh\"auser, 2000).

\bibitem{Szabo:2001kg}
R.~J.~Szabo,
Quantum field theory on non-commutative spaces,
Phys.\ Rept.\  {\bf 378}, 207 (2003)
[arXiv:hep-th/0109162].

\bibitem{Douglas:2001ba}
M.~R.~Douglas and N.~A.~Nekrasov,
Non-commutative field theory, 
Rev.\ Mod.\ Phys.\  {\bf 73}, 977 (2001)
[arXiv:hep-th/0106048].

\bibitem{history}
\textit{Letter of Heisenberg to Peierls (1930)},   Wolfgang Pauli,
Scientific Correspondence, Vol. II, p.15, Ed. Karl   von Meyenn,
Springer-Verlag, 1985; 

\textit{Letter of Pauli to Oppenheimer   (1946)}, Wolfgang Pauli,
Scientific Correspondence, Vol. III,   p.380, Ed. Karl von Meyenn,
Springer-Verlag, 1993.

\bibitem{Groenewold:kp}
H.~J.~Groenewold,
On The Principles Of Elementary Quantum Mechanics,
Physica {\bf 12}, 405 (1946).

\bibitem{Snyder:1946qz}
H.~S.~Snyder, 
Quantized Space-Time, 
Phys.\ Rev.\  {\bf 71}, 38 (1947);
The Electromagnetic Field in Quantized Spacetime, 
Phys. Rev.{\bf 72} (1947) 68;

\bibitem{Yang:ud} C.~N.~Yang,
On Quantized Space-Time,
Phys.\ Rev.\  {\bf 72}, 874 (1947).

\bibitem{Moyal:sk}
J.~E.~Moyal,
Quantum Mechanics As A Statistical Theory,
Proc.\ Cambridge Phil.\ Soc.\  {\bf 45}, 99 (1949).

\bibitem{Jackiw} R. Jackiw, 
Physical Instances of Noncommuting Coordinates,
Nucl.Phys.Proc.Suppl. \textbf{108}, 30
(2002) [hep-th/0110057];

\bibitem{topology1} A. P. Balachandran, G. Marmo, B. S. Skagerstam,
  A. Stern, {\it Classical Topology and Quantum States},
World Scientific, Singapore, 1991.

\bibitem{moskalev} D.~A. Varshalovich, A.~N. Moskalev and V.~K. Khersonsky, 
{\it Quantum Theory of Angular Momentum}, World Scientific, New
Jersey, 1998.

\bibitem{Holstein} T.~Holstein and H.~Primakoff,
Field Dependence Of The Intrinsic Domain Magnetization Of A Ferromagnet,
Phys.\ Rev.\  {\bf 58}, 1098 (1940).

\bibitem{Sen} D. Sen, 
Quantum-spin-chain realizations of conformal field theories
Phys. Rev. B 44, 2645(1991)

\bibitem{Lipkinbook} H. J. Lipkin {\it Lie Groups for Pedestrians },
  Dover Publications, 2002.

\bibitem{Goddard}  P.~Goddard and D.~I.~Olive,
Kac-Moody And Virasoro Algebras In Relation To Quantum Physics,
Int.\ J.\ Mod.\ Phys.\ A {\bf 1}, 303 (1986).

\bibitem{defqu}
F.~Bayen, M.~Flato, C.~Fronsdal, A.~Lichnerowicz and D.~Sternheimer,
Deformation Theory And Quantization. 1. Deformations Of Symplectic
Structures, Annals Phys.\  {\bf 111}, 61 (1978); 

F.~Bayen, M.~Flato, C.~Fronsdal, A.~Lichnerowicz and D.~Sternheimer,
Quantum Mechanics As A Deformation Of Classical Mechanics,
Lett.\ Math.\ Phys.\  {\bf 1}, 521 (1977).

\bibitem{weyl} H. Weyl {\it Gruppentheorie und Quantenmechanik
The theory of groups and quantum mechanics}, New York, Dover Publications, 1950;

H.~Weyl, Quantum Mechanics And Group Theory,
Z.\ Phys.\  {\bf 46}, 1 (1927).

\bibitem{Klauder} J. R. Klauder and B. S. Skagerstam, {\it Coherent States:
Applications in Physics and Mathematical Physics}, World Scientific (1985).

\bibitem{perelomov} A. M. Perelomov {\it Generalized Coherent States
  and their Applications}, 
Springer-Verlag (1986).

\bibitem{GarnikAlSasha} G.~Alexanian, A.~Pinzul and A.~Stern, 
Generalized Coherent State Approach to Star Products and Applications to
the Fuzzy Sphere, Nucl.\ Phys.\ B {\bf 600}, 531 (2001), [arXiv:hep-th/0010187].

\bibitem{Voros} A. Voros, 
Wentzel-Kramers-Brillouin method in the Bargmann representation
Phys. Rev. {\bf A40}, 6814 (1989).

\bibitem{Haag} R. Haag, {\it Local quantum physics : fields,
  particles, algebras.} Berlin, Springer-Verlag (1996).

\bibitem{kontsevich}
M.~Kontsevich,
Deformation quantization of Poisson manifolds, I,
Lett.\ Math.\ Phys.\  {\bf 66}, 157 (2003)
[arXiv:q-alg/9709040].

\bibitem{chiral} 
P.~Pre\v{s}najder, 
The origin of chiral anomaly and the non-commutative geometry,
J.\ Math.\ Phys.\  {\bf 41}, 2789 (2000)
[arXiv:hep-th/9912050];

\bibitem{regulator} H.~Grosse, C.~Klimcik and P.~Pre\v{s}najder,
Towards finite quantum field theory in non-commutative geometry,
Int.\ J.\ Theor.\ Phys.\  {\bf 35}, 231 (1996)
[arXiv:hep-th/9505175].
  
\bibitem{GP1} 
H.~Grosse and P.~Pre\v{s}najder,
The Construction on non-commutative manifolds using coherent states,
Lett.\ Math.\ Phys.\  {\bf 28}, 239 (1993);

\bibitem{bal} A.~P.~Balachandran and G.~Immirzi, 
Fuzzy Nambu-Goldstone physics,
Int. J. Mod. Phys. {\bf A 18} (2003) 5981,arXiv:hep-th/0212133.

\bibitem{berryphase}M.~V.~Berry,
Quantal Phase Factors Accompanying Adiabatic Changes,
Proc.\ Roy.\ Soc.\ Lond.\ A {\bf 392}, 45 (1984).

\bibitem{wegge-olsen}  N.E. Wegge Olsen, 
{\it K-theory and $C^*$-Algebras-a Friendly Approach}, 
Oxford University Press, Oxford, 1993.

\bibitem{watamura} U.~Carow-Watamura and S.~Watamura,
Chirality and Dirac operator on noncommutative sphere,, 
Commun.\ Math.\ Phys.\  {\bf 183}, 365 (1997)
  [arXiv:hep-th/9605003].

\bibitem{jr} R.~Jackiw and C.~Rebbi,
Spin From Isospin In A Gauge Theory,
Phys.\ Rev.\ Lett.\  {\bf 36}, 1116 (1976).

\bibitem{Hasenfratz} P.~Hasenfratz and G.~'t Hooft,
A Fermion - Boson Puzzle In A Gauge Theory,
Phys.\ Rev.\ Lett.\  {\bf 36}, 1119 (1976).

\bibitem{baletal} 
A.~P.~Balachandran, A.~Stern and C.~G.~Trahern,
Nonlinear Models As Gauge Theories, Phys.\ Rev.\ D {\bf 19}, 2416 (1979). 

\bibitem{trg} T.~R.~Govindarajan and E.~Harikumar, 
O(3) sigma model with Hopf term on fuzzy sphere,
Nucl.\ Phys.\ B {\bf 655}, 300 (2003) [arXiv:hep-th/0211258].

\bibitem{chan} Chuan-Tsung Chan, Chiang-Mei Chen, Feng-Li Lin, Hyun Seok Yang,
`${\mathbb C}P^n$ Model on Fuzzy Sphere', {\it Nucl.Phys.}  B625 (2002) 327
 and {\tt hep-th/0105087};
 
Chuan-Tsung Chan, Chiang-Mei Chen, Hyun Seok Yang,
``Topological $Z_{N+1}$ Charges on Fuzzy Sphere'', {\tt hep-th/0106269}.
    
\bibitem{dabrowski}
Ludwik Dabrowski, Thomas Krajewski, Giovanni Landi, `Some Properties
of Non-linear $\sigma$-Models in Noncommutative Geometry',
{\it Int. J. Mod. Phys.} B14 (2000) 2367 and {\tt hep-th/0003099}. 

\bibitem{wojciech} W.J. Zakrzewski, `Low dimensional sigma models', 
Adam Hilger, Bristol 1997.

\bibitem{mignaco} 
J. A. Mignaco, C. Sigaud, A. R. da Silva and F. J. Vanhecke,
`The Connes-Lott program on the sphere', 
{\it Rev. Math. Phys.} {\bf 9} (1997) 689 and  {\tt hep-th/9611058};
 
J. A. Mignaco, C. Sigaud, A. R. da Silva and F. J. Vanhecke,
`Connes-Lott model building on the two-sphere', 
{\it Rev. Math. Phys.} {\bf 13} (2001) 1 and {\tt hep-th/9904171}.

\bibitem{landi} G. Landi, Projective Modules of Finite Type over
the Supersphere $S^{2,2}$, Differ. Geom. Appl. {\bf 14} (2001) 95-111 and math-ph/9907020.

\bibitem{yangsigma} A.~M.~Polyakov,
Interaction Of Goldstone Particles In Two-Dimensions. Applications To
Ferromagnets And Massive Yang-Mills Fields,
Phys.\ Lett.\ B {\bf 59}, 79 (1975); 

A.~M.~Polyakov and A.~A.~Belavin,
Metastable States Of Two-Dimensional Isotropic Ferromagnets,
JETP Lett.\  {\bf 22}, 245 (1975) [Pisma Zh.\ Eksp.\ Teor.\ Fiz.\
  {\bf 22}, 503 (1975)].

\bibitem{GKP2} H.~Grosse, C.~Klimcik and P.~Pre\v{s}snajder,
Topologically nontrivial field configurations in non-commutative geometry,
Commun.\ Math.\ Phys.\  {\bf 178}, 507 (1996) [arXiv:hep-th/9510083].

\bibitem{monopole} 
S.~Baez, A.~P.~Balachandran, B.~Ydri and S.~Vaidya,
Monopoles and solitons in fuzzy physics,
Commun.\ Math.\ Phys.\  {\bf 208}, 787 (2000)
[arXiv:hep-th/9811169].

\bibitem{instanton1}A.~P.~Balachandran and S.~Vaidya,
Instantons and chiral anomaly in fuzzy physics,
Int.\ J.\ Mod.\ Phys.\ A {\bf 16}, 17 (2001)
[arXiv:hep-th/9910129].

\bibitem{GP2}
H.~Grosse and J.~Madore,
A Non-commutative version of the Schwinger model,
Phys.\ Lett.\ B {\bf 283}, 218 (1992).

H.~Grosse and P.~Pre\v{s}najder,
A non-commutative regularization of the Schwinger model,
Lett.\ Math.\ Phys.\  {\bf 46}, 61 (1998).

\bibitem{chaichian1} M.~Chaichian, P.~Presnajder, M.~M.~Sheikh-Jabbari and A.~Tureanu,
Noncommutative standard model: Model building,Eur.\ Phys.\ J.\ C {\bf 29}, 413 (2003)
[arXiv:hep-th/0107055];

M.~Chaichian, P.~Presnajder, M.~M.~Sheikh-Jabbari and A.~Tureanu,
Noncommutative gauge field theories: A no-go theorem,
Phys.\ Lett.\ B {\bf 526}, 132 (2002)
[arXiv:hep-th/0107037].

M.~Chaichian, A.~Kobakhidze and A.~Tureanu,
 Spontaneous reduction of noncommutative gauge symmetry and model
 building,arXiv:hep-th/0408065.

\bibitem{nonabelianmonopole}
 A.~P.~Balachandran, G.~Marmo, N.~Mukunda, J.~S.~Nilsson, E.~C.~G.~Sudarshan and F.~Zaccaria,
Monopole Topology And The Problem Of Color, Phys.\ Rev.\ Lett.\  {\bf
50}, 1553 (1983);

A.~P.~Balachandran, G.~Marmo, N.~Mukunda, J.~S.~Nilsson, E.~C.~G.~Sudarshan and F.~Zaccaria,
Nonabelian Monopoles Break Color. 1. Classical Mechanics, Phys.\ Rev.\
D {\bf 29}, 2919 (1984);

A.~P.~Balachandran, G.~Marmo, N.~Mukunda, J.~S.~Nilsson, E.~C.~G.~Sudarshan and F.~Zaccaria,
Nonabelian Monopoles Break Color. 2. Field Theory And Quantum
Mechanics, Phys.\ Rev.\ D {\bf 29}, 2936 (1984).

\bibitem{sachin} S.~Vaidya, Scalar multi-solitons on the fuzzy sphere,
  JHEP {\bf 0201}, 011 (2002)
  [arXiv:hep-th/0109102]. 

\bibitem{Nair-Poly}  D.~Karabali, V.~P.~Nair and A.~P.~Polychronakos,
  Spectrum of Schroedinger field in a noncommutative magnetic monopole,
  Nucl.\ Phys.\ B {\bf 627}, 565 (2002)
  [arXiv:hep-th/0111249].

\bibitem{ginsparg} P.~H.~Ginsparg and K.~G.~Wilson,
A Remnant Of Chiral Symmetry On The Lattice,
Phys.\ Rev.\ D {\bf 25}, 2649 (1982).

\bibitem{giorgio}
A.~P.~Balachandran and G.~Immirzi,
The fuzzy Ginsparg-Wilson algebra: A solution of the fermion doubling problem,
Phys.\ Rev.\ D {\bf 68}, 065023 (2003)
[arXiv:hep-th/0301242];

\bibitem{fujikawa} K.~Fujikawa,
Path Integral Measure For Gauge Invariant Fermion Theories,
Phys.\ Rev.\ Lett.\  {\bf 42}, 1195 (1979);

K.~Fujikawa,
Path Integral For Gauge Theories With Fermions,
Phys.\ Rev.\ D {\bf 21}, 2848 (1980)
[Erratum-ibid.\ D {\bf 22}, 1499 (1980)].

\bibitem{BalNair} A.~P.~Balachandran, G.~Marmo, V.~P.~Nair and C.~G.~Trahern,
A Nonperturbative Proof Of The Nonabelian Anomalies,
Phys.\ Rev.\ D {\bf 25}, 2713 (1982).

\bibitem{luscher} S.~Randjbar-Daemi and J.~A.~Strathdee,
  On the overlap formulation of chiral gauge theory,
  Phys.\ Lett.\ B {\bf 348}, 543 (1995)
  [arXiv:hep-th/9412165];

S.~Randjbar-Daemi and J.~A.~Strathdee,
  Gravitational Lorentz anomaly from the overlap formula in two-dimensions,
  Phys.\ Rev.\ D {\bf 51}, 6617 (1995)
  [arXiv:hep-th/9501012];

S.~Randjbar-Daemi and J.~A.~Strathdee,
  Chiral fermions on the lattice,
  Nucl.\ Phys.\ B {\bf 443}, 386 (1995)
  [arXiv:hep-lat/9501027];

S.~Randjbar-Daemi and J.~A.~Strathdee,
  Consistent and covariant anomalies in the overlap formulation of chiral
  gauge theories,
  Phys.\ Lett.\ B {\bf 402}, 134 (1997)
  [arXiv:hep-th/9703092];

M.~Luscher,
  Exact chiral symmetry on the lattice and the Ginsparg-Wilson relation,
  Phys.\ Lett.\ B {\bf 428}, 342 (1998)
  [arXiv:hep-lat/9802011];

M.~Luscher,
  Weyl fermions on the lattice and the non-abelian gauge anomaly,
  Nucl.\ Phys.\ B {\bf 568}, 162 (2000)
  [arXiv:hep-lat/9904009];

H.~Neuberger,
  Chiral symmetry outside perturbation theory,
  arXiv:hep-lat/9912013;

W.~Kerler,
  Dirac operator normality and chiral fermions,
  Chin.\ J.\ Phys.\  {\bf 38}, 623 (2000)
  [arXiv:hep-lat/9912022];

J.~Nishimura and M.~A.~Vazquez-Mozo,
  Noncommutative chiral gauge theories on the lattice with manifest
  star-gauge invariance,
  JHEP {\bf 0108}, 033 (2001)
  [arXiv:hep-th/0107110].

\bibitem{fermion} 
A.~P.~Balachandran, T.~R.~Govindarajan and B.~Ydri,
The fermion doubling problem and noncommutative geometry,
Mod.\ Phys.\ Lett.\ A {\bf 15}, 1279 (2000)
[arXiv:hep-th/9911087].

A.~P.~Balachandran, T.~R.~Govindarajan and B.~Ydri,
The fermion doubling problem and noncommutative geometry. II,
arXiv:hep-th/000621.

\bibitem{bassetto} A. Bassetto and  L. Griguolo,  
{\it Journ. of Math. Phys.} {\bf 32} (1991) 3195. 

\bibitem{aoki} H.~Aoki, S.~Iso and K.~Nagao,
Chiral anomaly on fuzzy 2-sphere,
Phys.\ Rev.\ D {\bf 67}, 065018 (2003)
[arXiv:hep-th/0209137].

\bibitem{wess} J.~Madore, S.~Schraml, P.~Schupp and J.~Wess,
 Gauge theory on noncommutative spaces,
  Eur.\ Phys.\ J.\ C {\bf 16}, 161 (2000)
  [arXiv:hep-th/0001203];

X.~Calmet, B.~Jurco, P.~Schupp, J.~Wess and M.~Wohlgenannt,
  The standard model on non-commutative space-time,
  Eur.\ Phys.\ J.\ C {\bf 23}, 363 (2002)
  [arXiv:hep-ph/0111115].

\bibitem{seiberg}N.~Seiberg and E.~Witten,
String theory and noncommutative geometry,
JHEP {\bf 9909}, 032 (1999)
[arXiv:hep-th/9908142].

\bibitem{ydrithesis}
B.~Ydri, Ph.D. Thesis, Syracuse University, NY,2001, arXiv:hep-th/0110006;

B.~Ydri,
Noncommutative chiral anomaly and the Dirac-Ginsparg-Wilson operator,
JHEP {\bf 0308}, 046 (2003)
[arXiv:hep-th/0211209].

\bibitem{nahm1} 
M.~Scheunert, W.~Nahm and V.~Rittenberg,
Graded Lie Algebras: Generalization Of Hermitian Representations,
J.\ Math.\ Phys.\  {\bf 18}, 146 (1977).

\bibitem{nahm2} M.~Scheunert, W.~Nahm and V.~Rittenberg,
Irreducible Representations Of The Osp(2,1) And Spl(2,1) Graded Lie Algebras,
J.\ Math.\ Phys.\  {\bf 18}, 155 (1977).
 
\bibitem{pais} A.~Pais and V.~Rittenberg, 
Semisimple Graded Lie Algebras,
J.\ Math.\ Phys.\  {\bf 16}, 2062 (1975)
[Erratum-ibid.\  {\bf 17}, 598 (1976)].

\bibitem{Dewitt} B. Dewitt, {\it Supermanifolds}, Cambridge University Press, Cambridge (1985); 

M. Scheunert, {\it The Theory of Lie Superalgebras}, Springer-Verlag, Berlin (1979).

\bibitem{cornwell} J. F. Cornwell, {\it Group Theory in Physics Vol. III}, Academic Press, San Diego (1989).

\bibitem{sorba} L. Frappat, A. Sciarrino, P. Sorba, Dictionary on Lie
  Superalgebras, hep-th/9607161.

\bibitem{chaichian} 
M.~Chaichian, D.~Ellinas and P.~Pre\v{s}najder,
Path Integrals And Supercoherent States,
J.\ Math.\ Phys.\  {\bf 32}, 3381 (1991).

\bibitem{gradechi} A. El Gradechi and L. M. Nieto, 
Supercoherent states, superKahler geometry and geometric quantization,
Commun. Math. Phys.,{\bf 175} (1996) 521, and hep-th/9403109; 

A.~M.~El Gradechi,
On the supersymplectic homogeneous superspace underlying the OSp(1/2) coherent states,''
J.\ Math.\ Phys.\  {\bf 34}, 5951 (1993).

\bibitem{bordemann} 
M. Bordemann, M. Brischle, C. Emmrich and S. Waldmann, subalgebras
with convering star products in deformation quantization: An algebraic
construction for ${\mathbb C}P^n$, J. Math. Phys., {\bf 37} (1996)
6311; q-alg/9512019; 

M. Bordemann, M. Brischle, C. Emmrich and S. Waldmann, Lett. Math. Phys., {\bf 36} (1996) 357;

S. Waldmann, Lett. Math. Phys., {\bf 44} (1998) 331.

\bibitem{Berezin:jz}
F.~A.~Berezin and V.~N.~Tolstoi,
The Group With Grassmann Structure Uosp(1,2),
Commun.\ Math.\ Phys.\  {\bf 78}, 409 (1981).

F. A. Berezin, {\it Introduction to Superanalysis}, D.Reidel Publishing Company, Dordrecht, Holland (1987). 

\bibitem{marmo} A. P. Balachandran, G.Marmo, B. S. Skagerstam and A. Stern, 
Supersymmetric Point Particles And Monopoles With No Strings,
{\it Nucl. Phys.} {\bf B164} (1980) 427; 

G. Landi and G. Marmo, {\it Phy. Lett.} {\bf B193} (1987) 61-66.
Extensions Of Lie Superalgebras And Supersymmetric Abelian Gauge Fields,

\bibitem{Fronsdal}  C. Fronsdal {\it Essays on
  Supersymmetry}. Mathematical Physics Studies Volume 8, Editor:
  Fronsdal, C. , Dordrecht, Reidel Pub.Co. 1986.

\bibitem{Wess} J. Wess, J. Bagger {\it Princeton series in physics:
  supersymmetry and supergravity}, Princeton, Princeton University
  Press, 1983.

\bibitem{susybreaking} A.~P.~Balachandran, A.~Pinzul and B.~Qureshi,
SUSY anomalies break N = 2 to N = 1: The supersphere and the fuzzy
supersphere, arXiv:hep-th/0506037.

\bibitem{klimcik1} 
C.~Klimcik,
A nonperturbative regularization of the supersymmetric Schwinger model,
Commun.\ Math.\ Phys.\  {\bf 206}, 567 (1999)
[arXiv:hep-th/9903112];

C.~Klimcik,
An extended fuzzy supersphere and twisted chiral superfields,
Commun.\ Math.\ Phys.\  {\bf 206}, 587 (1999)
[arXiv:hep-th/9903202].

\bibitem{seckin2}
S.~Kurkcuoglu,
Non-linear sigma models on the fuzzy supersphere,
JHEP {\bf 0403}, 062 (2004)
[arXiv:hep-th/0311031].

\bibitem{witten} E.~Witten, 
A Supersymmetric Form Of The Nonlinear Sigma Model In Two-Dimensions,
Phys.\ Rev.\ D {\bf 16}, 2991 (1977).

\bibitem{divecchia} 
P.~Di Vecchia and S.~Ferrara,
Classical Solutions In Two-Dimensional Supersymmetric Field Theories,
Nucl.\ Phys.\ B {\bf 130}, 93 (1977);

\bibitem{sweedler} M. E. Sweedler {\it Hopf Algebras}, W. A. Benjamin, New York, 1969. 
A. A. Kirillov, {\it Elements of the Theory of Representations}, Springer-Verlag, Berlin, 1976.

\bibitem{Mach-Schomerus} 
G.~Mack and V.~Schomerus, QuasiHopf quantum symmetry in quantum theory, Nucl.\ Phys.\ B {\bf 370}, 185 (1992);

G. Mack and V. Schomerus 
in {\it New symmetry principles in Quantum Field Theory}, Edited by J.Frohlich et al., Plenum Press, New York, 1992;

G.~Mack and V.~Schomerus, Quantum symmetry for pedestrians, preprint, DESY-92-053.

\bibitem{seckin3}
A.~P.~Balachandran and S.~Kurkcuoglu,
Topology change for fuzzy physics: Fuzzy spaces as Hopf algebras,
arXiv:hep-th/0310026.

\bibitem{Figari} R.~Figari, R.~H\" {o}egh-Krohn and C.~R.~Nappi,
Interacting relativistic boson fields in the de Sitter universe with two space-time dimensions,
Commun.\ Math.\ Phys.\  {\bf 44}, 265 (1975).

\bibitem{steinacker} J.~Pawelczyk and H.~Steinacker, 
A quantum algebraic description of $D$-branes on group manifolds,
Nucl.\ Phys.\ B {\bf 638}, 433 (2002) [arXiv:hep-th/0203110].

\bibitem{Bal-Trahern} A. P. Balachandran and C. G. Trahern {\it Lectures on Group Theory for Physicists},  
Monographs and Textbooks in Physical Science, Bibliopolis, Napoli, 1984.

\bibitem{BalPaulo} A.~P.~Balachandran, E.~Batista, I.~P.~Costa e Silva and P.~Teotonio-Sobrinho,
Quantum topology change in (2+1)d,
Int.\ J.\ Mod.\ Phys.\ A {\bf 15}, 1629 (2000) [arXiv:hep-th/9905136]; 

A.~P.~Balachandran, E.~Batista, I.~P.~Costa e Silva and P.~Teotonio-Sobrinho,
The spin-statistics connection in quantum gravity,
Nucl.\ Phys.\ B {\bf 566}, 441 (2000) [arXiv:hep-th/9906174];

A.~P.~Balachandran, E.~Batista, I.~P.~Costa e Silva and P.~Teotonio-Sobrinho,
A novel spin-statistics theorem in (2+1)d Chern-Simons gravity,
Mod.\ Phys.\ Lett.\ A {\bf 16}, 1335 (2001)
[arXiv:hep-th/0005286]; 

\bibitem{dascalescu} S. Dascalescu, C. Nastasescu, S. Raianu {\it Hopf
  algebras : an introduction}, New York, Marcel Dekker, 2001 \,.

%%%%%%%%%%%%%%%%%%%%%%%%%Fuzzy Physics and string Theory%%%%%%%%%%%%%%%%%%%%%%%%
{\it Relation of Fuzzy Physics to Brane physics have been
investigated. Some articles on this subject are:}

\bibitem{Alekseev:1999bs}
A.~Y.~Alekseev, A.~Recknagel and V.~Schomerus,
Non-commutative world-volume geometries: Branes on SU(2) and fuzzy spheres,
JHEP {\bf 9909}, 023 (1999)
[arXiv:hep-th/9908040]. 

A.~Y.~Alekseev, A.~Recknagel and V.~Schomerus,
Open strings and non-commutative geometry of branes on group manifolds,
Mod.\ Phys.\ Lett.\ A {\bf 16}, 325 (2001)
[arXiv:hep-th/0104054].

\bibitem{Klimcik:1996hp}
C.~Klimcik and P.~Severa,
Open strings and D-branes in WZNW models,
Nucl.\ Phys.\ B {\bf 488}, 653 (1997)
[arXiv:hep-th/9609112].

\bibitem{Alekseev:1998mc}
A.~Y.~Alekseev and V.~Schomerus,
D-branes in the WZW model,
Phys.\ Rev.\ D {\bf 60}, 061901 (1999)
[arXiv:hep-th/9812193].

\bibitem{Gawedzki:1999bq}
K.~Gawedzki,
Conformal field theory: A case study,
arXiv:hep-th/9904145.

\bibitem{Garcia-Compean:1999uw}
H.~Garcia-Compean and J.~F.~Plebanski,
D-branes on group manifolds and deformation quantization,
Nucl.\ Phys.\ B {\bf 618}, 81 (2001)
[arXiv:hep-th/9907183].

\bibitem{Myers:1999ps}
R.~C.~Myers,
Dielectric-branes,
JHEP {\bf 9912}, 022 (1999)
[arXiv:hep-th/9910053].

\bibitem{Trivedi:2000mq}
S.~P.~Trivedi and S.~Vaidya,
Fuzzy cosets and their gravity duals,
JHEP {\bf 0009}, 041 (2000)
[arXiv:hep-th/0007011].

\bibitem{Das:2000ab}
S.~R.~Das, S.~P.~Trivedi and S.~Vaidya,
Magnetic moments of branes and giant gravitons,
JHEP {\bf 0010}, 037 (2000)
[arXiv:hep-th/0008203].

%%%%%%%%%%%%%%Fuzzy Physics and Numerical Studies%%%%%%%%%%%%%%%%%%%%%%%%%
{\it Quantum field theories on fuzzy spaces are also studied via numerical
  methods. Some articles on this subject are:}  

\bibitem{Martin:2004un}
X.~Martin,
A matrix phase for the phi**4 scalar field on the fuzzy sphere,
JHEP {\bf 0404}, 077 (2004)
[arXiv:hep-th/0402230].

\bibitem{Azuma:2004zq}
T.~Azuma, S.~Bal, K.~Nagao and J.~Nishimura,
Nonperturbative studies of fuzzy spheres in a matrix model with the Chern-Simons term,''
JHEP {\bf 0405}, 005 (2004)
[arXiv:hep-th/0401038].


\end{thebibliography}
\end{document}